# THE EVOLUTION OF ESPORTS:

*An analysis of its origin and a look at its prospective future growth as enhanced by Information Technology Management tools.*

Submitted by:

# Anders Hval Olsen

**Master in Science (M.Scs.) At Coventry University**

Management of Information Technology

September 2014 - September 2015

**Supervised by:** Stella-Maris Ortim

**Course code:** ECT078 / M99EKM

**Student ID:** 6045397

# DECLARATION OF ORIGINALITY

COVENTRY
BUSINESS
SCHOOL

**Student surname:** OLSEN

**Student first names:** ANDERS, HVAL

**Student ID No:** 6045397

**Course:** ECT078 – M.Scs. Management of Information Technology
**Supervisor:** Stella-Maris Ortim

**Second marker:** Owen Richards

**Dissertation Title:** The Evaluation of eSports: *An analysis of its origin and a look at its prospective future growth as enhanced by Information Technology Management tools.*

**Declaration***: I certify that this dissertation is my own work. I have read the University regulations concerning plagiarism.*

*Anders Hval Olsen* *15/08/2015*

# ABSTRACT

As the last years have shown a massive growth within the field of electronic sports (eSports), several questions emerge, such as how much is it growing, and will it continue to grow? This research thesis sees this as its statement of problem, and further aims to define and measure the main factors that caused the growth of eSports. To further enhance the growth, the benefits and disbenefits of implementing Information Technology Management tools is appraised, which additionally gives an understanding of the future of eSports.

To accomplish this, the thesis research the existing literature within the project domain, where the literature is evaluated and analysed in terms of the key research questions, and further summarised in a renewed project scope. As for methodology, a pragmatism philosophy with an induction approach is further used to understand the field, and work as the outer layer of the methodology. The inner layer is rather based on qualitative data, whereas a thematic analysis serves as the foundation of the later chosen business analytical tools.

Through observation of the data collection, the themes ‘games played’, ‘games genre’, and ‘single-player or team-player’ was deemed most important. A statistical analysis was additionally done on the numerical data, which showed significant correlation between the themes ‘prize pool’, ‘participants’ and ‘participating countries’ towards live audience. This indicates that the mentioned themes affect how many people attends live eSport event. The analysis also consists of a trend analysis, which presents that nearly every theme analysed will continue to grow in a five-year period. The trend analysis further expects a 305.51% growth in the live audience from 2014-2019, while the virtual streaming audience is expected to experience a growth of 140.61%. To enhance this growth, the research thesis recommends using the presented trends and patterns found in the short-term, to conduct successful eSport events. However, in the long-term an implementation of a Data-driven Decision Support System can be done, which will periodically gather data to know the best possible way to deliver eSport events to the eSport community. This can further be a part of a knowledge management system, and an Information Strategy, both of which are believed to enhance the growth seen in the field of eSports.

# ACKNOWLEDGEMENTS

I want to express my sincere gratitude to my supervisor, Stella-Maris Ortim, for all the help she has given during the months it has taken to construct this master thesis. Without her guidance, as well as feedback, the final product would not be the same.

I additionally express my deepest thanks to Petter E. Kristiansen, whom have been a huge asset to me in regards of ensuring a correct language throughout the thesis.

I would also thank my friends and family, whom supported me and made it possible for me to write this master thesis.

# TABLE OF CONTENTS

# TABLE OF FIGURES

## TABLE OF TABLES

# GLOSSARY

**Competitive Gaming:** A scenario or activity that bases on that two or more persons plays video games against each other at a competitive level.

**Data mining:** Defined by Oracle (2015) as "a practice of automatically searching large stores of data to discover patterns and trends that go beyond simple analysis"

**Data warehouses:** Defined by Oracle (2002) as a "relational database that is designed for query and analysis […]". It is further based on historic data, and works as a means of data storage.

**FPS:** A game genre. Short for First Person Shooter

**Information Technology Management tools**: Tools that is included within the field of IT management, where the most common is Knowledge Management and Information Strategy.

**MOBA:** A game genre. Short for Multiplayer Online Battle Arena

**MMOA**: A game genre. Short for Massively Multiplayer Online Action Game

**OCCG:** A game genre. Short for Online Collectible Card Game

**PVG:** A game genre. Short for Platform Video Game

**RTS:** A game genre. Short for Real-Time Strategy

**Spectating Sport**: A sport that can be viewed live over internet through internet services, such as YouTube and Twitch.tv.

**Twitch.tv:** A live streaming service, making it possible for people all over the world, with connection to internet, to broadcast themselves either playing video games or talking about video games or relevant content, while chatting with their viewers (Dredge 2014).

**Virtual Audience (Also known as spectating or streaming audience):** Can be understood as people using services such as YouTube and Twitch.tv to watch various eSport events through their own devices, rather than attending to the events physically.

# CHAPTER 1 INTRODUCTION

## 1.1 Chapter 1 Introduction

The first chapter of this research paper will be focused upon giving an introduction on the specific field chosen. This includes a brief background, as well as the statement of problem. This is followed by the motivation of the researcher, and the relevance to the degree. Chapter 1 ends with defining the problem domain, the target audience, the aim, its objectives and its Research Questions. These research questions and objectives will work as the core for every chapter, as this is what the research paper will seek to answer.

## 1.2 Background

The field of eSport is a rapidly growing community, both in size and consumption. However, the actual awareness of eSports can be understood as not optimal, and therefore a brief explanation is given. Esports (also known as Competitive Gaming or Electronic Sports), is an entertaining sport based upon individuals or teams playing video games against each other for prize money and glory (Heaven 2014a). Today, these games are normally played on a Personal Computer, but there are tournaments played upon nearly any console existing. The eSport term existed decades ago, while pinball tournaments were held throughout the world, and the electronic entertainment field was slowly starting to spread throughout the world (Hope 2015). Today it is something completely different, seeing 5-digit live audience numbers, and even more spectating through internet (Heaven 2014b). This is only a brief explanation of eSports, so a further explanation of the state of problem can be given. A deeper understanding of eSports will be brought through the later chapters, as a common definition is not easily done, as for some eSport is unheard of. Nevertheless, with the brief explanation given, a state of problem of this research paper is given.

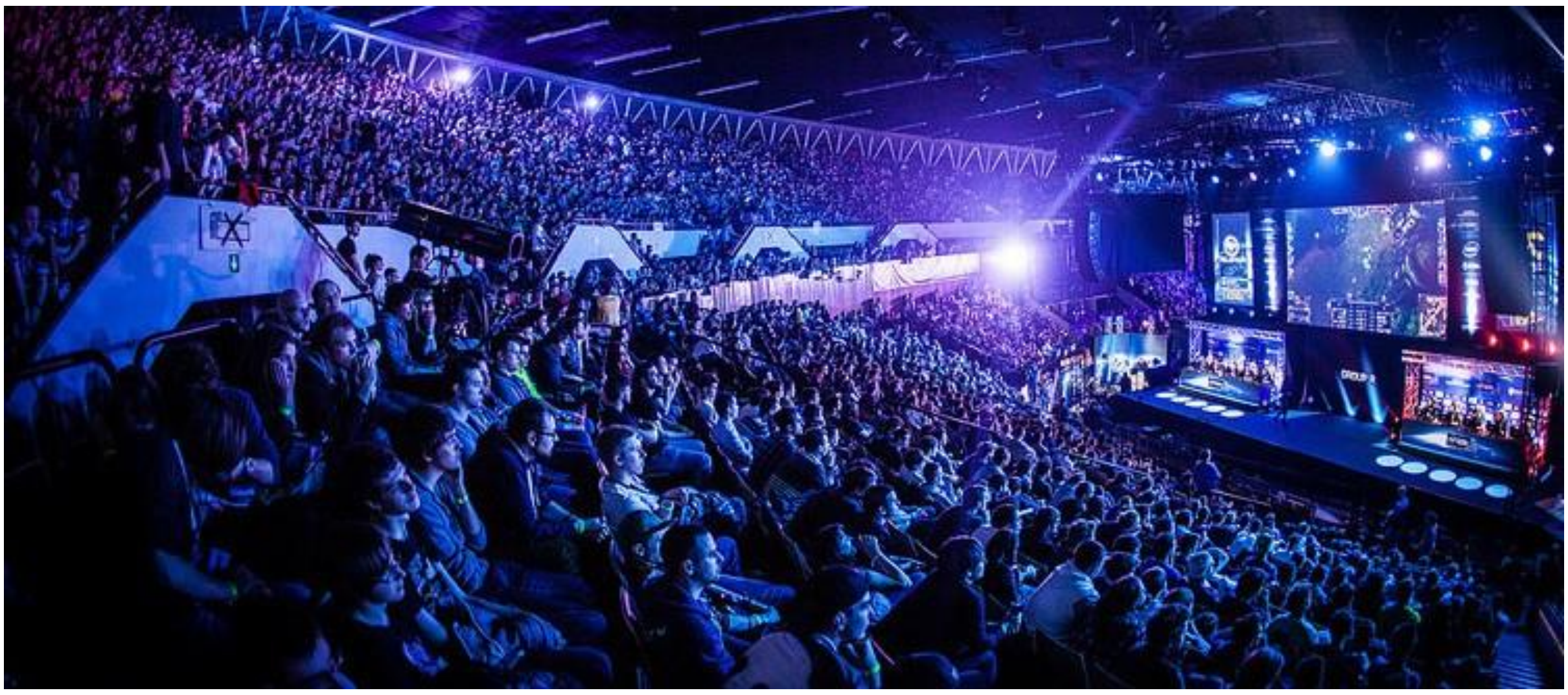

*Figure 1.1 eSport event: ESL One Katowice (Ingamers 2015)*

### 1.2.1 Statement of Problem

The statement of problem is essential for the research thesis overall, as it delineates the essence of its content, and will therefore be further discussed. The last years have changed the community of eSports.  It has experienced massive growth, and one cannot say for certain why (Hope 2014; Hewitt 2014). This brings forth the first statement of problem: why is the growth happening now, and what is triggering it? Although this question is interesting by itself, it also triggers additional questions, which brings forth the second state of problem: will the business continue to grow, or has it reached its peak? To fully understand and analyse this problem, a set of data collection tools must be utilized to get valid and reliable data on the topic. Such data would involve various factors, which are causing the growth seen recently in eSports. This will not only give an understanding of its past, but additionally give indications of its future by shedding some light on why eSports is seeing such a growth today. To tie this to a business approach, these factors will be implemented by using various Information Technology Management (Further named ITM) tools, which will work towards achieving an understanding of the future of eSports. This, in total, explains the state of problem, and a discussion around the domain can now be given.

### 1.2.2 Problem Domain and Project Domain

As presented in the introduction, this research thesis will use evolve around the eSports business. The eSport domain can be broken down into various organisations and communities, which will be explained by examples throughout the research. However, it can be understood as a part of both the electronic entertainment field and the event entertainment field. Nevertheless, this thesis will focus on the field of events, as eSport events will have a big influence on the results of this thesis. Considering the preceding statements, this research will be divided two different areas; firstly, this dissertation will use business analytics tools to critically analyse what momentums and factors that have led to the increase of eSport consumption. This specific analysis will be based upon data gathered from various events since the beginning of eSports. These results will further strive to give a short-term recommendation for the eSport community as a whole. Secondly, this thesis will analyse the effect of implementing ITM tools and theories to both uphold and enhance the growth of eSports. This will focus on giving a long-term recommendation for the eSport community as a whole.

## 1.3 Target Audience

This research is understood to be beneficial for two parts: Firstly, for any company wanting to understand the role of eSports in the future and how it can influence the entertainment sector. Secondly, this research will be beneficial for eSport event companies, such as Turtle Entertainment and Riot Games, as it seeks to find interesting data on trends within the field. Furthermore, as a study of the future of sports bearing an IT perspective, it may be a somewhat interesting read for any person with an interest in Information Technology. Nevertheless, the research paper is mainly written towards the field of eSports, as the field is still new and lack a lot of academic work within its field.

## 1.4 Motivation

There exist two main motivations for this research to be constructed: firstly, why it interesting for the researcher, and secondly, why it is interesting in the broad aspect. Each motivation factor will further be discussed in the subsequent points.

### 1.4.1 Self-motivation

I, the researcher, have since a child loved gaming, and everything around it. I also have experience from being a store manager at GAME, and my bachelor is within Culture and Leadership, where I specified myself under cultural entertainment. During my undergrad I started following on different eSports shows and channels, and I have since then been a consumer of the entertainment they bring, and watched it grow every day. I want to keep chasing my passion within gaming in a business aspect, and that is why I ended up writing this thesis in the field of eSport. ESport is a fresh, unique but massively growing business, which will need people with a higher education — especially within IT, mainly because it needs to be done profitable. The personal aim of this research thesis is towards a futuristic dream job, where I can analyse and implement different strategies from a business perspective to a passion I hold dearly.

### 1.4.2 Broad Motivation

The broad motivation aspect refers mainly to the growth eSport is seeing in the entertainment field. Today, there is a huge discussion if eSport is even a sport, or if it is just a competition scenario such as chess — as people such as the ESPN boss declares it is not (Tassi 2014). However, if Heaven's (2014b) number is correct, it will be a 275% increased consumption of eSport by 2018, which will make the growth of eSports undeniable. Nevertheless, the lack of business aspects in the eSport community is lethal, and there needs to be feasibility studies, marketing strategies, business strategies, organisational strategies, and IT/IS strategies to further improve eSports awareness and overall condition (Yong 2010). Lastly, the data collected can be used and compared with other entertainment fields as well, as the implementation of strategy techniques in eSports can give knowledge to e.g., the movie industry or the music industry.

## 1.5 Relevance to the Degree

This research is taken under the course of M.Scs. in Management of Information Technology at Coventry University. The university itself defines the courses aim as to *"develop an understanding of information systems and database systems, the technology available to support management decision-making, the role of the internet and communications technology, and the issues facing IT managers in small, medium and large organisations"* (Coventry 2015). This correlates with this research thesis in three aspects:

I. ESports: eSports is an emerging entertainment platform, and a door opener for different internet driven spectator platforms such as Twitch.tv and YouTube.

II. ITM: ITM tools will be used and juxtaposed to give an earnest understanding on how the available technology might be able to support the future of eSports, and the decision-making at the managerial level.

III. Organisations: This research also intend to gather primary data to gain an understanding on how the eSport community is doing in an organisational aspect.

## 1.6 Project Scope

To narrow the scope of the study and provide structure, an overall aim has been defined for this thesis. The aim is broken down into a various objectives, which define specific goals that have to be reached to achieve the aim. These objectives are further broken down into deliverables, which is a set of tasks that have to be done to complete the objectives. The aim, objectives and deliverables are further discussed in the subsequent points. However, before these can be defined, an understanding of the aforementioned points must is summarised as a statement of purpose for this study.

### 1.6.1 Purpose of this study

The purpose of this mixed methodology study is to analyse the past of eSports to gain an understanding of its key performance factors, relying mainly on qualitative data. These factors will further be used to gain an understanding of the future of eSports, by both implementing business analytics and ITM tools. Initially, data will be obtained through semi-structured interviews at locations convenient to the participants, as well as observation of previous eSport events, understood as case studies gathered through secondary and primary sources. The data gathered will be managed and further analysed in Microsoft Office and SPSS, where various business analytics will be performed to get an understanding of the past, present and future of eSports.

### 1.6.2 Aim

To interpret what has to be gathered throughout the research, an aim with it underlying objectives is composed. These are further defined into the Research Questions, which serve as the very basis of this research paper. Considering the project domain and the purpose of the study, the following aim has been defined for this research paper: *This study aims to define and measure the key factors that caused the growth of eSports, and further appraise the benefits and disbenefits of implementing Information Technology Management tools to maintain, and enhance its growth.*

### 1.6.3 Objectives and deliverables

The following objectives is based upon Robinson's (2014) 'SMART' method, meaning that each objective should be Specific, Measurable, Attainable, Relevant and Time-bounded. Considering this, eleven objectives is created based upon either knowledge, application or evaluation. These objectives is further explained by its deliverables, which is certain tasks that has to be done to achieve the certain objective.

**Objective 1: Discover previous research within the field, and summarize their importance for this study**

- *(D1) A complete literature review by week 19*
- *(D2) A summary of the major findings within the eSport community by week 20*

**Objective 2: Define what eSport is, and outline its major historic milestones**

- (D3) A Complete and synthesised definition of eSports by week 16.
- (D4) A summary with the milestones of eSport history by week 19.
- (D5) A visual timeline of eSports historic data by week 20.
- (D6) A summary of the most important series and leagues within the eSport community by week 20.
- (D7) A report on the major eSport events by week 20

**Objective 3: Arrange and compare the main financial factors within eSport**

- (D8) A list of the main financial factors/trends within eSports by week 21
- (D9) A comparison of financial factors to differentiate their impact by week 22
- (D10) A brief explanation of the financial factors by week 22

**Objective 4: Review previously successful eSport events, and use business analytics tools to visualize the common patterns**

- (D11) A critical analysis of the most important success factors in previous eSport events by week 25.
- (D12) A time line of the most important eSport event by week 25.
- (D13) A statistic correlation test to see if there is any common patterns between the events by week 25.

**Objective 5: Discover the features of eSport that defines the essence of the eSport business**

- (D14) An interview with some of the important actors within eSports by week 25
- (D15) A synthesised report of the comparison of the financial factors by week 26
- (D16) A juxtaposed summary of the most important features within eSport by week 26
- (D17) A juxtaposed summary of the causing factors of the recent growth within eSport by week 27

**Objective 6: Examine the overall awareness of eSports**

- (D18) An in-depth analysis on how many people watches eSports and it overall awareness by week 20.
- (D19) A report based upon the overall viewership of eSport, both consisting of live audience and streaming audience by week 27.

**Objective 7: Predict the financial future of eSports**

- (D20) An in-depth analysis of the economic future of eSports based upon both statistical data and trend lines made from business analytics, as well as the literature review by week 28
- (D21) A set of prediction of the finical situation of eSport based upon findings by week 28
- (D22) A brief recommendation on how eSport can improve its financial value by week 30

**Objective 8: Evaluate the benefits and disbenefits of implementing ITM tools in the eSport community and synthesise which ITM tools would be most effective**

- (D23) A complete definition of Information technology, and its tools, by week 21
- (D24) A critical evaluation of the financial effect of the implementation of Information Technology Management tools by week 22
- (D25) An evaluation of which ITM-tools implementations that can be done within eSport by week 23

**Objective 9: Evaluate the effect of implementing well-justified ITM tools to increase the growth of eSport**

- (D26) A list of benefits and disbenefits of the implementation of ITM-tools in eSports by week 26
- (D27) A critical evaluation of how various ITM tools might improve the growth of eSport, and its effects by week 27

**Objective 10: Recommend alterations the eSport business might commence to uphold and increase its growth.**

- (D28) A summary of the data gathered in the dissertation by week 31
- (D29) A list of recommendation based upon the data gathered by week 31
- (D30) A evaluation of the cause and effect of the recommendations made by week 32

***Objective 11: Evaluate and validate the findings, results and recommendations***

- (D31) A Project Management and Quality Assurance of conduct of project by week 32
- (D32) A conclusion of the credibility of the data found by week 33
- (D33) A critical evaluation of all the findings and writings by week 33

## 1.7 Research Questions

Based upon the findings in previous points, the research questions were conducted to define what major points this research paper seek to answer. In total, six research questions were constructed based upon the aim, objectives and deliverables:

I. What is eSports, and how aware are people of the eSport culture?
II. What are the major milestones of eSports history?
III. Why is eSport seeing its growth today, and which factors is pushing it upwards?
IV. Will the growth of eSport continue, and what can the eSport business do to ensure a continuous growth?
V. Can Information Technology Management Tools be used beneficially for the field of eSport?
VI. What outcomes would the most beneficial Information Technology Management tool give in the eSport business?

## 1.8 Procedures and Methodology

To answer the research questions within the study, various historic data and observations is needed. In addition, numeric numbers is essential to understand the financial future. Based upon the expected data, both qualitative and quantitative methods will be used, which makes the research thesis a mixed-methodology. However, additional knowledge is needed of the ITM tools, which makes the thesis dependent on qualitative data. The procedures to collect this data will rely on historic case studies of eSport events, as well as interviews to gain a valid understanding of the eSports field. This will further be analysed by business analytic tools, to gain both a statistical and visual understanding of the financial future of eSport

## 1.9 Structure of the study

This research paper is divided into ten different chapters. Each chapter is seen as a step towards answering the overall aim. Each Chapter is briefly explained in the following points:

1. Chapter 1 is the introduction, where the background, objectives and Research Questions serve the most important role.
2. Chapter 2 is the literature review, which includes the collection of previous data in this research field, as well as a renewed scope based upon the findings in the literature review.
3. Chapter 3 is the methodology, which includes an explanation on how the data is collected to answer the research questions.
4. Chapter 4 is the Interpretations, which explains how and why the various data collection tools were conducted.
5. Chapter 5 is the data collection and analysis, which includes several discussions focused upon answering the research questions based upon the data collection findings.
6. Chapter 6 is the conclusion, which includes the final answers for the research questions, as well as the short-term and long-term recommendations and the future works.
7. Chapter 7 is the project management and quality assurance chapter, which includes a critical evaluation of how well the paper over all is written, and if it reaches its aim, objectives, deliverables and research questions.
8. Chapter 8 is the reference list, which is all the references used for this written thesis. An additional reference list for the thematic analysis is attached in Appendix B, which is further explained in the later chapters.
9. Chapter 9 is the bibliolatry, which is the various reference used to gain knowledge or understanding, but were not used in this thesis.
10. Chapter 10 is the appendix, which consist of 9 Parts:
    a. Thematic analysis: excel file (444 events)
    b. Thematic analysis: Reference List
    c. Various findings in the thematic analysis and the business analytics that were not used.
    d. Business analytics: Trend Lines

e. Results of the pilot study
f. Gant Chart
g. Ethical documents for ethical consideration
h. Time line for the most important eSport events (23 events)

### 1.10 Limitation of the Study

Within this study, the following is seen as limitations, which further is discussed in 6.4.1 Future works and 7.4 Lessons learnt. The severity is high for each limitation, and they are therefore discussed individually.

#### 1.10.1 Information Technology Management Tools

The use of ITM tools in this research study has been limited due of both the size of the paper, as well as the time limitation. Due of the limitation, the following ITM tools will be studied in this research paper: Decision Support Systems, Knowledge Management and Information Strategy. These tools are additionally seen in a general approach, as where the most of their underlying tools is not discussed individually, as that would be too broad for this thesis. Each tool is therefore looked with a holistic approach, so a generic understanding can be conducted.

#### 1.10.2 Global Aspect

The size of eSport is viewed in a global aspect as the existing data is mostly looked upon globally. This limits the availability to perform various tasks, such as surveys. Conducting global surveys with different participating countries is not feasible in the time limit for this study, and therefore affect the possible methodologies used in this thesis.

### 1.11 Summary of Chapter 1

The introduction chapter involved the main reasoning for the chosen topic for this research thesis, as well as stating its purpose. As introduced in the background, this thesis will regard the field of eSports, which is understood as an entertainment field where professional players play video games competitively, either in front of a live audience or a spectating virtual audience. Due of the recent growth, this thesis will further focus on understanding the past of eSports, and what has made it suddenly grow in the recent years. It will further focus on the future, and what patterns of the past can be used to enhance the future of eSport business. Additionally, the chapter defines a specific aim, which the thesis seeks to answer. The aim is broken down into eleven objectives and thirty-three deliverables, which will work as milestones towards achieving the aim. This information is summarised into six research questions, which will stand as an essential point in every chapter – as it is these questions the research paper ultimately seeks to answer. Before these questions can be answered, an understanding of the existing literature needs to be accumulated, which is why a literature review is amassed in the subsequent chapter.

# CHAPTER 2 LITERATURE REVIEW

## 2.1 Introduction to Chapter 2

In Chapter 1, an introduction was given as well as the aim, objectives and research questions within this research project. This chapter, however, will examine the topic as well as the research questions and get an understanding of what is the current thinking around the topic, and what research has been done. This is done by conducting a literature review, which is defined by Randolph (2009:5) as the "means of demonstrating an author's knowledge about a particular field of study, including vocabulary, theories, key variables and phenomena, and its methods and history". After the literature review is conducted, a renewed project scope is defined based on the findings in the literature review, which ensures that this research thesis will not reproduce previous academic work within the field.

In this research paper, the literature review is created by following Randolph's (2009: 9) five research stages in conducting a literature review, as well as by seeking inspiration from Cooper's (1988: 109) taxonomy of Literature Reviews. Randolph's five stages are named problem formulation, data collection, data evaluation, analysis and interpretation, and public presentation. However, for this research thesis these steps are named differently to give a better understanding to the reader, and will therefore be named literature problem formulation, literature collection, literature evaluation and validity, analysis and Interpretation, and literature presentation. The first step will therefore introduce the literature problem formulation, which is the first step towards the completion of the literature review.

## 2.2 Literature Problem Formulation

The literature problem formation is the first out of five steps in the literature review, and it is built towards finding which questions will be asked and answered in the literature review (Randolph 2009). For the literature review, the research questions will stand in the essence of what evidence that should be included in the literature review. To get a better understanding of what data to look for, each questions in connected to various objectives, and then further given a specific goal including the expected data collection, to get an understanding of what data is needed to explain the different research questions. The specific goal is further linked to the expected data collected, which is the various deliverables named in chapter 1. This information is further presented in table 2.1.

| *Research Questions* | Objectives | Literature review Goal | Deliverables (Expected data collected) |
|---|---|---|---|
| *[1] What is eSports, and how aware are people of the eSport culture?* | Objective 2: Define what eSports is, and outline its major historic milestones<br>Objective 6: Examine the overall awareness of eSports | To gain an understanding of what eSport is, and the overall awareness of esports in a global aspect | • (D3) A complete and synthesised definition of eSports by week<br>• (D18) An in-depth analysis on how many people watches eSports and it overall awareness by week 20.<br>• (D19) A report based on the overall viewership of eSport, both consisting of live audience and streaming audience by week 27. |
| *[2] What are the major milestones of eSports history* | Objective 2: Define what eSports is, and outline its major historic milestones<br>Objective 4: Review previously successful eSport events, and use business analytics tools to visualize the common patterns | To gain an understanding of which events as well as factors that has led to the definition of eSport known today. | • (D4-10) A summary of the most important milestones in eSport history, which include a list of the major eSport events through time and a time line.<br>• (D11-13) A understanding of the most important success factors in previous eSport events |
| *[3] Why is eSport seeing its growth today, and which factors/patterns is pushing it upwards?* | Objective 3: Arrange and compare the main financial factors within eSport<br>Objective 5: Discover the features of eSport that defines the | To gain an understanding of which factors that has increased the consumption of eSports | • (D8-10) A summary and a comparison of the most important financial factors within eSports<br>• (D14-D17) A summary of the most important factors, events and cases that has led to the specific growth seen today |
| | essence of the eSport business | | |
| *[4] Will the growth of eSport continue, and what can the eSport business do to ensure a continuous growth?* | <u>Objective 5:</u> Discover the features of eSport that defines the essence of the eSport business<br><br><u>Objective 7:</u> Predict the financial future of eSports | To gain an understanding of the close future of eSports, and which main factors that can be used to ensure the growth. | • (D17) A juxtaposed summery of the most important features and the causing factors of the growth within eSports<br>• (D20) An in-depth analysis of the economic future of eSports<br>• (D21-22) A list of predictions of the future of eSports |
| *[5] Can Information Technology Management Tools be used beneficially for the field of eSport? ?* | <u>Objective 8:</u> Evaluate the benefits and disbenefits of implementing ITM tools in the eSport community and synthesise which ITM tools would be most effective | To gain an understanding of what Information Management Tools can do to ensure any growth in the eSport community. | • (D23) A Definition of Information Technology and its tools<br>• (D24-25) A list of benefits and disbenefits of various Information Technology tools that can be adopted in the eSport community |
| *[6] What outcomes would the most beneficial Information Technology Management Tool give in the eSport business? ?* | <u>Objective 9</u>: Evaluate the effect of implementing well-justified ITM tools to enhance the growth of eSport<br><br><u>Objective 10</u>: Recommend alterations the eSport business might commence to uphold and increase its growth. | To gain an understanding of which ITM that would be most beneficial within the field of eSport based on the earlier findings. | • (D26-D27) A brief discussion on how these benefits might affect eSports<br>• (D28-D31) A list of recommendations to further enhance the future of eSports |

I. Deliverable 1 and 2 is under objective 1 which is to conduct the literature review, and is therefore not used in the table

II. Deliverable 31, 32 and 33 is under objective 11, which concerns evaluating and validating the findings, and is therefore not used in this table.

Table 2.1 Literature review goals and expected data

## 2.3 Literature Collection

Advancing to the second step is the literature collection. This step include finding various journals, articles and similar documents that relates to the previous expected data collection written in table 2.1 (Randolph 2009). In the field of eSports, there exist an umpteen number of literature, mostly existing of newspaper articles, forum posts and blog posts. To elucidate the fundamental findings within the chosen articles, table 2.2 is created to summarise the main findings. The table additionally includes the researcher's aim/purpose of the study of the paper conducted, to gain an understanding if their research is applicable to this thesis' objectives.

| *(*) Additional information about the specific literature is given under the left column if applicable* | | |
|---|---|---|
| **Research Author and additional information (*)** | **Researcher's aim/purpose** | **Data Collection (Main Findings)** |
| ***Alavi and Leidner (1999): Knowledge Management Systems: Issues, Challenges, and benefits*** | *A journal article aimed towards understanding Knowledge Management Systems, and how it can be used to improve organisations.* | *Benefits of knowledge management:*<br>• *Communication benefits (Enhanced communication, faster communication, more visible opinions of staff and increased staff participation)*<br>• *Efficiency benefits (Reduced problem solving time, shortening proposal times, faster results, faster delivery to market and greater overall efficiency)*<br>• *Financial benefits (increased sales, decreased cost and higher profitability)*<br>• *Marketing benefits (Better service, improved customer focus, better targeted marketing and better proactive marketing)*<br>• *General benefits (Improved project management and reduce number of personal needed)* |
| ***Bergeron (2003): Essentials of Knowledge Management*** | *A book intended to learn what knowledge management it, its implementation issues, management challenges, and* | Definition of knowledge management:<br>• Knowledge Management (KM) is a deliberate, systematic business optimization strategy that selects, distils, stores, organizes, packages, and communicates information essential to the business of a company in a manner that improves employee performance and corporate competitiveness |
| | *show the best proven knowledge management practices* | |
| ***Borowy and Jin (2013): Pioneering ESport: The experience economy and the marketing of Early 1980 Arcade Gaming Contest*** | *A journal article set out to historicize the development of eSport in the early 1980s, by examining its origins as both a marketised event and experiential commodity.* | • The first video game tournament ever staged is in 1980, held by Atari and evolved around the video game Space Invaders.<br>• The second big video game tournament is named the National Video Game Championship, held at the Chicago Exposition Centre in 1981. The tournament is based upon a game named Centipede, which they used $240,000 to finance and promote.<br>• The third video game tournament is the National Video Game Master Tournament held in 1984, using more than 60 different arcade games.<br>• The fourth milestone according to Borowy and Jin would be the Electronic Circus, a traveling video game circus going across North America throughout 1983.<br>• The article additionally points other factors that led to the existence of eSports back in 1980s.<br>  ○ Firstly is the article in LIFE magazine released in November 1981, about the best video game players in the United States.<br>  ○ Secondly, there was various TV-Show, such as 'That's amazing!' and 'Starcade' that all together raised the awareness of eSports as a competition sport. |
| ***The Business Dictionary (2015a): Definition of Information Technology*** | *A written definition used to define Information Technology by the Oxford Dictionary* | Definition of Information Technology:<br>"Set of tools, processes, and methodologies (such as coding/programming, data communications, data conversion, storage and retrieval, systems analysis and design, systems control) and associated equipment employed to collect, process, and present information" |
| ***Casselman (2015): Resistance is futile: eSports is massive … and growing***<br><br><u>Additional Info (*)</u><br>*Some data is based* | *A Newspaper article made to understand how big eSport actually is, and how it can be* | • *Time spent watching eSports:*<br>  ○ *2012: 1.3B*<br>  ○ *2013: 2.4B*<br>  ○ *2014: 3.7B*<br>• *eSport audience:*<br>  ○ *2012: 58 Million*<br>  ○ *2013: 74 Million* |
| *upon Newzoo's (2015) report and is not used* | *compared to other sports* | ◦ *2014: 89 Million*<br>• *Twitch is the most important contributor to eSports' recent growth, and has the following important data:*<br>◦ *55M active users*<br>◦ *1M Peak concurrent Viewers Per month*<br>◦ *100M Unique Viewers Per month*<br>◦ *11M Broadcasts Per month*<br>• *Demographical data of eSports fans:*<br>◦ *Age 10—20 (16%)*<br>◦ *Age 21-35 (56%)*<br>◦ *Age 36-65 (28%)*<br>◦ *Female (38%)*<br>◦ *Male (63%)*<br>• *Both the League of Legends and the Dota 2 Championship in 2014 has more viewership than the NBA Finals*<br>◦ *Super Bowl 111.2M*<br>◦ *LOL Championships: 27M*<br>◦ *Masters (Golf): 25M*<br>◦ *DOTA 2 International: 20M*<br>◦ *NBA Finals: 15.5M*<br>• *What people follows on YouTube:*<br>◦ *Music: 85M*<br>◦ *Gaming: 79M*<br>◦ *Sports: 78 M*<br>◦ *News: 35M*<br>◦ *Popular: 27M*<br>◦ *Spotlight: 22M*<br>◦ *Movies: 18M*<br>• *Prize pools in eSports:*<br>◦ *World Series of Poker Champions: $10M*<br>◦ *Super Bowl Champion (2015):*<br>◦ *$5.1M*<br>◦ *Dota 2 Champions (2014): $5M*<br>◦ *NBA Finals Champion (2014): $4.1M*<br>◦ *Stanley Cup Champion (2014: $3.8M*<br>◦ *Smite World Champions (2015): $1.3M*<br>◦ *League of Legends World Champions (2014): $1M* |
| ***Chobopeon (2012): A History of eSports***<br><br>Additional Info (*) | *A forum post meant to explore the whole history* | *The following events is the major findings defined as the important milestones in eSport history by Chobopeon:*<br>• *The Creation of electronic games in the 1950s*<br>• *The Arcade games and tournaments that was introduced in the 1980s.* |
| *The post it a text translation of various podcasts, split in centuries* | *of competitive gaming* | • *Technology such as ARPAnet and DWANGO network, which made it possible to play against each other in the 1990s.*<br>• *The rise of multiplayer games, such as Doom and Doom 2 in the early 1990s.*<br>• *The increase of eSport events in 1995, where Deathmatch '95 and Microsoft's Judgement Day '95 is seen as the biggest.*<br>• *The increase of modem technology in mid-2000, making it easier to play against each other globally.*<br>• *The creation of the video game Warcraft, which introduced RTS in the competitive field in 1994.*<br>• *The creation of the Cyberathlete Professional League in 1997, and the FRAG tournaments that followed*<br>• *The World Cyber Games Challenge (WCGC), which were held in 2000, working as an Olympic inspired tournament for eSports worldwide.* |
| ***Davenport, Long and Beers (1990): Successful Knowledge Management Projects*** | *A Journal article on how a company can create, share and use knowledge effectively* | Knowledge can be defined din two ways:<br>• Tacit knowledge (subjective knowledge)<br>• Explicit knowledge (documented knowledge)<br><br>Factors that lead to successful knowledge projects?<br>• Link to economic performance or industry value<br>• Technical and organizational infrastructure<br>• Standard, flexible knowledge structure<br>• Knowledge friendly culture<br>• Clear purpose and language<br>• Change in motivational practices<br>• Multiple channels for knowledge transfer<br>• Senior management support |
| ***D.Devil (2011): A Short Story of nearly everything*** | *A forum post meant to make a summary of what lead esports to be where it is today* | *The following events is the major findings defined as the important milestones in eSport history by D.Devil:*<br>• *(1997) eSports is born in 1997, with the Red Annihilation Quake tournament*<br>• *(1997) The CPL is created in 1997*<br>• *(2000) The 2000 CPL Razer tournament, with an overall prize poll of $100 000*<br>• *(2000) The WCG tournament that took place in South Korea in the year 2000, having a prize pool of over $200 000. The WCG tournaments following the other years is also important seen in the historic aspect.*<br>• *(2000) Turtle Entertainment is created in 2000, as one of the first professional leagues within eSports*<br>• *(2001) The dot.com bobble setbacks eSports in 2001* |
| | | • *(2002) MLG is created in 2002, bringing the professional leagues in the America*<br>• *(2002) The fighting game tournament EVO is born in 2002*<br>• *(2003) SK Gaming, one of the first gaming communities, is created in 2003.*<br>• *(2003) The ESCW (Electronic Sport World Cup) Takes place in 2003, and is the later years a big tournament within the global eSport market.*<br>• *(2005) The World eSports Games are born in the year 2005, giving a prize pool of around $700 000.*<br>• *(2006) MLG secures $35M in venture capital funding, making it one of the biggest leagues in eSports*<br>• *(2006) Turtle Introduces the Intel Extreme Masters for the first time, an event that will increase for every year in the field of eSports.*<br>• *(2008) The International eSport Federation is founded*<br>• *(2010) GOM TV announces the Global StarCraft II League (GSL) In South Korea – the first television league known in history*<br>• *(2011) League of Legends and DOTA makes it mark in the competitive gaming, with two of the most viewed championship seen so far in eSport history.* |
| ***Garfield (2014): 15 Knowledge Management benefits*** | *A blog entry intended to the various benefits which can result from knowledge management and enterprise social networks* | • Knowledge management enables better and faster decision making<br>• Knowledge management makes it easier to find relevant information and resources<br>• Knowledge management can be used to reuse ideas, document and expertise.<br>• Knowledge management can be used to avoid redundant effort<br>• Knowledge management can be used to avoid making the same mistakes twice<br>• Knowledge management can be used to take advantage of existing expertise and experience<br>• Knowledge management can be used to promote standard, repeatable process and procedures<br>• Knowledge management can be used to provide methods, tools, templates, techniques, and examples.<br>• Knowledge management can be used to make scarce expertise widely available<br>• Knowledge management can be used to show customers how knowledge is used for their benefit |
| | | • Knowledge management can be used to accelerate delivery to customers<br>• Knowledge management can be used to enable organisations to leverage its size<br>• Knowledge management can be used to make the organisations best problem-solving experiences reusable<br>• Knowledge management can be used to stimulate innovation and growth |
| ***Heaven (2014a): Rise and Rise of eSports*** | *A journal article that were constructed to discuss the popularity of eSports* | • eSport are getting big all over the world – not only Asia<br>• G3 was just help in the UK, with 4000 live audience, $140,00 prize pool and 8.5 Online watchers<br>• Competitive players that perform well can be watch as superstars as the same level as football and baseball stars<br>• The rise of eSports is based upon both better video-streaming technology and faster internet connections. |
| ***Heaven (2014b): eSports by numbers*** | *A journal article that were constructed to give a quick and summarised numbers of the field of eSports* | • 2.4 billion Hours of esports video watched last year<br>• 6.6 billion hours is expected to be watched by 2018, according to market research firm IHS Technology<br>• 15,000 Seats in the world's first dedicated esports stadium, due to open in 2017 in Hengqin, China<br>• 1972 Year of the first known video game contest. On 19 October, Stanford University hosted students fighting it out in Spacewar. The prize? A year's subscription to Rolling Stone<br>• Newbee shared the highest winning price of $5 million during the international contest in 2014 (DOTA 2) |
| ***Hewitt (2015): Will eSports Ever Become Widely Accepted as Official Sports and How Will They Affect the Way We Entertain Ourselves If they do?*** | *A journal article meant to discuss how the sport industry and the gaming industry affect each other* | • The competitive field of eSports did not 'blow out of the water' before recently, especially with the introduction of games as StarCraft II and League of legends<br>• A Major threat of the promotion of eSport is that it lack of physicality, and many are afraid of promoting a lifestyle of sitting in front of a computer. Hewitt further discuss that maybe the implementation of Virtual Reality might help the weakness that exists within the field of eSports.<br>• Another threat that especially is seen lately in Korea is that many teenagers are leaving their education to train to become eSport players. As the sport itself is very restricted in the sense of size, leaves the chances of actually becoming recognized low, hence doing so it |
| | | quite risky. This might lead to negative attitude towards the field of eSports.<br>• The future of eSports is looking bright, especially as it just only has really been commercialised. Hewitt further write that the future of eSports are most likely going to be an enormous contributor to the entertainment industry in the years to come.<br>• Another issue within the future of eSports is the legal considerations. There exist to one 'owning' the sport of eSports, it is only various events that shows tournaments in different games – and this might show problems in the future. |
| ***Heyoka (2011): Before eSports: Nintendo World Championship*** | *A forum post meant to give an insight on the early Nintendo eSport events* | • Heyoka names the Nintendo World Championship, which was held in 1990, as one of the biggest milestones in eSport history.<br>• Nintendo continued their eSport success with the Nintendo Campus Challenge and Nintendo PowerFest '94 |
| ***Holsapple and Sena (2005): ERP plans and decision-support benefits*** | *A journal article created to explore connections between ERP systems and Decision support based on the perception of 53 ERP system adopters.* | *Benefits of using a DSS system:*<br>• *Better knowledge processing*<br>• *Better Cope with complex problems*<br>• *Reduced Decision time*<br>• *Reduced decision cost*<br>• *Greater reliability*<br>• *Better communication*<br>• *Greater exploration/discovery*<br>• *Stimulates fresh perspective*<br>• *Better coordination*<br>• *Greater satisfaction*<br>• *Decisional improvement*<br>• *Competitive advantage* |
| **Hope (2014): The Evolution of the Electronic Sports entertainment Industry and its popularity** | *A journal article meant to analyse and gain a understanding of the most important historic events within the field eSports, and a brief discussion* | There are in total 9 events that together is seen as the major milestones in eSports history:<br>• The Space Invades tournament held in 1981<br>• The Starcade TV program held in the mid-1980s<br>• The Creation of IRC (Internet Relay Chat) in 1988<br>• The release of the game Doom in 1993<br>• The release of Doom II in 1994, which included the multiplayer option through DWANGO (Dial Up Wide Area Network Gaming Operations)<br>• The WCG event held in 2000 and 2001<br>• The Increase of gaming organisation seen in the beginning of 2000 (IEM, MLG) |
| | *of what the future holds* | • The Creation of the streaming service Justin.tv in 2007<br>• The increase in mainstream promotion seen in the end of 2000<br>• The future of eSport lies within live TV, and looks bright. |
| ***Jackson (2013): The rise of eSports in America*** | *A newspaper article made to take a look at how strategy games, MOBAs, and fighters have become a legitimate sports in America* | • *eSport has been growing immensely after Blizzard Entertainment released the game StarCraft II in 2010*<br>• *Major League Gaming (MLG) has become one of the most prestigious eSports organizations in the world, pioneering eSports in America and leading to its growth*<br>• *MLG Pro Circuit Champions had 11.7 million unique viewers in 2012, compared to 3.5 million in 2011.*<br>• *The Evolution Championship Series (EVO) has led to an increase of viewership in the fighting side of eSports, hitting 1.7 million unique online viewers at their tournament in 2012.*<br>• *Blizzards StarCraft II World Championship series (WCS) had a prize pool for $1.6 million dollars which is massive for the field of eSports*<br>• *Valve and Riot Games had both caused an increase eSport consumption with their games and events*<br>• *The Rise of the streaming platform twitch, reached over 35 million viewers that watch gaming-related content. This is to be understand as one of the main factors of the increase in eSports in America.*<br>• *The rise of eSports commentators and casters has also lead eSport to be a more likeable sport* |
| ***Jarvis (2014): eSports: Behind the next billion-dollar industry***<br><br><u>Additional Info (*)</u><br>*Includes findings from previous stated literatures and therefor all findings is not concluded* | *A Newspaper article from The market for computer & video games (MCV) based upon eSport statistics today, and potential growth factors* | The following data is considered major findings within the newspaper:<br>• Prize pool in 2013: $25 000 000<br>• Prize money is a major step forward<br>• The International 2014 is the golden example of prize money, with $10.9 Million is prize pool<br>• One of the biggest factors that pushes eSport is the game genre MOBA, with the underlying games DOTA and League of Legends. But there is still a huge interest for FPS games such as Counter-Strike: Global Offensive and offensive RTS's such as StarCraft II |
| ***Kreiswirth (2015): ESPN The Magazine's First-Ever eSports Issue*** | *A newspaper article from ESPN based upon the* | ESPN Recognizes eSport as a sport by making an own issue. This is a big push as ESPN is one of the leading names within the sport industry |
| ***on Newsstand Friday*** | *upcoming magazine which bases itself on eSports* | |
| ***Lolesports (2015): Teams \| Lol Esports*** | *A webpage made to show various eSport teams performing at the seasonal League of Legends championships in Korea* | The following companies is represented in the Korean league:<br>• Samsung Galaxy<br>• SK Telecom T1 |
| ***McCatterty (2014): How companies benefit from ITSM***<br><br>Additional Info (*)<br>IT service is seen as a tool within Information System Management, and is therefore further considered | *A Survey and report conducted by Enterprise Management Associates (EMA) to understand the efficiencies and improved user experiences that is possible through the increased use of IT Service Management (ITSM).* | • 52% of the organisations that implemented IT Service Management (ITSM) initiatives, describes the result as either a "extremely successful" or "very successful"<br>• Nearly 50% of the respondents believe that ITSM is poised for growth<br>• Several post show huge growth potential while using ITSM tools, which further make adaptations to cloud and other technologies easier. |
| ***McCool (2013): A Current Review of the Benefits, Barriers, and Considerations for Implementing Decision Support Systems*** | *A Journal article based upon the challenges and benefits when implementing a DSSs as a database for the CPGs in the health sector* | Implementation of DSS effects:<br>• Increased education to the users<br>• Improve patient (costumer):<br>  ○ Outcomes<br>  ○ Cost efficiency<br>  ○ Safety<br>  ○ Diagnosis<br>  ○ Mortality<br>  ○ Morbidity |
| ***McWhertor 2014: The International Dota 2 tournament watched by more*** | *A newspaper article from Polygon based upon the* | • More than 20 million in streaming audience<br>• 2 Million Peak concurrent viewership<br>• Prize pool exceeded $10,000,000 |
| ***than 20M viewers, Valve says*** | *success of the recent International eSport event* | |
| ***Newzoo (2015): Global Growth of eSports Report: Trends, Revenues & Audience Towards 2017***<br><br><u>Additional Info (*)</u><br>(1) The full report is not available for academic reading – so the free 11-page overview was used.<br><br>(2) Collected data is based over 25 countries | *An annual statistical trend report (dataset) created to provide the industry with a global overview of the eSport economy, viewership and participants as well as giving a realistic estimate of the futuristic potential of eSports.* | • There are 205 million people globally that watch eSports<br>• The viewers can be divided in four segments:<br>    ○ Regular viewers / participants (13 Million)<br>    ○ Occasional Viewers / Participants (19 Million)<br>    ○ Regular Viewers (56 Million)<br>    ○ Occasional Viewers (117 Million)<br>• 40% of the esports viewers do not play the games themselves.<br>• Esports audience has a growth existing of 50,7% from 2012 to 2014, and is expected to grow 63,4% from 2014 to 2017<br>• The esports economy is based upon teams, events, publishers and leagues, which is paid by brands and consumers.<br>• The report concludes that the amount of esports enthusiasts is about the same number of Ice hockey fans throughout the world. |
| ***North et al.(2004): The benefits of Knowledge Management: Some Empirical Evidence*** | *A Empirical study conducted to show empirical evidence on the benefits of knowledge management* | *Benefits of using knowledge management:*<br>• *More acceleration of process*<br>• *Reduction in redundancies*<br>• *Heavily increase in re-use of internal knowledge*<br>• *Leads to increased process transparency*<br><br>*Benefits in an employee perspective:*<br>• *Lead to an increase in motivation*<br>• *Enhance personal knowledge base*<br>• *Improving the teamwork*<br><br>*Benefits in a customer perspective:*<br>• *Increase the quality of products and services* |
| ***Oxford Dictionaries (2015): Definition of Information Technology*** | *A written definition used to define Information Technology by the Oxford Dictionary* | "[Information Technology is] *"the study or use of systems (especially computers and telecommunications) for storing, retrieving, and sending information"* |
| ***Paulsen and Saunders (2013):*** | *A book intended to* | Information Strategy Definition: |
| ***Strategic Management of Information Systems*** | *provide the reader with a foundation of basic concepts relevant to using and managing information.* | • "The plan an organization uses to provide information services".<br>• A function to give a strategy on how they will use the coming technology and communication to their advantage in the overall strategy. |
| ***Pereria (2013): League of Legends Infographic Highlights Eye-popping Numbers*** | *A newspaper article from IGN based upon various findings in the usage of the game League of Legends* | League of Legends has<br>• 70 Million Users (2013) (15 Million in 2011)<br>• 32.5 Million Active Users 2013) (1.4 Million in 2011)<br>• 1.3 Billion Hours of gameplay a month worldwide<br>Season 2 World Finals statistics:<br>• 8,000 Attendance (Live audience)<br>• $5,000,000 Total Prize Pool (over the whole season)<br>• 8,282,000 Unique Viewers<br>Season 3 World Finals Statistics<br>• 13,000 Attendance (Live audience)<br>• $8,000,000 Total Prize Pool (over the whole season)<br>• 32,000,000 Online viewers |
| ***Pick (2008): Benefits of Decision Support Systems*** | *A chapter within a book aimed towards surveying the decision support system field, defining it and further discuss its benefits and challenges.* | Using a decision support system will give you the following benefits:<br>• Improvement in decision quality<br>• Subtle improvement in the decision process |
| ***Power, Sharda and Burstein (2008): Decision Support Systems*** | *An journal article constructed towards defining Decision Support Systems, and its underlying methods* | DSS can be broken down into five different system types:<br>• Communication-driven<br>• Data-driven<br>• Document-driven<br>• Knowledge-driven<br>• Model-driven<br><br>DSS Definition:<br>• "A class of computerized information system that support decision-making activities" |
| ***Raisinghani (2004): Business Intelligence in the digital economy: opportunities, limitation and risks*** | *A book aimed towards defining Business Intelligence, and its underlying concepts.* | Business Intelligence can be defined as a Decision Support System, as both is defined as a umbrella term including architecture, tools, databases, analytical tools, applications and methodologies. |
| ***Rouse (2015): What is IT Management?*** | *A webpage made to define IT management and the processes within it* | IT management is the process of overseeing all matters related to information technology operations and resources within an IT organization. |
| ***Statisa (2015): Most Played PC Games on Gaming Platform Raptr in February 2015*** | *A monthly, scientific dataset based upon which computer games are played the most* | The following computer games is the most played:<br>• League of Legends (25%)<br>• World of Warcraft (8.03%)<br>• DOTA 2 (7.17%)<br>• Counter-Strike Global Offensive (5.15%)<br>• Smite (2.5%)<br>• Hearthstone (1.6%)<br>• Heroes of the Storm (1.09%) |
| ***SuperDataResearch (2015): eSports: Market Brief 2015***<br><br>Additional Info (*)<br>(1) *The data is global, but no information is given regard the participant people nor countries* | *A scientific dataset constructed to analyse the eSport industry with 'big data', track the key performs across digital markets and find various trends and values in the eSport field.* | • There are 71 million people globally that watch competitive gaming<br>• The main companies within competitive gaming are Riot Games, Wargaming, Valve and Ubisoft<br>• Companies like Intel and Coca-Cola uses eSports to reach out to their audience<br>• The League of Legends Season 3 World Championship had 32 Million viewers, which is bigger than BCS National Championship and the NBA Finals.<br>• The report include information provided by Newzoo in 2013, which say that 70% of the esports viewers in U.S. is men.<br>• Concludes that the major games within eSport is League of legends, StarCraft II and Dota 2. They also mention that fighting games light Street Fighter 4 and Super Smash Brothers Melee have a long and loyal stay within the eSport community |
| ***Turban et al. (2006): Decision support and*** | *A book constructed to introduce the* | *Benefits of using a DSS system:*<br>• *81% Faster and more accurate reporting*<br>• *78% improved decision making within the company* |
| ***Business Intelligence Systems*** | *management support systems (MSS), and its underlying branches.* | <ul><li>*56% improved customer service*</li><li>*49% increase of total revenue for a company*</li></ul> |
| ***Twitch.tv (2015): Twitch Annual Report 2014*** | *An annual report constructed to give an overview of its streaming numbers in 2014.* | <ul><li>Twitch.tv had 16 billion minutes watched per month</li><li>Twitch.tv had 100 million unique viewers per month</li><li>Twitch.tv had 1.5 million unique broadcasters per month</li><li>Twitch.tv had 11 million videos broadcast per month</li><li>Twitch.tv had 10 000 partnered channels</li></ul> |
| ***Urban Dictionary (2015): Definition of eSports*** | *A dictionary entry of various definition used to explain eSports* | "Pretty much another term for competitive video games. They are not sports, but only considered sports by nerds that want to be athletes. » |
| ***VNS (2011)): 3 Benefits of Having an Information Strategy*** | *A blog entry meant to understand the importance of aligning an Information Strategy with the business objectives, and how it further can give benefits for a company.* | <ul><li>Information Strategy initiates options that leads to overall reduced cost for the company. This can be seen as e.g., IT outsourcing or cloud computing.</li><li>Information Strategy open up various opportunities that leads to increased customer satisfaction. This is mostly done by increased the efficiency, which is increased by shortening the order cycle which is made possible by an IS.</li><li>Lastly, Information Strategy can increase the overall revenue generation. This is done by making it possible to reach a bigger audience with its technology, as well as making it possible to find and cut less efficient workflows.</li></ul> |
| ***Wagner (2006): On The Scientific Relevance Of Esports*** | *A conference paper created to lay a foundation for a proper academic treatment of eSports,* | Generic findings:<ul><li>One of the earliest uses of the term 'eSports' was in 1999, at the press release of the Online Gamers Association</li><li>ESport seems to be routed deeply in digital your culture.</li></ul><br>Historic milestones in eSport history: |
| | *including a historic overview and a definition* | • (1993) Release of the game Doom<br>• (1994) Release of the game Warcraft<br>• (1996) Release off the game Quake<br>• (1997) The creations of several gaming leagues, including CPL.<br>• (1997) The CPL: FRAG event, which made eSport considered as an emerging spectator sport<br>• (1998) The release of the game StarCraft<br>• (1999) Release of the game Counter-Strike<br><br>Definition of eSport:<br>• "ESports" is an area of sport activities in which people develop and train mental or physical abilities in the use of information and communication technologies. |
| ***Williams (2015): eSports are Growing Up*** | *A newspaper article made to understand the changes that is experienced in the field of competitive gaming* | • *eSports is slowly but surely growing*<br>• *eSports is getting more organized*<br>• *More channels make it possible to reach a wider audience for the field of eSports*<br>• *There exist a huge debate whether eSports is a sport or not.* |
| ***Witowski et al. (2013): eSport on a rise?: Critical consideration on the Growth and Erosion of Organized Digital Gaming Competitions*** | *A journal article on how to perform a written panellists debate where the topic of discussion is the present and the future of eSports, and the impact it will have on sporting events* | • At a recent Reddit AMA (Ask Me Anything), Emmett Shear, CEO of Twitch.tv, the leading live games, claimed that in a decade eSports will be bigger than athletic sport<br>• ESport represents a rare opportunity to observe the historical emergence of interactive gaming in a sporting 'skin', as well as new forms of sports-like competition realised through interactive gaming platforms.<br>• Hutchins argues that while eSport reproduces some features of modern sport, it also currently lacks the spaces and routines that enabled a key feature of sport to emerge – rituals and myth-making |
| ***Yong (2010): Korea's Online Gaming Empire***<br><br><u>Additional Info (*)</u><br>*Literature is based on data from Korea* | *An e-book aimed to contextualize stories of gaming culture in Korea, and what the culture means for the political economy of Korea.* | • The online game market was valued $4.96 Billion in 2006, and expected to reach $11.88 Billion in 2011.<br>• Korea constituted 49% of the global online game market in 2006<br>• The first TV eSport league was seen in Korea in 1998, where a league for the game StarCraft was created.<br>• WCG was hosted in Korea in 2001, dubbed the "Olympics of Video games"<br>• Oxford defines eSports as "a computer game played in professional competitions, especially when it is watched by fans and broadcast on the Internet or on television." Michael Wagner (2006) defines eSports as "an area of sport activities that includes sport activities in which people develop and train mental or physical abilities in the use of information and communication technologies."<br>• In 2000, 21.4% of the internet users of Korea used their computers for online gaming. This number increased to 55.5% in 2006.<br>• The first television eSport event was a*mazing, a program in Australian where children played Nintendo games against each other.<br>• The Dot-com boom between 1991-2000 accelerated the growth of eSports<br>• ESport market was worth $26.7 million in 2004, and grew to $39.5 million in 2005. It is expected to grow to $77.4 Million in 2007 and $120.7 Million by 2010.<br>• The growth of eSports correlates with the growth of number of games:<br>  ○ (1999) 72 Games and $1.5 Million was awarded<br>  ○ (2000) 82 Games and $2.0 Million was awarded<br>  ○ (2001) 93 Games and $3.0 Million was awarded<br>  ○ (2002) 187 games and $3.5 Million was awarded<br>  ○ (2003) 144 games and $4.0 Million was awarded<br>  ○ (2004) 98 games and $4.5 Million was awarded |
| | | ○ (2005) 278 Games and $5 Million was awarded<br>○ (2006) 124 Games and $3.3 Million was awarded<br>○ (2007) 70 Games and $1.8 Million was awarded<br>○ (2007) there were 21 online games being played compatibly.<br>• Audience are primarily 18-34 years old (74) |

Table 2.2 Collected Literature

## 2.4 Literature Evaluation and Validity

Advancing to the third step, an evaluation of the literature validity is created. Moreover, this step evaluate the collected literature in the sense of its relevance to the research question and the objectives (Randolph 2009). Considering the findings table 2.2, an addition table is conducted to review the relevance of each literature to this research study. This is done by taking each reviewed literature, and measuring up against the relevance of both the research question and the research objectives, which then summarises if the literature is valid or invalid. This is done to increase the research significance and the fact that it will lead to new knowledge (Hofstee 2006).

| Literature | Research Question Relevance | Research Objective Relevance | Valid or Invalid |
|---|---|---|---|
| *Alavi and Leidner (1999)* | Yes – RQ [5,6] | Yes – OB [7,8] | Valid |
| *Bergeron (2003)* | Yes – RQ [5,6] | Yes – OB [7,8] | Valid |
| *Borowy and Jin (2014)* | Yes – RQ [1,2] | Yes – OB [1,2] | Valid |
| *The Business Dictionary (2015a)* | Yes – RQ [5,6] | Yes – OB [7,8] | Valid |
| *Casselman (2015)* | Yes – RQ [1,2,3,4] | Yes – OB [2,4,5,7] | Valid |
| *Chobopeon (2012)* | Yes – RQ [2] | Yes – OB [2] | Valid |
| *D.Devil (2011)* | Yes – RQ [2] | Yes – OB [2] | Valid |
| *Davenport et al. (1998)* | Yes – RQ [5,6] | Yes – OB [7,8] | Valid |
| *Garfield (2014)* | Yes – RQ [5,6] | Yes – OB [7,8] | Valid |
| *Heaven (2014a)* | Yes – RQ [1,3,4] | Yes – OB [2,5,10] | Valid |
| *Heaven (2014b)* | Yes – RQ [2,4] | Yes – OB [2,3,7] | Valid |
| *Hewitt (2015)* | Yes – RQ [1,2,3,4] | Yes – OB [2,4,5,7] | Valid |
| *Heyoka (2011)* | Yes – RQ [2] | Yes – OB [2] | Valid |
| *Holsapple and Sena (2005)* | Yes – RQ [5,6] | Yes – OB [7,8] | Valid |
| *Hope (2015)* | Yes – RQ [1,2,3,4] | Yes – OB [2,4,5] | Valid |
| *Jackson (2013)* | Yes – RQ [1,2,3,4] | Yes – OB [2,4,5,7] | Valid |
| *Jarvis (2014)* | Yes – RQ [1,3,4] | Yes – OB [3,4,7] | Valid |
| *Kreiswirth (2015)* | Yes – RQ [1,2,3,4] | Yes – OB [2,3,4,7] | Valid |
| *McCattery (2015)* | Yes – RQ [5,6] | Yes – OB [7,8] | Valid |
| *McCool (2013)* | Yes – RQ [5,6] | Yes – OB [7,8] | Valid |
| *McWhertor (2014)* | Yes – RQ [1,2,3] | Yes – OB [2,3,4] | Valid |
| *Na.lolesport.com (2015)* | Yes – RQ [1] | Yes – OB [2] | Valid |
| *Newzoo (2015)* | Yes – RQ [3,4] | Yes – OB [3,5,6,7] | Valid |
| *North et al. (2004)* | Yes – RQ [5,6] | Yes – OB [7,8] | Valid |
| *Oxford Dictionaries (2015)* | Yes – RQ [5,6] | Yes – OB [7,8] | Valid |
| *Paulsen and Saunders (2013)* | Yes – RQ [5,6] | Yes – OB [7,8] | Valid |
| *Pereria (2013)* | Yes – RQ [1,2,3] | Yes – OB [2,3,4] | Valid |
| *Pick (2008)* | Yes – RQ [5,6] | Yes – OB [7,8] | Valid |
| *Power, Sharda and Burstein (2015)* | Yes – RQ [5,6] | Yes – OB [7,8] | Valid |
| *Raisinghani (2004)* | Yes – RQ [5,6] | Yes – OB [7,8] | Valid |
| *Rouse (2015)* | Yes – RQ [5,6] | Yes – OB [7,8] | Valid |
| *Statista (2015)* | YES – RQ [3] | Yes – OB [3] | Valid |
| *SuperDataResearch (2014)* | Yes – RQ [1,3,4] | Yes – OB [3,5,6,7] | Valid |
| *Turban (2008)* | Yes – RQ [5,6] | Yes – OB [7,8] | Valid |
| *Twitch.tv (2015)* | Yes – RQ [1,3,4] | Yes – OB [5,6] | Valid |
| *Urban Dictionary (2015)* | Yes – RQ [1] | Yes – OB [2] | Valid |
| *VNS (2015)* | Yes – RQ [5,6] | Yes – OB [7,8] | Valid |
| *Wagner (2006)* | Yes – RQ [1] | Yes – OB [2] | Valid |
| *Wagner (2006)* | Yes – RQ [1] | Yes – OB [2] | Valid |
| *Witowski, Hutchins and Carter (2013)* | Yes – RQ [1,4] | Yes – OB [1,7] | Valid |
| *Yong (2010)* | Yes – RQ [1,2,3,4] | Yes – OB [2,3,4,5,6] | Valid |

Table 2.3 Literature validity and relevance

As table 2.3 shows, every literature reviewed has some connection to either the research questions or objectives, making them valid to this thesis. The high percentages of validity is caused by early removing none relevant literature, which was done before presenting the literature findings. These valid literatures will therefore be used further in the next step, where they are analysed up against the research questions.

### 2.5 Literature Discussion, Analysis and Interpretation

The literature discussion, analysis and interpretation, which is the fourth step, will use the valid literature that were previously gathered and evaluated to answer the research questions. This is done to give a clearer understanding of the various literatures importance in this project paper, and further how well the research questions can be answered by the existing literature. The discussion will be used to see what literature is sufficient, what literature is lacking, and what can be done better. The analysis of the discussion will give a final indication the reviewed literature, making it possible to answer what this research thesis should further focus on to answer the research questions in the best possible way.

#### 2.5.1 Research Question 1: Literature Discussion

*‘What is eSports, and how aware are people of the eSport culture?’*

An introduction were previously given on what eSport (also known as Electronic sports or competitive gaming) is in chapter 1, but something that was already noticed in this introduction was the lacking of one specific definition. The literature points out that a common definition of eSport seize to exist, as the collected literature has multiple definitions. Nevertheless, there seem to be some correlation between the various definitions. According to Yong (2010), Oxford defines eSport as “a computer game played in professional competitions, especially when it is watched by fans and broadcast on the Internet or on Television”. Nevertheless, this definition does not exist through the online dictionaries, making it a less likely used definition. Urban Dictionary (2015) on the other hand, rather defines eSport as another term for competitive video games, and further points out that it should not be compared to a sport. In a more academic matter, Wagner’s (2006) definition is more commonly used, where he define eSport as “an area of sport activities that includes sport activates in which people develop and train mental or physical abilities in the use of information and commination technologies”. Heaven (2014a) however, defines eSport as a newly acknowledged, but still debatable sport, which is rising from the gaming community. He further states that it is based upon competitive gaming, where people play various computer games with or against each other. These competitive matches can be viewed either as a live participant, audience, or through streaming services (also known as spectator services) through internet services.

Wagner's (2006) definition includes the findings from the others, correlates with Heaven's (2014a) literature, and is therefore deemed as well constructed definition, which this project thesis deem usable.

A deeper understanding of competitive playing can be given by an example. In 2013, Riot Games conducted their annual World Championship Finals in their computer game 'League of Legends'. This event was held in Los Angeles' Staple Centre in front of a 13,000 live fans, as well as a virtual audience consisting of 32,000,000 spectators (virtual viewers) through various platforms on the world wide web (Casselman 2015; Jackson 2013; Redbeard 2013). This event is seen as the biggest to this data, but it might not be long until a bigger event is constructed.

The size of eSports is essential in defining the field of eSports. The financial size of eSport is according to Newzoo (2015) set to be around $190 000 000, which is higher than various known sports, whereas ice hockey is the clearest example. Casselman (2015) further states that a total of 3.7 billion hours was spent watching eSports in 2014, and this is according to Heaven (2014b) predicted to increase to 6.6 billion by 2018.

The overall awareness can be understood by looking at the audience numbers, which interestingly is different in the various literature. SuperDataResearch (2015) states that 71 million watch eSports, while Casselman (2015) states the number is higher with 89 Million watchers. This does not correlate with Newzoo's (2015) data, as they state that the eSport audience is over 205 million people. This can however be deemed less likely, as such a high number does not correlate with other collected literature. Newzoo (2015) further difference the audience in four segments: The occasional viewers, the regular viewers, the semi-professional regular viewers (people who occasionally play competitively within eSports) and the professional viewers (people who fully play competitively within eSports). The occasional viewers is seen as the biggest segment, filling 58.5%, while the semi-professional viewers consist of 28%. This gives an indication that the eSport lack fully engages viewers, but the awareness do exist. The number of viewers is predicted to continue growing, whereas the number of total viewers is believed to increase by 67.5% by 2017. The growth can further be claimed genuine, as the increase from 2012 to 2014 alone was 49.25% (Newzoo 2015).

### 2.5.2 Research Question 2: Literature Discussion

*What are the major milestones of eSports history?*

The very question of the major milestones of eSport history correlate with anything relating to the history of eSports. As for existing literature, there exist quite few published works related to this topic. However, a discussion in various newspapers, forums and blogs do exist.

The first eSport event held is considered as a major event, which was according to Hope (2015) an event held in 1981, for the arcade game 'Space Invaders'. Borowy and Jin (2013) raise the question if this can be named as an eSport event, as this is one of many events based upon arcade games that existed in the early stages of the 1980s. Chobopeon (2012) points out that other games existed long before this, but the events held (if any) were never large or prevalent. One of these events was the 'Space war' competition held in 1972, which is according to Heaven (2015a) the first eSport event ever held as it included a prize. Nevertheless, the majority of people, including Borowy and Jin (2013), conclude that the Space Invaders tournament is considered the first video game tournament ever staged, although they say it were to exist in 1980, not 1981.

The second milestone can be seen as the Seattle-based tournament Games held in October 28, 1981 (Borowy and Jin 2013). This event spent over $200,000 to finance and promote the tournament, and the field of eSports have never seen so much money put into one single event at that time. Nevertheless, the year of 1981 had many downfalls, where tournaments such as the Atari Coin-Op $50,000 World Champions held with the capacity of 10,000 competitors, only attracted 250 competitors (Chobopeon 2012).

The future of 1980s future include many arcade game tournaments. Borowy and Jin (2013) names the National Video Game Master Tournament and the Electronic Circus held in the US in 1983-84 as a major milestones, mainly because of its public promotion. Annual tournaments was also created in the end of 1980s, where the American Video Game Challenge were one of the biggest (Borrow and Jin 2013).

The next major milestone for the eSports is seen by both Borrow and Jin (2013) and Hope (2015) as getting the competitions on television. In-between 1982 and 1984, a programme named Starcade was broadcasted, which features various competitive gaming tournaments and interviews.

Hope (2015) explains further that the eSports continued existing in a small scale until the start of 1990, mainly due lack of technology. In the 1990s some technology was introduced which made the multiplayer functionality of eSports easier, which lead to an increase of tournaments and players. The technology also made it possible to play over the internet, with was one of the reason that led to the creation of two well-known game genres used to day: Real Time Strategy (RTS) and FPS (First Person Shooter) (Chobopeon 2012). D.Devil (2011) name two major events that used this technology to the fullest, which is the Red Annihilation Tournament and the Cyberathlete Professional League's (CPL) FRAG tournament, both held in 1997. D.Devil name these events as the birth of eSports known today, not involving any arcade games. The success of the Red Annihilation Tournament in 1997 also created a chain-reaction that made various companies construct annual tournaments, where QuakeCon is one of the first examples (Chobopeon 2012)

Nintendo, a well-known game distributer, had their first tournament as well in the beginning of the 1990s, known with the Nintendo World Championship, which attracted people from all around the world (Chobopeon 2012; Heyoka 2011). The number of events continued to grow slightly until 2000, which is seen as one of the strongest peaking years within eSports.

The next milestone is seen to be The WCG (World Cyber Games), who got involved within the field of eSports in the year 2000, creating events involving over 35 countries and having prize pools consisting of up to $600,000 (Hope 2015). WCG showed a great form of success, which led to the foundation of major eSports companies known today as IEM (Intel Extreme Masters) and MLG (Major League Gaming) (Hope 2015).

Additionally, Wagner (2006) states the creation of the games of Doom (1993), Warcraft (1994), Quake (1996), StarCraft (1998) and Counter-Strike (1999) as important milestones of the growth of eSports in the beginning of 2000 century.

D.Devil (2011) also name the mid-2000s importance for the history of eSports, as it includes the creation of the first eSport communities known today, where SK Gaming is the first example, created in 2003.

Another milestone in the history of eSports is the launch of live streaming, which appeared in 2007 by a website named Justin.tv, which today is known as Twitch.tv (Hope 2015). This made it possible for all professional players, and all tournaments to show the matches online rather having staged shows with large television – making it much easier to reach out to the consumers. This is not the creation of streaming in the sense of spectating however, as being able to watch games online through spectator modes was available before the launch of Justin.tv.

The increase of eSport grew drastically the following years, and tournaments were held everywhere, focusing globally on various game genres. Hope (2015) name the release of the game League of Legends by Riot Games in the end of 2009 as the next milestone of eSports. The game itself is today seen as one of the leading games played in the field of eSports, attracting players and viewers from all continents of the world. This can further be seen in 2013, where the League of Legends world finals made the biggest eSport event yet, which had 32,000,000 online viewers throughout the tournament (Pereira 2013). The growth of eSports can also be seen in the sense of prize pool, whereas The International tournament in 2014 reached a total prize pool of over $10,000,000 (McWhertor 2014)

The next milestone according to Hope (2015) is finding a way into television around the world. This is already quite popular in Korea, where channels are showing eSports 24/7, but the eSport on television has yet not reached Europe and North America in the same scale. Nevertheless, Hope (2015) name this as the future of eSports, which is discussed further in research question 4.

### 2.5.3 Research Question 3: Literature Discussion

*<u>Why is eSport seeing its growth today, and which factors/patterns is pushing it upwards?</u>*

The previous question regarded what is the most important events that has existed in the history of eSports, and that question leads into a new one: why are eSports seeing it major growth today, and not earlier? Both Hope (2015) and Borrow and Jin (2013) touch this question by naming the necessary access of internet not existing until the end of 1990, which lead up to an opportunity to increase the eSport community. Nevertheless, this is not the only momentum that lead to the growth of eSports.

Jackson (2013) name an important factor for the growth is the increase of various leagues, especially MLG (Major League Gaming) and EVO (Evolution Championship Series). This correlates with Williams (2015) thoughts, as he point outs that the scene of eSports lacked a sense of organisation rules, which can be seen today.

The launching of internet streaming services in 2007 was also seen as a major historic milestone, which made it possible to view live streaming viewership (Hope 2015). Williams (2015) debates that this was only the beginning of something that plays a big part of the growth seen today. However, Justin.tv, which is now known as Twitch.tv is only one channel; there exist a broad aspect of different channels such as HLTV and MLG.TV, who were some of leading providers of streaming services in the early 2000s.

The amount of games that exist today can also be seen as a pattern pushing it upwards. Hope (2015) previously named League of Legends as an important milestone, but as Jarvis (2014) debates in his interviews; it is the game genres that pushes the field of eSport upwards. Jarvis (2014) interviews mostly point out two genres, which is FPS and MOBA, which is the games that is mostly viewed. This correlates with Statista's (2015) continuously analyse, which show the most common played games today is FPS and MOBA, but also MMOA and OCCG. SuperDataBase research (2015) report on digital trends in the eSport market differentiate itself some however, as it includes RTS as a major genre, as well as Fighting games. Nevertheless, SuperDataBase (2015) states that this is mainly due the game genres has existed longer in the eSport community, and therefore have a bigger loyal fan base than other genres.

Yong (2010) has a different approach to the growing factors, as his data is especially from Korea. However, his literature shows that the growth in the field of eSports heavily increased in 1991-2000, which is the dot-com boom. Yong (2010) further points out a correlation between the prize pool and number of games, which can be understood as two important factors within the field of eSports.

Additionally, Newzoo's (2015) annual eSport report show some interesting data regarding who watches eSport. The earlier years were heavily depended that only people playing the game itself. However, Newzoo (2015) state that 40% of the eSport viewership does not play the game themselves, which can be a factor of the overall increase of eSports consumption.

#### 2.5.4 Research Question 4: Literature Discussion

*Will the growth of eSport continue, and what can the eSport business do to ensure a continuous growth? (Future of eSports)*

Heaven (2014a) expect to see a sufficient growth in the field of eSport. In the sense of viewership, the hours watched in 2013 equalled 2.4 billion hours of eSport video consumption. This growth is believed to grow to 6.6 billion hours, which is a 275% video consumption increase in only 5 years. He further bases his opinion on the increase of eSports arenas, where the eSport stadium due to open in 2017 with 15,000 seats in Hengqin, China is the golden example.

Newzoo's (2015) data also show signs of even further growth within the field of eSports the coming years. Their data show an increase in audience by 50% from 2012. The data further indicates that the size of audience is expected to grow even further 63.4% by 2017. This means that the total audience number today, which is 205 million according to Newzoo (2015), is expected to grow to nearly 334 million people.

Jackson (2015) also believe in the growth of eSports, and point out various examples. Firstly, there is the growth seen in twitch.tv, which further makes an increase of eSports commentators and castors. Secondly, more professional teams are sprouting all over the world, and can be created through pure will – not necessary companies buying professional players to increase their name recognition, which have been a trend the last years. Thirdly, Jackson believe that the new eSports Association (TeSPA) is a great network that will increase the gaming-community especially at college campuses. This again might increase the consumption and enhance the growth of eSports.

Williams (2015) also mention a bright future for eSports, although it might come slowly. Williams further believes that the success of eSports today is mainly caused by four factors: Better-organised structure within the field of eSports, New gaming channels, more knowledge in the sense of more casters that are experienced, communities and management, and lastly the higher achievements in the sense of getting time on TV programs and being named an actual sport. This beliefs correlates highly with the previous discussed literature.

ESPN, the heavily acknowledge sport network, has also lately recognized the size of eSports, and that it is still growing (Kreiswirth 2015). In June 2015, ESPN released an own issue focusing on eSports in their own magazine. This affects the future of eSports, as the discussion of whether eSports is a sport or not has been one of the original momentums holding eSports back. The discussion whether it should be a sport or not is something that might have an impact of the future of eSports, as today only a few people are acknowledging the likelihood.

Witowski et al. (2013) however does not strongly agree in this sense, as they point out that eSports is lacking spaces and routines that enables various key features so that a sport genre can emerge.

Hope (2015) answer the question if eSports will be viewed more in the future with a simple "Yes!" He further point outs that the different varieties of showing eSports is increasing, making gaming no longer a looked-down hobby, but rather something that involves serious dedication. He further concludes that there is no specific answer for the future that holds for eSports, as it will be based upon the advances of technology itself.

### 2.5.5 Research Question 5 and 6: Literature Discussion

*Can Information Technology Management Tools be used beneficially for the field of eSport? & what outcomes would the most beneficial Information Technology Management Tool give in the eSport business?*

To this date (10.07.2015), no previous research could be found where ITM tools were used within the field of eSports. This indicates that it is not possible to assiduously answer research question 5 and 6 with existing literature. However, it does exist literature that discusses the effect ITM tools have on organisations and businesses in general. Considering this, the tools limited by the study in chapter 1 will be deliberated, where their benefits and disbenefits is further discussed.

#### *2.5.5.1 Information Technology Management*

ITM tools involves specific tools, techniques and strategies commonly used within the field of Information Technology. Before getting an understanding of which tools could be most beneficial, it is firstly important to give a definition on what ITM is. Although various definition exists, the most common one exist within the Oxford dictionaries (2015): "[Information Technology is] the study or use of systems (especially computers and telecommunications) for storing, retrieving, and sending information". The business Dictionary (2015a) further define it as a set of tools, processes, and methodologies that is used to collect, process and present information. ITM on the other hand, can be understood as the management process of these processes, as Rouse (2015) further define ITM as "the process of overseeing all matters related to information technology operations and resources with an IT organization". These three definitions give an understanding of the field, which further indicated the choice of the selected ITM tools.

### *2.5.5.2 Information Technology Management Tools*

Within the field of ITM, there are various areas, which involves set of techniques and processes. Some examples of these would be Knowledge Management within the field of IT, Project Management within the field of IT, Decision Support Systems Management, Enterprise Resource Planning Management and Information Technology Strategy Management. By the factors gathered from the business analysis, as well as previous literature reviewed later in this chapter, the most suited theories will be used to implement the factors to sustain and continue the growth of eSport. If available, all existing tools would be considered. However, due the limitation of the paper stated in chapter 1, only three specific tools will further be discussed. These tools is chosen based upon belief of what might be the most influential, as well as which tool correlates with the previous definitions given. To gain a better understanding of the chosen ITM tools an introduction is given followed by a discussion of its benefits and disbenefits.

#### 2.5.5.2.1 Decision Support Systems

A Decision Support System (DSS) is an ITM tool used to gather data to help decision making at a managerial level. Generally, it is understood within Power, Sharda and Burstein (2015) definition: "[A DSS is a] class of computerized information system that support decision-making activities". However, Turban et al. (2008: 21) additionally defines DSS as an umbrella term, as it can be understood in numerous possibilities. That is why Power, Sharda and Burstein (2015) further breaks it down into either a Communication-driven DSS, Data-driven DSS, Document-driven DSS, Knowledge-driven DSS or a Model-driven system DSS. These types of DSS's differentiate in how they gather data that is used towards making decisions for a managerial level. DSS can also be defined as a Business Intelligence, which is another umbrella term, which according to Raisinghani (2004) combines architectures, tools, databases, analytical tools, applications and methodologies. Based upon the given definitions, the next matter discuss the various benefits.

#### 2.5.5.2.1.1 Benefits of Decision Support Systems

A DSS exist in various forms as stated in the definition, and each of these will have its different benefits in the field of eSports. However, as this is seen in a holistic approach, the existing literature give some indications of its benefits. Pick (2008) states that a DSS firstly improve the decision quality. He further states that a DSS helps making the decision for the management, and makes the decision more secure and improved. This together improves the decision process overall. These sort of benefits is very holistic, and look at DSS in the macro perspective. Holsapple and Sena (2005) however looks at the benefits using a DSS system in a micro perspective, giving some additional literature. Their findings include the facts such as DSS makes it possible to better cope with complex problems, it reduces the decision time and cost, it improves the communication and coordination within the company, and overall gives a competitive edge to the users. These benefits can further be understood in accurate numbers, as Turban et al. (2008) concludes the following numerical benefits of using a DSS:

- 81% Faster and more accurate reporting
- 78% improved decision making within the company
- 56% improved customer service
- 49% increase of total revenue for a company

#### 2.5.5.2.2 Knowledge Management

Knowledge management is an abiding term, which includes multitudinous amount of various definitions, as the term itself can be understood differently. It is however understood as an ITM tool, and as this research paper is done within a business aspect, Bergeron's (2003: 8-9) definition can be used to get an understanding of the term: "Knowledge Management (KM) is a deliberate, systematic business optimization strategy that selects, distils, stores, organizes, packages, and communicates information essential to the business of a company in a manner that improves employee performance and corporate competitiveness". *K*nowledge itself is differentiated by Davenport et al. (1997) in either tacit knowledge (subjective knowledge) or explicit knowledge (documented knowledge). This can be understood as the knowledge itself can either be easily transferable (explicit) through a sense of medium, or that the knowledge is hard to explain (tacit), and thereby is not easily transferable. With a definition of Knowledge management given, the following beneficial matter can be discussed.

###### 2.5.5.2.2.1 Knowledge Management Benefits

With the definition of Knowledge management in mind, a knowledge management system can bring various benefits to organisations and businesses based upon using already existing knowledge. Alavi and Leidner (1999) break these benefits into five categories: Communication benefits, efficiency benefits, financial benefits, marketing benefits and general benefits. They further point out that these benefits includes an enhanced and faster communication, a faster delivery focused and cost-efficient system, a higher profitability, a better customer focus and an increased utilisation of workers.

These benefits correlates with Garfield's (2014) 15 benefits of using knowledge management, although he argues there exist more other benefits. These benefits includes that knowledge management enables better and faster decision making, as well as it makes it easier to find relevant information and resources. Garfield (2014) further states that knowledge management can be used to take advantage of exiting knowledge, both tacit and explicit. Garfield (2014) lastly discuss that knowledge management can be used to stimulate innovation and growth, which correlates greatly with the research questions. These benefits is also mentioned in Davenport's et al. (1997) literature.

North et al. (2004) data differentiate itself from the others, as it looks at the benefits of knowledge management in a business perspective, an employee perspective and a customer perspective. In the sense of implementation, North et al. (2004) defines the following benefits of Knowledge Management:

- Benefits of using knowledge management
    - More acceleration of process
    - Reduction in redundancies
    - Heavily increase in re-use of internal knowledge
    - Leads to increased process transparency
- Benefits in an employee perspective
    - Lead to an increase in motivation
    - Enhance personal knowledge base

- Improving the teamwork
- Benefits in a customer perspective
    - Increase the quality of products and services

Davenport et al (1997) further stress the importance of having a technical and organisational infrastructure, as well as a knowledge friendly culture and a clear purpose and language for this to be achieved.

There is none numerical numbers of the exact effect, as it would different itself on the different projects. However, it is commonly acknowledge that the previous stated benefits is legitimately found in various projects.

#### 2.5.5.2.3 Information Strategy (IS)

A large organisations strategy is often broken down into three different strategies, which is the business strategy, the organizational strategy and the Information Strategy. Paulson and Saunders (2013) refers this as the Information Systems Strategy Triangle, as each strategy has an equal effect on the organisation that is conducting a new strategy. Information Strategy can be understood as a vague term; it can be defined and broken down to either an Information System Strategy, an Information Technology Strategy or simply a Technology Strategy. The management of these strategies can additionally vary, as IT management can be broken down into e.g. IT service management or IT systems management. As stated in the limitations, all of these terms are understood as one, which is Information Strategy. This is done to have a more holistic approach to the research thesis, and rather focus on the main objectives and research questions. Paulson and Sanders (2013: 36) defines a IS as "the plan an organization uses to provide information services". They further explain it function to give a strategy on how they will use the coming technology and communication to their advantage in the overall strategy. With a holistic approach, the following benefits and disbenefits is discussed in the subsequent point.

##### 2.5.5.2.3.1 Information Strategy Benefits

An Information Strategy is defined as an extensive landscape of tools, where their benefits vary. Additionally, there is nearly none literature written about the benefits of Information Strategy with a holistic approach. The literature rather focus on initiatives such at IT Service

Management, which is a branch in effect under the Information Strategy. However, to get an understanding of the benefits Information Strategy bring, these branches is included in the findings. As for the usage of Information Strategies, VNS (2015) name following benefits:

- Leads to overall reduced cost for the company
- Increases customer satisfaction
- Increases the efficiency within the company
- Shortens the order cycle
- Increase overall revenue generation
- Makes it possible to reach a bigger audience and cut less efficient workflows

These benefits can further be seen in practice, where IT Service Management is used within companies. According to McCatterty (2015), ITSM bring growth to the company in 50% of the cases. His findings also state a huge growth potential within companies where Information Strategies exist.

Paulsen and Saunders (2013) mention that an Information Strategy can be to implement a Knowledge Management System or a Decision support system, which makes it possible to see Information Strategy as something that might involve the benefits stated in the other tools. However, these tools are discussed in a holistic approach, making the implementation of an Information Strategy and the implementation of a KM system two completely different areas. There is therefore no further information on the benefits within usage of Information Strategy.

## 2.6 Literature Presentation

The last step according to Randolph's (2009) model is literature presentation, which summarises what information should be included further in the thesis, and which literatures should be prioritised. Table 2.3 can be used yet again; to see the data used as well as its validity. However, this step will rather focus on how well the research question is answered with the existing literature, and if the literature itself answer the objectives and deliverables. This is done by conducting a discussion on what can be done better, or differently. This will thereby conclude the further focus on this thesis, and renew the project scope.

### 2.6.1 Research Question 1: Presentation

In the first step, the expected data in the first research question involved a definition of what eSport is as well as a summary of the eSport awareness. There exist a lot of literature in regards of the definition of eSports, and the definition used by Wagner (2006) is often used. As for the awareness, there are many numbers throughout various articles and surveys, giving a clear indication of what both the objectives and deliverables ask for. The existing data is also gathered in a global aspect, making it suitable to these projects aim. As table 2.3 shows, more than 16 literature involve research question 1, proving that it exist literature for the question. In conclusion, the literature existing answers the research questions well, as both objective number 2 and 6 is answered. Considering this, little additional research can be done. However, a lack of significance was found in the literature regarding the overall awareness, which is something that could be researched more on.

### 2.6.2 Research Question 2: Presentation

The second research question expected to include a summary of the most important milestones in eSport history, which included a list of the major eSports events through time as well as understanding of what made these events important. The existing literature of this is for the most based upon personal experience from single persons posted at forums, and lack highly academic documentation. Holistically, the literature gives an indication of the most important events, but it lacks specifically numbers and explanations of its major roles. As table 2.3 shows, more than 14 literature involve research question 2, proving that there do exist relevant literature. However, the literature lacks either the academic approach or depth in their findings. In other words, the existing literature is not sufficient to answer research question 2, as objective number 4 cannot be fully answered.

### 2.6.3 Research Question 3: Presentation

The third research question regards the factors and patterns that are pushing eSports upwards. The existing literature answer this to some extent**,** but alike research question 2, these findings is more directed at specific milestones. The literature show various milestones which is majorly important for the growth seen today, but there is few specific trends and patterns. As table 2.3 shows, more than 14 literature involve research question 3, proving that existing relevant literature exist. However, the literature gave some indication of various patterns, such as total number of games at events, event prize pool, game genres and some specific games. These patterns and trends had an indication of the growth seen in eSport. Additionally, all the found patterns were mainly in eSport events. Nevertheless, the findings briefly touch the research questions, and additionally lack the academic depths to answer sufficiently, making objective 3 unanswerable.

### 2.6.4 Research Question 4: Presentation

The fourth research question expected to get an understanding of the close future of eSports, as well as what can be done to enhance the growth. As for the close future, there are various literature give both economic and consumption predictions, making it sufficient literature to answer objective 7, which involves the future of eSports. As for how it can continue to grow, next to none literature exist. Alike the previous research question, very little factors and patterns are found and used within the existing literature. Without these patterns, little can be said about using it to the future. Nevertheless, table 2.3 shows, more than 13 'literatures' involve research question 4, proving that literature regarding this question exist. However, the existing literature gives only an understanding of the future of eSport, which is objective 7, but lacks factors to further academically discuss it. It is therefore not fully possible to answer objective seven and objective 5, which means research question 3 cannot fully be answered with the existing literature.

### 2.6.5 Research Question 5 and 6: Presentation

Both Research question 5 and 6 involves ITM tools that would be deemed beneficial for the eSport business to use. However, both questions are deemed not possible to be answered by the existing literature, as there exist none. Nevertheless, literature on various general benefits of using ITM tools is used instead, to get an indication of what it might bring the field of eSports. The general benefits gives some indication on how it might affect, but specific tools, factors and patterns are needed to answer the research questions. As table 2.3 shows, more than 17 'literatures' involve RQ5 and RQ6 in the generic sense. However, the lack of specific literature makes it not possible to answer both research question five and six, as their underlying objectives eight, nine and ten, require an understanding of the benefits, which it might bring into the field of eSports.

## 2.7 Summary of Randolph's Five Steps

With Randolph's (2009) steps finished, an understanding of what research has been done in eSports is conducted. As discussed in the literature presentation, some of the research questions could be answered to some extent. However, some questions lacked any literature at all, or the literature were not qualified for academic presentation. This changes the focus of the research thesis, as the findings reveals what is actually needed to answer the research questions in the best possible way. The previous presentation is therefore taken into consideration, when the future scope of the thesis is defined.

### 2.7.1 Renewed Project Scope Based Upon Literature Findings

To debate what will stand in the essence of this thesis based upon the literature findings, a project scope is created. The project scope is defined as "The work that needs to be accomplished to deliver a product, service, or result with the specified features and functions" (Project Management Institute 2001). This was previously defined as the aim and objectives in chapter 1, but a renewed scope is done based on the finding in the literature review. The project scope is further based upon both understanding of what should be done better, and what does not need additional research. Considering this, the following points were conducted:

I. The various growth factors and patterns were numerous and hard to define. Rather than focusing on every specific milestone in the eSport field, the research thesis will rather focus on eSport events. This is done by using analytical tools to get a better understanding of which factors are important when creating an eSport event. It will further arrange these factors, and seek a correlation with the growth eSports sees today.

II. The literature did include important eSport events as milestones, but these findings did not always correlate with the existing literature. Additionally, the existing literature was in the most cases not academic correct in the sense of content. Considering this, the project thesis will focus on constructing an analysis consisting of an abounding number of events, where the successfulness of the event is decided by the size of its audience. The analysis will further include various themes, to gain an understanding of various patterns within eSport events.

III. This research thesis will use the analytical findings consisting of various factors and patterns to enhance future growth of eSport. This is done by using specific ITM tools, chosen from both theory and the findings in the analysis. This split the research paper in two halves: What can be done to improve eSport events immediately, and what can be done to improve eSport events in the future.

IV. This research paper will further predict and recommend certain actions the eSport business can do to enhance the growth, both in short-term and long-term, based upon the chosen ITM tools previously found.

In addition to understanding what the research will base itself on, it is also important to state what it will not answer. The following points therefore state what this research thesis will not seek to answer, as existing literature is found sufficient:

I. The research thesis will not try to define what eSport is, as this is already done in the existing literature. It will rather focus upon the awareness to answer research question 1.

II. The research thesis will further not consider futuristic technological factors in eSports, such as improved gaming platforms and gaming technology. This is purely due to size, and therefore the focus is directed at eSport events. This causes the focus on IS strategy a chosen ITM to be less likely, which will be discussed in the impending chapters.

III. The research thesis will neither try to answer which segment watches eSports, as the existing literature answer this question well.

## 2.8 Summary of Chapter 2

This Chapter was dedicated to the literature review, where various literature in the project domain were gathered, discussed and analysed. An understanding on how the research questions relates to the objectives was firstly constructed which formulated the problem foundation, showing what is expected to find in the literature review. In total 41 various literatures were gathered, and were concluded valid for this research thesis as it had relevance to both research questions and objectives. The analysis of the literature further concluded that research question 1 had sufficient literature, while research question 2, 3 and 4 had some literature but lacked academic depths. Research question 5 and 6 however, had no existing literature that had direct relevance, so the questions was looked upon in a general matter. To further gain an understanding of how well the existing literature is, and what can be done better, the research thesis construct a renewed project scope. Supervening the literature review, the project scope states that the further focus will be to use eSport events to answer the various research questions. The findings in the analysis will further be used to implement an ITM tool, which would give a better answer for research question 5 and 6.

# CHAPTER 3 METHODOLOGY

## 3.1 Chapter Introduction

The Previous chapter examined and discussed the literature formerly done within the project domain. The valid literature was further used to try answering the research questions, to give an indication of what needs to be done further too fully answer the questions. In conclusion, the literature review created new unique goals and objectives for the upcoming chapters, which was summarised as the renewed project scope. This chapter, however, will discuss how the aim will be achieved in the sense of methodology and research. As stated previously, the aim for this research is to define and measure the key factors that caused the growth of eSports, and further appraise the benefits and disbenefits of implementing ITM tools to maintain, and enhance its growth. To achieve this aim, a set of both primary and secondary data is required. However, before this can be discussed, an understanding of research philosophy, research methodology and research methods is required.

### 3.1.1 Defining Research Philosophy and Research Approach

A model repeatedly used within the research methodology field is the research onion model presented in figure 3.1 (Saunders et al. 2009: 109). The model gives an understanding of which factors that together defines a research, and further what defines data collection and data analysis. The supervening points will focus on the two outermost layers referred as philosophies and approaches. However, an understanding of the term ‘research’ is needed afore discussing its multiple layers.

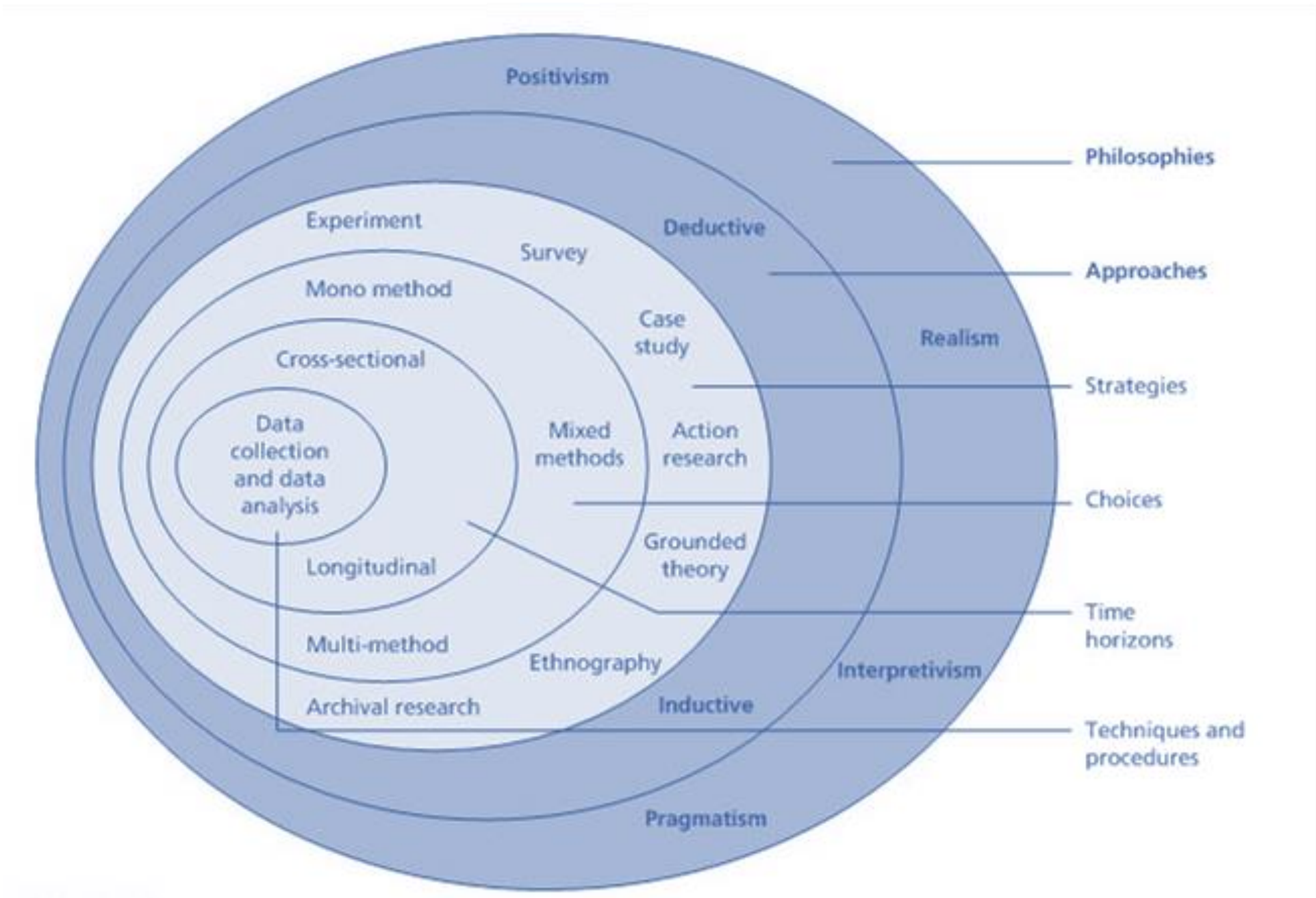

*Figure 3.1 the Research Onion (Saunders et al. 2009: 109)*

### *3.1.1.1 Research definition*

Research is defined by Saunders et al. (2009: 5) as "an original contribution to the existing stock of knowledge making for its advancement". This can been seen in a broader aspect, as Kothari (2004: 1-2) defines research as a pursuit of truth with the help of study, observation, comparison and experiment. This correlate greatly with the purpose of this thesis, as it strives to achieve its aim and objectives. There are three further phrases used within research that differentiate from each other; firstly, there is the research philosophy, which defines why the research is done. Secondly, there is the methodology, which defines and discuss the various methods to use within the research, and further define the correct method for the research or study to use. Thirdly, there is methods and tools, which define which tools to be used while collecting the required data. All these three aspects will further be discussed within this chapter.

## 3.2 First Layer: Research Philosophy

A research philosophy is used to give and understanding of why the data is collected, and what the research questions desires to answer (Saunders et al. 2009). These philosophies could be dwelled upon for pages, but due the research extent, their only briefly considered.

Within business research, the most common philosophy approaches is ontology, epistemology and axiology, which is used to consider the correct philosophy for this research (Saunders et al. 2009: 110-119). The ontology approach is used when working towards the nature of reality or being, while the epistemology approach is used to consider what acceptable knowledge is. Axiology on the other hand, consider what the values role is within the research in total. In addition to these there is the most common philosophies within business, which is positivism, realism, interpretivism and pragmatism – which all have a very different aspect on how the researcher view the data. These philosophies is considered within the different approach to give an understanding of which philosophy is most suited for this research.

To define the research philosophy approach, the aim and the research questions must be considered. This creates an understanding of what eSports is, and why it is experiencing a recent growth. To answer this, a set of qualitative data is needed. In addition, a scenario must be created to give an understanding of how ITM tools could be implemented, which is further shown with business analytic tools. The first part of the aim could be understood as realism, as the existing data need to be understood and the culture and experience within eSports stand as an essence. However, the research questions is more essential than the culture within eSports, and therefore the pragmatism philosophy is more suited for this research. Pragmatism can be understood as a school of philosophy that correlates greatly with research's aim, which according to Mastin (2008), can be understood as "[Pragmatism] considers practical consequences or real effects to be vital components of both meaning and truth". Most of the research questions is additionally centres around value, which is essential for the pragmatism philosophy in an axiology approach. This could be discussed further, but this is the final reasoning for having a pragmatism philosophy as the outer layer of this research.

### 3.3 Second Layer: Research Approaches

Introducing the second layer of the research onion is the research approaches for the study. As Figure 3.1 indicates, there exist two different approaches to use within a research study, which is the deduction approach and the induction approach. Saunders et al. (2009: 127) defines deduction as a scientific approach, which often moves from theory to data and is often based upon quantitative data. The induction approach however, is used to collect qualitative data and is used to gain a greater understanding of the data. Trochim (2006) further defines the two approaches in four steps, which gives a greater understanding of each approach. The deduction approach is formulated as: 1. Theory, 2. Hypothesis, 3. Observation and 4. Confirmation (Trochim 2006). This is according to Kothari (2004: 11) the most used approach, as he further uses this approach to define the seven steps in a research paper. The induction approach however use a button-up approach and turns everything around: 1. Observation, 2. Pattern, 3. Tentative Hypothesis, 4. Theory (Trochim 2006). With the research aim in mind, the induction approach is believed to be more efficient. This is mainly due the research questions as well as the aim, is firstly constructed to get a understanding of what eSports is, and which factors that have made it grow – this can be seen as observation and pattern. Furthermore, these factors are implemented with ITM tools and theory, which can be seen as tentative hypothesis and theory. It is with this reasoning the induction approach is selected, and the second layer of the research onion is chosen.

### 3.3.1 Defining Primary and Secondary Data

Before continuing the layers of the research onion, it is first important to define how this research paper will use and understand various data. For this research, both primary and secondary data will be understood within Malhotra and Birk's definition (2007: 85):

- Primary data: "Data originated by the researcher specially to address the research problem"
- Secondary data: "Data that have already been collected for purposes other than the problem at hand"

This can further be understood as library research (secondary data) and field research (primary data), which gives an easier understanding of the methods. This also includes a third method, which is Laboratory Research; this is based upon studying patterns of random behaviour, often within the phycology field, but will however not be used within this thesis (Kothari, 2004: 7).

Reminiscing the objectives, a high amount of both primary and secondary data is needed to succeed defining eSports, understanding its historic milestones and analysing a pattern of success factors in previous eSports events. Additionally, primary data is essentially needed to ensure data on various objectives, where the awareness of the eSport culture stand as the highest importance. To achieve this, sets of specific research methods were chosen to obtain the primary and secondary data. Kothari (2004: 2-3) differentiate the research methods that can be used against each other, which is descriptive versus analytical, applied versus fundamental, conceptual versus empirical and quantitative versus qualitative. For this research thesis, the quantitative versus qualitative is seen most beneficial, as the objectives and aim refers to data requirements such as quality in its findings. Kothari (2004: 3) further define these those methods as:

- *Qualitative data:* based upon the phenomena relating to or involving quality.
- *Quantitative data:* based upon the measurement of quantity or amount.

This correlates with Sharma's (2007:121) definition of qualitative methods in sports, where she differentiate these two by defining qualitative methods as data used to get insightful information, while quantitative methods are merely used to gather data for checking hypotheses.

## 3.4 Third Layer: Research Data

The third layer of the research onion focuses on how the primary and secondary data will be gathered. However, there exist several tools to be used to collect the necessary qualitative and quantitative data. Kothari (2007: 19) names the main ones, which is observation, personal interviews, telephone interviews, questionnaires, and schedules. These are not every tool that exists within the research universe – they rather work as the first level of both qualitative and quantitative research tools that can further be broken down into two digit numbers. Each of the tools could be beneficial to use to answer the research questions, as some of the data is based upon history. As there exist numerous set of tools, the research paper further differentiate the tools that is to be used within the qualitative data analysis, and the quantitative data analysis. To make the understanding of needed tools easier, table 3.1 is constructed to consider the relevance of quantitative data and qualitative data to the research questions.

| # | Research Questions | Qualitative or Quantitative |
|---|---|---|
| **1** | What is eSports, and how aware are people of the eSport culture | Qualitative<br>*Quantitative* |
| **2** | What are the major milestones of eSports history | *Qualitative* |
| **3** | Why are eSport seeing its growth today, and which factors/patterns is pushing it upwards | *Quantitative*<br>Qualitative |
| **4** | Will the growth of eSport continue, and what can the eSport business do to ensure a continuous growth? | *Quantitative*<br>*Qualitative* |
| **5** | Can Information Technology Management Tools be used beneficially for the field of eSport? | *Qualitative* |
| **6** | Which Information Technology Management Tool would be most beneficial for the eSport business to use? | Qualitative |

Table 3.1 Research questions relevance to qualitative and quantitative data

### 3.4.1 Qualitative Data

The research aims to both predict the financial future for eSports and defining it, and will therefore depend profoundly on qualitative data, as shown in Table 3.1. Auerbanch and Silverstein (2003: 4) further breaks down the tools and techniques that can be used within qualitative data, and it is these tools and definitions that will be further discussed within the research thesis:

- Participant observation
- Fieldwork
- Ethnography

- Unstructured interviews
- Life histories
- Textual analysis
- Discourse analysis
- Critical cultural history

Auerbanch and Silverstein (2003) elucidates that this is again only a brief list of the tools that exist within the qualitative data universe, but these are the mainly used ones. Fieldwork, ethnography and critical cultural history could be beneficial within a broad aspect, but due of the research limitations, they are not considered. Interviews on the other hand would be greatly beneficial to gather the qualitative data, as there exist some companies that both create the eSports events and successful video games. To gather this data, either semi-structured, structured or unstructured interviews could be done to achieve a great discussion around the research questions within this research thesis. This also includes life histories and participant observation – as they are both elements within a 1-on-1 interview.  A discourse analysis could also be considered, as some data on previous eSports events and historic milestones is needed. However, a discourse analysis is more aimed towards the actual communication and language usage, and is therefore not seen that well correlated with this research thesis (Schegloff 1999). A textual analysis could also be useful to get an understanding of how individuals make sense of eSports, and what it means for them (McKee 2003). However, this does not answer the research questions or the purpose of the thesis very well, and will therefore not be includes. Nevertheless, other quantitative tools can be used to get numerical data rather than a deeper verbal understanding.

In addition to Auerbanch and Silverstein's qualitative data tools, case studies, content analysis and meta-analysis is additionally considered. Using case studies of previous events can give a great understanding of the main historic milestones within the eSports timeline. However, these set of tools can both be considered to collect qualitative and quantitative.

### 3.4.2 Quantitative Data

Alike qualitative data tools, various quantitative tools such as observation and interviews can be done; the difference is that the collected data is based upon numbers rather than text (Kothari 2004). However, within quantitative data tools there are more specific tools meant to collect quantitative data, where online surveys is considered the most used tool (Sharma 2007). Within this research, quantitative data tools could be used to answer research question 1, regarding the overall awareness of eSport. However, as this question look at eSports in a global factor, this could not be done within this research due limitations, as previous introduced in Chapter 1 under point 1.10. To get data that is both reliable and valid in a global aspect would take months, something that was not seen feasible in this research. Considering this, a decision is done that most of the quantitative data in this thesis is taken through secondary sources, to give a more valid answer to the research questions. Nevertheless, as table 3.1 indicates some quantitative data will have relevance for research question 3 and 4, which is further discussed in the chosen tools for this research thesis.

## 3.5 Concluding the Research Onion: Chosen Methodology for this Thesis

Introducing the fourth, fifth and six layers of the research onion is the overall methodology chosen for the research thesis, and its underlying tools. The methodology and its tools differentiate itself from the research data and its approaches, as it defines how the data is collected. As stated previous, a qualitative vs quantitative approach is used within this research paper, and it is these approaches that is deliberated when the research instruments are introduced.

### 3.5.1 Qualitative Research Instruments

As table 3.1 indicate, the qualitative research will both be important in the primary and secondary data and it is therefore crucial to deliberate its research instruments. With the previous discussion in mind, the subsequent tools will be set as ideal tools within this thesis.

#### *3.5.1.1 Semi-Structured interviews*

As briefly discussed, interviews is one of the most used collection tools within the field of qualitative data. Kothari (2007: 98-100) discuss this within his book, where he differentiate the sort of interviews that can be made. Firstly, there is personal interviews, which is further broken down into focussed interviews, clinical interviews or non-directive interviews. Secondly, there is telephone interview; this includes not meeting the person itself, but rather using a phone or another communication service such as skype. The interview can further be structured, unstructured or semi-structured, meaning that the questions can be strictly followed and have a systematic approach, or be more orderly – leaving the communication to be more openly and natural (Kitchin and Tate, 2000)

The purpose of the interviews is to get qualitative data that will help to critical analyse all the objectives within this research. The data gathered will consist of insightful information that regards the operation of eSports business, as well about its future. It is with this reasoning a semi-structured approach has been chosen, which is the most common interview technique used within qualitative methods (Kitchin and Tate, 2000: 213). Wengraf (2001: 4-5) elucidates that a semi-structured interview includes designing questions alike a structured interview, but these questions must be designed to be sufficiently open as the response of the interviewer cannot be planned in advanced, and therefore addition questions must be improvised in a careful and theorised way. This leads to a more openly interview, which is essential to gather information for the research questions within the research. To achieve this, the semi-structured interviews require much preparation, more discipline and creativity during the interviews, and more time for analysis and interpretation than the other interview techniques (Wengraf 2005: 5). Additionally, both personal interviews set out to be telephone interviews, due few adequate persons near the researcher's current location. The techniques used will further be based upon participant observation, life history and the understanding of eSports overall.

##### 3.5.1.1.1 Interview Representatives

Considering the necessary data that is needed to answer the research questions, only a handful of people and companies might have the required knowledge to answer the upcoming questions. The well-suited respondents will therefore be titled as first tire participants, while those who might not answer the questions in full extent will be titled second tire participants. These representatives is further shown and discussed in table 3.4

#### *3.5.1.2 Observation*

To get an understanding of the past of eSports, and its underlying factors, a set of qualitative data will be collected. This data itself could be understood as observation of case studies, as Kothari (2004: 113) defines a case study as a popular qualitative analysis that place emphasis on the full analysis of a limited number of events or conditions and their interrelations. This correlate with Gillham's (2008: 1) definition, where he underline that a case study is used to investigate a specific case and get a deeper understanding of it so new knowledge can be created. Gillham (2008: 1-2) further specifies that a case could exist in many variations: it could be an individual, a group, an institution, a community, a town, an industry, and the list goes on. These observations will include both qualitative data about historic eSport events, and quantitative data such as prize pools and number of attendees. This is possible by using an analysis named thematic analysis.

##### 3.5.1.2.1 Thematic Analysis

An observation of case studies might be referred to as content analysis, which is defined as systematic qualitative analysis that brings out generic information of the data analysis (Marks and Yardley 2003: 56-58). However, a more specific result is needed and therefore a thematic analysis can be conducted. A Thematic analysis does not only go through the content, but it specifically looks for patterns that is predefined as themes. It further uses a technique named coding, which is a process that analyses the raw data and identify its implicit and explicit contents (Boyatzis 1998). The thematic analysis strives to both give businesses actionable insight, performance measurements and value measurements, as well as future insight based upon its themes, which makes it highly applicable in this thesis (Stubbs 2013). These themes can however be based on both qualitative and quantitative data, and is therefore suitable to be named both although it is considered as a qualitative technique. To understand the patterns found in the thematic analysis, various analytical tools will be used to give a more effective understanding to the reader. This kind of analysis is named business analytics, which is further discussed.

##### 3.5.1.2.2 Business Analytics

To understand what business analytics is, a definition of analytics as a whole is needed. Cooper (2012) define analytics as "[…] the process of developing actionable insights through problem definition and the application of statistical models and analysis against existing and/or simulated future data”.  These definitions is still too extensive to define the analytics to the thesis aims and objectives, as there are different subgroups of analytics. Stubbs (2013) further explain that within analytics there are some common examples such as; reporting which concludes a summarization of historical data; trending where time series of data is identified; and segmentation where the similarities of the data is identified and weighted of each other. These examples are mainly based on data, or various mathematical techniques that transform the data and add value, so it might be transformed to information for the company, which again can give them knowledge. Stubbs (2013) further defines the subsequent business analytic tools as either advanced analytics, business analytics or visual analytics. Advanced analytics can be understood as the form of analytics constructed to predict what will happen in the future. It bases itself on advanced statistical and mathematical techniques, such as algorithmically based predictive models or nonparametric statistics (Stubbs 2013). Visual Analytics is rather a more basic set of analytics, which Keim et al. (2008) defines as a “combined, automated analyses technique with the interactive visualization for an effect. Business analytics however, is something that strives to achieve various business outcomes, such as growth.

Both advanced analytics, business analytics and visual analytics has beneficial use within this study. Nevertheless, the research thesis will further refer these tools together as just business analytics, which is the umbrella term for this kind of analysis. This can be seen in Garner’s (2005) definition of business analytics, which is “[Business Analytics is] comprised of solutions used to build analysis models and simulations to create scenarios, understand realities and predict future states”. It is with this definition the research thesis wish to learn and create scenarios, so an understanding of the main themes of eSports can be found.

One example of business analytics could be understood as a trend line, which is a visualisation of the certain trends in the collected data. It can further be defined as done by the business dictionary (2015b) as a “Straight or curved line in a trend chart that indicates the general pattern or direction of a time series data (information in sequence over time). It may be drawn visually by connecting the actual data points or (more frequently) by using statistical techniques such as 'exponential smoothing' or 'moving averages.' This is one of the chosen tools existing under the umbrella term of business analytics, and is used in this thesis to get an indication of the future of eSports. This is further discussed and introduced in 5.4.5.

### 3.5.2 Quantitative Research Instruments

The quantitative data used as the first step within the induction step in mind, can be understood as observation, which further creates a pattern. The instruments to accomplish this is often referred to as a survey, sampling, observation, modelling or correlation tests (Kothari 2004: 33-40). As stated in 3.2.2, the quantitative data is gathered by a secondary source, due of the time limitations. This source is mostly Newzoo`s (2015) annual report of the eSports growth, launched in February 2015. However, this data is critically analysed with relevant quantitative data that might correlate. The annual report include surveys, samples, observation and models of the overall awareness of eSport, and defines viewing segments. However, to use business analytics in the best possible way, quantitative data will additionally be gathered in the thematic analysis, shown in chapter 4. Although this is not seen as a quantitative research instrument, it will include numerical findings that further can be statistical analysed.

## 3.6 Primary and Secondary Data within this Study

With the instruments in mind, a further discussion has to be conducted to get an understanding of these tools will be used on secondary or primary data. Table 3.2 further shows a summary of the expected primary and secondary data based upon the methodology:

| Instrument/Tool | Definition/Summary |
|---|---|
| *Semi-structured Interviews* | The semi-structured interviews will include only primary data, as all the information gathered from this instrument is solely collected for this research. |
| *Thematic Analysis* | The thematic analysis will be based upon data gathered by either other persons or the company themselves. As this is considered as secondary data, it effects the validity of the analysis tremendously. This is further discussed under 3.10 limitations. |
| *Business Analytics* | The data gathered in the thematic analysis will be used to get statistical data, which help predicting the future of eSports by implementing Business analytics tools, together with ITM tools. This in an essence can be understood as observation, as the data is observed to get both an understanding of the history of eSports, and find a common pattern that can help understanding the future. However, although the creation of new patterns can be seen as primary data, it is still based on the data collected in the thematic analysis, and is therefore considered as secondary data. |
| *ITM Tools* | Secondary data will be used to get an understanding of the outcome of implementing ITM tools to sustain the growth eSports is partaking. In addition to this, case studies will be used to see similar scenarios where ITM tools are implemented. |

Table 3.2 Primary and secondary data within the research thesis

## 3.7 Credibility, Reliability and Validity

When estimating how good the collected data is, different measurements can be used. The most common ones are reliability and validity, which together creates the credibility of the study (Bashir et al. 2008). Joppe (2000) defines reliability as "*The extent to which results are consistent over time and an accurate representation of the total population under study*". It is important to estimate the reliability when evaluating quantitative research, at it is mostly based upon numbers (Bashir et al. 2008). Validity can however be understood as "whether the research truly measures that which it was intended to measure or how truthful the research results are" (Joppe 2000).  To ensure an understanding of this research's reliability and validity, it is important to discuss these terms. This is done by following Patton's (2001) three questions to ensure credibility of the research, which is further shown in table 3.3.

| Patton's (2001) questions concerning the thesis credibility | What can be done in this study to ensure credibility |
|---|---|
| **What techniques and methods were used to ensure the integrity, validity and accuracy of the findings?** | • Piloting the studies (4.5 Piloting the data)<br>• Measuring the collect data to the literature review (6.2.1-6.2.6 Conclusion of Research Question 1-6)<br>• Discussion of flaws and strength in the thematic analysis (5.2 Credibility of the thematic analysis and (7.2 Critical evaluation of project conduct) |
| **What does the researcher brings to the study in terms of experience and qualification?** | • A vast experience of the eSport community from a consumers point of view<br>• Experience from working within the gaming community |
| **What assumptions undergrid the study?** | • That eSports is seeing a undergoing growth in consumption<br>• That there exist various trends in the eSport events conducted since it first began in 1980 |

Table 3.3 what have been done to increase the credibility based upon Patton's (2001) questions

Whether the data collected is reliable or not is further discussed in the later chapters. However, following Patton's (2001) three questions gives an indication that the data collectable show signs of being credible.

## 3.8 Ethical Procedures

As this study is based upon a global aspect, there was some ethical consideration to partake. However, every theme or questions made were very specifically made to be as generic as possible, so no ethical issues would arise. To ensure no ethical problematic would occur, an ethical approval form were constructed and approved at the 4th of June at medium risk (Appendix G III). The ethical approval also includes a Participate Information Sheet as well as an Information Consent Form for the interview representative, which is further discussed in chapter 7 (Appendix G I-II).

## 3.9 Limitation of methodology

Considering the methodology, two limitations affected the research thesis, which further caused implications on the subsequent chapters. Due of the size of these implications, they are discussed individually in the subsequent points.

### 3.9.1 Limitations of the Thematic Analysis

As briefly mentioned previously, the thematic analysis is purely based upon data that is collected by others (secondary data). This means that this research can neither take credit on the data, nor ensure that the data is correct. However, the analysis is based on over 400 sources of information, making it as valid as it could be considering that it is based only on secondary references. The findings are therefore referred to the conducted references list, which is found in Appendix B. To further increase the validity of this thesis as a whole, the interviews were to correlate their answers with the collected data, to see if they answered the same. This would have proved somewhat how well the thematic analysis answered the research questions, as the interviews findings would have been based on primary data. However, this was not deemed possible due a lack of respondents, which is further discussed under 3.9.2.

#### *3.9.1.1 Thematic Analysis: Streaming audience*

During the analysis a lack of data were found under the theme streaming audience. This is mainly due the streaming service is still new, and not many companies or sites share their viewing numbers. It is with this reasoning that the business analysis choose to concentrate the streaming data on one specific theme, which is the unique viewership per event. This is further discussed in chapter 5.

#### *3.9.1.2 Thematic Analysis: Excluding data from 2015*

In the thematic analysis, little of the data from 2015 will be used although it is analysed. It would be unrealistic to include the findings from 2015, as some of the main events has not begun, as well as the analysis bases upon a year-to-year basis. However, the data will be concluded if it is found useful in the sense of average values and discussion, but this will always be mentioned if done.

### 3.9.2 Limitations of the Semi-structured Interview

As shown in Appendix I, the questions formulated for this thesis are very specific, making the possible candidates quite few. However, a total of 25 different companies and individuals were asked to attend. These potential representatives were contacted firstly by telephone if the number was available. If this was not the case, their preferred contact form were used. During the thesis production, a total of two months were used trying to reach out to various persons, starting in June when the project achieved an ethical approval. However, as shown in table 3.4, none of the persons that could bring the desired information was available or responded. Three people firstly agreed to attend an interview, but the contact was suddenly lost and no email or telephones were responded to, making it impossible to conduct the semi-structured interviews within the time limit. This limits the validity of the findings in this research paper. However, this is compensated for by an increased amount of events and references used in the thematic analysis, so as to help improve the validity of this thesis, although this data is secondary.

| | *Company / Contact person* | *Contacted* | *Responded* | *No* | *Maybe* | *Yes* | *Contact form* |
|---|---|---|---|---|---|---|---|
| | **Tire 1 Interviews** | | | | | | |
| 1 | Turtle Entertainment | Y | N | X | | | *email* |
| 2 | Riot Games | Y | N | | | | *email* |
| 3 | Fnatic | Y | Y | X | | | email |
| 4 | Blizzard Entertainment | Y | N | | | | *email* |
| 5 | Dreamhack | Y | Y | X | | | email |
| 6 | QuakeCon | Y | N | | | | *Contact form and email* |
| 7 | Valve (Dota 2 team) | Y | N | | | | *email* |
| | **Tire 2 Interviews** | | | | | | |
| 8 | SK Gaming | Y | N | | | | *Twitter and email* |
| 9 | TSM (Community) | Y | N | | | | *email* |
| 10 | TSM (Website) | Y | N | | | | *contact form* |
| 11 | CLG | Y | N | | | | *e-mail* |
| 12 | Team Dignitas | Y | N | | | | *Facebook* |
| 13 | Ram “Broken Shard” Djemal | Y | Y | | | X (*) | Facebook and email |
| 14 | Xarly “Ocelote” Santiago | Y | N | | | | Facebook |
| 15 | Gamers 2 | Y | N | | | | *email* |
| 16 | Red Bull eSports | Y | Y | X | | | telephone and email |
| 17 | Jesse Cox | Y | N | | | | *email* |
| 18 | Copenhagen Wolves | Y | N | | | | *contact form* |
| 19 | Mouseesports | Y | Y | | | X (*) | *contact form and email* |
| 20 | Reason Gaming | Y | N | | | | *contact form* |
| 21 | Fusion Gaming | Y | N | | | | *contact form* |
| 22 | Cloud 9 | Y | N | | | | *email* |
| 23 | Team Liquid | Y | N | | | | *email* |
| 24 | GGBeyond | Y | N | | | | *YouTube contact message* |
| 25 | Esportearnings.com | Y | N | | | | *email* |

(*) Initially stated interested but did not reply to the second email regarding information about the project

Table 3.4 Contacted companies for the research paper

## 3.10 Summary of Chapter 3

Chapter 3 includes a discussion of which methodologies is best suited for this research thesis. It concludes that this thesis will use a pragmatism philosophy with an inducation approach, which correlates greatly with research questions. Qualitative data will have a major role within this research thesis, although some quantitative data will also be used. The instruments used within this study, and its correlations with objectives and research questions, is further summarised in table 3.5. The research instruments mentioned is additionally sorted based upon its relevance.

| # | **Research Questions** | **Collation Tools** | **Qualitative or Quantitative** |
|---|---|---|---|
| *1* | What is eSports, and how aware are people of the eSport culture | Existing literature<br>Thematic Analysis<br>Semi-structured interviews | Both |
| *2* | What are the major milestones of eSports history | Thematic Analysis<br>Semi-structured interviews | Qualitative |
| *3* | Why are eSport seeing its growth today, and which factors/patterns is pushing it upwards | Thematic Analysis<br>Business Analytics<br>Semi-structured interviews | Both |
| *4* | Will the growth of eSport continue, and what can the eSport business do to ensure a continuous growth? | Thematic Analysis<br>Business Analytics<br>Semi-structured interviews | Both |
| *5* | Can Information Technology Management Tools be used beneficially for the field of eSport? | Business Analytics (of the thematic analysis)<br>Existing literature | Qualitative |
| *6* | What outcomes would the most beneficial Information Technology Management Tool give in the eSport business? | Existing literature<br>Business Analytics (of the thematic analysis) | Qualitative |

Table 3.5 Summary: Collection tools for each research question

Due of limitations, none semi-structured interviews will be conducted and further analysed, and therefore the interviews is coloured red in the previous mentioned table. The thesis will continuously use the mentioned methods to hold both a high credibility, and a high ethical formality. This is done by using the various methods shown in this chapter, and is further evaluated in chapter 7.

# CHAPTER 4 INTERPRETATIONS

## 4.1 Introduction Chapter 4

The previous chapter discussed the methodology of the thesis, including an introduction to the proposed research tools and instruments that is chosen to answer the research questions within this thesis. This chapter will further introduce the chosen instruments, in the sense of interpretation, and further define how the tools will be used towards collecting the data. As no semi-structured interviews were conducted due of the limitation of the methodology, this chapter will mainly focus on the thematic analysis and its subsequent business analytic tools used. However, the interview guide along with the information of the interviews is further attached in appendix I, although it is not analysed**.** Furthermore, this chapter will additionally include the piloting of the data, which is an efficient tool used to ensure the quality of the data collection. Nevertheless, to fully explain the interpretations, an explanations of the software used in this thesis must first be given, which is discussed as the first point in this chapter.

## 4.2 Software Used in the Data Collection

To perform the thematic analysis, as well as the business analysis, two different software were used. The first software is SPSS, (Statistical Package of the Social Science), which is a statistical software devolved by IBM (IBM 2015). Firstly, SPSS is used in this research to accomplish analytical processes, such as finding patterns by looking at descriptive findings and various tests. Secondly, SPSS is used towards constructing mathematical formulations used to accomplish the business analytics. The second software, Microsoft Office Excel, is a spreadsheet software developed by Microsoft (Office 2015). It is used in this research as a platform to store all the collected data. It is also used to visualise the findings in various graphs, as well as performing the trend analysis. Each software were chosen based upon its quality and professional recognition to gain the best possible results for this research thesis.

## 4.3 Thematic Analysis

The thematic analysis includes various themes found in the literature review, which is further focus on finding trends, patterns and overall historic milestones within the consisting data of

eSport events. The thematic analysis data, which were further used to perform business analytics, is attached as Appendix A. Each themes used within the thematic analysis is defined in the subsequent points, before table 4.2 summaries each theme's relevance to the research questions.

### 4.3.1 Year

The titled 'year' is created around all the research questions, so an understanding of the patterns in various years could be seen. By having this as an own theme, it becomes possible to analyse trends in specific years.

### 4.3.2 Event (Name)

The theme titled 'event' is created to get an overview of the name of the event, so it may be referred to in the analysis. It is therefore somewhat important to research question 2 and 4, as the name of the event might be of interest.

### 4.3.3 Series

The theme titled 'series' is created to find out the trend is specific leagues, tournaments or series throughout eSport events. This will further help to get an understanding on research question 2, 3 and 4.

### 4.3.4 Where

The theme titled 'where' is created to get an understanding of where the eSports events is normally held. This further helps to get an understanding of which places that can be named as the capital of eSports history, which will be beneficial to answer research question 2 and 3.

### 4.3.5 Participants

The themes titled 'participants' is used to get an understanding of how many professionally players attend to different events. It is further broken down to how many participants attended in total, how many finalists attended and how many different countries attended. This data will be used to mainly answer research question 2, 3 and 4, as it will be used to see if it is any correlation between how many participating and the success of an eSports event.

### 4.3.6 Continent? (NA, EU, Global or Asia?)

The theme titled 'continent' is used to see where the event is held, so an understanding of the most common places to construct eSport events is gained. The data will be used to answer research question 2, 3 and 4.

### 4.3.7 Audience

The themes titled 'Audience' is used to get an understanding of the size of the events, in the sense of how many people watched it. This was done by collecting numbers of the live audience, as well as how many people watched it virtually through streaming. To get a more realistic understanding of the data, the audience were broken into four themes: Live audience, Streaming (total), Streaming (Unique viewers), and Streaming (Peak concurrent viewers). The audience numbers will have a high influence on all the research questions.

#### *4.3.7.1 Live Audience*

The theme titled 'Live Audience' is used to get an understanding of how many people attended to various physically.

#### *4.3.7.2 Streaming Audience*

Alike the 'Live Audience' theme, the 'Streaming' theme is used to get an understanding of how many people watched the events in total. Streaming however, as written in the glossary and the literature review, is not an easily defined word. To get a more realistic understanding of the streaming numbers, three different measurements were used. These three measurements together answer all the research questions.

- The theme titled 'streaming (total views)' is used to get an understanding of how many virtually views in total the event got.

- The theme titled 'Unique viewers' is used to get an understanding of how many individuals watched the event virtually; this is measured on how many different IP addresses that was the event virtually.
- The theme titled 'Peak concurrent viewers' is how many different IP addresses watched the event virtually at a specific time which defines the peak.

### 4.3.8 Prize pool

The theme titled 'prize pool' is used to get an understanding of how much money was awarded to the attending players throughout the event. As it is said to be a very important theme in the literature review, it is deemed important for all the research questions. The currency used is American Dollars ($), due it was the most common currency used in the secondary data.

### 4.3.9 Games

The themes titled 'Game', 'Total Number of Games', 'Games Genre' 'Single player or Multi Player' are four different themes made to get a understanding of what games were played at various events, as well as how many games were played and its name and genre. This was used to get an understanding of how the usage of games has changed over the years, and which games that might be used in eSports in the future. Therefore, this theme includes data for all the research questions.

### 4.3.10 Duration - Length (Months)

The themed titled 'Length' is used to get an understanding of how long the eSport events is, and how this can have an impact on future eSports events. This data will therefore include information regarding research question 2, 3 and 4.

### 4.4.11 Deleted Themes

In addition to these themes, various others were consider as can be seen in the excel document attached in Appendix A. These themes were not used in the analysis due either lack of content, validity, reliability or not responding efficiently to the research questions. The following themes were removed:

- Television Coverage
    - Lack of data
- Paper Coverage
    - Lack of data
    - Lack of validity
- Streaming: Sessions
    - Lack of data
- Extraordinary events and Extraordinary events (Yes or No)
    - Lack of validity
    - Lack of data
- Sponsors
    - Lack of relevance

### 4.4.12 Relevance of Themes

Table 4.1 further discuss the relevance for each theme within the four first research questions of this thesis. Although the findings is used to decide research question 5 and 6, it is rather the analytical findings in the thematic analysis that is used towards answering the last two research questions. These questions is therefore not included in the following table.

| **Research Question** | **Themes with none importance for the research question** | **Themes with some importance for the research question** | **Themes with great importance for the research question** |
|---|---|---|---|
| **<u>Research Question 1?</u>**<br>**What is eSports, and how aware are people of the eSport culture?** | • Event (Name)<br>• Where<br>• Series<br>• Games<br>• Prize pool<br>• Duration<br>• Continent | • Year<br>• Participants<br>• Audience<br>• Streaming viewership | |
| **Research Question 2:**<br>**What are the major milestones of eSports history** | | • Event (Name)<br>• Series<br>• Participants<br>• Continent<br>• Games<br>• Duration | • Year<br>• Where<br>• Audience<br>• Streaming viewership<br>• Prize pool |
| **Research Question 3:**<br>**Why are eSport seeing its growth today, and which factors/patterns is pushing it upwards** | • Event (Name)<br>• Series | • Year<br>• Where<br>• Continent Games<br>• Duration | • Participants<br>• Audience<br>• Streaming viewership<br>• Prize pool |
| **Research Question 4:**<br>**Will the growth of eSport continue, and what can the eSport business do to ensure a continuous growth** | | • Where<br>• Event (Name)<br>• Series<br>• Continent<br>• Duration | • Year<br>• Audience<br>• Streaming viewership<br>• Prize pool<br>• Participants<br>• Games |

Table 4.1 Relevance of themes in the thematic analysis

## 4.4 Primary and Secondary Sources within the Thematic Analysis

To get an indication of the past and the future of eSports, data was collected on 444 eSports events from its origin to today (1980-2015). As stated in the methodology, this data were mainly collected from secondary sources, but also some primary sources were used such as well. This should not be misunderstood as primary data however; primary sources in this sense means that the data was taken from the company or the event itself, while secondary sources is collected through another company or person who claim to have the correct data. The collected data as a whole consist of over 1,000 references, and is therefore given an own reference list to give a better overview for the reader, which can be seen in appendix B. As stated in the limitations, all the data is gathered from various sources, making the results based upon their findings. In the sources used, there were some more commonly used that others, which is further in table 4.2

| | |
|---|---|
| **Primary sources frequently used** | • Major League Gaming (MLG)<br>• Intel Extreme Masters (IEM)<br>• Electronic Sports League (ESL)<br>• Cyber athlete Professional League (CPL)<br>• World Championship Series (WCS)<br>• BlizzCon<br>• QuakeCon<br>• DreamHack |
| **Secondary sources frequently used** | • E-Sports Earnings |

Table 4.2 Sources frequently used in the thematic analysis

### 4.4.1 Raw Data: Choosing the Events

When considering which events to include in the analysis, some specific qualifications had to be met. The number of eSports events that is held or being is held is immeasurable, and including every event in the analysis is just not feasible. The amount of events chosen in the thematic analysis is therefore shortened to 444 events. The events is chosen by an extraordinary number of audience, prize pool, or by being an extraordinary event in the sense of including other forms over entertainment. These 444 events is seen as the most successful eSport events in history, based on the findings.

## 4.5 Piloting of the Data

Within this study, two pilot studies were conducted; one for the case study data collection, and one for the semi-structured interview. Tejlingen and Hundly (2001), defines a pilot study in two various way, it can either be different trial runs done to ensure that everything will be understood during the interview, or it can be to understand the feasibility of the study. For this research, pilot studies were conducted to ensure the quality of the data collection and making sure that no questions could be misunderstood. The pilot interview were completed on several fellow students at Coventry University, which can be seen as undesirable as the student may lack an understanding of some of the questions. However, the pilot concerns if the questions were understood or not – not the actual content of the answer – and therefore this is not seen as a major problem.  As for the thematic analysis, different business analytic tools were used to get an understanding if the collected data is measurable, and if it is possible to get the wanted result. The results of the piloting can be viewed under Appendix E I (Thematic analysis) and Appendix E II (Semi-structured interview).

### 4.5.1 Pilot studies: Thematic Analysis and Business Analysis (Appendix E I)

The pilot of the case studies performed well. Although the data were not fully collected, it gave some succinct figures on how the result might look alike. However, one problem that will influence the data collection were found throughout the pilot, which is that some of the graphs will consist of to many entries. This is however resolvable by removing some of the less used entries, to get a more befitting and understanding figures, which can further be analysed. An example of this is under the theme 'games', where some games were only used once throughout the analysis, and is therefore not deemed important for the analysis. These entries will be chosen after the data collection, and are further discussed under point 5.4.1 Outliners.

### 4.5.2 Pilot studies: Semi-structured Interviews (Appendix E II)

The pilot of the semi-structured interviews performed well, and most of the questions were understood by the persons participating. However, some changes had to be done based upon the findings in appendix E II, as there were some confusion in my questions:

- Question 1a had to be redefined into several questions, as it was confusing for the respondent to answer several questions at once.
- Some questions, such as question 3c, consisted of too much information, making it hard to answer if not prepared for the questions. It will therefore be important to specify that these questions need to be prepared to the respondent. before the interview takes part,
- The pilot included some findings of what the respondent might answer, which will be beneficial for the actually interview, as it is a semi-structured interview and additional questions will be asked from the researcher.

## 4.6 Summary of Chapter 4

Chapter 4 includes the interpretations of the chosen instruments in this research thesis. The chapter concludes that the chosen software's for this study will include Microsoft Excel and SPSS, which will perform the business analysis of the thematic analysis. As for the instrumental tools used in the study, the semi-structure interviews is not included, but its relevant data is attached in appendix I. The thematic analysis however, will further consist of 11 themes: Year, Event, Series, Participants, Continent, Audience, Prize Pool, Games, Games Genre and Duration. These themes further hold sub themes, where live audience, streaming audience, single player or multi player and game genre is deemed the most important. To summarise the expected data collection, table 4.3 has been created, which outlines the data collection expectation for each research question. To ensure that these expectations is met, the interpretation chapter perform a set of pilots of the instruments, to ensure that the collected data is what it is set out too. Both pilots performed well, and gave several indications, which is taken further in the data collection chapter.

| **Research Questions** | **Expected findings in the thematic analysis** | **Themes** |
|---|---|---|
| Research Question 1:<br>What is eSports, and how aware are people of the eSport culture? | • How many attend to eSport events (both participants and audience)<br>• Where is the eSport events held | • Where<br>• Participants<br>• Audience<br>• Streaming Audience |
| Research Question 2:<br>What are the major milestones of eSports history | • Which event is seen as the biggest events in eSports history, based upon:<br>  ○ Audience<br>  ○ Participants<br>  ○ Prize pool<br>• Why are these events deemed important? | • Year<br>• Event<br>• Series<br>• Where<br>• Participants<br>• Continent<br>• Audience<br>• Streaming Audience<br>• Prize pool<br>• Games<br>• Duration |
| <u>Research Question 3:</u><br>Why are eSport seeing its growth today, and which factors/patterns is pushing it upwards | • How much growth has there been since eSports origins in the 1980s?<br>• What seems to be the causes of growth of successful events? | • Year<br>• Event<br>• Series<br>• Where<br>• Participants<br>• Continent<br>• Audience<br>• Streaming Audience<br>• Prize pool<br>• Games<br>• Duration |
| <u>Research Question 4:</u><br>Will the growth of eSport continue, and what can the eSport business do to ensure a continuous growth? | • How does the future of the various themes look based upon the collected data?<br>• Which themes seems to be the most influential factors for the future? | • Year<br>• Event<br>• Series<br>• Where<br>• Participants<br>• Continent<br>• Audience<br>• Streaming Audience<br>• Prize pool<br>• Games<br>• Duration |

Table 4.3 Expected findings for the data collection

# CHAPTER 5 DATA COLLECTION AND ANALYSIS

## 5.1 Chapter 5 Introduction

The previous chapter gave an interpretation of how the data will be collected within this research thesis, and what it will consist of. It further concluded with the expected findings of the thematic analysis, which can be understood as an introduction to this chapter. This chapter will firstly focus on the collection of the required data previously stated. To do so, a presentation of the findings will be made. This is followed by an analysis, which state the relevance of the findings towards the research questions. Subsequently, the specific research questions will be answered by the newly collected data, and conclude which factors that have a significant influence on eSport events, as previously done with the collected literature in chapter 2. However, before the presentation of the data collection can be done, a discussion needs to be held in regards of the credibility of the collected data.

## 5.2 Credibility of the Thematic Analysis

According to Bashir et al. (2008), the overall credibility can be understood in various ways. However, the reliability and validity stands as the most common terminologies within this field. Joppe (2000) argues that in addition to this, the data collection should also considered if the data collection gather the expected data, as well as the overall truthfulness of it. For this research thesis, a pilot were done for both the interview and the thematic analysis, which made sure that the expected findings would be gathered through the chosen instruments, as presented in the previous chapter. As for the overall truthfulness, this is as declared before somewhat lacking. Even though the thematic analysis consists of over 1,000 references, it is still mostly websites and blogs from primary and secondary sources. This means that not every reference is collected from the companies themselves (primary sources), which reduces the validity heavily. As intentioned, the analysis of the interviews would be measured against the findings in the thematic analysis to increase the validity, but this was not possible due of the limitations of the study as stated in the methodology chapter. That is why so many events is included in the analysis, and each event chosen has approximately 2-4 references each, which increases the truthfulness in the numbers – even though it is not optimal, it is the best data available for this research. An addition flaw in the thematic analysis is that some data is missing from some of the themes, as either the needed data could not be found nor the time did not limit itself to collect it, and this affect the credibility. The overall credibility of the collected data is therefore further summarised in table 5.1, which is based upon how many events include the specific data in the theme, divided on the total number events.

| Theme | Data Included (%) | Credibility summary |
|---|---|---|
| Year | 444 (100%) | The data collected to the theme 'year' is included on every event, and is therefore not losing any credibility due of lack of data. |
| Series | 427(96%). | The data collected to the theme 'Series' is nearly included on every event, and is therefore not losing any credibility due of lack of data. |
| Where | 435 (98%) | The data collected to the theme 'Series' is nearly included on every event, and is therefore not losing any credibility due of lack of data. |
| Participants Total | 76 (17%). | The data collected to the theme 'Participants total' is lacking, and is therefore losing credibility in its findings due of lack of data. |
| Participants Finalist | 403 (91%) | The data collected to the theme 'Participants Finalist' is nearly included on every event, and is therefore not losing any credibility due of lack of data |
| Participants countries | 402 (91%) | The data collected to the theme 'Participants countries' is nearly included on every event, and is therefore not losing any credibility due of lack of data. |
| Continent | 427 (96%). | The data collected to the theme 'Continent' is nearly included on every event, and is therefore not losing any credibility due of lack of data. |
| Audience: Live Audience | 33 (7%) | The data collected to the theme 'Live audience' is lacking, and is therefore losing credibility in its findings due of lack of data. |
| Audience: Streaming (Total views) | 28 (6%) | The data collected to the theme 'Streaming (Total Views)' is lacking, and is therefore losing credibility in its findings due of lack of data. |
| Audience: Streaming (Unique views) | 33 (7%) | The data collected to the theme 'Streaming (Total Views)' is lacking, and is therefore losing credibility in its findings due of lack of data. |
| Audience: Streaming (Peak concurrent views) | 18 (4%): | The data collected to the theme 'Streaming (Total Views)' is lacking, and is therefore losing credibility in its findings due of lack of data. |
| Prize Pool | 420 (96%) | The data collected to the theme 'prize pool' is nearly included on every event, and is therefore not losing any credibility due of lack of data |
| Game | 434 (98%) | The data collected to the theme 'Game' is nearly included on every event, and is therefore not losing any credibility due of lack of data. |
| Total number of games | 434 (98%) | The data collected to the theme 'Game' is nearly included on every event, and is therefore not losing any credibility due of lack of data. |
| Team based, single player or both | 421 (97%) | The data collected to the theme 'TB, SP, Both' is nearly included on every event, and is therefore not losing any credibility due of lack of data. |
| Duration | 421 (95%) | The data collected to the theme 'Duration' is nearly included on every event, and is therefore not losing any credibility due of lack of data. |
| Total (Average) | 296 (67%) | In total, 67% of the data is missing which is mostly caused the missing data under the live audience and streaming themes. Nevertheless, this is worth mentioning in further discussions as it decreases the credibility of the thematic analysis. |

Table 5.1 Credibility of the thematic analysis

## 5.3 Data Collection Findings within the Thematic Analysis

Business analytic tools, as explained in the project scope, is further used on the data collected from the thematically analysis to give a better understanding to the reader. Using business analytics tools gives additionally both more useable and visual data. The subsequent points is the finding for each theme within this study. However, to gain a better understanding of the various events, most analytics excluded information from the year 2015, as briefly mentioned in the limitations. This is due that the year is only halfway during the writing point of this thesis, so the data collected cannot be measured against the previous years as either the total value is not high or the biggest events is yet not completed. However, if seemed reasonable, the 2015 numbers is used to gain an understanding on the futuristic trend. The years used is further stated on either the name of the figures or defined in the description. Additional findings that were not presented, is further found in Appendix C.

### 5.3.1 Theme: Single Player or Team Based

The collected data showed a high frequency of both including team based games and single player games, whereas 47.9% of the games used both. Overall, the single player games had a slightly higher tendency with an overall percent of 26.5% while the team based was at 25.6%. If this is looked in a yearly perspective, the single player trend has existed since eSports first began in 1980, and peaked both in 1999 and 2013. The team based did not come through until the year 2000, but today it is the leading factor in this sense. The visualisation of the data gives a slight indication that team based games might have an importance for the future.

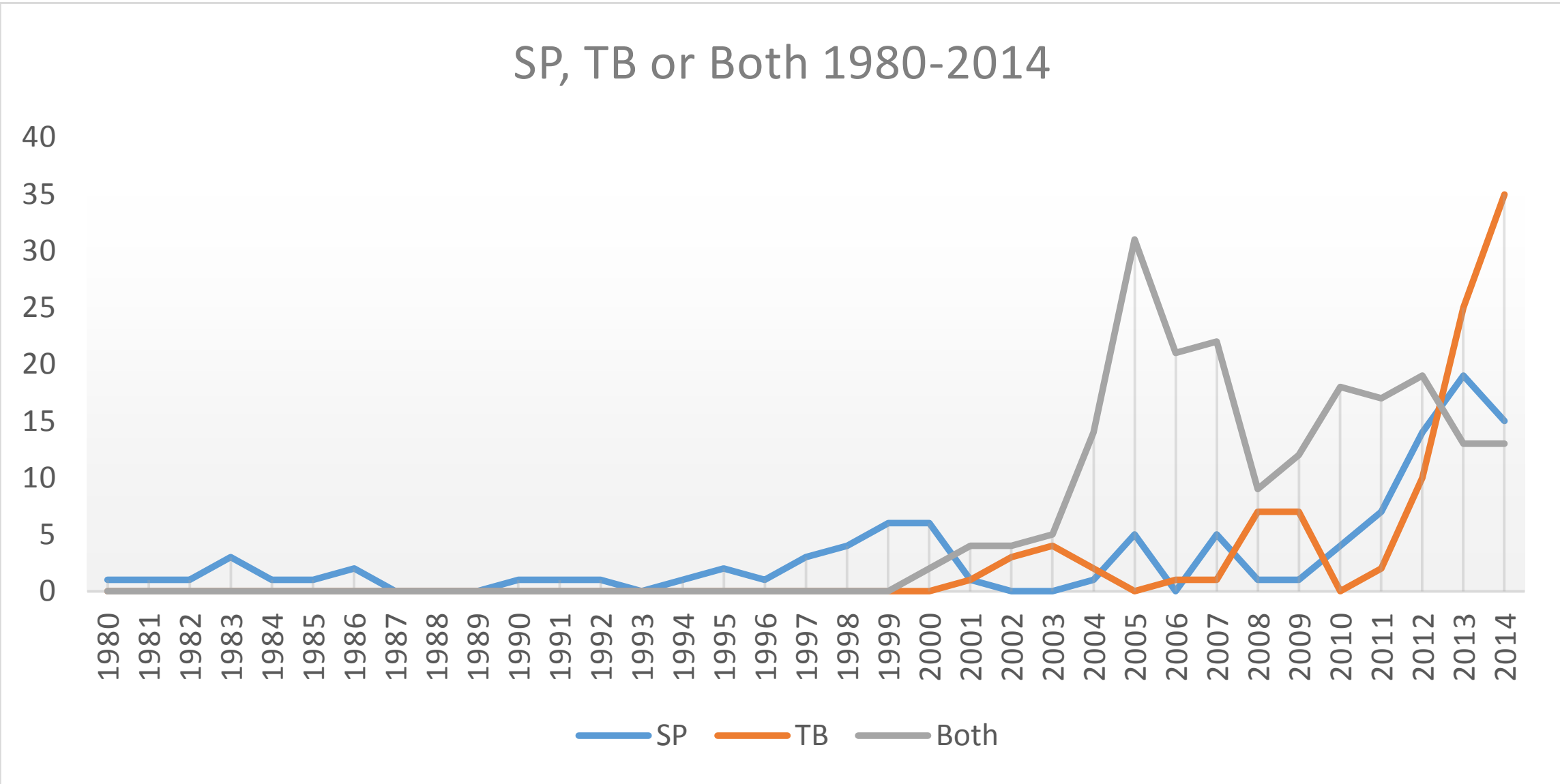

*Figure 5.1 SP, TB or Both 1980-2014 (Appendix B)*

### 5.3.2 Theme: Event Location

The collected data gives an indication of that USA is the capital of eSports, whereas over 40% of the events in this analyses took place in USA. Other countries that were highly used to make eSports events were Korea, China, Sweden, Germany, Malaysia and France. ESports events that were just held online and were only streamed with no live audience was ranked number 5 out of the collected data.

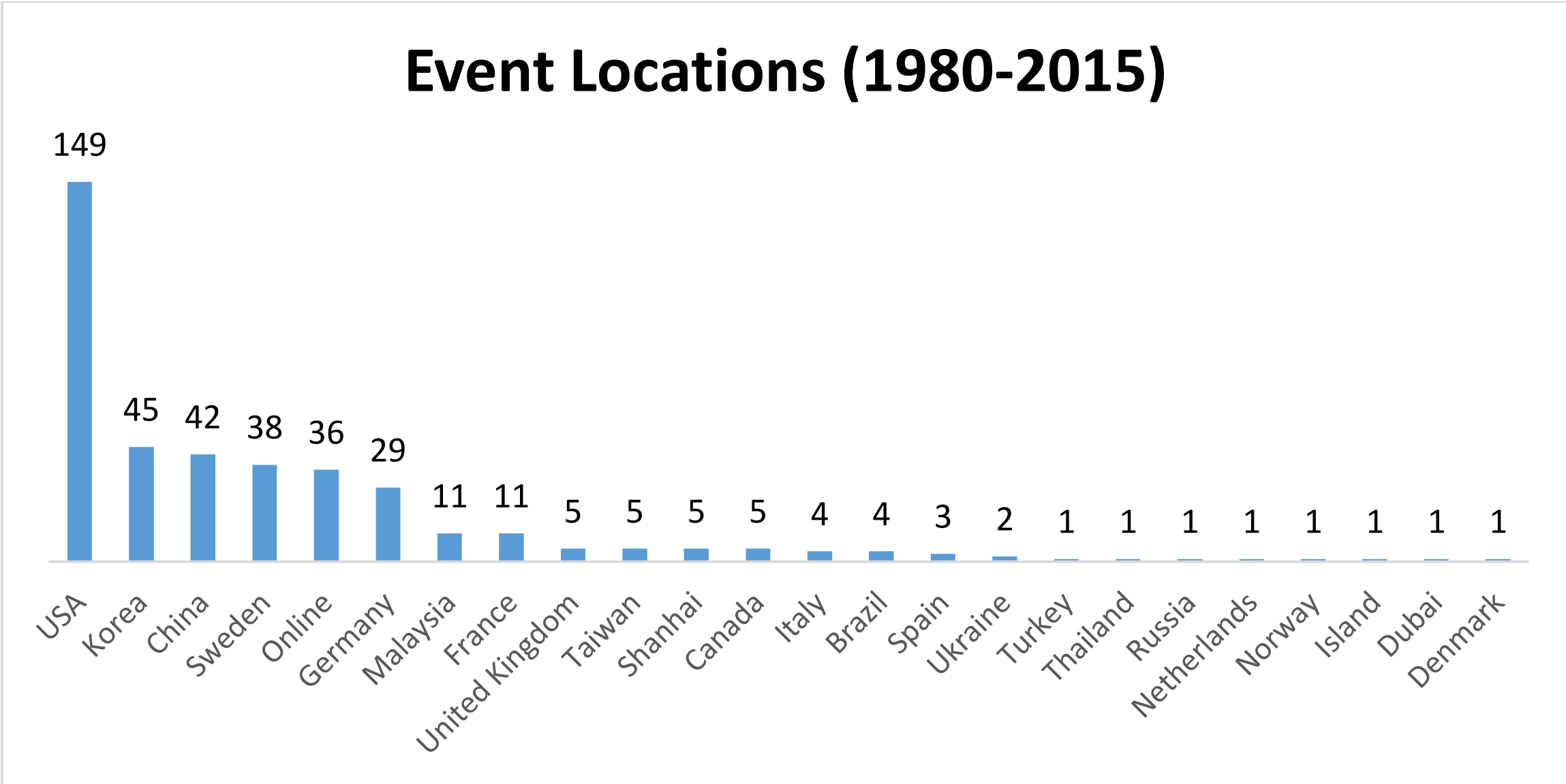

*Figure 5.2 Event Locations 1980-2015 (Appendix B)*

### 5.3.3 Theme: Series/Tournament/Event Name

By analysing the events that were used in the thematic analysis, a large variation on the data was seen. Due to the large variation, the biggest percentages lie within 'others', as there existed numerous hosts. Nevertheless, the data indicates that the biggest events/tournaments/leagues throughout history is MLG (16%), IEM (ESL) (11%), CPL (9%) and DreamHack (9%).

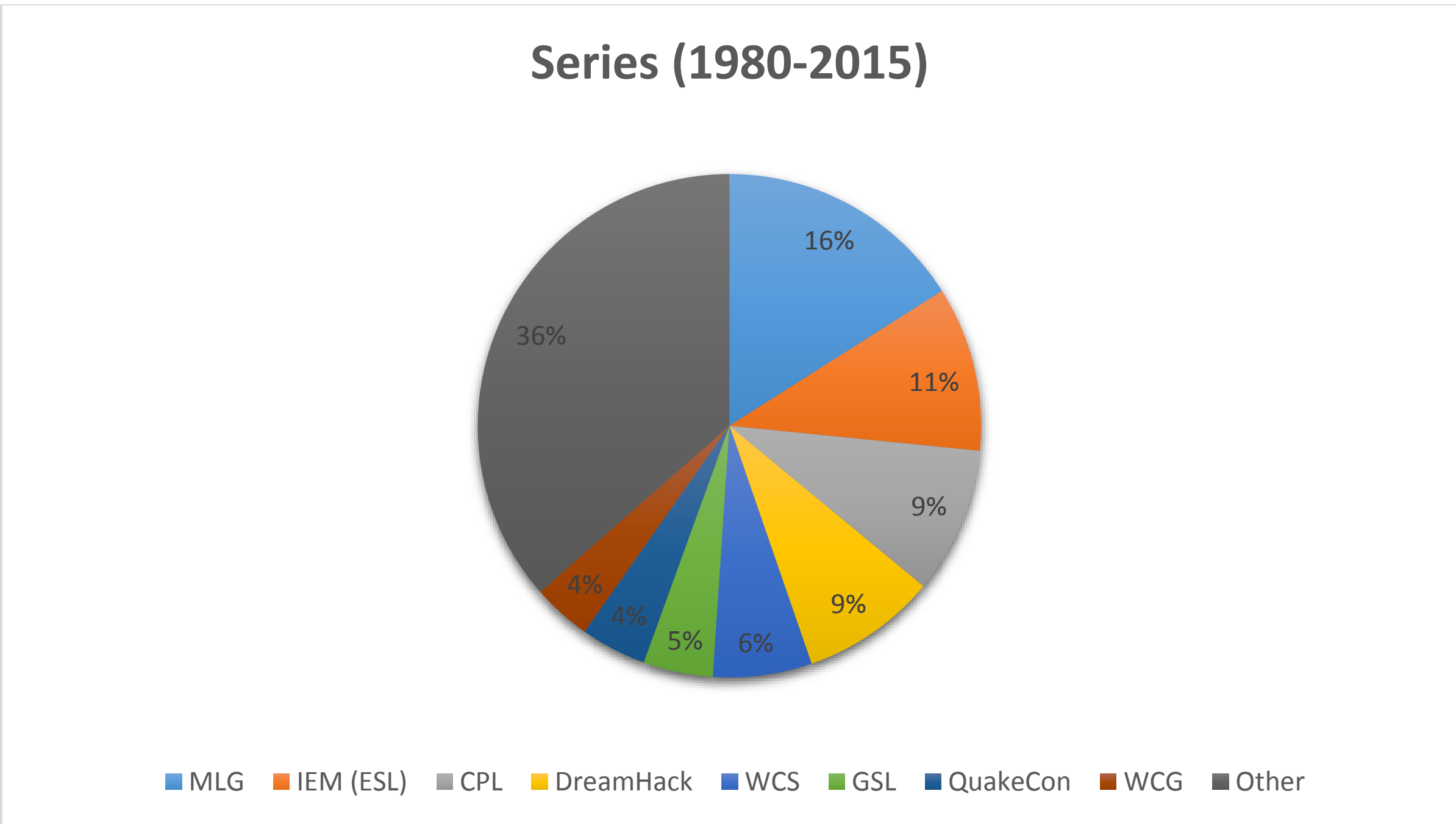

*Figure 5.3 Tournaments/Series 1980-2015 (Appendix B)*

### 5.3.4 Theme: Games (Name)

By analysing the theme games, the data gives an indication on that the StarCraft series (17%) is the most played game in the eSports events, followed by Counter-Strike (14%) and League of Legends (9%). This show an interesting result, as they all is different game genres (RTS, FPS and MOBA). It is understandable that the StarCraft series, the Quake series and the Counter-Strike series is amongst the highest, as the series has existed nearly since eSports began. Considering this makes the findings of League of Legends more interesting, which will be discussed further in the analysis.

| Games played at eSport events 1980-2015 | | | | | |
|---|---|---|---|---|---|
| Game Name | # | Game Name | # | Game Name | # |
| **StarCraft series (*)** | **162** | Track mania Series (*) | 13 | Various Arcade Games | 6 |
| **Counter-Strike Series (*)** | **133** | PainKiller | 13 | Gran Turismo Series (*) | 6 |
| **League of Legends** | **85** | Super Street Fighter Series (*) | 12 | Doom Series (*) | 5 |
| Quake Series (*) | 74 | RBS Series (*) | 12 | Age of Empires (*) | 5 |
| Warcraft series (**) | 73 | Point Blank | 12 | Battlefield Series (*) | 4 |
| Halo Series (*) | 68 | Unreal Tournament Series (*) | 10 | World in Conflict | 4 |
| DOTA series (* & ***) | 48 | Hearthstone | 10 | F.E.A.R | 4 |
| World of Warcraft | 31 | Heroes of Newearth | 9 | Bloodline Champions | 4 |
| Super Smash Bros. Series (*) | 30 | NFS series (*) | 9 | Dead or Alive series (*) | 3 |
| FIFA series (*) | 24 | Crossfire | 7 | PES Series (*) | 3 |
| Call of Duty Series (*) | 21 | Project Gotham series (*) | 7 | Shadowrun | 3 |
| Gears of War Series (*) | 16 | World of Tanks | 6 | Forza MS Series (*) | 2 |
| Tekken Series (*) | 16 | SoulCalibur Series (*) | 6 | Heroes of the Storm | 2 |
| Total number of games: 958 | | | | | |
| ***(i) Includes every game under the franchise name since its initial release.*** | | | | | |
| ***(ii) Excluding World of Warcraft and DOTA*** | | | | | |
| ***(iii) DOTA that originated as a mod from Warcraft III*** | | | | | |

Table 5.2 Games Played in eSport events (Appendix B)

### 5.3.5 Theme: Game Genre

By analysing the theme 'Game genre', the data gives an indication that FPS (32%) is the most common genre in eSports events from 1980-2015 . Just alike the top games, the genres RTS (26%) and MOBA (17%) follows on the top three list.

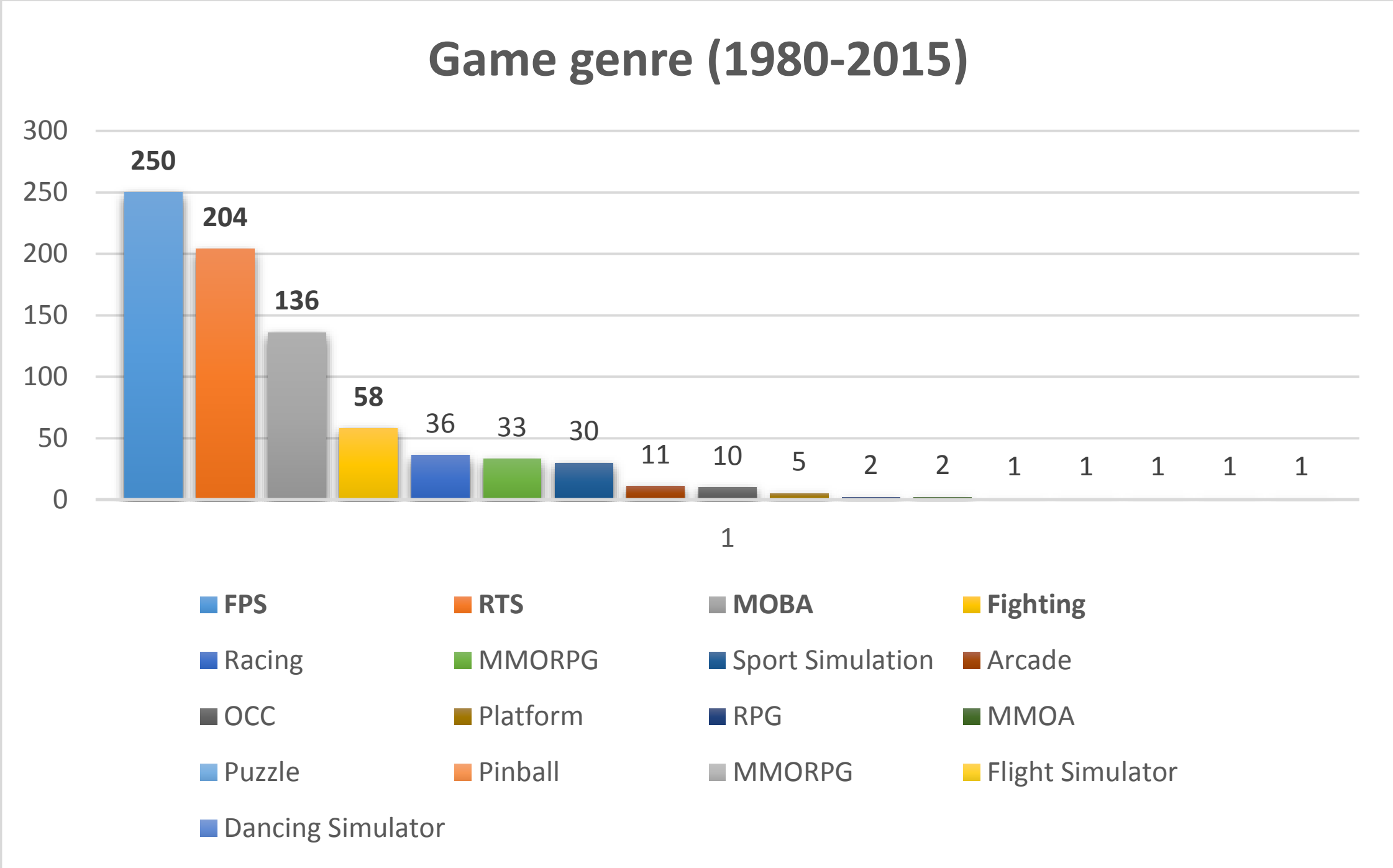

Figure 5.4 Game Genre 1980-2015 (Appendix B)

### 5.3.6 Theme: Prize Pool

By analysing the theme prize pool, the data give an indication of a growth that started in the year 2000. The total prize pool total went down again in 2006, which is caused by an uneven number of events from year to year, and therefore the average figure is used. The findings also shows a high rise in both total prize pool and average prize pool in the last five years, which might show promise for the future of eSports.

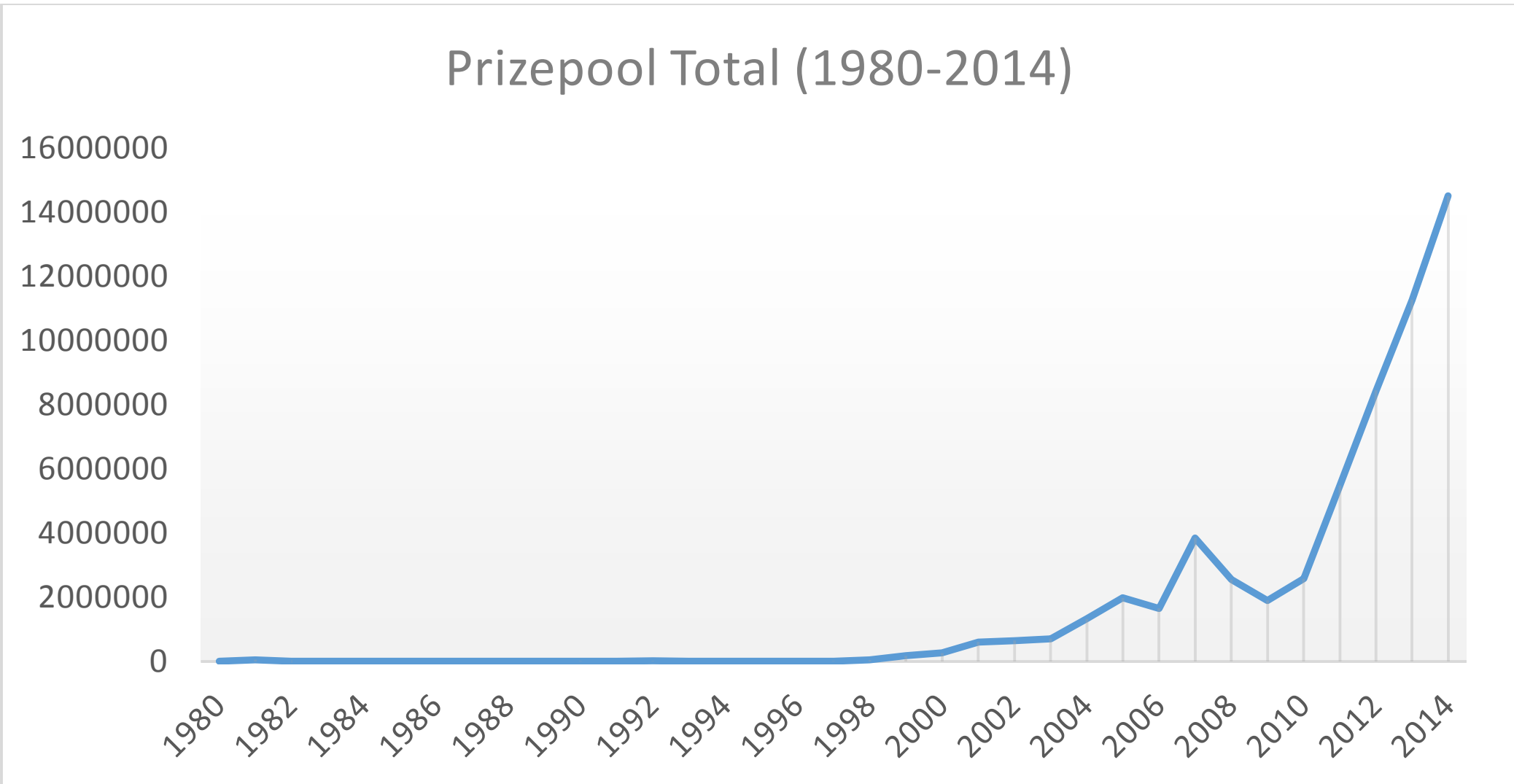

Figure 5.5 Prize pool Total 1980-2014 (Appendix B)

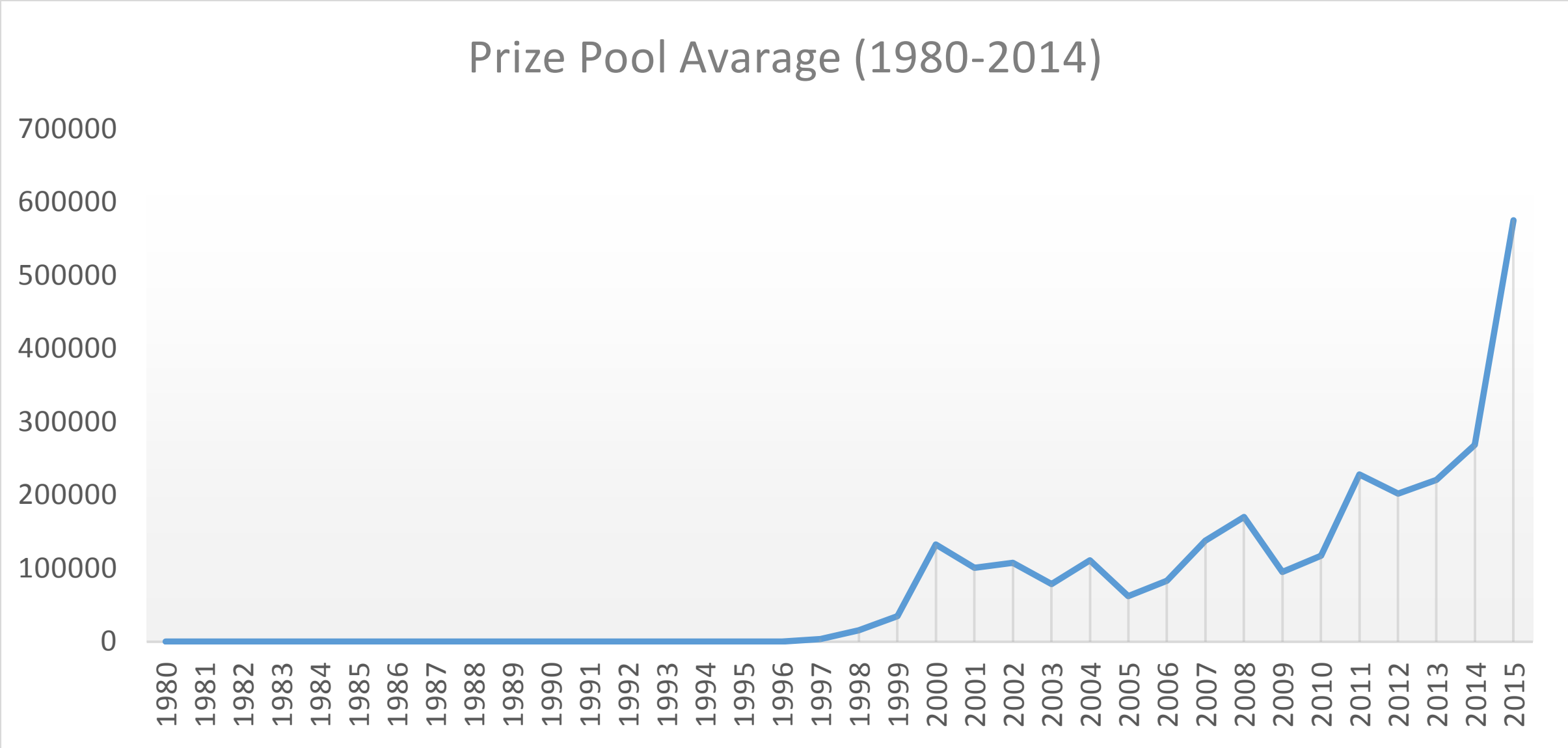

*Figure 5.6 Prize pool Average 1980-2014 (Appendix B)*

### 5.3.7 Theme: Event Duration

By analysing the theme Duration, the data give an indication that 72% of the events last less than one month. The data further shows that the events rarely last more than four months (10%). Those events that last more than one month are often seen as a series that plays matches every weekend for a quarter of a year.

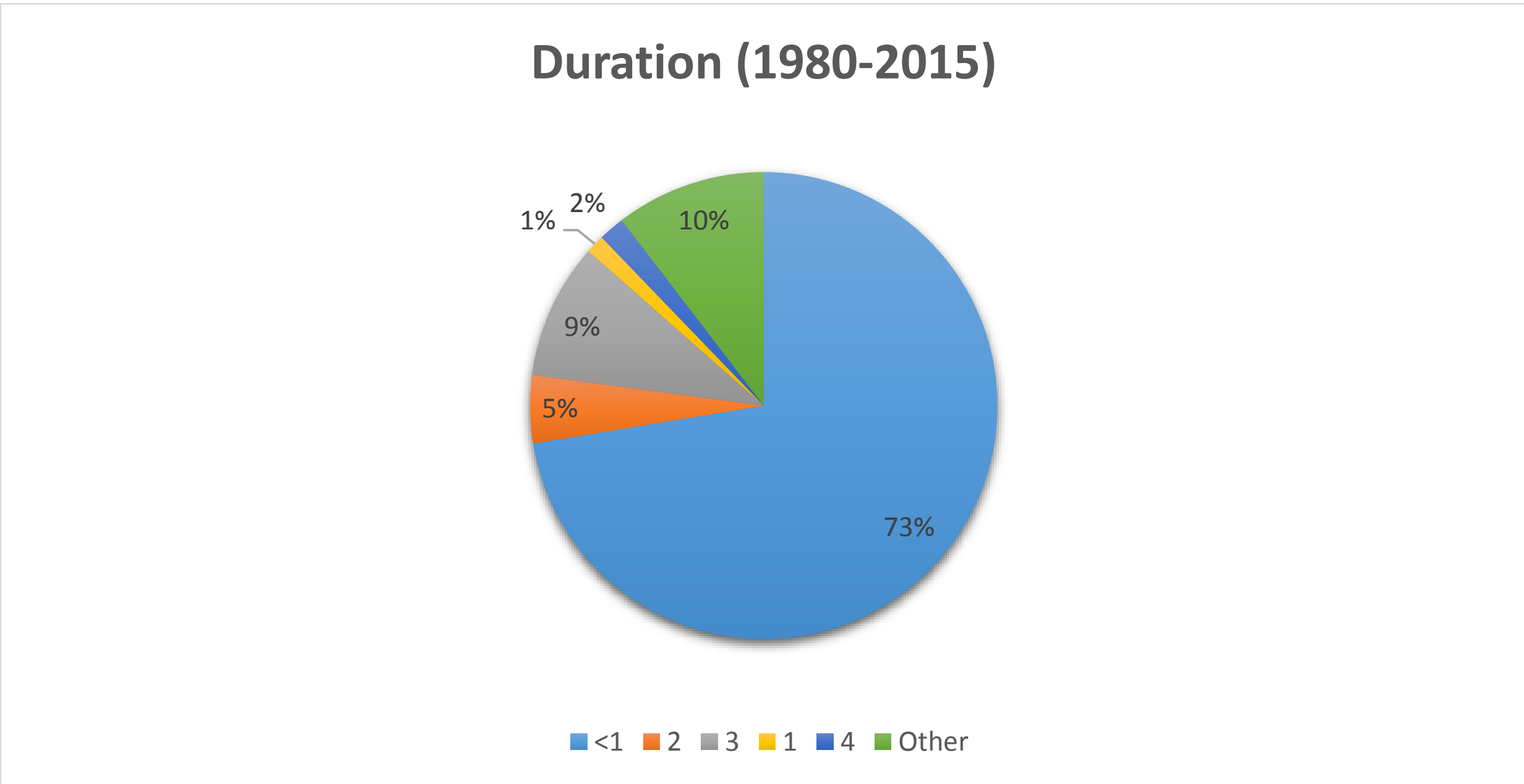

*Figure 5.7 Duration of events 1980-2015 Number 1(Appendix B)*

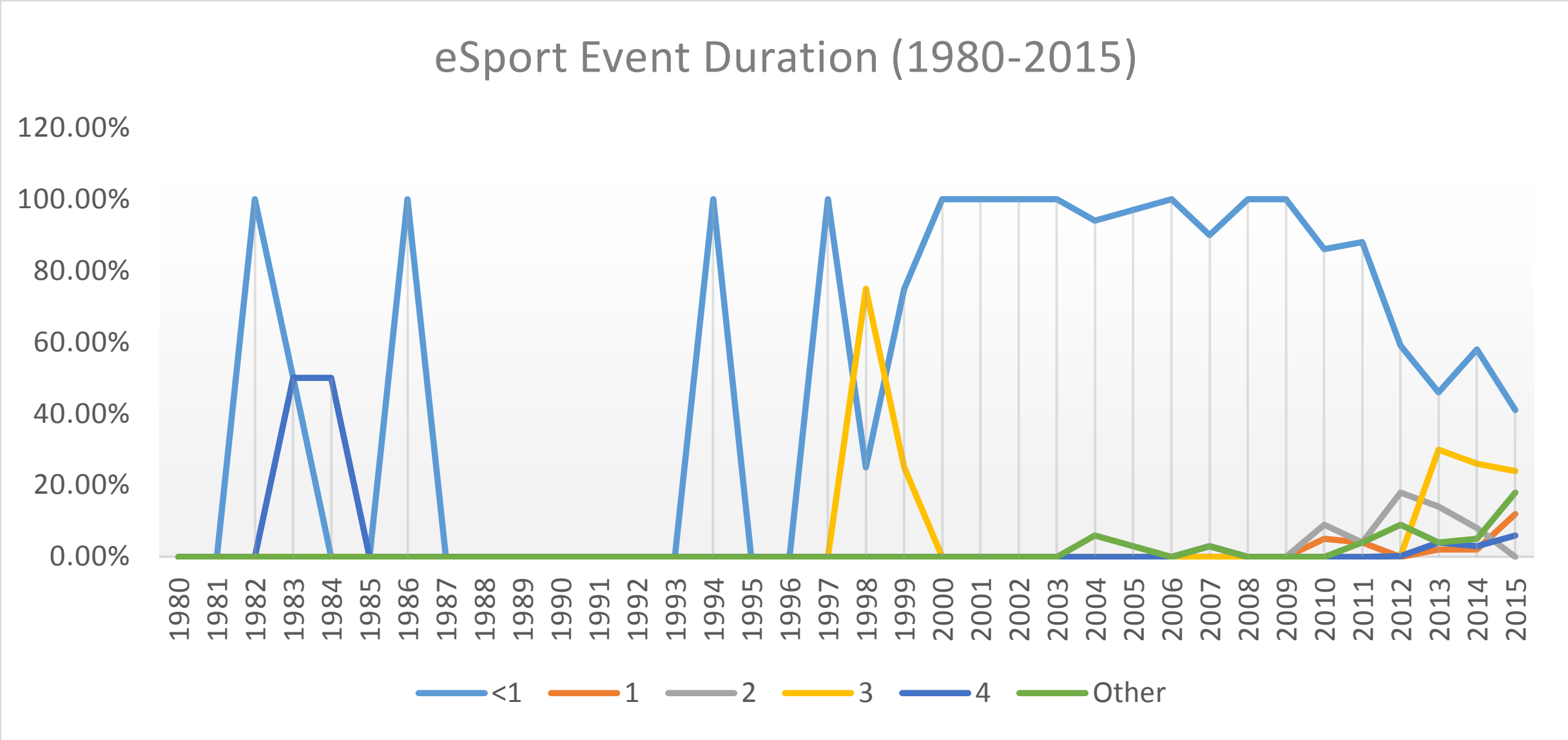

*Figure 5.8 Duration of events 1980-2014 Number 2 (Appendix B)*

### 5.3.8 Theme: Live Audience

By analysing the theme Live Audience, the data give an indication that the number of live audience is been somewhat stable until the year 2013, were a heavy rise started. A peak also hit 2005, caused by one single event held in Korea. This event is seen as an outliner, which is discussed in 5.4.1, and an additional figure is created excluding this event.

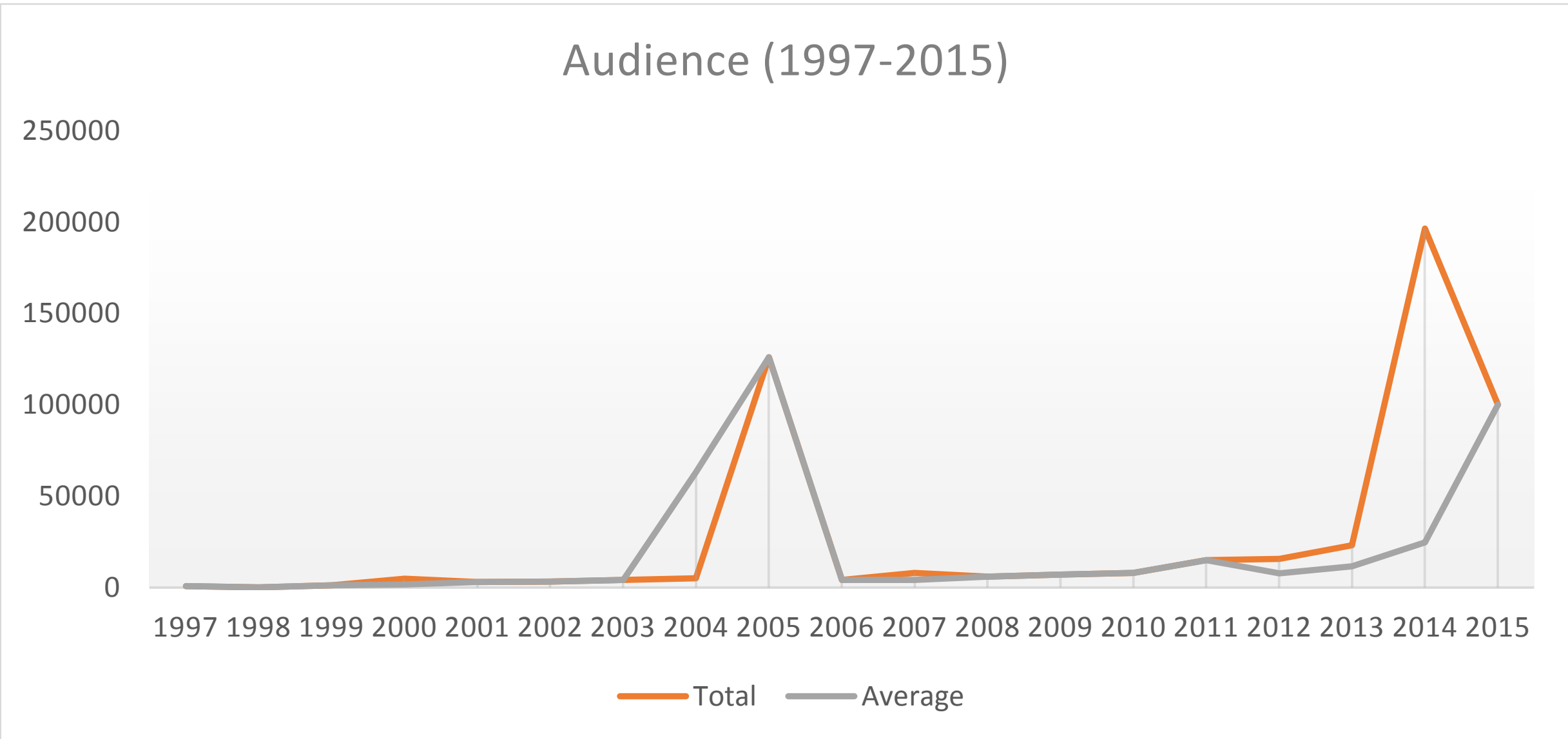

*Figure 5.9 Live Audience Total and Average 1997-2015 (Appendix B)*

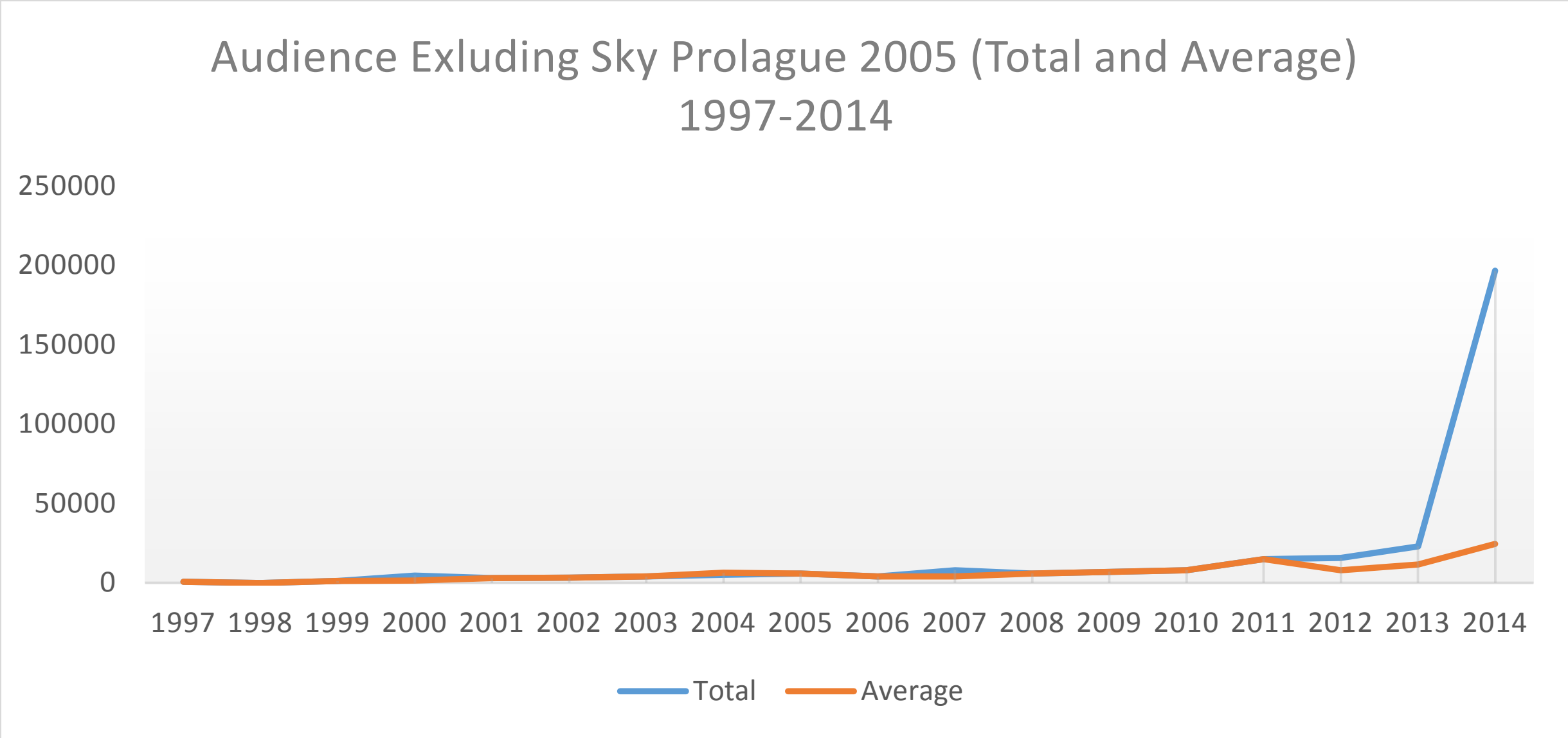

*Figure 5.10 Live Audience Total and average 1997-2014 Excluding SKY Prologue 2005 (Appendix B)*

### 5.3.9 Theme: Streaming (Overview)

As mentioned in the limitation of the methodologies, there were a lack of data in the field of streaming. Analysing the data collected was revealed nearly impossible, as no relevant data could be gathered. This is largely due the number difference between total views and the concurrent peak is very diverse, as seen in figure 5.10. Therefore, to get a better understanding of the data, the most credible streaming audience point were chosen, which is unique views. The definition of unique views also correlate greatly with the definition of live audience, as it is based upon how many unique persons watches the various eSport events, which was the main reasoning for further focusing on unique views. However, the virtual audience data consist of numbers existing from 2010, as it was not possible to found data in this case before these years.

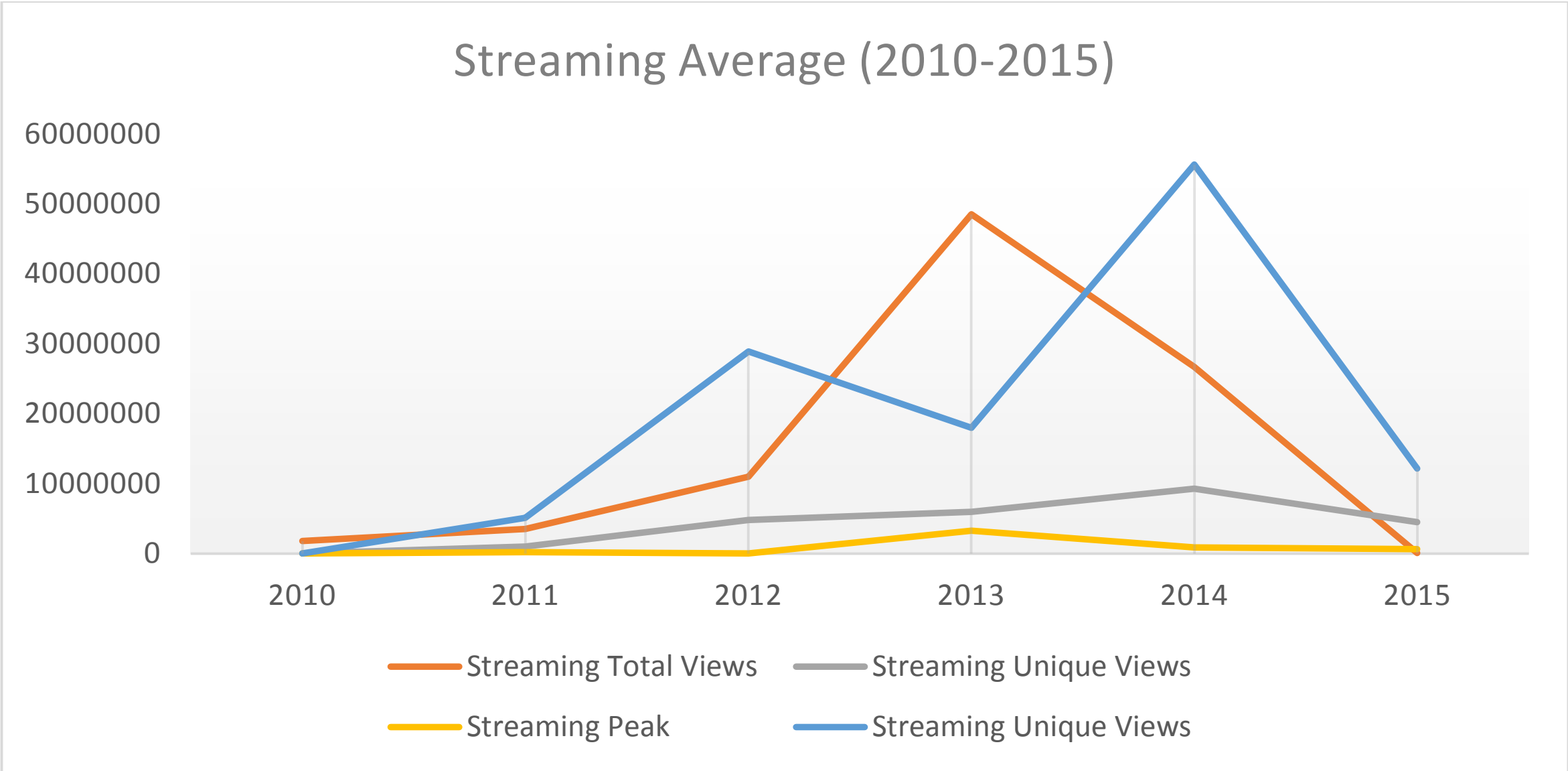

*Figure 5.11 Streaming Average 2010-2015 (Appendix B)*

### *5.3.9.1 Theme: Streaming: Unique viewers*

By analysing the theme Streaming Unique Views, an indication of a rise seems to be existing. Although consisting on data of the last 5 years, a rapid growth is indicating, peaking an average at over 9,000,000 in 2014.

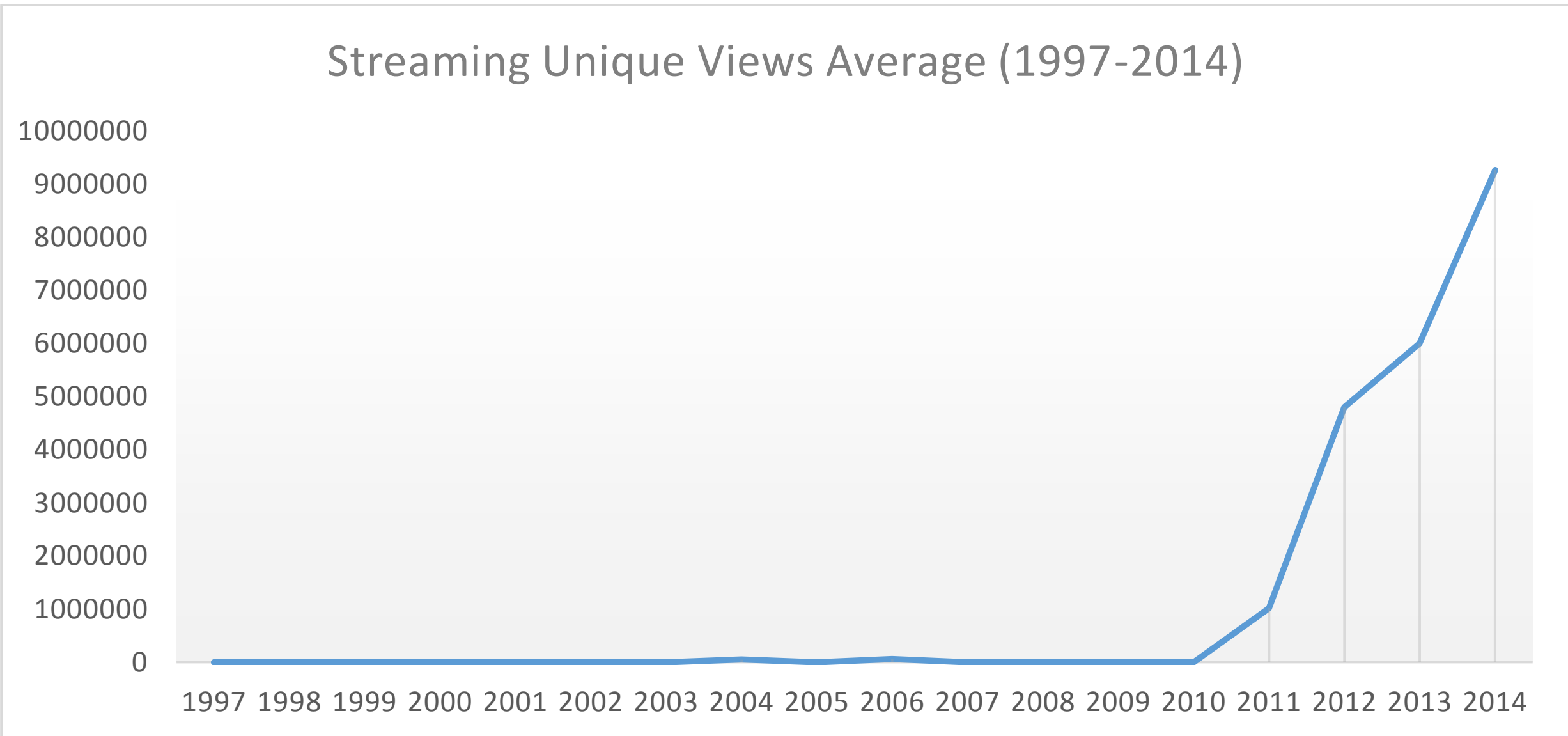

*Figure 5.12 Streaming Unique viewers Average 1997-2014 (Appendix B)*

### 5.3.10 Theme: Participants Total

By analysing the theme Participants Total, the data shows a steady growth year for year. However, it peaked very highly in 2003, mainly because of one event: ESCW 2003. This event is seen as an outliner, and is therefore removed as explained in 5.4.1. Considering this, an addition figure is created without the ESCW 2003 event to get a better visualisation of the participant total`s growth.

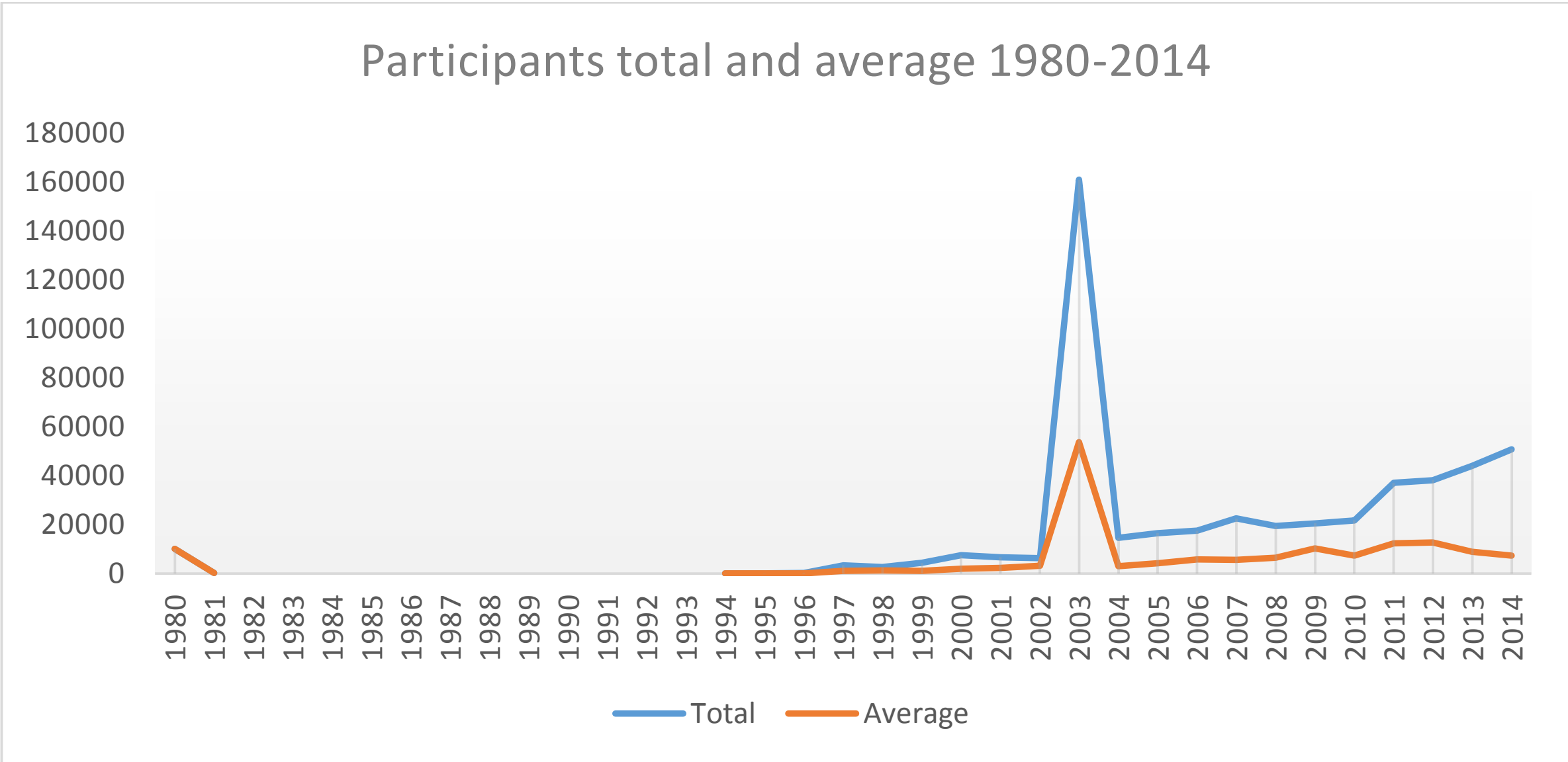

*Figure 5.13 Participants total and average 1980-2014 (Appendix B)*

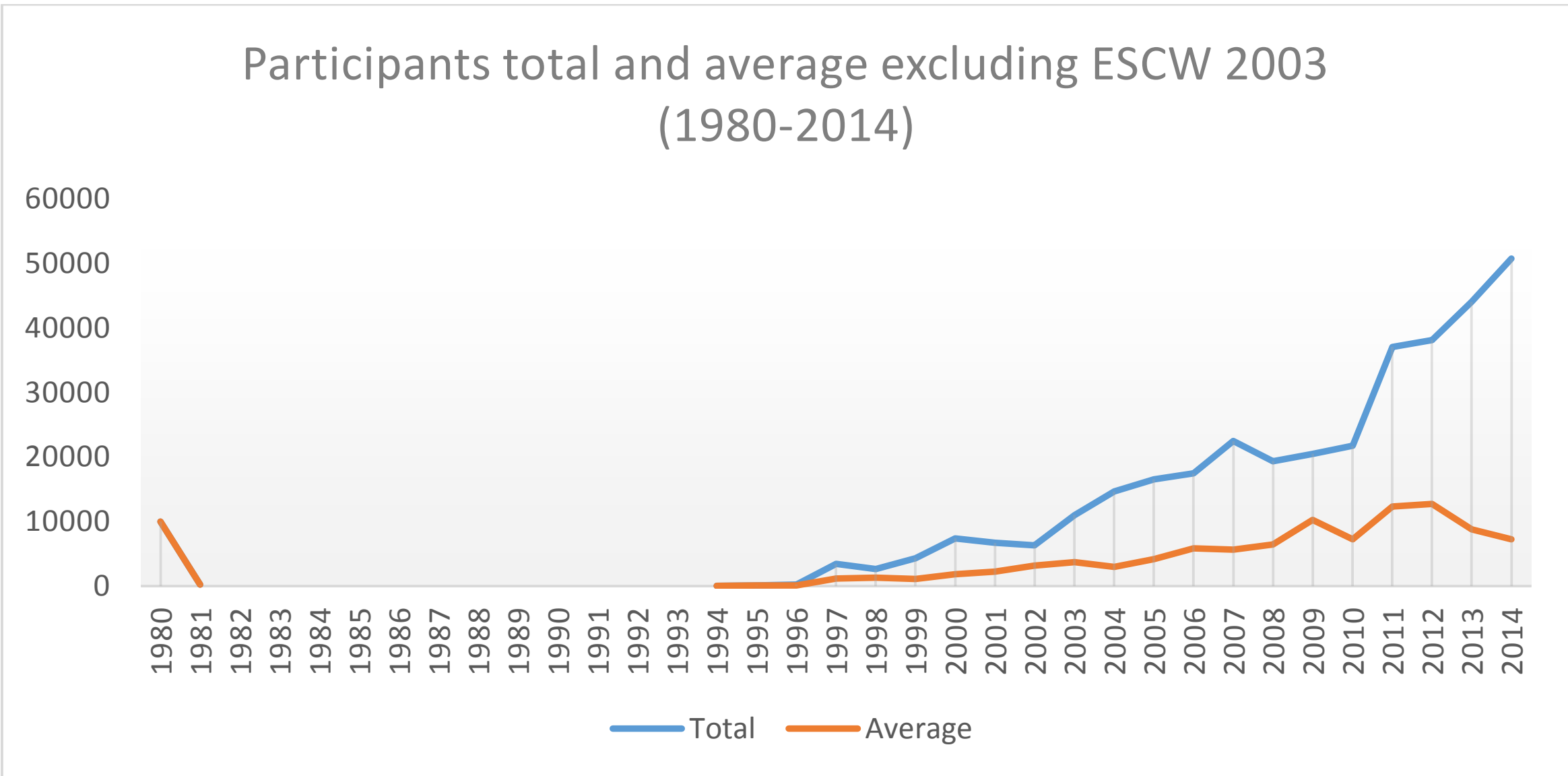

*Figure 5.14 Participants total and average excluding ESWC 2003 1980-2014 (Appendix B)*

### 5.3.11 Theme: Participants Finalists

By analysing the theme' Participants Finalist', the data indicates that no specific growth in the last 10 years. However, a high rises and high losses occurred between the periods 1990-2002, which is something that can be analysed more in depth.

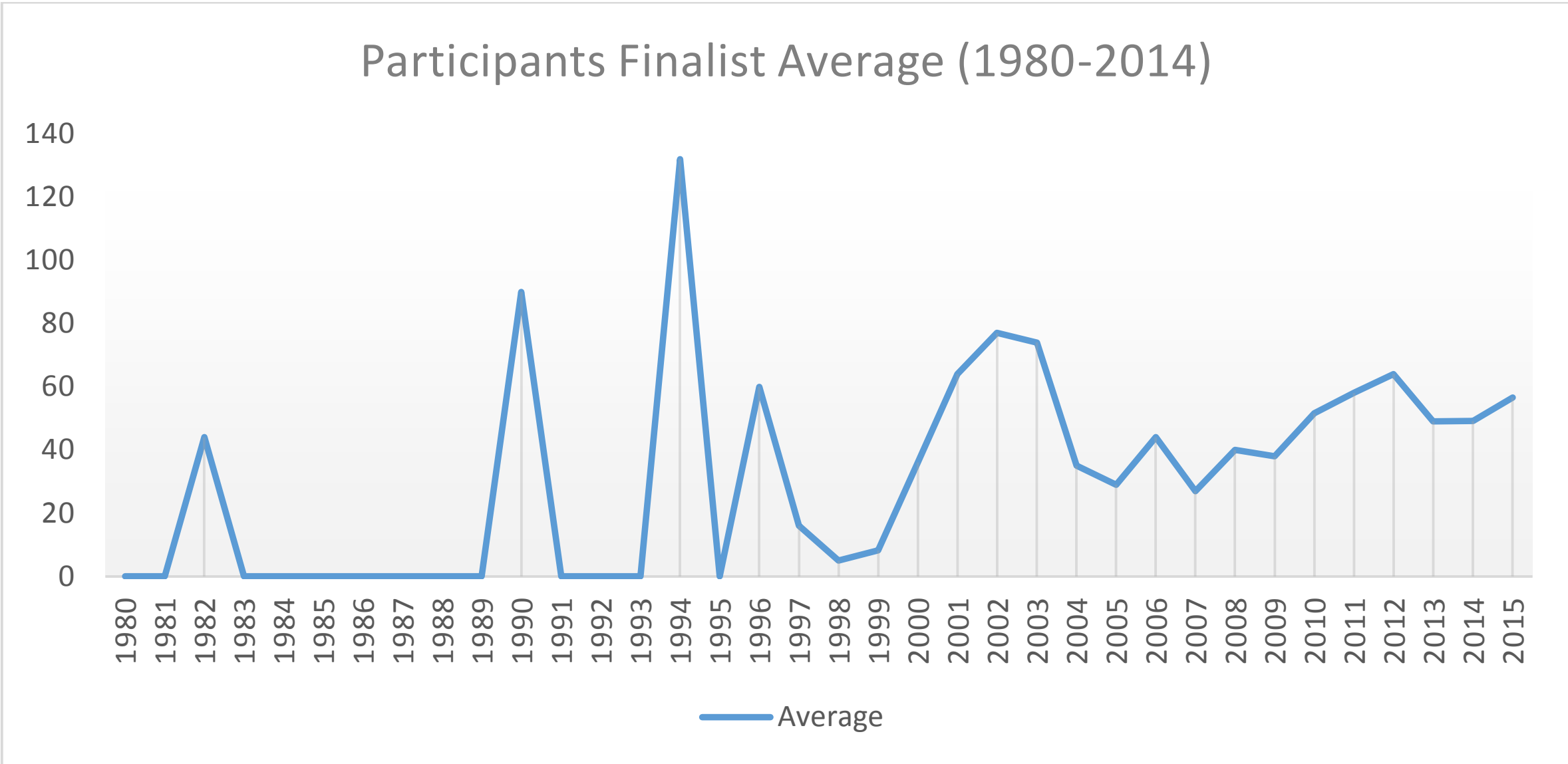

*Figure 5.15 Participants Finalist 1980-2014 (Appendix B)*

### 5.3.1.12 Theme: Participants Countries

By analysing the theme 'Participants Countries', the data show us an indication of growth since 1996. This means that the average countries that participants increases, which might be interesting to analyse further in the sense if this has any impact on an events success or not.

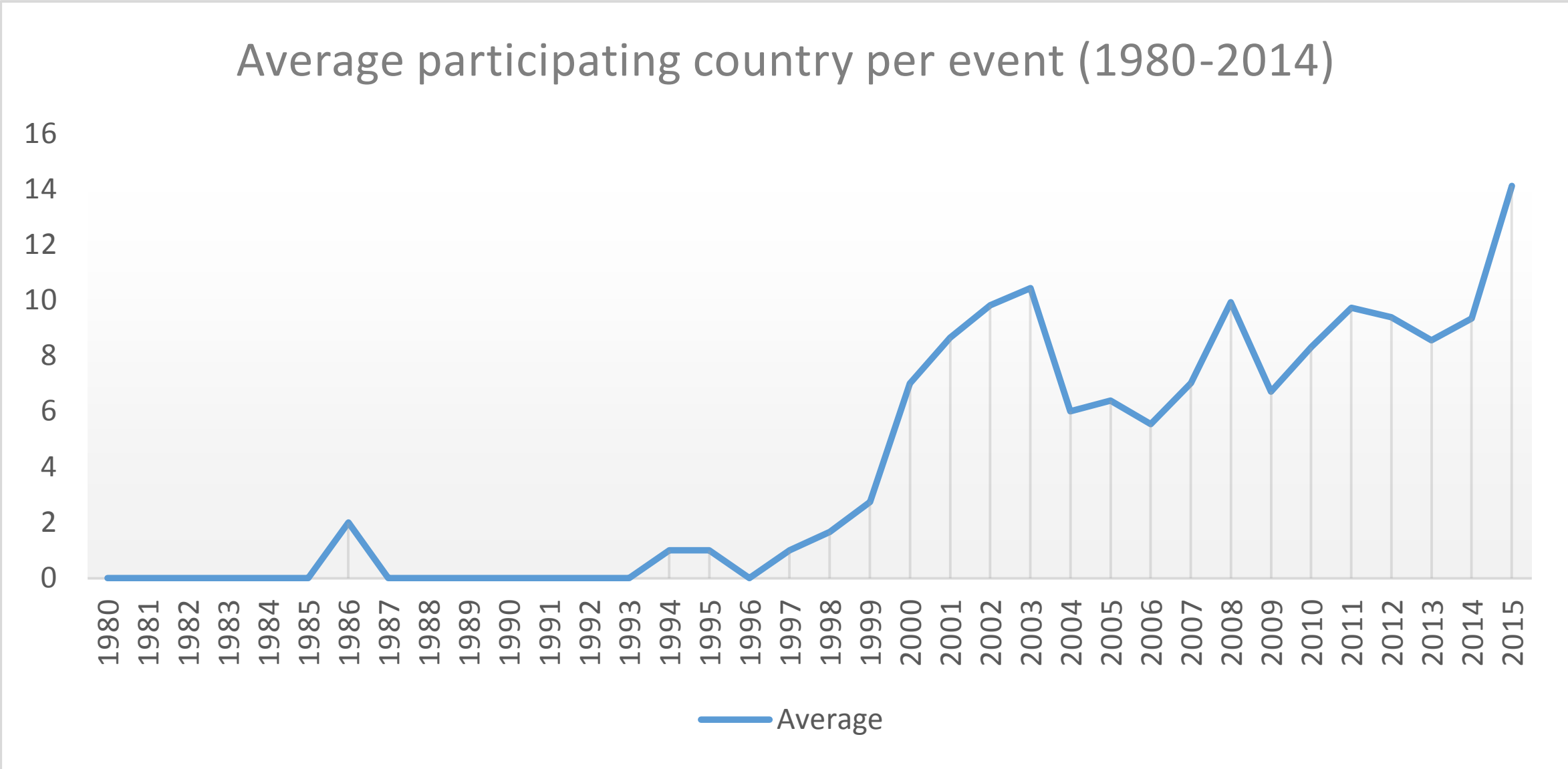

*Figure 5.16 Average participating country per event 1980-2014 (Appendix B)*

## 5.4 Analysis

The presented data collection and findings is further analysed by business analytical tools to answer the research questions. However, before performing the analysis, a measurement of the data collection towards the expected data collection has to be completed. This is done by comparing the presented data with the conclusion of the interpretation given in table 4.2. The reasoning for doing this is to ensure the quality of the findings, which is normally done by various statistical analysis. However, not all the data is numerical, making techniques such as Cronbach Alpha not possible to conduct. Therefore, this is the preferred and chosen solution to continuously ensuring the validity. However, a brief explanation is given on the outliners in this study, before each research question is further analysed in the subsequent points.

### 5.4.1 Outliners

Through the presentation of the data, two specific outliners were mentioned. An outliner is defined by Aguinis et al. (2013) as “data points that deviate markedly from others”, and that is exactly what the events Sky Prologue and ESCW does. Based upon Aguinis et al. (2003) definitions and findings on how to handle outliners, a decision is made to omit these events in the analysis. In addition to this, some of the entries in the thematic analysis were removed due of lack of entries, as mentioned in the pilot of the study under point 4.5.1. These entries is also defined as outlines, as they lack impact to the overall data collection goals.

### 5.4.2 Research Question 1: Data Collection Findings

As for research question 1, the collected data grants an indication of a growing awareness of eSports. The data further indicate that events are firstly held in USA, Korea and China, making it possible that the eSport awareness is higher in these countries compared to others. The attendance in both live audience and streaming audience has grown the past years, giving an indication of the people are getting more aware of the eSports community. The overall attendance in both streaming and live audience is further analysed in research question 4 at point 5.4.5.

### 5.4.3 Research Question 2: Data Collection Findings

As mentioned before, the number of eSport events is umpteen. The analysis based upon the 444 most historic and extraordinary events based upon the collected data, and these are seen as the most important events through history. However, to further answer the questions and meet the expected data, the main events from the thematic analysis is further analysed. Based upon the themes live audience, streaming audience, extraordinary events and prize pool, the following 83 events in table 5.4 are considered as important eSport events that together serve as milestones in the eSport history. Each event that is deemed important has its own reasoning theme, which is firstly explained in the table. If required or necessary, additional information are used to elucidate why this event is seen as an important milestone, including definitions of the reasoning in table 5.3.

| Reasoning | Definition/explanation |
|---|---|
| ***Extraordinary events:*** | An event seen to bring something out of the ordinary in the concept of eSports. |
| ***Extraordinary events (origins):*** | Although not necessary big in its started year, this is an important event created, as it will serve as the origin of one of the biggest eSport events today |
| ***Prize pool, Live audience, streaming audience and participants:*** | Having an extreme or unordinary number for that specific year |

Table 5.3 Reasoning definitions for table 5.4

| # | Year | Event Name | Reasoning | Additional Information |
|---|---|---|---|---|
| *1* | 1980 | First National Space Invaders competition | Extraordinary event | One of the first events held to ever be named as a eSport event |
| *2* | 1990 | Nintendo World Championship | Prize pool | The first big prize pool awarded event, including $10,000 cash, a convertible and a television |
| *3* | 1992 | Nintendo Campus Challenge | Price pool | |
| *4* | 1994 | DreamHack 1994 | Extraordinary Event (Origin) | |
| *5* | 1996 | QuakeCon 1996 | Extraordinary Event (Origin) | |
| *6* | 1997 | Red Annihilation Quake Tournament | Extraordinary Event and Prize pool | The CEO of the creator of DOOM, gave away his Ferrari 328 GTS Cabriolet as first prize |
| 7 | 1997 | CPL: FRAG | Extraordinary Event (Origin) and prize pool | First event ever held by the CPL, a league that is going to be important for the eSport culture. |
| 8 | 1998 | PGL Season 1-4 | Prize pool | One of the first events that took the same season concept seen in other sports, stretching out over 12 months. Also included a high prize pool. |
| 9 | 2000 | QuakeCon 2000 | Participants | |
| 10 | 2000 | World Cyber Games Challenge 2000 | Prize pool, Extraordinary Event, and Extraordinary event (origins) | With over $150,000 in prize pool and a high number of participating countries, WCG became an important milestone. |
| 11 | 2001 | DreamHack 2001 | Participants and prize pool | |
| 12 | 2001 | World Cyber Games (WCG) 2001 | Prize pool and participants | |
| 13 | 2001 | CPL World Championship | Prize Pool and Extraordinary Event | Finals for the long CPL World Tour |
| 14 | 2002 | World Cyber Games (WCG) 2002 | Prize pool and participants | |
| 15 | 2003 | Electronic Sport World Cup 2003 | Participants | |
| 16 | 2003 | World Cyber Games (WCG) 2003 | Prize pool | |
| *17* | 2004 | World Cyber Games (WCG) 2004 | Prize pool | |
| *18* | 2004 | CPL Summer 2004: Extreme World Championship | Prize pool and Streaming | One of the first events to have the tournament/series format. |
| *19* | 2004 | DreamHack Summer and Winter 2004 (2 events) | Participants | World Largest LAN Record |
| *20* | 2004 | MLG USA Tours (9 events) | Prize pool and Extraordinary event (origins) | The start of a very successful set of event, created by the Major League Gaming Community. |
| *21* | 2005 | SKY Proleague 2005 | Live Audience | |
| *22* | 2005 | World eSport Games I, II and III (3 events) | Prize Pool | |
| *23* | 2005 | CPL World Tour Finals | Prize Pool | |
| *24* | 2005 | BlizzCon 2005 | Extraordinary (Origin) | |
| *25* | 2005 | QuakeCon 2005 | Live Audience | |
| *26* | 2005 | DreamHack Summer and Winter 2005 (2 events) | Participants | World Largest LAN Record |
| *27* | 2006 | Electronic Sport World Cup 2006 | Prize Pool | |
| *28* | 2006 | DreamHack Summer 2006 | Participants, Extraordinary event | Included the WSCG 2006 |
| *29* | 2006 | DreamHack Winter 2006 | Participants | |
| *30* | 2007 | Championship Gaming Series 2007 by DirectTV | Prize pool Extraordinary event | One of the first big events shown through Television |
| 31 | 2007 | MLG 2007 (6) Events | Prize pool | |
| 32 | 2007 | DreamHack Winter 2007 | Participants | |
| 33 | 2007 | IEM I Finals (Friday Night Games) | Extraordinary Event (Origins) | |
| 34 | 2008 | Championship Gaming Series 2008 | Prize pool | |
| 35 | 2009 | QuakeCon 2009 | Participants | |
| 36 | 2009 | IEM III and IV (6 events) | Participating countries | |
| 37 | 2009 | DreamHack Summer and Winter 2009 | Participants | |
| 38 | 2010 | MLG 2010 (5 Events) | Streaming | One of the first events showing the tournament through streaming services with a high audience |
| 39 | 2010 | Global League Open (3 Events) | Extraordinary Event, Extraordinary Event (origins) | A Blizzard owned seasonal tournament grown colossal in size the latest year, specially focus on one game. |
| 40 | 2010 | BlizzCon 2010 | Live Audience | |
| 41 | 2010 | DreamHack Summer and Winter 2010 (2 events) | Participants, Extraordinary event | World Largest LAN record |
| 42 | 2011 | DreamHack 2011 Summer | Extraordinary Event (Origins) | The origin of the League of Legends World Championships. |
| 43 | 2011 | DreamHack 2011 Winter | Participants, Extraordinary Event, | World Largest LAN record |
| 44 | 2011 | The International 2011 | Extraordinary Event (Origins), Prize pool and Streaming audience | |
| 45 | 2011 | MLG 2011 (7 Events) | Prize pool and streaming audience | |
| 46 | 2011 | Call of Duty: Experience 2011 | Extra ordinary event, Extraordinary event (origin) , prize money | Considered as an extraordinary event with live music performance from Kanye West and Dropkick Murphys, making it a multi-culture event. |
| 47 | 2011 | League of Legends Continental Series | Extraordinary event (Origin) | Started sessional tournaments in the game League of Legends, which were split in various continents such as EU, NA and Korea. |
| 48 | 2011/ 2012 | IEM VI (7 events) | Prize pool, participants | |
| 49 | 2012 | MLG 2012 (6 Events) | Streaming Audience | |
| 50 | 2012 | DreamHack Summer 2012 | Participants | |
| 51 | 2012 | DreamHack Winter 2012 | Participants and | World Largest LAN Record and including |
| | | | Extraordinary Event | matches for the League of Legends Season 2 |
| 52 | 2012 | League of Legends World Championship Season 2 | Extraordinary Event, Prize pool, Live audience and Streaming audience | |
| 53 | 2012 | The International 2012 | Prize pool | |
| 54 | 2012 | QuakeCon 2012 | Live Audience | |
| 55 | 2012 | WCS 2012 (4 events) | Prize pool and participating countries | |
| 56 | 2013 | MLG 2013 | Streaming Audience | |
| 57 | 2013 | DreamHack Winter 2013 | Prize pool and participants | |
| 58 | 2013 | Turbo Racing League $1,000,000 ShellOut | Extraordinary event and prize pool | Used a competitive game with a high prize money reward for marketing a newly released film |
| 59 | 2013 | Call of Duty Championship 2013 | Prize pool | |
| 60 | 2013 | The International 2013 | Prize pool and streaming audience | |
| 61 | 2013 | QuakeCon 2013 | Live Audience | |
| 62 | 2013 | League of Legends World Championship Season 3 | Live Audience, Streaming Audience, Prize | Largest viewership seen in a eSport event seen today |
| | | | pool, Extraordinary event | |
| 63 | 2014 | DreamHack 2014 Summer and Winter | Participant | |
| 64 | 2014 | Garena Premier League 2014 (3 events) | Prize pool | |
| 65 | 2014 | OGN 2014 (2 events) | Streaming Audience | |
| 66 | 2014 | Smite Launch Event | Prize pool and Extraordinary event (origins) | Launch of a new game named Smite, a game that show promise for future eSport events |
| 67 | 2014 | World Cyber Arena 2014 | Prize pool | |
| 68 | 2014 | Call of Duty Championships 2014 | Prize pool | |
| 69 | 2014 | BlizzCon 2014 | Prize pool and extra ordinary event | Including a musical event with the highly recognized band Metallica. |
| 70 | 2014 | WCS 2014 (9 Events) | Prize pool | |
| 71 | 2014 | The International 2014 | Prize pool, Audience, Streaming audience, Extraordinary Event | Included television coverage on the highly acknowledge sport channel ESPN |
| 72 | 2014 | ESL One 2014: Frankfurt | Extraordinary Event, Live Audience and | First eSport event to be held on a World Cup Stadium |
| | | | Streaming Audience | |
| 73 | 2014 | ELS One 2014: New York | Streaming Audience | |
| 74 | 2014 | QuakeCon 2014 | Live Audience | |
| 75 | 2014 | IEM VIII World Championship | Live Audience and Streaming Audience | |
| 76 | 2014 | IEM IX San José | Streaming Audience | |
| 77 | 2014 | League of Legends World Championship S4 | Live Audience, Streaming Audience, Prize pool and Extraordinary event | Including an own song written by the band Imagine Dragons, a top-chart band during 2014, dedicated to the world finals (who additionally played live at the event, held at a Seoul World Cup Stadium) |
| 78 | 2014 | League of Legends All Star Event 2014 | Streaming Audience | |
| 79 | 2015 | Call of Duty Championship 2015 | Prize pool | |
| 80 | 2015 | ESL ONE: Katowice | Prize pool, Live audience and Streaming audience | |
| 81 | 2015 | Heroes of the Dorm | Extraordinary Event, Prize pool | Included a prize pool paying the winning team a full scholarship at any |
| | | | | chosen university. The event finals was also shown live at ESPN 2, being one of the first live shows ever streamed at ESPN. |
| *82* | 2015 | Smite World Championship | Streaming Audience | |
| *83* | 2015 | Dota 2 Asian Championship | Prize pool | |

Table 5.4 The 83 most important eSport events in eSport history (Appendix A; Appendix B)

### 5.4.4 Research Question 3: Data Collection Findings

To present the expected data collection shown in table 4.3, as well as to answer research question 3, an understanding of which factors is pushing towards the growth of eSports is needed. In the thematic analysis, the growth is based on the audience size, both in live and streaming – although there could be other measuring factors, this is the strongest tool measurable due of its numeric values as well as relations to awareness and consumption. As the thematic analysis consist of both numbers and text, the correlation between the themes can both be observed and calculated, this is further defined in table 5.5. Therefore, to answer research question 3, both a summary in the sense of observation, and a statistical based summary will be conducted.

| *Themes* | **Numerical Findings summary** | **Tools** |
|---|---|---|
| *Single Player and Multi Player* | Excel Chart (Observation) | Diagrams, Scatter Plots and Trend lines |
| *Event Location* | Excel Chart (Observation) | Diagrams |
| *Series* | Excel Chart (Observation) | Diagrams |
| *Games Played* | Excel Chart (Observation) | Diagrams |
| *Game Genre* | Excel Chart (Observation) | Diagrams |
| *Prize Pool* | Excel Chart (Observation) and SPSS Data (Statistical data) | Diagrams, Scatter Plots, Trend lines and Correlations |
| *Duration* | Excel Chart (Observation) | Diagrams |
| *Live Audience* | Excel Chart (Observation) and SPSS Data (Statistical data) | Diagrams, Scatter Plots, Trend lines and Correlations |
| *Streaming Audience* | Excel Chart (Observation) and SPSS Data (Statistical data) | Diagrams, Scatter Plots, Trend lines and Correlations |
| *Participants: Total and Average* | Excel Chart (Observation) and SPSS Data (Statistical data) | Diagrams, Scatter Plots, Trend lines and Correlations |
| *Participants Countries* | Excel Chart (Observation) and SPSS Data (Statistical data) | Diagrams, Scatter Plots, Trend lines and Correlations |

Table 5.5 Correlation between themes and tools to measure significance

#### 5.4.4.1 Observation

The visual data previously presented in the data collection can further be used to give various indications of which factors that has influenced the growth seen in the field of eSports. This can be done by observing any changes seen in the recent year, and see if they correlate with the growth seen in audience. Centred on the collected data, the following observed findings at table 5.6 is considered as growth factors within eSports.

| Theme (Growth Factor) | *Observed reasoning for being a growth factor* |
|---|---|
| **Single player or Team player** | • An increase of Team based eSport events, increasing from 2011<br>• An increase of single player based events, increase from 2009.<br>• A slight decrease in 'both' eSport events, decreasing from 2005. |
| **Event Location** | The most events is held USA, making it the most popular place to construct eSports events. |
| **Series** | MLG, IEM, CPL, and DreamHack has shown to be the most important series in eSport history. The observation of the data shows that these series brings a more structuralised view on the eSport organisation, making them more likely to succeed. Even though some series have already gone bankrupt, the strongest series is growing at the same rate as eSports, or even higher. |
| **Games played** | The following games seems to have an effect on the popularity of the event:<br>• StarCraft<br>• Counter-Strike<br>• League of Legends<br>• Quake series<br>• Warcraft Series<br>• Halo series<br><br>Based upon the recent five years, StarCraft, Counter-Strike, League of Legends, Call of Duty and Hearthstone seems to have an impact of the success of the eSport event. |
| **Games genre** | It seems to be quite important to use the FPS games, the RTS games as well as MOBA games. Analysing the previous years, MOBA and FPS are growing at the same rate as the growth of eSports |
| **Duration** | Observing the data consisting information of durations of eSport events, it shows that almost 75% of the events last less than 1 month, making it plausible to recommend events not lasting longer. |

Table 5.6 Observed findings summary of the thematic analysis (Appendix A; Appendix B)

#### *5.4.4.2 Statistical Data*

As table 5.5 indicates, only four of the thematic data can further be analysed in SPSS, while the others is used for historic observation and trend lines. Nevertheless, considering the numerical data, two hypotheses is created to discuss how prize pool and participants affect streaming audience (Unique viewers) and live audience. The hypotheses is based upon the average values of each year, to have a more thoroughly analysis as the number of events used differ from year to year.

**Hypotheses 1:**

$H_0$: Audience = Prize pool, Participants and participating countries.

(Audience has no correlation with the selected themes).

$H_1$: Audience ≠ Prize pool, Participants and participating countries.

(Audience has a significant positive or negative correlation with the selected themes).

**Hypothesis 2:**

$H_0$: Streaming (Unique viewers) = Prize pool, Participants and participating countries.

(Streaming unique viewers has no correlation with the selected themes).

$H_1$: Streaming (Unique viewers) ≠ Prize pool, Participants or participating countries.

(Streaming unique viewers has a significant positive or negative correlation with the selected themes).

To analysis these hypothesis, both Pearson correlation and Spearman correlation is conducted. The difference between the two techniques is that Pearson mainly focuses on linearity in statistics, while Spearman focuses on the monotonic similarity (Hauke and Kossowski 2011). The reasoning for using both is to achieve a richer understanding of the significant correlation of the findings. Using both will give an understanding of the linear correlation and the monotonic correlation. This is deemed important, as the numerical data shown in the previous data collection can be both. Both techniques uses a 2-tailed test of significance, as the various themes can affect the audience themes negatively or positively.

Alike the other analysis, the data used is the average values rather than the total values; this is due some years have more events than others do, and the total value varies heavily due of this. The correlation test can further be find in table 5.7 and 5.8, which was conducted through SPSS.

**Correlations**

| | | Live Audience (Average) | Prize Pool (Average) | Streaming UW (Average) | Participants (Average) | Participating Countries (Average) |
|---|---|---|---|---|---|---|
| Live Audience (Average) | Pearson Correlation | 1 | .918** | -.029 | -.058 | .598** |
| | Sig. (2-tailed) | | .000 | .963 | .813 | .007 |
| | N | 19 | 19 | 5 | 19 | 19 |
| Prize Pool (Average) | Pearson Correlation | .918** | 1 | -.024 | .213 | .772** |
| | Sig. (2-tailed) | .000 | | .970 | .382 | .000 |
| | N | 19 | 19 | 5 | 19 | 19 |
| Streaming UW (Average) | Pearson Correlation | -.029 | -.024 | 1 | -.338 | -.197 |
| | Sig. (2-tailed) | .963 | .970 | | .578 | .750 |
| | N | 5 | 5 | 5 | 5 | 5 |
| Participants (Average) | Pearson Correlation | -.058 | .213 | -.338 | 1 | .429* |
| | Sig. (2-tailed) | .813 | .382 | .578 | | .037 |
| | N | 19 | 19 | 5 | 24 | 24 |
| Participating Countries (Average) | Pearson Correlation | .598** | .772** | -.197 | .429* | 1 |
| | Sig. (2-tailed) | .007 | .000 | .750 | .037 | |
| | N | 19 | 19 | 5 | 24 | 36 |

**. Correlation is significant at the 0.01 level (2-tailed).

*. Correlation is significant at the 0.05 level (2-tailed).

Table 5.7 Pearson Correlation Test

**Correlations**

| | | | Live Audience (Average) | Prize Pool (Average) | Streaming UW (Average) | Participants (Average) | Participating Countries (Average) |
|---|---|---|---|---|---|---|---|
| Spearman's rho | Live Audience (Average) | Correlation Coefficient | 1.000 | .795** | -.100 | .688** | .551* |
| | | Sig. (2-tailed) | . | .000 | .873 | .001 | .014 |
| | | N | 19 | 19 | 5 | 19 | 19 |
| | Prize Pool (Average) | Correlation Coefficient | .795** | 1.000 | -.100 | .530* | .681** |
| | | Sig. (2-tailed) | .000 | . | .873 | .020 | .001 |
| | | N | 19 | 19 | 5 | 19 | 19 |
| | Streaming UW (Average) | Correlation Coefficient | -.100 | -.100 | 1.000 | -.200 | -.800 |
| | | Sig. (2-tailed) | .873 | .873 | . | .747 | .104 |
| | | N | 5 | 5 | 5 | 5 | 5 |
| | Participants (Average) | Correlation Coefficient | .688** | .530* | -.200 | 1.000 | .497* |
| | | Sig. (2-tailed) | .001 | .020 | .747 | . | .013 |
| | | N | 19 | 19 | 5 | 24 | 24 |
| | Participating Countries (Average) | Correlation Coefficient | .551* | .681** | -.800 | .497* | 1.000 |
| | | Sig. (2-tailed) | .014 | .001 | .104 | .013 | . |
| | | N | 19 | 19 | 5 | 24 | 36 |

**. Correlation is significant at the 0.01 level (2-tailed).

*. Correlation is significant at the 0.05 level (2-tailed).

Table 5.8 Spearman correlation test

#### 5.4.4.2.1 Pearson correlation findings (Table 5.7)

The Pearson correlation test show that there is a positive correlation between audience and prize pool. The number is also significant, showing that increased prize pool effect the number of live audience. This is not the case with the streaming audience, as there exist nearly none correlation, which is nothing close to significant. The Participant seems to have a negative effect on the streaming audience, but this is not significant and does not compare with the expectations of the collected data, and is therefore not analysed. The number of countries participating also has a positive correlation with audience. The number is significant, proving that more participating countries causes an increase of live audience.

#### 5.4.4.2.2 Spearman Correlation Findings (Table 5.8)

The Spearman correlation is somewhat different from the Pearson correlation test, as the results includes numbers that are more significant. Considering hypotheses 1, both prize pool and participating countries has a significant positive outcome. A difference in this test is that the participating average also has a significant positive effect, making all the themes significate affecting the total audience. The streaming audience however has no significant correlations.

#### 5.4.2.2.3 Correlation summary

Considering the information previously discussed, as well including the information gathered from table 5.7 and 5.8, the following conclusion is made and presented in table 5.9.

| **Hypothesis** | **Summary** |
|---|---|
| Hypotheses 1:<br>$H_0$: Audience = Prize pool, Participants and participating countries.<br><br>$H_1$: Audience ≠ Prize pool, Participants and participating countries. | True, in a linear correlation both participants and prize pool has a positive correlation. While in a monotonic correlation all three themes has a positive correlation. |
| Hypothesis 2:<br>$H_0$: Streaming (Unique viewers) = Prize pool, Participants and participating countries.<br><br>$H_1$: Streaming (Unique viewers) ≠ Prize pool, Participants or participating countries. | Untrue, none themes has a significant effect on the streaming audience, which is perhaps caused by the lack of data, something which is discussed 6.2.3, the evaluation of research question 3, and 7.2, the critical evaluation of project conduct. |

Table 5.9 Conclusion of Hypothesis 1 and 2 within the correlation of the chosen data

### 5.4.5 Research Question 4: Data Collection Findings

The previous research question included findings of the significance of various themes related to the audience. Research question four rather focuses on the future of eSports, and if the trends seen will continue the coming years. To understand this, several trend lines is constructed to get an indication of the future of the themes. As defined in the methodology, a trend line is a branch under business analytics used to get an understanding of how a specific situation is going, and how it might change towards the future. As eSports is rapidly growing as indicated by the purpose of study and the literature review, the trend line analysis is based upon only five years in the future, as seeing more towards the future is deemed less realistic. To measure the significance of the various trends, $R^2$ (coefficient of determination) is used. This means the higher $R^2$ (closer to 1), the more significant the number shown is. All themes included the audience is further analysed, to give an indication of both the size and the factors existing within eSports events. As each theme have 2-4 figures each, they were attached to the appendix for a better reading experience. The further findings and visualisations of the trend lines is therefore attached in appendix D.

#### *5.4.5.1 Audience: Streaming Unique Views*

**Visualisation on the data is found under Appendix D: I.**

The trend lines indicate a continuous growth in the unique viewers streaming audience. Considering the data collected, the growth seen today is seen out of the ordinary, but not by much. The trend line further expends the growth to nearly 16,000,000 unique viewers per event, which is an increase of nearly 7,000,000 viewers. The $R^2$ is however close to 0.80, and is therefore considered quite reliable, but not fully. This is most possible due some of the earlier years are missing data. Therefore, an additional trend line is made based upon the last four years with data. This trend line shows a higher increase, to a viewership of nearly 22,500,000 on an average. This is additionally done on a $R^2$ consisting of 0.96. Nevertheless, this is just considering the last years making it less possible than the previous figure. However, both indicates a growth in unique streaming viewers.

#### *5.4.5.2 Audience: Live Audience*

**Visualisation on the data is found under Appendix D: II.**

The trend line of Live Audience is considered without using the ESCW 2003 event, as it is consider an outliner as explained in 5.4.1. The first trend line that bases upon the data consisting from 1997, show little valid data to use due a low $R^2$. The second model however bases on data consisting from 2009, and indicate a growth in the total audience but very little change in the average audience. Considering the first model, the growth seen in 2014 were so extreme that the actual trend line does expect this number before 2019, the $R^2$ is however quite low, so this indication is not very significant. The average value however seems only to grow by very few numbers, meaning that there will be more audience all over, as there will be more events, but the number of people attending the event will not necessarily increase by that much. Considering the last year, the audience indication is increased 6 times it number in the total value, but alike the previous model the average value only rise by very little.

#### *5.4.5.3 Trends in eSport events: Single or Multi player*

**Visualisation on the data is found under Appendix D: III.**

The trend lines indicate a growth seen in all genres. Based upon the data from 1980, single player events will still have biggest role for the future years, where it will actually take more than five years for the multiplayer events to surpass. However, if this were rather look upon the last 5 years, the results would be different. In this perspective, the team-based events will highly increase while single player events will slightly increase. Big events including both however, will decrease. The trend analyses based upon the last five years has a highest $R^2$ consisting of 0.67-0.97, making it the most likely scenario.

### *5.3.5.4 Trends in eSport events: participants average*

**Visualisation on the data is found under Appendix D: IV.**

The trend lines indicate a continuous growth of participants, exceeding to approximately 68 000 by 2019 in total. The total value has a high $R^2$, making it reliable. The average value however only got a $R^2$ on 0.71, making it less reliable. If only the last five years is looked on, a high decrease in $R^2$ is found – making it purposeless to discuss its findings. However, as the average value is deemed most truthful, it is concluded that a growth in average participants will go from approximately 8,000 to 18,000 by 2019.

### *5.4.5.5 Trends in eSport events: Participants Countries*

**Visualisation on the data is found under Appendix D: V.**

The trend line indicate a growth seen in average countries participating at eSport event. In 2014, the number were hovering around nine, but the trend line indicate a growth to nearly 14 countries participating at eSport events at an average by 2019. The $R^2$ is 0.78, meaning the data is some-what reliable, but not fully. If studied with just conserving the last 5 years, no significant data is found so the first figure is deemed most significant.

### *5.4.5.6 Trends in eSport events: Participants Finalist*

**Visualisation on the data is found under Appendix D: VI.**

The data for participants average is vary scattered, making it nearly impossible to perform a trend line on the data. Considering all the data, the highest $R^2$ available is 0.25, which is nearly not significant at all. The table further indicate that the future of finalist is somewhat stabilizing. Considering just the last years, a trend line show a high risk of a shortage of Average participants. Although the $R^2$ is 0.81 making somewhat significant, this cannot be seen as that realistic due the basis of 5 years. However, the decrease is something worth mentioning in future discussions.

#### *5.4.5.7 Trends in eSport events: Duration*

**Visualisation on the data is found under Appendix D: VII.**

Alike participants finalist, the event duration data is somewhat scattered, making it nearly impossible to predict the future as there is few common patterns. All of the trend lines exist of a $R^2$ that is less than 0.50, making them less significant. Nevertheless, the trend lines indicate a much similar event duration seen the last years, where events lasting less than one is the highest populated. However, considering the duration in a smaller time scale based upon the last five year, the results changes drastically. Due the rise of events going 3 months, the trend line indicates that this duration will surpass the <1 duration by 2016. The second trend line also include higher $R^2$ than the previous model, making it more realistically although the numbers are not significantly high.

#### *5.4.5.8 Trends in eSport events: Prize Pool*

**Visualisation on the data is found under Appendix D: VIII.**

In the matter of prize pool, the trend line indicate a slow but notable growth. Adding just 5 years made the trend line predict the same value seen in 2014, however, if additionally 5 years is added a further indication of growth is seen. The average trend line is rather based upon data from 1997, as none data existed before it. The trend line here has a $R^2$ consisting of 0.80, making it give a better indication the growth. Again, the growth is not much considering its five years, but it is nevertheless a growth. As average is the most realistically data, an additional trend line were created based upon the last five years. The data here has a higher $R^2$, consisting of 0.88. However, the answer is nearly exactly same as seen in the previous trend line test.

### 5.5 Research Question 5 and 6: Data Collection Findings

As stated previous, the thematic analysis have no clear findings to answer either Research Question 5 or 6. However, since this is not a possible, the chosen ITM tool is based upon literature and various theories, giving a theoretical approach on answering the last two research questions. Nevertheless, the previous findings will be used to give an indication of how the various ITM might use the various themes in their systems and strategies, but before doing this a chosen tool must firstly be decided – which is the research question 5. Research question 6 will additionally not be analysed further in this chapter. It rather be focused upon under recommendation in chapter 6, where the strength and weakness is measured of the chosen ITM tool. Nevertheless, as none data is collected – no data can further be analysed. The theoretical answering of the research questions will rather find place in the next chapter.

### 5.6 Summary of Chapter 5

Chapter 5 includes all the essential findings of the thematic analysis, followed by analysis using business analytical tools to answer the research questions. The credibility of the data collection is deemed slightly lacking, as the thematic analysis only included 67% of all the data intended to be gathered, which is caused by a lack of shared data in the field of eSports. As for the collection, 17 different figures and tables were constructed to visualise the gathered data. These findings is further discussed as each research question is for a second time considered. For the first research question, an indication of growth is seen in the sense of audience, which relates to the overall awareness of eSports. Additionally, the data indicates that USA is commonly used to hold eSport events. For the second research questions, 83 events is chosen to define the major milestones in eSport history. Research question 3 shows various correlations between the themes by observation and by using both a Pearson correlation test and a Spearman correlation test. The correlations indicates that live audience and prize pool is the most significant themes, while participating countries and participates total additionally have significant value. As for trends within the field, each numerical theme in the thematic analysis is analysed. Nearly every theme that were analyses showed indications of positive trends in the future. These findings included high $R^2$ values, which makes it possible to predict that the growth will continue, including which themes will grow the most based on the thematic analysis. Research question 5 and 6 however is not included in this chapter, as none data were collected to answer them. They will rather be based on theory in the next chapter, where the various themes is used to get a understanding of which ITM tool would be most sufficient.

# CHAPTER 6 CONCLUSION AND RECOMMEDATIONS

## 6.1 Chapter Introduction

The previous chapter based on both the data collection from the thematic analysis, and the analysis constructed with business analytical tools. It briefly discussed the findings, before analysing the data to find both the significant numbers as well as trend lines of the data. This data was further used to indicate the growth factors seen in eSport events today, so an understanding of both the past and the future of eSports could be made. However, the previous chapter did not include findings on research question 5 and 6, which is something this chapter will focus on answering. Additionally, this chapter will focus on concluding the findings, and endeavour to answer the aim, objectives and the research questions. This is done by firstly disputing the differences of the findings in the literature review and the thematic analysis. Then, when fully understanding the questions that were to be answered, a recommendation will be made, which will further answer research question 6. Finally, there is a conclusion of the work, which includes what can be done further within the chosen problem domain. Nevertheless, before the recommendations can be done, a summary of the research question is done based on both the findings in the literature review and in the data collection.

## 6.2 Conclusion of the Research Questions

To conclude each research question, a brief discussion is done with the collected data and the literature. As both a discussion is done in the literature review, and in the data collection, only a brief summary is constructed.

### 6.2.1 Research Question 1: Conclusion

*'What is eSports, and how aware are people of the eSport culture?'*

As for the question regarding what eSport is, the literature review answered the question quite well by defining it as a competitive sport. The literature further defined it as an activity requiring either mental or physical abilities, although these were a quite debatable topic. It was with this reasoning eSport events were further analysed, as they are the essence of the definition of eSports. As for the awareness of eSport, the literature review confirmed an audience consisting of a number between 71 million people to 205 million people in a global aspect, where it was deemed closer to 71 million people rather than 205 million people. The thematic analysis did not give any clear indication on how many people watches eSport globally, but it did give an indication of how many watches eSport events live, and through streaming services. In 2014, eSport events had an average live audience close at 24,549 (+309% from 2013), and an average unique viewership on streaming services at 9,268,000 (+854% from 2013). This is the first indication that people are getting more aware of eSports, as the consumption growth is sufficient.

The thematic analysis further indicated that the awareness is greatest in USA, followed by Korea and China. These differ somewhat from the literature review, as it states Korea to be the number one country in regards to eSport – which is not the case in the thematic analysis. In conclusion, an understanding of eSports is made, and both the literature review and the data collection indicate that the overall awareness of eSport is growing.

### 6.2.2 Research Question 2: Conclusion

*What are the major milestones of eSports history?*

As for the major historic milestones in the history of eSport, there were some differences in the findings. Firstly, the literature review reviewed every important factor, such as the introduction to technology devices and the dot.com bobble. However, this was not deemed feasible as mentioned in the limitations of methodology, and therefore the thematic analysis based upon eSport events rather than every factor possible. For the thematic analysis, the first constructed list consisted of 444 events, as shown in Appendix A. To further answer research question 2, the list was shortened to 83 events, as explained in 5.3. To further answer research question 2, a visualisation of the timeline has been made, based on the summary findings in the literature review as well the 83 events mentioned in thematic analysis discussion. This timeline is further attached in Appendix H, due to its size. The events coloured in orange indicates that they are based upon the thematic and business analytics findings. The events coloured in green are based on the findings in the literature review, while the events coloured in blue indicate that they are based on findings from both.

.

### 6.2.3 Research Question 3: Conclusion

*Why are eSport seeing its growth today, and which factors/patterns is pushing it upward?*

The various factors and patterns found in the literature review were intended to be analysed in the thematic analysis, in order to find answers that are more significant. However, the thematic analysis ended up finding several more factors, which were used – although some were deleted as explained in the interpretation chapter. The following factors and patterns found are further summarised in table 6.1. The significance of the findings is discussed in the right aligned row, where a brief explanation of its meaning is given in the beginning of the table. However, it is worth mentioning that as for the data collected none factor showed significance to the virtual streaming audience. This was briefly touched upon in table 5.9, where it was stated that this could be caused by a lack of data. After analysing the matter, this is the only conclusion found reasonable, as the trend analysis expect a high growth as will be presented in subsequent points. Therefore, this factor causes a lack of credibility in research question 3, which is further discussed and presented in 7.2.2. Table 6.1 will therefore focus on the significance with the live audience.

## Growth Factors in eSport events

*[1] Observation equals findings in the thematic analysis*

*[2] Literature equals findings in the literature review*

*[3] Statistical significant equals findings in the correlation tests based upon live audience*

| Factor / Pattern | **Literature review factors/patterns** | **Thematic analysis factors/patterns** | **Observation or Significant** |
|---|---|---|---|
| **Gaming Series** | Increase in gaming series and leagues. | A high increase in gaming series, where MLG, IEM, CPL and DreamHack is shown as the most important | Literature and Observation |
| **Games played** | Some games had additionally a high correlation to the | Some games are more often played in successful eSport events | Literature and Observation |
| | growth of eSports, where league of legends where the brightest example | than others are, where the highest examples is League of Legends, StarCraft and Counter-Strike | |
| **Game Player-mode** | | A high increase in team based eSport games | Observation |
| **Game Genre** | A correlation between the increase of FPS, MOBA, MMOA and OCCG to the growth | Based upon the events analysed, FPS, RTS and MOBA games are the most played genres. | Literature and Observation |
| **Streaming services** | the release of Justin.TV in 2007, making it easier for streaming services around – although this existed earlier with examples such as HLTV at MLG.TV | An increase of streaming audience from 2007-2010. | Literature And observation |
| **Dot.Com bobble** | The dot.com bobble that existed 1991-2000, although not a factor it is believed to be one of the causing factors of the growth of eSports | | Literature |
| **Event Duration** | | Of the events analysed, events lasting less than a month seems tend to be more popular. | Observation |
| **Prize pool** | | Prize pool has both a linear and monotonic similarity on the attending live audiences | Statistical significant |
| **Participating countries** | | Participating countries has both a linear and monotonic similarity on the attending live audiences | Statistical significant |
| **Participating players** | | Participants on eSport events has a monotonic similarity on the attending live audiences | Statistical significant |

Table 6.1 Summary of factors and patterns that is enhancing the growth of eSport

### 6.2.4 Research Question 4: Conclusion

*Will the growth of eSport continue, and what can the eSport business do to ensure a continuous growth? (Future of eSports)*

The fourth research question regards the future of eSports, where there existed numerous literature. Both the literature review and the analysis show significant numbers, observation and trends that eSport will continue to grow. Three tables are further used to indicate the growth potential in the consumption of eSports. This is based upon the two audience factors, streaming (unique viewers) and live audience, as well as the most significant correlating theme, prize pool. Prize pool is chosen to give an indication on how a trend might affect the consumption and awareness of eSport with a futuristic approach. These tables can be viewed under table 6.2, 6.3 and 6.4. Lastly, a summarisation is made of the themes observed with growth, stating the most important factors based on the findings, on how to create a successful eSport event. This is discussed in the subsequent point.

# Growth of eSports: Live Audience

| Year | **Literature review** | **Business Analysis and thematic analysis** | **Technique** |
|---|---|---|---|
| **2000** | | +42.45% | Literature and Observation |
| **2001** | | +91.44% | Literature and Observation |
| **2002** | | +8.33% | Literature and Observation |
| **2003** | | +23.07% | Literature and Observation |
| **2004** | | +57.50% | Literature and Observation |
| **2005** | | -4.76% [I*] | Literature and Observation |
| **2006** | | -33.33% [I*] | Literature and Observation |
| **2007** | | 0.00% | Literature and Observation |
| **2008** | | +50% | Literature and Observation |
| **2009** | | +16.67% | Literature and Observation |
| **2010** | | +14.26% | Literature and Observation |
| **2011** | | +87.50% | Literature and Observation |
| **2012** | | -47.82% | Literature and Observation |
| | | | |
| **2013** | +27.58% | +46.95% | Literature and Observation |
| **2014** | +20.27% | +113.47% | Literature and Observation |
| **2015** | +21.13% [IV] | +61.10% [I*, II*] | Literature and Observation and trend line analysis |
| **2016** | +21.13% [IV] | +61.10% [I*, II*] | Literature and trend line analysis |
| **2017** | +21.13% [IV] | +61.10% [I*, II*] | Literature and trend line analysis |
| **2018** | | +61.10% [I*, II*] | Literature and trend line analysis |
| **2019** | | +61.10% [I*, II*] | Literature and trend line analysis |

*[I]: Based on findings from 2009-2019, where the total growth is expected to be 305.51% from 2014-2019. This number is further divided on 5 years to get a linear understanding of the growth*

*[II]: Based upon the trend line from 2009-2014*

*[III]: Based upon the trend line from 1997-2014*

*[IV]: Based on the expected difference from 2014-2017 which is 63.4%. It is further dived on the three years to get a linear understanding of the growth*

Table 6.2 Expected growth of live audience based upon findings (Appendix B; Casselman 2015; Newzoo 2015; Heaven 2014b; Hope 2014)

## Growth of eSports: Streaming Audience (Unique Viewers)

| **Year** | Literature Review | Business Analysis and thematic analysis | Technique |
|---|---|---|---|
| 2011 | - | - | - |
| 2012 | - | +371.67% | trend line analysis |
| 2013 | - | +24.90% | trend line analysis |
| 2014 | - | +54.57% | trend line analysis |
| 2015 | - | +28.32% (I*) | trend line analysis |
| 2016 | - | +28.32% (I*) | trend line analysis |
| 2017 | - | +28.32% (I*) | trend line analysis |
| 2018 | - | +28.32% (I*) | trend line analysis |
| 2019 | - | +28.32% (I*) | trend line analysis |
| [I] *Based on the difference from 2014-2019 which is 140.61%. It is further dived on the five years to get a linear understanding of the growth.* | | | |

Table 6.3 Expected growth of live audience based upon findings (Appendix B)

## Growth in Prize pool

| | Thematic analysis Total | Thematic analysis Average | Literature review |
|---|---|---|---|
| **1998** | 660% | 440% | - |
| **1999** | 376.22% | 225.86% | - |
| **2000** | 152.46% | 380.93% | - |
| **2001** | 228.44% | -23.85% | - |
| **2002** | 106.74% | 106,74% | - |
| **2003** | 109.45% | -27.03% | - |
| **2004** | 188.01% | 141.01% | - |
| **2005** | 149.06% | -44.10% | - |
| **2006** | -16.43% | 133.70% | +22.22% |
| **2007** | 232.51% | 166.08% | +38.64% |
| **2008** | -33.81% | 123.55% | +4.92% |
| **2009** | -25.55% | -44.16% | -45.31% |
| **2010** | 135.88% | 123.52% | +48.57% |
| **2011** | 212.53% | 194.82% | +86.54% |
| **2012** | 154.06% | -11.65% | +35.05% |
| **2013** | 133.43% | 109.50% | +51.14% |
| **2014** | 128.73% | 121.59% | +81.81% |
| **2015** | - | +26.87% [I*] | +97.22% |
| **2015** | - | +26.87% [I*] | - |
| **2016** | - | +26.87% [I*] | - |
| **2018** | - | +26.87% [I*] | - |
| **2019** | - | +26.87% [I*] | - |

[I] *Based on the difference from 2014-2019 which is 134.36%. It is further dived on the five years to get a linear understanding of the growth.*

Table 6.4 Expected growth in the theme prize pool (Appendix B; Newzoo 2015)

#### 6.2.4.1 What can be done to ensure growth?

The results in the matter of growth enhancement differ between the thematic analysis and the literature review. Other than a wish to get more focus on television rights of eSports events, there were little literature saying what the field of eSports should do further. The thematic analysis however analysis the trend within eSport events, showing what trends and pattern it should use to achieve the best possible event. These recommendations is seen as a part of the short-term recommendations, discussed in point 6.3.1. The following trends and patterns is further recommended to follow to ensure growth on eSport events:

I. The event should be based upon Team Based games
II. The event should last less than a month
III. The event has a higher chance to be successful if part of an organisation or a series
IV. The games included should be either FPS, RTS or MOBA.
V. The games played in the eSport should include StarCraft, Counter-Strike and League of Legends, Hearthstone and DOTA.
VI. The event should have a high prize pool, as it improves the amount of audience attending.
VII. The event should mainly try to attract players from the entire world, as this increases the live audience. Additionally, having more participants in total is also shown to have a positive effect on the total number of audience.
VIII. Holding the event in USA might attract more audience.

### 6.2.5 Research Question 5: Conclusion

*Which Information Technology Management Tool would be most beneficial for the eSport business to use?*

Research question 5 regards various ITM tools, which were only discussed in the literature review. The findings on these tools is, as mentioned in the limitations-section, based upon theory in the sense of literature, since no interviews were conducted. Research Question 5 is further discussed based upon exactly this, while the 6th research question will be based upon the recommendation itself, as it concerns the effect of implementing the chosen ITM tool. However, the previous collected data give an indication of how an ITM tool can be used, which is further discussed in the subsequent points.

#### 6.2.5.1 *Information Technology Management Tools*

Previously in the literature review, three different ITM tools were briefly discussed to implement the findings of the data collection. The decision regarding which ITM tool, were meant to both be based upon which factors are deemed important in the data collection, as well as information gathered from the semi-structured interviews. However, as mentioned earlier in the limitations-section, no interviews were conducted, which leaves the ITM tool decision based purely on the findings in the thematic analysis and the literature. Nevertheless, the themes in the analysis are not considered internal factors, they are rather factors focused towards growth. This makes it harder to discuss an Information Strategy, as the growth factors are not sufficient to construct such a strategy. However, what the thematic analysis has is trends based upon historic events, which can be understood as knowledge.

This type of knowledge is in essence beneficial in the sense of all three chosen ITM tools. For example, a decision support system can be created to collect more data around the field of eSports; a knowledge management system can be constructed based on the knowledge gathered; and these two systems can be a part of an overall Information Strategy of either an eSport company or the business as a whole. To further answer research question 5, a brief discussion based upon the literature review and the collected data is constructed to understand how ITM tools can be used.

#### *6.2.5.2 Using Information Technology Tools in the Field of ESports*

As defined in the literature review, Knowledge Management is understood as an essential part to either a business or a company in the manner of employee performance and corporate competiveness. For the benefit of eSport, a specific system must be implemented to collect necessary knowledge, so as to facilitate an understanding of what is pushing it forward, alike the factors found in the data collection. In addition to the definition given in the literature review, Frost (2015) argues that a system as a Knowledge Management System makes it possible to "[…] store and retrieve knowledge, improve collaboration, locate knowledge sources, mine repertories for hidden knowledge […]". As this research thesis mainly looks at eSports events, the knowledge gathered would be somewhat similar to what was collected in the thematic analysis; however, that collected data for this thesis is only used to give an indication. Nevertheless, the data that is involved within the thematic analysis is understood as explicit knowledge, which can easily be found by automated systems and translated into text.

The actual collection of the data highly correlates with the Decision Support System (DSS) tool. Although both collection and storing knowledge can be called a Knowledge Management tool, the techniques are rather based upon DSS theory. Storing knowledge is mainly possible by using 'data warehouses' (see glossary), and further by using various DSS techniques to automatically collect the data.

An Information Strategy however, would serve a holistic plan for the future of eSports, giving it direct guidelines on how to operate the next five-ten years. Within the Information Strategy, there would be a plan to construct a Knowledge Management system, which is based upon an automated collection of knowledge to gain a futuristic understanding of the impactful factors in eSport events. To collect such knowledge, various tools based upon 'data mining' (see glossary) and 'data warehouses' would be required, which leads to the required system. These concepts are known as Decision support systems, which can work as the underlying core of the suggested ITM approach.

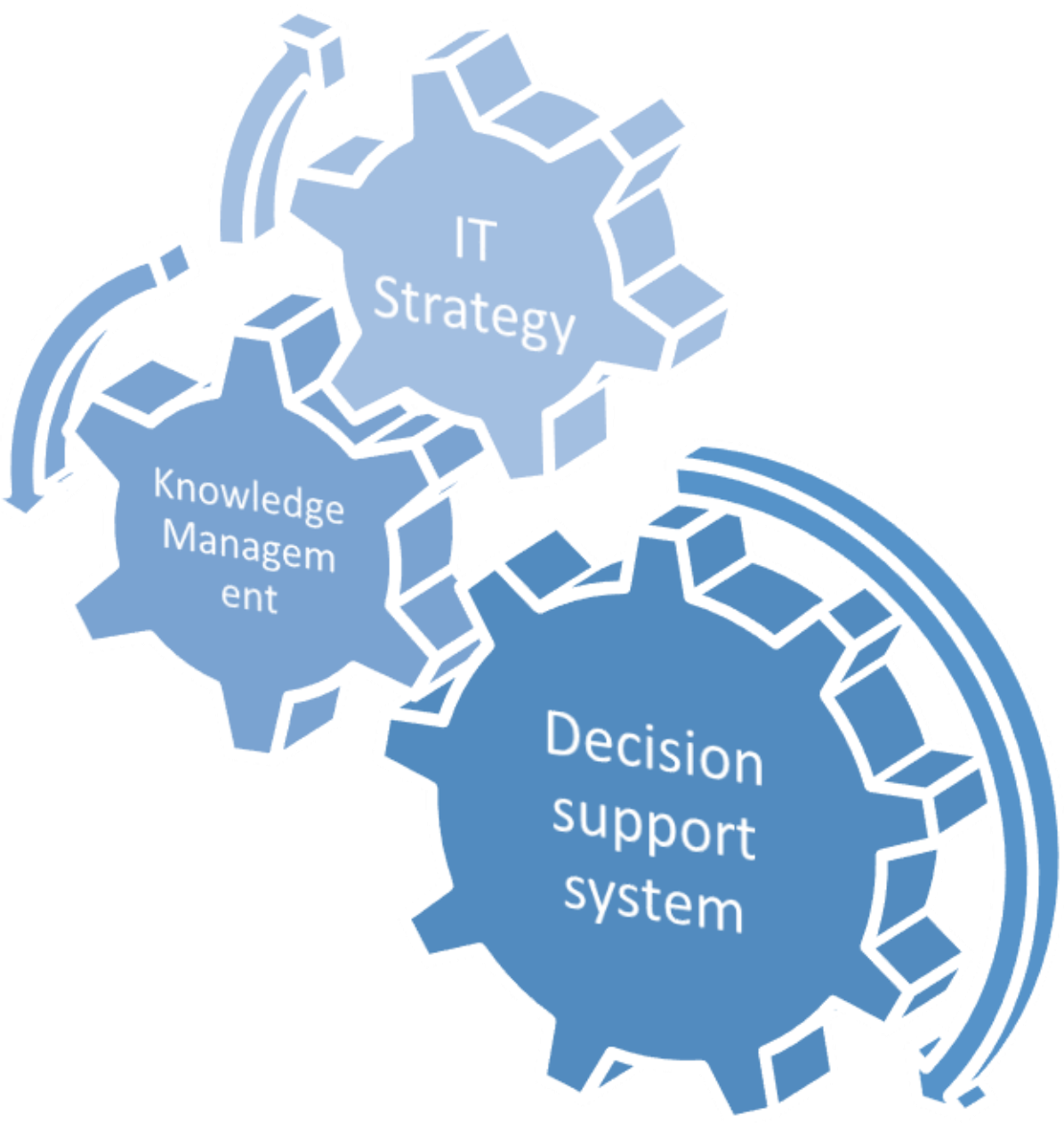

*Figure 6.1 Decision Support Systems as a working core*

With the limitations considered, as well the size of the paper, the DSS is further used to answer research question 5 and 6. The decision is based on three essential factors, which will work as the introduction for the last research question:

- The literature collected indicated significant improvement possibilities by implementing a DSS, where 81% faster and more accurate reporting is considered the most essential one (Turban et al. 2008). This stands in high relevance in the wanted system, as specific reports on trends need to be analysed to get an understanding on the market.
- Although both an Information Strategy and a Knowledge Management System is considered beneficial, they cannot be done without underlying internal and external factors. A DSS can however help extract these factors, making it possible to fully make decision and construct systems based upon the findings of such a proposed system.
- The findings and recommendations that is presented in table 6.5 highly correlate with the expected findings of a DSS. Using a system to gather this information, rather than gathering them solely, makes it possible to increase the number of data gathered, which might further increase the validity of the data.

To conclude research question 5, the theory argues that it is possible to gain benefits from using ITM tools. However, of the three tools considered tools, a DSS should be constructed first so it can serve as a main substance for further techniques. The actually outcomes of this is further discussed in research question 6.

### 6.2.6 Research Question 6: Conclusion

*Which outcomes would the most beneficial Information Technology Management tools give in the eSport business?*

As research question 5 stated, the DSS could be seen as the most beneficial ITM tool based upon the gathered data and theory. However, before the outcomes can be discussed, an understanding of how the DSS would function is required.

### *6.2.6.1 Implementation of a Decision Support System*

A DSS is implemented around solving one specific problem, before designing an actual concept. In total, a system is constructed around four steps based upon a theory, which originated from Simon (1977). This theory is further analysed and deliberated by Gidel et al. (2008), and their updated model is further used as showed in figure 6.2. Each step consisting within this model is further discussed to answer research question 6.

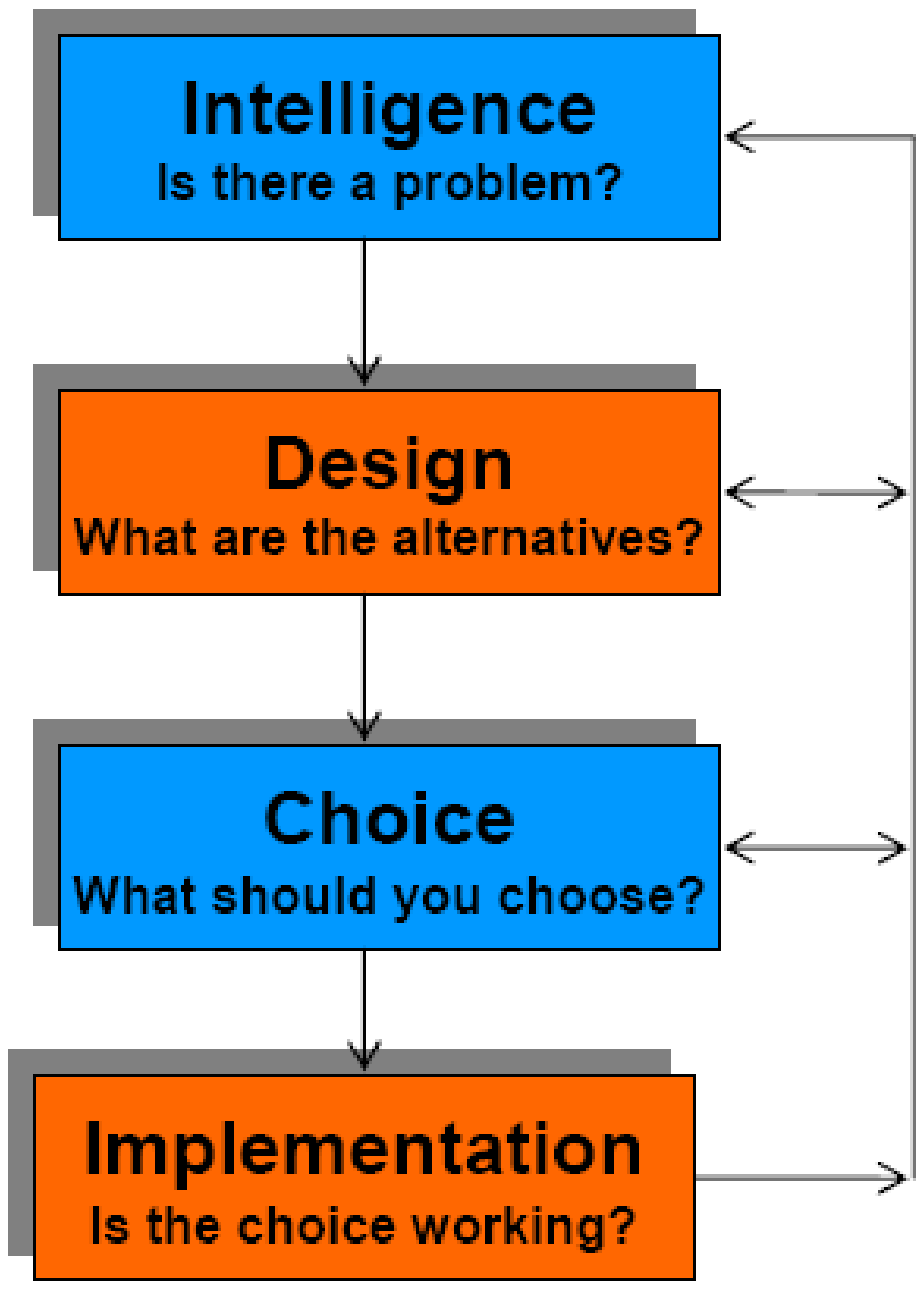

*Figure 6.2 Simons 4-phase model of decision making process (Blogwind 2004)*

#### 6.3.6.1.1 Phase 1: Intelligence

The problem exists in the sense that eSport is experiencing an intense growth, both in audience and its leading themes. The problem will revolve around building a system that enhances the growth seen in the best possible way, using past history to gain an understanding on what eSport events should focus upon in a futuristic approach.

#### 6.2.6.1.2 Phase 2: Design

The Design phase can be understood in Turban et al. (2008) definition, which is where the possible actions are developed. As previously touched upon, the action for this DSS is to collect various knowledge to give clear information to the decision-makers. Within a DSS, there are five specific methods to use, as briefly touched upon in the literature review: Model-Driven DSS, Data-Driven DSS, Knowledge-Driven DSS, Documents-Driven DSS and Communication driven DSS. The main difference of these DSS's is what source of knowledge the systems is based upon, as well as how it will be collected. This can further be seen under Kopáčková and Škrobáčková's (2006:102) model shown in figure 6.3, which briefly explains how various Business Intelligence tools, is used in the mentioned DSS.

#### 6.2.6.1.3 Phase 3: Choice

The aforementioned phases show that historic data will stand in the essence of the chosen DSS, meaning that data mining will be influential. In addition to the previous definitions, data-mining tool (also known as knowledge discovery) is defined by Turban et al (2008:305) as “a process that uses statistical, mathematical, artificial intelligence and machine learning techniques to extract and identify useful information and subsequent knowledge from large databases”. A thematic analysis is a mathematical example of this, but a DSS rather does this automatically rather than manually, as done in this research thesis. Based upon the need for data mining, only two Decision support systems are possible to choose from, according to Kopáčková and Škrobáčková’s (2004: 102) model, which is either a Data-Driven DSS or a Knowledge-Driven DSS.

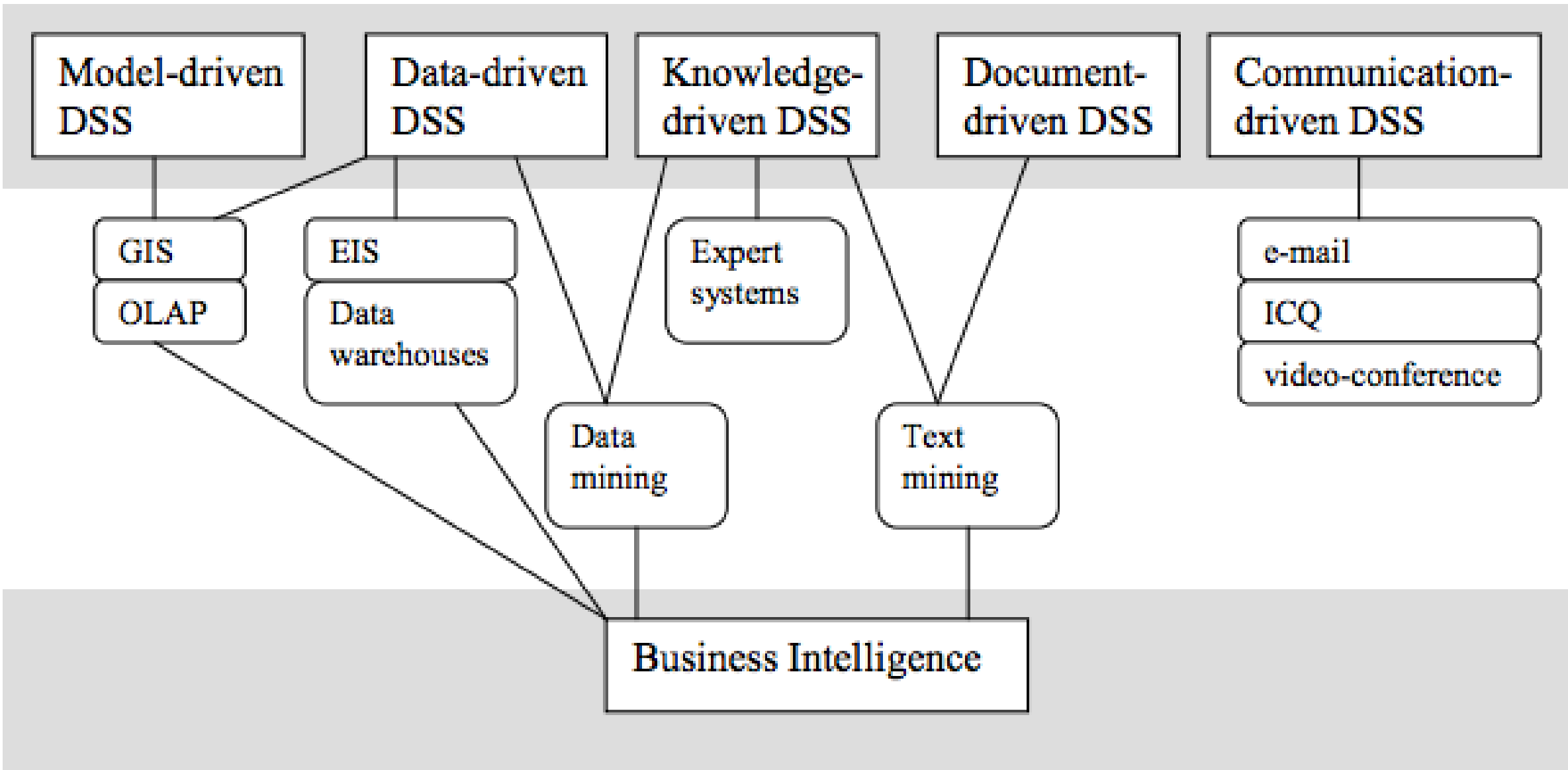

*Figure 6.3 Components of BI and DSS (Kopáčková and Škrobáčková 2004: 102)*

Knowledge-Driven DSS refers to a specific expertise in one particular domain (Powel 2015a). An example of this could be just focusing on one specific theme, and search for hidden patterns within it. Data-Driven DSS however, can be understood as the name suggest, as a system driven by various data. Powel (2015b) defines it as a DSS that “emphasizes access to and manipulation of a time-series of internal company data and sometimes external data”. It uses data existing mainly in data warehouses, and use it for analysis. Powel (2015b) further argues one essential point in the decision step for this specific problem, which is that Data-Driven DSS together with On-line Analytical processing (OLAP) delivers unsurpassed functionality when it comes to analysis of historical data – which is the essence of the Intelligence problem. Considering the definitions, as well as the connections to OLAP, Data warehouses and data mining, the Data-Driven DSS is the chosen system deemed most beneficial for the chosen problem.

##### 6.2.6.1.4 Phase 4: Implementation

The last step includes Implementation, which regards how well such as system might perform. As this is only a discussion based upon theory, no actual proven benefits can be summarised, other than the overall benefits of using a DSS discussed in the literature review. Nevertheless, if a Data-Driven system were to be constructed as a part of a knowledge management system, striving towards understanding the field of eSport based upon previous events, the possible outcomes are many, and this is exactly what research questions six concerns. To understand the possible outcomes of implementing a DSS, its strengths and weaknesses are further discussed.

#### *6.2.6.2 Strength and Weakness of Implementing a DSS System*

When analysing the strengths and weaknesses of implementing a DSS system, there exists various techniques and tools to aid the analysis. These tools is often referred to as either macro-analysis or micro-analysis, as they seek to understand either the internal or the external factors of a company or a product (Bähr and Lindar 2013). In most cases, a SWOT analysis is used, which briefly look at the strengths, weaknesses, opportunities and threats (Srdjevic et al. 2012). Techniques that focus more globally could be used as well, such as PESTLE or McKinsey 7S's, however; the external factors they consider are not deemed that important in the implementation of a DSS. Considering the theory and information gathered in this research thesis, a SOAR analysis was deemed the best to understand the positive aspects of implementing a DSS (Stavros and Hinrichs 2009). Like SWOT, SOAR analyses the external attributes of opportunities and the internal attributes of strengths. However, instead of looking at harmful attributes such as threats and weaknesses, it rather looks at these in a more positive light, naming them aspirations and results. Nevertheless, to further look upon the possible threats and weaknesses, a PESTLE analysis is made. The PESTLE analysis is used to look at a specific idea or company in a bird's-eye view, analysing the external factors on a macro level (Srdjevic et al. 2012). Each analysis is further completed in the subsequent points, as shown in figure 6.4 and 6.5 before research questions 6 is further discussed.

### *6.2.6.3 SOAR analysis*

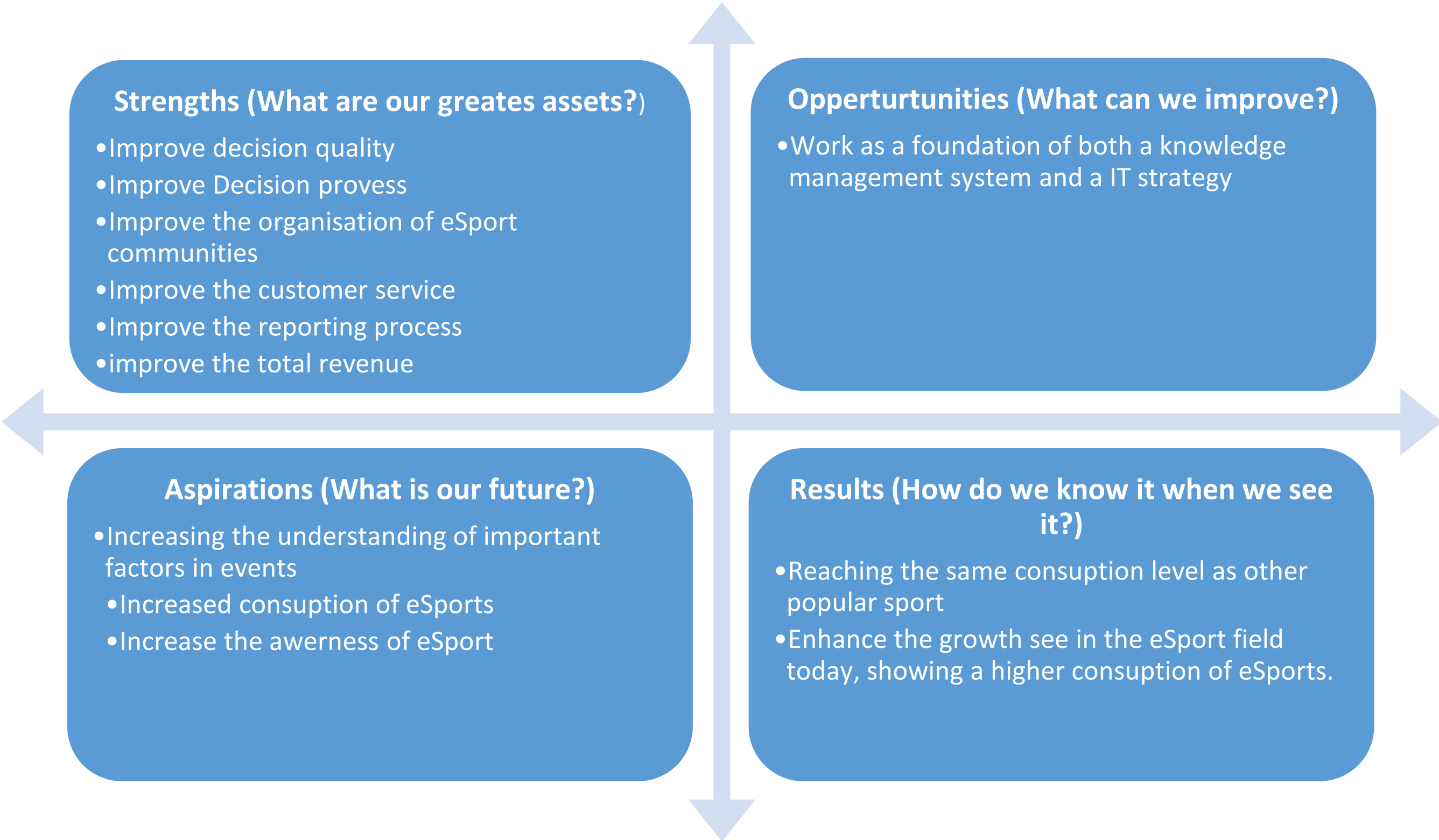

*Figure 6.4 SOAR Analysis of the implementation of a DSS system (Davenport et al. 1997; Bergeron's 2003; Turban et al. 2008; Holsapple and Sena 2005; Pick 2008)*

#### 6.2.6.3.1 Strengths

The strengths of a DSS system will involve the set of attributes it will give the field of eSports. This includes the previous findings, where the most important finding is the improved decision quality and diction process within the eSport community. Secondly, a DSS in theory will make the eSport community more organised, knowing both the pasts mistake and the futuristic trends and patterns. Lastly, a DSS system works towards improving the customer service, the reporting process, and the total revenue seen within the eSport community.

#### 6.2.6.3.2 Opportunities

The opportunities of implementing a DSS revolves around the possibility to further build on its findings. The patterns and factors found within the DSS can further be used into a knowledge management system or an Information Strategy, further improving the organisational level of the eSport community.

#### 6.2.6.3.3 Aspiration

The Aspiration of a DSS system involves an increased understanding of the important event factors, making it possible for the managerial level to construct successful eSport events. This will further lead to an increase of eSport consumption, and an overall growth in the eSport community

#### 6.2.6.3.4 Results

A DSS will aspire to give sufficient knowledge to improve the decision progress when making eSport events. This will further work towards the growth of eSport, making it possible to have the same audience as other popular sports seen today at the brightest, or at least enhance the growth.

### *6.2.6.4 PESTLE*

The PESTLE analysis is further used to see the possible negative aspects of implementing a DSS in a macro perspective, based upon the data gathered in this research paper. This includes consideration of the political aspects, the environmental aspects, the social aspect, the technologies aspects, the legal aspects as well as the economic aspects (Srdjevic et al. 2012). Each aspect is further discussed individually in the subsequent points.

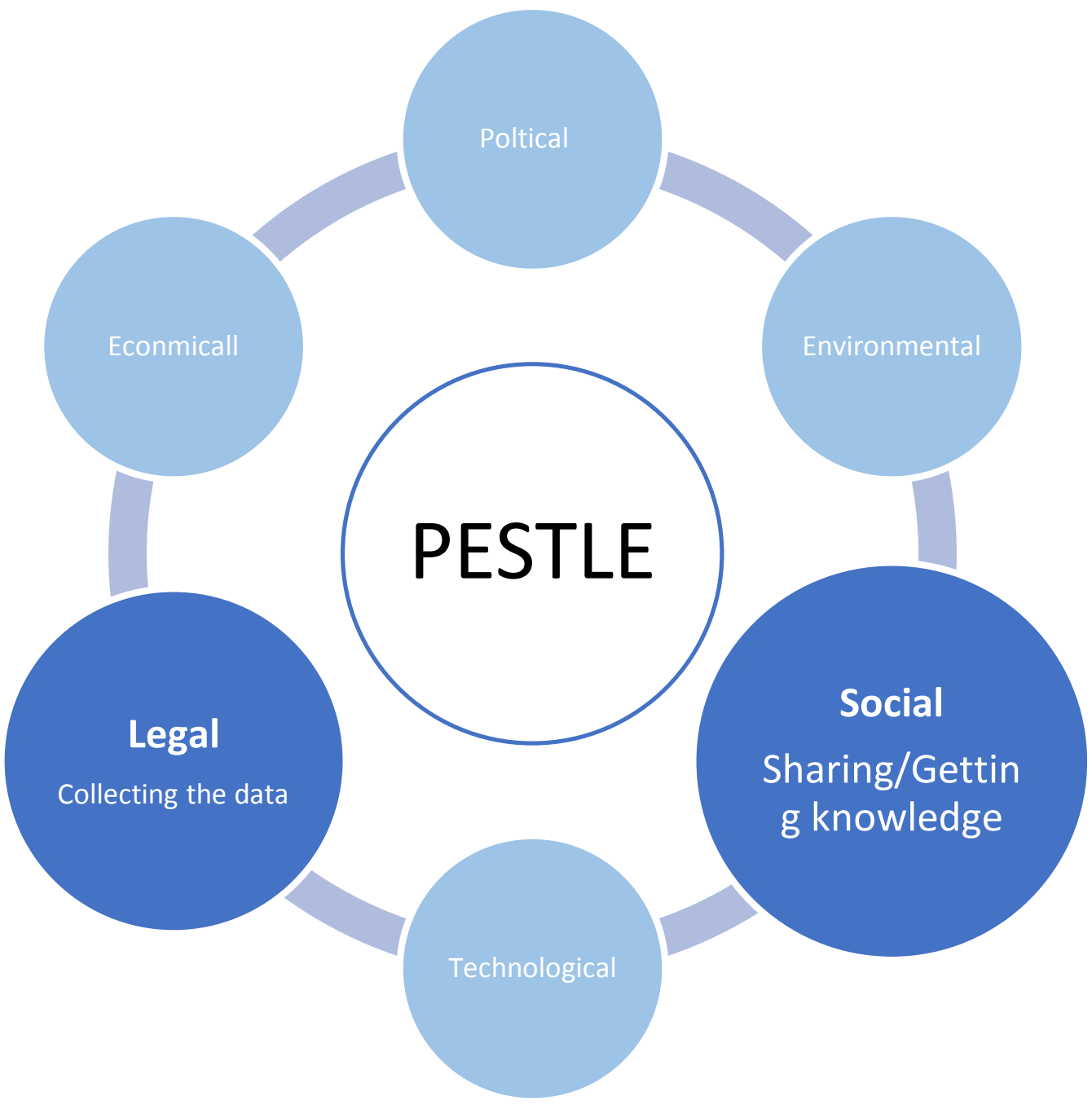

*Figure 6.5 PESTLE analysis (Srdjevic et al. 2012)*

##### 6.2.6.4.1 Political:
None political factors is deemed important when implementing a DSS.

##### 6.2.6.4.2 Environmental:
None economic factors is deemed important when implementing a DSS.

##### 6.2.6.4.3 Social:
As the literature review briefly mentioned during the discussion of knowledge management, people tend to not share their tacit or explicit knowledge due various reasons. Nevertheless, as this is financial data it is deemed less like for it to occur, but it is a threat.

##### 6.2.6.4.4 Technological:
Although a DSS is based upon technology, none technological factors is deemed important when implementing a DSS. An improved system might however be invented, but this will rather enhance the effect of the DSS, rather to work as a threat.

##### 6.2.6.4.5 Legal:
The Legal factor is deemed somewhat important when implementing a DSS. This is mainly caused by the collection of the data necessary to define the patterns in the eSport; this data might be held by various events not willing to share the data – making it impossible to collect the data. However, as the collection of the data is believed to be in the best interest for all this is not likely to happened, but is considered a threat.

##### 6.2.6.4.6 Economical
None Environmental factors is deemed important when implementing a DSS.

#### *6.2.6.5 Summary of Research Question 6*
Although the question is analysed in a high detail, it is not possible to producing a precise answer of how beneficial it will be to implement an ITM tool due of lacking numerical data. However, the various findings give a clear indication that using a tool such as DSS might provide various strengths and opportunities, enhancing the growth seen in the field of eSports. The strengths will be based upon the decision-making, which makes it easier for the managerial level to construct successful eSport events, which eventually might cause an increase of both consumption and size within the field of eSports. However, as pointed out in the PESTLE analysis, this will only be possible if all the organisations within the eSport community co-operate. Nevertheless, the theory and findings indicate that a Decision support system might improve, even heavily improve, the field of eSports if implemented. A further example of the findings in the prior examples is presented in table 6.4, where a fictional event is created based on the trends previously analysed. This sort of results is exactly what a data-driven DSS can provide, if sufficient data is given.

### 6.3 Final Recommendations

Based upon the previous findings and discussion, a set of recommendations can be done. Before doing so, it is important to reiterate this research paper's aim, which is to "define and measure the key factors that caused the growth of eSports, and further appraise the benefits and disbenefits of implementing Information Technology Management tools to maintain, and enhance its growth". The final recommendation can further be understood as a short-term recommendation and a long-term recommendation. The short-term recommendation will focus on the first four research questions, while the long term further focus on the last 2 research questions.

#### 6.3.1 Short-Term Recommendation

The first recommendation is based upon how eSport organisations and communities can create events based upon the findings in this research thesis. Research questions 3 and 4 concluded that the leading key factors found were prize pool, participating countries and participating players. The remaining factors were also observed as important, all though not deemed equally significant. The trends is summarised in table 6.5, which is according to the data collection, the best possible way to construct an eSport event. The table conclude the short-term recommendation, which is how use the found trends to increase the consumption of eSport events.

| Theme | Recommendations |
|---|---|
| **Games played** | The following games sorted by popularity is essential to include in a eSport event:<br>1. StarCraft (Single Player, RTS)<br>2. Counter-Strike (Team Based Game, FPS)<br>3. League of Legends (Team Based, MOBA)<br>4. DOTA (Team Based, MOBA)<br>5. Call of Duty (Team Based, FPS)<br>6. Hearthstone (Single Player, OCCG)<br>7. Halo (Both, FPS) |
| **Event Duration** | The event duration should be less than 1 month, as this seems to be the standard within the eSport community. |
| **Prize Pool** | The prize pool should be at a minimum be $250,000 - $300,000. The higher prize pool the more audience will be attracted |
| **Focus Globally** | The tournament should invite player to compete from all over the world. This will firstly increase the competition, as well as increase the audience. |
| **Include as many as possible** | The event should invite as many players as possible, as this can help increasing the audience |
| **Location** | If possible, hold the event either in USA, South Korea or China, as these are seen as the capital cities within eSports. |
| **Series, League or tournament** | The event should be a part of a series, league or tournament, as this will attract more popularity |

Table 6.5 The ideal eSport event to be constructed in 2015, based upon both findings in the thematic analysis and the literature review, as well as futuristic trends in the business analysis (Appendix B; Appendix C)

### 6.3.2 Long-Term Recommendation

The trends and patterns presented in table 6.5 is understood as an example of how a DSS can be implemented, and its recommendations could be similar as these findings. However, the limits for an automated system is much more, making it possible to include more findings making the data more valuable, as well as including more themes deemed necessary. Based upon these findings, a further recommendation on implementing a Document-Driven DSS is completed. Making such a system will in theory help the eSport community construct better eSport events, knowing exactly what the leading factors and how to take advantage of these, at any time. It additionally makes it possible to use real-time data, meaning that it will be easier to the trends happening right at this very moment. The DSS can further be a foundation for other ITM tools such as a Knowledge Management System and an Information Strategy, that together work towards an enhancement of growth in the long-term aspect.

### 6.4 Conclusion of the Research Thesis

With the gathered data, this research thesis can conclude that the field of eSport is growing, and it will continue to grow in the future. The thesis summarised 444 events, which can be defined as the most important events in eSport history, where it all originated in 1981, when the first Space Invaders tournament was held. To this date the size of events and the overall consumption has increased profoundly, which can be shown by various events such as The International tournament, The League of Legends championships or the DreamHack digital festivals.

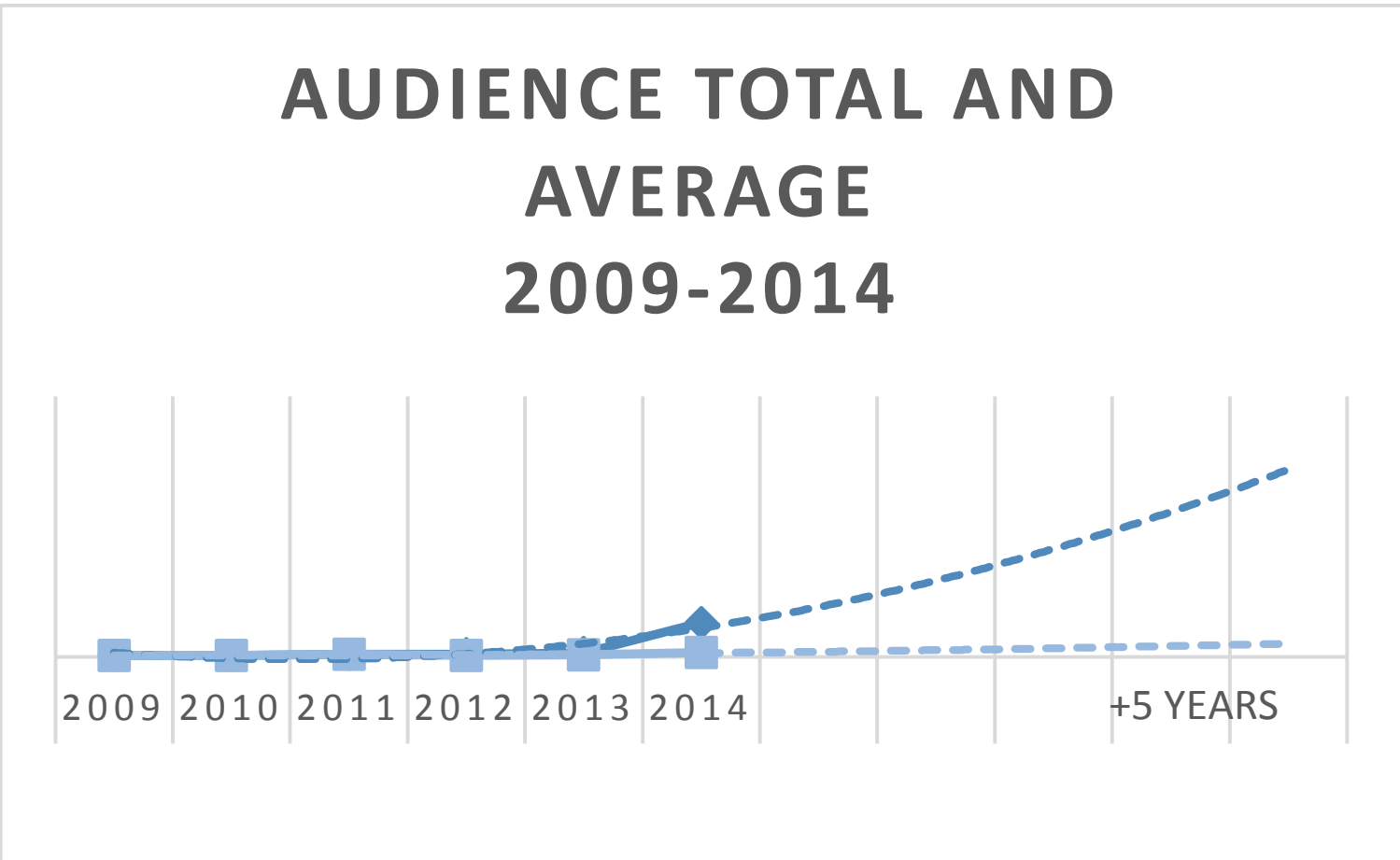

*Figure 6.6 Visual Trend line with highest $R^2$ of live Audience 2009-2014 (Appendix B)*

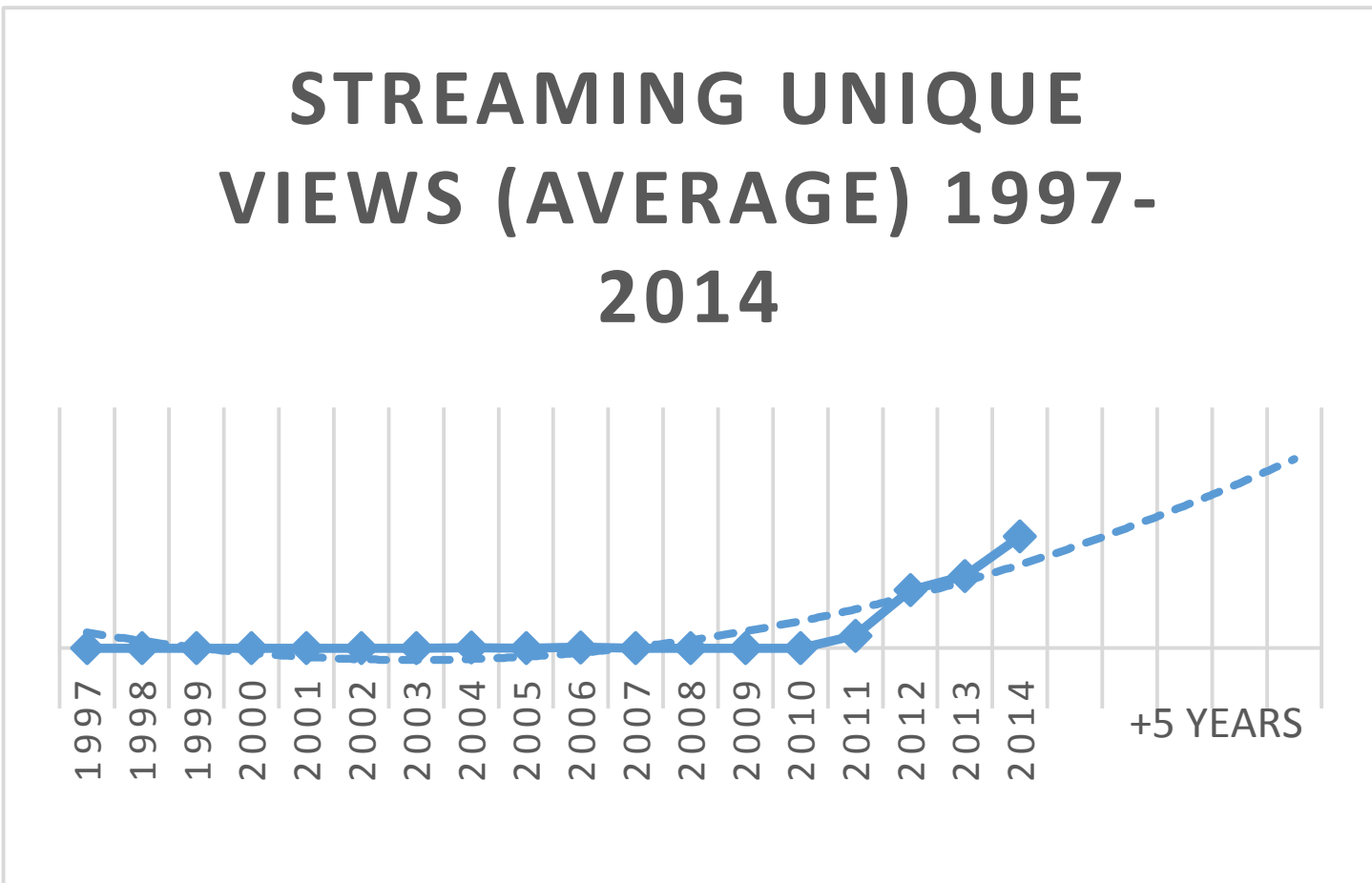

*Figure 6.7 Visual Trend line with highest $R^2$ of Unique viewers 1997-2014 (Appendix B)*

The data collection further proved there are various factors that push the success of the events, where prize pool and participation (in both persons and countries), were deemed important in the sense of live audience. Streaming however had no significant correlation, making it impossible to include any findings on that part. Figure 6.6 and 6.7 present some of the findings shown in appendix C and D, which are indications of the trend lines that are to be seen within the field of eSports. These trend lines indicate that the future of eSport events is bright, making the potential of the eSport community likely to grow. Although lacking some data and credibility within the collected data, the thesis concludes that the best method to ensure the growth in the future is to construct a data-driven decision support system, which will work on equal lines as the thematic analysis, just scaled immensely in the sense of size. This will increase the quality of the knowledge gathered, making the decisions and data more accurate, and therefore more likely to find the patterns within the field of eSports. This will further work as a foundation for a knowledge management system and an information strategy, the inclusion of which is highly recommended.

With the final recommendations given, the research thesis will further evaluate its findings and recommendations. This is done in chapter 7, where each objective, deliverable and research question is measured against the previous set requirements. However, before this is done an indication of what can be done further in this specific field will be given under the subsequent topic 'Future works'.

### 6.4.1 Future Works

The findings in this research thesis give various indications within the field of eSports, of which the biggest is the futuristic growth within the field. However, as indicated previous there exist a both lacking academic literature as well as lacking credibility on the collected data. This relates to first step that can be seen within the future works, which is to increase the amount of academic research in the field of eSports. This includes collecting additional primary data, which the planned semi-structured interviews can be shown as an example off (Appendix I). In addition to this, only three tools were discussed, due to limitations. For future work, various research on the effect of using different ITM tools would be beneficial to enhance the growth seen in the field of eSports. An additional future work would be to construct a more detailed plan for the implementation of a Decision Support System in the field, and getting more numeric data on how much it might enhance the growth seen in the field of eSports. All this will increase the knowledge in the eSport field, which further will increase the awareness as a whole, and expectantly enhance the growth of eSport.

### 6.5 Summary of Chapter 6

Chapter 6 concluded each research question based upon the findings of the literature review, the thematic analysis, and its underlying business analytic findings. The newly answered research questions firstly showed that the given eSport definition in the literature correlated with what could be found in the data. As for the awareness, it is best understood by the data collected from the literature review. Nevertheless, the thematic analysis also gave indications of an increase in awareness of the lifespan of eSports history. Question 2 and 3 indicates a list of 444, 83 and 23 events, which are seen as the most important eSport events in history. These events are further showed to have a pattern of success caused by either being a part of a gaming series, playing some specific games, game genres or team-playing games, having a certain number of participating players, or having a high prize pool for the event. These factors are further used to define what can be done to ensure growth, which concerns question 4. The growth of eSport seems existing, considering the data indicates a growth in live audience of 305% from 2014-2019, while in streaming a growth of 140.61% from 2014-2019 (Appendix B). The most significant factor, prize pool, was also shown to have a nearly equal growth as seen in the streaming services, although these are not significant correlative. These findings is further concluded as the long-term recommendation; creating an eSport event based on the stated findings in research question 1-4. Research question 5 and 6 however, concerns constructing a Data-driven Decision Support System, based upon the various findings. This system is examined, and a measure of consideration is put into how it might be implemented based upon Simon's (1977) 4-phase model of decision-making processes. This is also the final long-term recommendation for the field of eSport: to implement a DSS system that will work as a foundation for a Knowledge system and an Information Strategy, which further will work towards enhancing the growth in the field of eSports. Chapter 6 is however not critically evaluating the data used within the research thesis; this is performed in the next and final chapter, named Project Management and Quality Assurance.

# CHAPTER 7 PROJECT MANAGEMENT AND QUALITY ASSURENCE

## 7.1 Chapter 7 Introduction

The last chapter concluded the thesis with its recommendations, but lacked a critical evaluation of its findings. Chapter 7 intends to do just this, by evaluating the credibility of the collected data, existing by its reliability and validity. This is done by measuring how well the aim, the deliverables, the objectives and the research questions are answered. This chapter will therefore intend to discuss the various instruments employed, and to discuss the effectiveness of the results gathered. Chapter 7 also includes the project management tools used within this study, and further evaluates their functionality. The chapter is concluded by the overall lessons learned from the critical evaluation.

## 7.2 Critical Evaluation of Project Conduct

This research thesis started as a heavily self-motivated paper on the field of eSports. This field has a lack of academic writing, which firstly made it interesting to construct the thesis, but as noticed in the literature review it also got challenging. The thesis strove towards meeting the aim, which is "*to define and measure the key factors that caused the growth of eSports, and further appraise the benefits and disbenefits of implementing Information Technology Management tools to maintain, and increase its growth".* To evaluate how well the research paper answered the aim, several measurements will be done. Firstly, a discussion on how well the objectives, deliverables and key research questions is achieved, will be done. This is followed by a brief discussion of the chosen methods, the ethical approach and the chosen project management tools. Lastly, a conclusion is given based upon what was learned, as well as the effectiveness of the experience.

### 7.2.1 Critical Evaluation of Objectives and Deliverables

To attain an understanding on how well the research paper answered the aim, the objectives and deliverables are summarised in table 7.1. The table presents that only one deliverable could not be done, while four deliverables were only answered to some extent. These deliverables are further deliberated over, as they have an impact of the overall project conduct. However, firstly an understanding is needed on which deliverables have been achieved, and how and where they were achieved.

| ***Deliverable*** | ***Objective*** | ***Achieved*** | ***Where / How*** |
|---|---|---|---|
| *(D1) A complete literature review by week 19* | *OB 1* | *Yes* | • *Chapter 2: Literature review* |
| *(D2) A summary of the major findings within the eSport community by week 20* | *OB 1* | *Yes* | • *2.5 and its underlying points* |
| (D3) A Complete and synthesised definition of eSports by week 16. | OB 2 | Yes | • 2.5.1 Research Question 1: Literature discussion:<br>• 6.2.1: Research Question 1: Conclusion |
| (D4) A summary with the milestones of eSport history by week 19. | OB 2 | Yes | • Appendix A (Thematic Analysis) (444 Events)<br>• Appendix H (Time Line) (23 Events)<br>• Table 5.4 (83 Events) |
| (D5) A visual timeline of eSports historic data by week 20. | OB 2 | Yes | • Appendix H (Time Line) |
| (D6) A summary of the most important series and leagues within the eSport community by week 20. | OB 2 | Yes | • Appendix A (Thematic Analysis) |
| (D7) A report on the major eSport events by week 20 | OB 2 | Yes | • Appendix A (Thematic Analysis) |
| (D8) A list of the main financial factors/trends within eSports by week 21 | OB 3 | Yes | • Appendix (Thematic Analysis)<br>• Appendix D (Trend Analysis)<br>• 5.4.4 Analysis: Research Question 3 Findings<br>• 6.2.3 Research Question 3 Conclusion |
| (D9) A comparison of financial factors to differentiate their impact by week 22 | OB 3 | Yes | • 5.4.4 Analysis: Research Question 3 Findings |
| (D10) A brief explanation of the financial factors by week 22 | OB 3 | Yes | • 4.3 Thematic Analysis |
| (D11) A critical analysis of the most important success factors in previous eSport events by week 25. | OB 4 | Yes | • Appendix A (Thematic Analysis)<br>• Appendix B (Findings not used)<br>• Appendix C (Trend lines)<br>• 5.4.4 Analysis: Research Question 3 Findings<br>• 6.2.3 Research Question 3 Conclusion |
| (D12) A time line of the most important eSport event by week 25. | OB 4 | Yes | • Appendix H (Time Line) |
| (D13) A statistic correlation test to see if there is any common patterns between the events by week 25. | OB 4 | Yes | • 5.4.4.2 Statistical data |
| (D14) An interview with some of the important actors within eSports by week 25 | OB 5 | No | • Appendix I |
| (D15) A synthesised report of the comparison of the financial factors by week 26 | OB 5 | Yes | • 5.4.4 Research Question 3 Findings<br>• 5.4.5 Research Question 4 Findings<br>• 6.2.3 Research Question 3 Conclusion<br>• 6.2.3 Research Question 4 Conclusion |
| (D16) A juxtaposed summary of the most important features within eSport by week 26 | OB 5 | Yes | • 6.2.4.1 What can be done to ensure growth<br>• 6.3 Final Recommendations |
| (D17) A juxtaposed summary of the causing factors of the recent growth within eSport by week 27 | OB 5 | Yes | • 5.4.5 Research Question 4 Findings<br>• 6.2.4 Research Question 4 Conclusion |
| (D18) An in-depth analysis on how many people watches eSports and it overall awareness by week 20. | OB 6 | To some extent | • 5.4.2 Research Question 1 Findings<br>• 6.2.1 Research Question 1 Conclusion |
| (D19) A report based upon the overall viewership of eSport, both consisting of live audience and streaming audience by week 27. | OB 6 | Yes | • Appendix A (Thematic analysis) through theme 'Streaming Audience' and 'Live Audience' |
| (D20) An in-depth analysis of the economic future of eSports based upon both statistical data and trend lines made from business analytics, as well as the literature review by week 28 | OB 7 | Yes | • 6.2.4 Research Question 4 Conclusion |
| (D21) A set of prediction of the finical situation of eSport based upon findings by week 28 | OB 7 | To some extent | • Appendix D (Trend line analysis) |
| (D22) A brief recommendation on how eSport can improve its financial value by week 30 | OB 7 | Yes | • 6.3 Final Recommendations |
| (D23) A complete definition of Information technology, and its tools, by week 21 | OB 8 | Yes | • 2.5.5 Research Question 5 and 6 and its underlying points |
| (D24) A critical evaluation of the financial effect of the implementation of Information | OB 8 | To some extent | • 6.2.5 Research Question 5 Conclusion |
| Technology Management tools by week 29 | | | • 6.4 Conclusion of Research Question 6 |
| (D25) An evaluation of which ITM-tools implementations that can be done within eSport by week 26 | OB 8 | Yes | • 6.2.5 Research Question 5 Conclusion<br>• 2.4.5 Research Question 5 |
| (D26) A list of benefits and disbenefits of the implementation of ITM-tools in eSports by week 26 | OB 9 | Yes | • 2.5.5 Research Question 5 & 5 and its underlying points. |
| (D27) A critical evaluation of how various ITM tools might improve the growth of eSport, and its effects by week 27 | OB 9 | To some extent | • 6.2.5 Research Question 5 Conclusion |
| (D28) A summary of the data gathered in the dissertation by week 31 | OB 10 | Yes | • 5.6 Chapter 5 Conclusion<br>• 6.2 Conclusion of Research Questions |
| (D29) A list of recommendation based upon the data gathered by week 31 | OB 10 | Yes | • 6.3 Final Recommendations |
| (D30) A evaluation of the cause and effect of the recommendations made by week 32 | OB 10 | To some extent | • 6.2.6.2 Strength and Weaknesses of Implementing a DSS system |
| (D31) A Project Management and Quality Assurance of conduct of project by week 32 | OB 11 | Yes | • Chapter 7 |
| (D32) A conclusion of the credibility of the data found by week 33 | OB 11 | Yes | • 5.2 Credibility of the thematic analysis<br>• Chapter 7 |
| (D33) A critical evaluation of all the findings and writings by week 33 | OB 11 | Yes | • Chapter 7 |

Table 7.1 Summary of achievement in regard of the objectives

#### *7.2.1.1 Critical evaluation of the deliverables*

Each objective is further evaluated based upon the findings in table 7.1. These findings is used to gain a comprehensive understanding of the accomplishments of the Research Questions. However, before the discussion of the objectives is done, a further explanation of the not achieved deliverables is given in table 7.2.

| **Deliverable** | **Explanation** |
|---|---|
| *Deliverable 14*<br>An interview with some of the important actors within eSports by week 25 | As stated in the limitations of the methodology, no interviews were conducted due to a lack of response. This made it impossible to perform D14, as it specifically calls for the conducting of interviews. Additionally, this affects the Objective 5, as without the interviews it is not possible to discover every feature consisting within eSports. Nevertheless, the thesis strove to give the best results based on the data available, making Objective 5 still achievable. However, if a longer time limit existed, the first priority would be to conduct semi-structured interviews. |
| *Deliverable 18*<br>An in-depth analysis on how many people watches eSports and it overall awareness by week 20. | Although some data was gathered on the awareness of eSport, it related more towards the overall consumption. However, the literature review gave a very clear understanding of who watches eSport, as well as how many. As stated in the limitations, this was not studied further due to the size of the paper. This makes the understanding of the awareness of eSport existing, but lacking in primary data. |
| *Deliverable 24*<br>A critical evaluation of the financial effect of the implementation of Information Technology Management tools by week 29 | As for the financial effect of implementing ITM tools, only literature was used within the three chosen tools. This could have been analysed further, and more tools could have been included. However, this was deemed not possible due to the limitations. Nevertheless, a numeric finding on the financial effect was not possible to construct, and therefore this deliverable is only answered to some extent. |
| ***Deliverable 27***<br>A critical evaluation of how various ITM tools might improve the growth of eSport, and its effects by week 27 | Due to limitations, only three ITM tools were considered, and their underlying juxtaposes were brief. Therefore, this deliverable is only answered to some extent, as a lot more analysis and discussion could be done within this topic. This was however known early on, and therefore it has little effect of the achievement of objective 9. |
| ***Deliverable 30***<br>A evaluation of the cause and effect of the recommendations made by week 32 | An evaluation is done of its strengths and weaknesses, but like deliverable 24, no numerical effect calculation could be done. This is something that can be discussed further and analysed, but was named in the limitations as not possible due to the size of this thesis. |

Table 7.2 Summary of deliverables that was either not fully completed or were only answered to some extent

#### *7.2.1.2 Critical evaluation of the objectives*

Based upon the data in table 7.1 and 7.2, a conclusion is made regarding the objectives. It is conducted in table 7.3, to improve the readability of the information.

| **Objective** | ***Achieved*** | ***Credibility*** | ***Critical evaluation summary*** |
|---|---|---|---|
| Objective 1<br>Discover previous research within the field, and summarize their importance for this study | *Yes* | High | Objective 1 is based upon the literature review, which is fully conducted. This objective is therefore concluded fulfilled. |
| Objective 2<br>Define what eSport is, and outline its major historic milestones | *Yes* | High | This objective is based upon the definition given in research question 1, and the findings in the thematic analysis. Considering this, the objected is concluded fulfilled. |
| Objective 3<br>Arrange and compare the main financial factors within eSport | *Yes* | Medium | The limitation of just using eSport events can be considered as a lack of credibility within the thesis. However, as this thesis chose a global perspective, it was the only method deemed feasible. Therefore, this objective is concluded fulfilled. |
| Objective 4<br>Review previously successful eSport events, and use business analytics tools to visualize the common patterns | *Yes* | Medium | This objective is based on the findings in both the literature review and in the thematic analysis, which includes good summaries regarding the objective. However, as stated repeatedly, the data is not primarily gathered by the researcher, which causes a lack of truthfulness and validity of the data. However, the objective is concluded fulfilled. |
| Objective 5<br>Discover the features of eSport that defines the essence of the eSport business | *Yes* | *Low/Medium* | Due to the lack of achievement of deliverable 14, objective five lacks credibility. This objective also lack some credibility as there were only found significance correlations with the live audience and not the virtual audience, as discussed in 6.2.4. However, the research thesis gives an indication of the features of eSport events, making it possible to conclude that objective 5 is answered to some extent. |
| Objective 6<br>*Examine the overall awareness of eSports* | *Yes* | Medium | Affected by deliverable 18, Objective 6 strives to understand the awareness of eSports. The data collected in the thematic and business analytics only answer this question briefly. However, the secondary data used is deemed highly appropriate to define the overall awareness of eSports. Additionally, it was not |
| | | | possible do any more analysis due to the global aspect of this research thesis. It is therefore concluded that objective 6 is fulfilled. |
| <u>Objective 7</u><br>Predict the financial future of eSports | *Yes* | Medium | This objective is based upon the trend lines, and therefore has good content to be answered. However, like previous objectives, both the use of events and using others data affect the credibility of the answer. Nevertheless, based upon the limitations, this objective is still concluded fulfilled. |
| <u>Objective 8</u><br>Evaluate the benefits and disbenefits of implementing ITM tools in the eSport community and synthesise which ITM tools would be most effective | *Yes* | Low/ Medium | Affected by Deliverable 24, this objective is highly influenced by the limitations as previously discussed. However, this objective is believed to give a good answer based upon the data collected, although the credibility is lacking, as shown in deliverable 24. Nevertheless, it is concluded as fulfilled, although with low credibility. |
| <u>Objective 9</u><br>Evaluate the effect of implementing well-justified ITM tools to enhance the growth of eSport | *Yes* | *Low/ Medium* | Affected by deliverable 27, this objective, alike objective 8, lacks credibility. However, this objective accomplishes to be answered well based upon its limitation, and is therefore concluded fulfilled. |
| <u>Objective 10</u><br>Recommend alterations the eSport business might commence to uphold and increase its growth. | *Yes* | *Low/ Medium* | Affected by deliverable 30, this objective lacks numerical data that decrease the truthfulness as well as the validity of the final recommendations. However, considering the limitations, the objective is well discussed. Additionally, it gives recommendations that indicates a growth enhancement. It is |
| | | | therefore concluded that this objectives is fulfilled. |
| <u>Objective 11</u><br>Evaluate and validate the findings, results and recommendations | | High | Due of chapter 7, this objective is concluded to be fulfilled, whereas every underlying deliverable is discussed. |

Table 7.3 Critical evaluation of objectives

### 7.2.2 Critical Evaluation of Research Questions

The key research questions have played an imperative part for every chapter constructed. The previous chapter discussed each question, and concluded how well they are answered based upon both the literature review and the analyses constructed. However, the complete summary of the research questions can be based upon the previous discussion of the objectives. The credibility is additionally based on Patton's (2001) questions concerning the thesis credibility, as were summarised in 3.7. This together with the previous findings is further summarised in table 7.4.

| **Research Question** | **Objective** | **Achieved** | **Credibility** |
|---|---|---|---|
| [1] What is eSports, and how aware are people of the eSport culture? | OB 2 &OB 6 | Yes | Medium/High |
| [2] What are the major milestones of eSports history | OB 2 & OB 4 | Yes | Medium/ High |
| [3] Why is eSport seeing its growth today, and which factors/patterns is pushing it upwards? | OB 3 & OB 5 | Yes | Medium |
| [4] Will the growth of eSport continue, and what can the eSport business do to ensure a continuous growth? | OB 5 & OB 7 | Yes | Medium |
| [5] Which Information Technology Management Tool would be most beneficial for the eSport business to use? | OB 8 | Yes | Medium/Low |
| [6] What outcomes would the most beneficial Information Technology Management Tool give in the eSport business? | OB 9 & OB 10 | Yes | Medium/Low |

Table 7.4 Summary of achieved Research Questions

### 7.2.3 Critical Evaluation of Methodology

This thesis chose to look at the statement of problem from a mixed-methodology perspective, based in a pragmatism philosophy, and an indication approach. This was done owing to the fact that no hypothesis could be constructed early, as there was a lack of data on which trends and patterns that existed within the field. As the research aimed to find the effect of various factors in the field of eSport, the pragmatism is deemed the correct philosophy approach. The data collection depended heavily on qualitative data, which is essential to achieve the wanted deliverables and the aim. However, for future works a numerical approach would be preferable, making more quantitative analysis such as surveys a more appropriate method to use. This can be seen in the understanding of validity and reliability; as no interviews or surveys were conducted, no reliability measurements can be conducted for this thesis, making it overall decrease its credibility. As for the semi-structure interviews, this is something that should be done. The interview were intended to give data that relates to the thematic findings, which can further be discussed to increase the validity of the data immensely. A semi-structure is also believed to be important, as any follow-up questions are essential when understanding such a new and immense field as eSports. It is with this conclusion that conducting semi-structured interviews were part of the future works, as it overall affect the results of the chosen methodology.

### 7.2.4 Ethical Consideration

Various steps were conducted to ensure the ethical quality of the thesis. This included performing an ethical approval form, as well as constructing a participant information sheet and an informed consent form for the interviews. The ethical data is attached under Appendix G. However, it is important to state the importance of the thematic analysis existing of data collected by others. To ensure this is presented ethically correct, the Appendix B, which includes the references of the data collection, is conducted and used whenever their data were used in the thesis. Lastly, the global aspect is considered when talking about ethics, but as only internal factors is researched, no necessary analysis were constructed.

## 7.3 PM and Quality Assurance Tools

Throughout the research thesis deadline, two PM tools were used. The first one is the Gantt chart and the second one were the pilot study of both the thematic analysis and the semi-structured interview. In prudence, a risk management plan could be constructed to understand the risk of not constructing any interviews. However, it was not within this project as it was deemed not necessary, which is further stated in the lessons learnt.

### 7.3.1 Gantt Chart (Appendix F)

As shown in Appendix F, two Gantt charts were constructed. The first included a summary of the project, while the second included timelines of all the deliverables. This made it easy for the researcher to know what to do at which time, and what needed to be done to proceed to a next deliverable. These charts were used throughout the project period, and changed if any unexpected situations were to happen.

### 7.3.2 Piloting of Collection Tools (Appendix E)

As shown in Appendix E, piloting of data collection were done for both the interview and the semi-structured interview. This were done for two major reasons:

- To ensure that each question were understood within the interview
- To ensure that the collected data from both the interview and the analysis is to the expected standards.

In this research thesis, it greatly increase the understanding of the interview, as well as giving a good indication of how to use the results in the thematic analysis by using business analytical tools.

## 7.4 Lessons learnt

In retrospective, several lessons were learned based upon these previous discussions, which is further discussed as subsequent points:

- Having a realistic expectation of the project scope is understood to be imperative. The size of this thesis ended up more than expected, which have caused some deficiency in quality of the conduct. Research Question 1-4 and 5-6 differentiate quite much in the sense of literature and data collection, and each could be understood as a thesis of their own. However, the findings in research question 1-4 is the very reason a Decision Support System is chosen, so the question exists with good reason, although it caused an extended thesis paper in the sense of size.
- Having a global aspect limits the data collection heavily, and makes the required amount of data huge. If this thesis were to be re-done, a smaller area or simply a country would have been chosen, as including the global aspect is not deemed feasible when collecting numeric and quantitative data.
- Being early with the interviews does not necessarily mean that you will get interviews. If a risk management plan had been constructed within this thesis, it would have included the probability of not getting interviews and having an alternative plan. However, with over 25 people to interview, this was not seen as a probability in the start, but it is useful knowledge to have for the future.

## 7.5 Summary of Chapter 7

This chapter discusses how well this thesis answers the overall aim, and further dispute its credibility. It further showed that the overall credibility of the project conduct is slightly lacking, which is mainly due to the limitations of the study. However, every objective is concluded fulfilled, although this is not the case with the deliverables. In this study, one deliverable, which included interviews, was not fully answered, and a further four deliverables were only answered to some extent. Nevertheless, the research questions could still be answered fully, although Research Question 3, 4, 5 and 6 lack credibility, mostly caused by a lack of validity. The chapter further concludes that the methodology used within the research is well chosen, although for the future works more quantitative studies such as surveys could be used to construct a better understanding of the awareness of eSports. Considering this, three lessons are learned: weighing the global aspect, ensuring interviews, and having realistic expectations of project scope as the most important. However, the thesis is constructed well based on the available data, and gives a clear indication of the past and the future of eSports, and how a Decision Support System might benefit the enhancement of the growth, meaning the aim is fulfilled, and thus completing this research thesis.

# CHAPTER 8 REFERENCES

# CHAPER 9 BIBLIOLATRY

# CHAPTER 10 APPENDIXES

## Appendix A Thematic Analysis: Documents

### Appendix A I. Excel Document

| Number | Year | Event | Series | Where | Participants Total | Participants Finalists | Participant Countries | Global, EU, Asia, OC, Africa or NA? | Live Audience | Sessions | Streaming (Total views) | Streaming (Unique viewers) | Streaming (Peak concurrent viewers) | Television Coverage | Paper Coverage | Prize pool | Game | Total Number of Games | Game Genre | Single player (SP) or team based (TB)? | Sponsors | Length (Months) | Extraordinary events |
|---|---|---|---|---|---|---|---|---|---|---|---|---|---|---|---|---|---|---|---|---|---|---|---|
| 1 | 1980 | First National Space Invaders Competition | - | New York, USA | 10,000 | | | | - | | | - | | Yes | Yes | | Space Invanders | 1 | Arcade | SP | - | - | First ever eSport event; A lot of celebreties within the field |
| 2 | 1981 | National Video Game Championship (Atari Coin-Op $50 000 World Championship) | - | Chigago Exposition Center, Chicago, Illanouis, USA | 200 | | | | | | | | | | | $50,000 | Centipede | 1 | Arcade | SP | - | - | 200 (expected between 10,000 - 15,00) |
| 3 | 1982 | North America Video Game Olympics 1982 That's Incredible Ms. Pac-Man Tournument | NA VGO | Ottumwa, Iowa, Missisippi, USA | - | 44 | - | NA (USA) | - | - | - | - | - | - | - | $500 | Ms. Pac-Man | 1 | Arcade | SP | - | <1 | |
| 4 | 1983 | North America Video Game Olympics 1983 That's Incredible! | NA VGO | Ottumwa, Iowa, Missisippi, USA | - | - | - | NA (USA) | - | - | - | - | - | - | - | - | Super Pac-Man, Donkey Kong Jr., Milipede, Joust, Burger Time, Buck Rogers, and Frogger | 8 | Arcade | SP | - | <1 | Opening parada and a some live entertainment |
| 5 | 1983 | The Electronic Circus | - | Various | - | - | - | NA (USA) | - | - | - | - | - | - | - | - | 500 Different Arcade Games | 500 | Arcade | SP | - | 4 | Various, Including Live entertainment |
| 6 | 1983 | The National Video Game Masters Tournument '83 | Video Game MastersTournument | Various | - | - | - | NA (USA) | - | - | - | - | - | - | - | - | Joust, Defender, CongoBongo, Centipede, Ms Pac-manGalage, Q-bert | 7 | Arcade | SP | - | - | - |
| 7 | 1983 | North America Video Game Olympics 1983 | Ottumwa, Iowa, Missisippi, USA | - | - | - | - | NA (USA) | - | - | - | - | - | - | - | - | - | - | - | - | - | - | - |

| | | | | | | | | | | | | | | | | | | | | | | | | |
|---|---|---|---|---|---|---|---|---|---|---|---|---|---|---|---|---|---|---|---|---|---|---|---|---|
| 8 | 1984 | The National Video Game Masters Tournument '84 | Video Game Masters Tournument | Various | - | - | - | NA (USA) | - | - | - | - | - | - | - | - | Overall score in 60 different Arcade games | 60 | Arcade | | SP | - | - | - |
| 9 | 1984 | Coronation Day Championship 1984 | CDC | Ottumwa, Iowa, Missisippi, USA | - | - | - | NA (USA) | - | - | - | - | - | - | - | - | - | - | - | - | | - | - | - |
| 10 | 1985 | Coronation Day Championship 1985 | CDC | Los Angeles, California, USA | - | - | - | NA (USA) | - | - | - | - | - | - | - | - | - | - | - | - | | - | - | - |
| 11 | 1985 | The National Video Game Masters Tournument '85 | Video Game Masters Tournument | Various | - | - | - | NA (USA) | - | - | - | - | - | - | - | - | Overall score in 97 different Arcade games | 97 | Arcade | | SP | - | - | - |
| 12 | 1986 | 1986 Video Games Master Tournument '86 | Video Game Masters Tournument | Various | - | - | - | NA (USA) | - | - | - | - | - | - | - | - | Overall score in 129 different Arcade games | 129 | Arcade | | SP | - | - | - |
| 13 | 1986 | 1986 Video Games Master Tournument '87 | Video Game Masters Tournument | Various | - | - | 2 | NA | - | - | - | - | - | - | - | - | Overall score in 100 different Arcade Games | 100 | Arcade | | SP | - | <1 | - |
| 14 | 1990 | Nintendo World Championships | NWC | Various | - | 90 | - | - | - | - | - | - | - | - | - | $10,000, A convertible and a Television | Super Mario Bros, Rad Racers and Tetris | 2 | PVG and Racing | | SP | | <1 | |
| 15 | 1991 | Nintendo Campus Challenge 1991 | NWC | Various | - | - | - | NA (USA) | - | - | - | - | - | - | - | - | Super Mario Bros 3, Pin*Bot and Dr. Mario | 3 | PVG, Pinball and Puzzle | | SP | | | |
| 16 | 1992 | Nintendo Campus Challenge 1992 | NWC | Various | - | - | - | NA (USA) | - | - | - | - | - | - | - | $15,000 | Super Mario World, F-Zero and Pilot Wings | 3 | PVG and Flight Simulation | | SP | | | |
| 17 | 1994 | Nintendo PowerFest '94 | NWC | San Diego, California, USA | - | 132 | - | - | - | - | - | - | - | - | - | - | Super Mario Bros, Super Mario Kart and Ken Griffey Jr Presents Major League Baseball | 3 | PVG, Racing and Baseball (Sport Simulation) | | SP | | <1 | |
| 18 | 1994 | Blockbuster World Video Game Championships I | BWVG | Various | - | - | - | - | - | - | - | - | - | - | - | - | NBA JAM, Sonic the Hedgehog 3, Virtua Racing, TMNT, Clay Fighter Tournument Edition | 5 | Basketball (Sport Simulation), racing, PVG and fighting | | SP | | | |
| 19 | 1994 | DreamHack 1994 | DreamHack | Local School Cafeteria, Malung, Sweden | 40 | - | 1 | EU (Sweden) | - | - | - | - | - | - | - | - | - | - | - | | | | | |
| 20 | 1995 | Blockbuster World Video Game | BWVG | - | - | - | - | NA (USA) | - | - | - | - | - | - | - | - | Donkey Kong Country | 1 | PVG | | SP | - | - | - |

| | | | | | | | | | | | | | | | | | | | | | | | |
|---|---|---|---|---|---|---|---|---|---|---|---|---|---|---|---|---|---|---|---|---|---|---|---|
| | | Championships II | | | | | | | | | | | | | | | | | | | | | |
| 21 | 1995 | DreamHack 1995 | DreamHack | Local School Cafeteria, Malung, Sweden | 80 | - | 1 | EU (Sweden) | - | - | - | - | - | - | - | - | - | - | - | - | - | - | - |
| 22 | 1995 | Deathmatch '95 Judgement Day Deatmatch '95 Dwango's Deathmatch '95 | - | - | - | - | - | - | - | - | - | - | - | - | - | - | DOOM 2 | 2 | FPS | SP | Microsoft | - | - |
| 23 | 1996 | DreamHack 1996 | DreamHack | Local School Cafeteria, Malung, Sweden | 120 | - | 1 | EU (Sweden) | - | - | - | - | - | - | - | - | - | - | - | - | - | - | - |
| 24 | 1996 | QuakeCon 1996 | QuakeCon | Best Western in Garland, Texas | 100 | 60 | - | NA | - | - | - | - | - | - | - | - | Quake and Doom 2 | 2 | FPS | SP | - | - | - |
| 25 | 1997 | DreamHack 1997 | DreamHack | Kupolen, Börlange, Sweden | 750 | - | 1 | EU (Sweden) | - | - | - | - | - | - | - | - | - | - | - | - | - | - | - |
| 26 | 1997 | Red Annhilation Quake Tournument | E3 (ClanRing) | E3 Expo, Atlanta, Georiga, USA | 2,000 | 16 | - | - | - | - | - | - | - | - | - | Ferrari 328 GTS Cabriolet | QuakeWorld | 1 | FPS | SP | Microsoft | <1 | John Carmack (Co-founder of ID software gave away hes car) |
| 27 | 1997 | Cyberathlete Professional League (CPL): FRAG | FRAG | - | - | - | - | | - | - | - | - | - | - | - | Merchendise worth $4,000 | QuakeWorld | 1 | FPS | SP | Id software and Activision | <1 | |
| 28 | 1997 | QuakeCon 1997 | QuakeCon | Holiday Inn in Plano, Texas, USA | 650 | - | - | - | 650 | - | - | - | - | - | - | Hardware worth $3,000 | QuakeWorld | 1 | FPS | SP | | <1 | |
| 29 | 1998 | DreamHack 1998 | DreamHack | Kupolen, Börlange, Sweden | 1,800 | - | - | NA | - | - | - | - | - | - | - | - | | | | | | | |
| 30 | 1998 | Cyberathlete Professional League (CPL): FRAG 2 QuakeCon 2008 | Frag QuakeCon 2008 | Dallas, Texas, USA | 800 | - | - | - | - | - | - | - | - | - | - | $15,000 | Quake 2 | 1 | FPS | SP | | <1 | |
| 31 | 1999 | DreamHack 1999 | DreamHack | Kupolen, Börlange, Sweden | 3,000 | - | - | - | - | - | - | - | - | - | - | Harware and software | - | - | - | - | - | <1 | - |
| 32 | 1999 | Cyberathlete Professional League (CPL) FRAG 3 | Frag | Dallas, Texas | 500 | 12 | 3 | | - | - | - | - | - | - | - | $26,800 | Quake III Arena | 1 | FPS | SP | | <1 | |
| 33 | 1999 | Cyberathlete Professional League (CPL) Ground Zero | Ground Zero | New York City | 300 | 6 | 3 | | - | - | - | - | - | - | - | $25,000 | Quake III Arena | 1 | FPS | SP | | <1 | |
| 34 | 1999 | QuakeCon 1999 | QuakeCon | Mesquite Convention Center, Mesquite, Texas, USA | 500 | - | - | - | 1,100 | - | - | - | - | - | - | - | Quake III Arena | 1 | FPS | SP | ID software, Activision, AMD, Apple Computer, ATI Technologies, Logitech, Linksys, and Lucent Technologies. | <1 | |
| 35 | 1998 | PGL 1998: Season 1 | PGL | Seattle, Washington, USA | - | 8 | 1 | NA (USA) | - | - | - | - | - | - | - | $15,000 | Quake World | 1 | FPS | SP | AMD | 3 | |
| 36 | 1998 | PGL 1998: Season 2 | PGL | Atlanta, Georgia, USA | - | 4 | 2 | NA | - | - | - | - | - | - | - | $15,000 | Quake II | 1 | FPS | SP | AMD | 3 | |
| 37 | 1998 | PGL 1998: Season 3 | PGL | San Francisco, California, USA | - | 3 | 2 | NA | - | - | - | - | - | - | - | $16,200 | Starcraft: BW | 1 | RTS | SP | AMD | 3 | |
| 38 | 1999 | PLG 1998: Season 4 | PGL | XS New York, New York, USA | 128 | 8 | 2 | NA | - | - | - | - | - | - | - | $25,615 | Starcraft: BW | 1 | RTS | SP | AMD | 3 | |
| 39 | 1999 | Descent 3 $50,000 Tournament | | Las Vegas, Nevada, USA | 250 | - | - | - | - | - | - | - | - | - | - | $50,000 | Decent 3 | 1 | FPS | SP | - | < 1 | - |
| 40 | 1999 | 1999 Sport Seoul Cup | - | Seoul, South Korea | 64 | 7 | 3 | Global | - | - | - | - | - | - | - | $46,500 | Starcraft: BW | 1 | RTS | SP | | < 1 | |
| 41 | 2000 | DreamHack 2000 | DreamHack | Kupolen, Börlange, Sweden | 3000 | - | - | - | - | - | - | - | - | - | - | - | - | - | - | - | - | - | - |
| 42 | 2000 | Cyberathlete Professional League (CPL) FRAG 4 | FRAG | - | - | - | - | - | - | - | - | - | - | - | - | - | Quake 3 Arena | 1 | FPS | SP | - | < 1 | - |
| 43 | 2000 | QuakeCon 2000 | QuakeCon | Mesquite Convention Center, Mesquite, Texas, USA | 3,000 | - | - | - | 3,000 | | | | | | | - | Quake 3 Arena | 1 | FPS | SP | | < 1 | |
| 44 | 2000 | Razer CPL Tournument | CPL | Dallas, Texas | - | 32 | 9 | Global | 800 | | | | | | | $100,000 | Quake 3 Arena | 1 | FPS | SP | Razer | < 1 | |
| 45 | 2000 | World Cyber Games Challenge 2000 | WCG | Seoul, South Korea | - | 12 | 4 | Global | | | | | | | | $150,000 | Age of Empires II, Quake III Arena and StarCraft: BW | 3 | RTS and FPS | SP | | <1 | |
| 46 | 2000 | CPL Cologne 2000 | CPL | Cologne, Germany | - | 41 | 8 | Global | | | | | | | | $25,413 | Counter Strike, Quake III Arena | 2 | FPS | Both | | < 1 | |
| 47 | 2000 | Battletop Universal Challange | | Hollywood, California | - | 4 | 1 | NA (USA) | | | | | | | | $21,500 | Quake III Arena | 1 | FPS | SP | | <1 | |
| 48 | 2000 | Cyberathlete Professional League (CPL) FRAG 4 | CPL: FRAG | Hyatt Regency. Dallas, Texas | 400 | - | - | Global | 900 | | | | | | | $30,000 | Quake III Arena, Doom II | 2 | FPS | SP | | <1 | |
| 49 | 2000 | Babbage's CPL Tournament | CPL | Dallas, Texas, USA | 1,000 | 90 | 13 | Global | | | | | | | | $115,000 | Counter Strike, Quake III Arena | 2 | FPS | Both | | < 1 | |

| | | | | | | | | | | | | | | | | | |
|---|---|---|---|---|---|---|---|---|---|---|---|---|---|---|---|---|---|
| 50 | 2001 | DreamHack 2001 | DreamHack | Elmia, Jönköping, Sweden | 5,000 | - | - | - | | | - | - | - | | | | |
| 51 | 2001 | World Cyber Games 2001 | WCG | Everland, Seoul, Korea | 430 | 174 (35) | 17 (12) | Global | | | $300,000 | Counter-Strike, Quake 3 Arena, Unreal Tournament, Age of Empires II, FIFA 2001 and Starcraft: BW | 6 | 6 | FPS, RTS and Football | XSReality.com | <1 |
| 52 | 2001 | QuakeCon 2001 | QuakeCon | Mesquite Convention Center, Mesquite, Texas, USA | 1,250 | 3 | 1 | USA (NA) | 3,000 | | $45,000 | Quake III Arena | 1 | FPS | SP | | <1 |
| 53 | 2001 | CPL London 2001 | CPL | London, UK | - | 40 | 10 | EU NA | | | $26,394 | Counter-Strike and Quake III Arena | 2 | FPS | Both | | <1 |
| 54 | 2001 | CPL Berlin 2001 | CPL | Berlin, Germany | - | 41 | 7 | EU NA | | | $22,792 | Counter-Strike | 1 | FPS | TB | | <1 |
| 55 | 2001 | CPL Holland 2001 | CPL | Amsterdam, Netherlands | - | 47 | 7 | EU NA | | | $21,184 | Counter-Strike and Quake III Arena | 2 | FPS | Both | | <1 |
| 56 | 2001 | CPL World Championship 2001 | CPL | Dallas, Texas, USA | - | 78 | 10 | EU NA | | | $190,000 | Alien Versus Predator 2 and Counter-Strike | 2 | FPS | Both | Intel | <1 |
| 57 | 2002 | World Cyber Games 2002 | WCG | Daejeon, South Korea | - | 80 | 22 | Global | | | $300,000 | 2002 FIFA World Cup, Age of Empires II: The Age of Kings, Counter-Strike, Quake III Arena, StarCraft:BW and Unreal Torunument | 6 | Football, RTS and FPS | Both | | <1 |
| 58 | 2002 | DreamHack 2002 | DreamHack | Elmia, Jönköping, Sweden | 5,000 | - | - | - | | | - | Counter-Strike | 1 | FPS | TB | | <1 |
| 59 | 2002 | CPL Cologne 2002 | CPL | Cologne, Germany | - | 42 | 6 | Global | | 0 | $18,411 | Counter-Strike and Quake III Arena | 2 | FPS | Both | | <1 |
| 60 | 2002 | CPL Oslo 2002 | CPL | Oslo, Norway | - | 40 | 6 | Global | | | $9,791 | Counter-Strike | 1 | FPS | TB | | <1 |
| 61 | 2002 | CPL Summer Championship 2002 | CPL | Dallas, Texas, USA | - | 141 | 10 | Global | | | $100,000 | Counter-Strike | 1 | FPS | TB | | <1 |
| 62 | 2002 | CPL Winter Championship 2002 | CPL | Dallas, Texas, USA | - | 125 | 10 | Global | | | $125,000 | Counter-Strike and Unreal Tournament 2003 | 2 | FPS | Both | | <1 |
| 63 | 2002 | QuakeCon 2002 | QuakeCon | Mesquite Convention Center, Mesquite, Texas, USA | 1,300 | 35 | 5 | EU and NA | 3,250 | | $93,000 | Quke III Arena and Return to Castle Wolfenstein | 2 | FPS | Both | | <1 |

| | | | | | | | | | | | | | | | | |
|---|---|---|---|---|---|---|---|---|---|---|---|---|---|---|---|---|
| 64 | 2003 | Electronic Sports World Cup 2003 | ESCW | Futuroscope, Poitiers, France | 150,000 | 358 | 37 | Global | | $15,000 | Counter-Strike, Unreal Tournament 2003, Quake III Arena, WarCraft III | 4 | FPS and RTS | Both | BenQ, plantronics gaming, Nvidia and ubisoft | <1 |
| 65 | 2003 | CPL Cannes 2003 | CPL | Cannes, France | - | 37 | 10 | Global | | $20,000 | Counter-Strike and Warcraft III | 2 | FPS and RTS | Both | | <1 |
| 66 | 2003 | Cpl Copenhagen 2003 | CPL | Copenhagen, Denmark | - | 48 | 7 | Global | | $35,883 | Counter-Strike and Warcraft III | 2 | FPS and RTS | Both | | <1 |
| 67 | 2003 | CPL Summer 2003 | CPL | Dallas, Texas, USA | - | 80 | 10 | Global | | $200,000 | Counter-Strike | 1 | FPS | TB | | <1 |
| 68 | 2003 | World Cyber Games 2003 | WCG | Seoul, South Korea | - | 36 | 14 | Global | | $259,000 | Age of Mythology, Counter Strike, FIFA 03, StarCraft: BW, Unreal Tournament 2003 and WarCraft III | 6 | RTS, FPS and Football | Both | | <1 |
| 69 | 2003 | CPL Winter 2003 | CPL | Dallas, Texas, USA | - | 80 | 9 | Global | | $100,000 | Counter-Strike | 1 | FPS | TB | | <1 |
| 70 | 2003 | DreamHack Summer 2003 | DreamHack | Elmia, Jönköping, Sweden | 4,000 | 5 | 1 | EU (Sweden) | | $644 | Counter-Strike | 1 | FPS | TB | | <1 |
| 71 | 2003 | DreamHack Winter 2003 | DreamHack | Elmia, Jönköping, Sweden | 5,000 | 6 | 1 | EU (Sweden) | | $791 | Counter-Strike | 1 | FPS | TB | | <1 |
| 72 | 2003 | QuakeCon 2003 | QuakeCon | Mesquite Convention Center, Mesquite, Texas, USA | 2,000 | 18 | 5 | EU and NA | 4,000 | $76,000 | Quake III Arena and Return of Castle Wolfenstein | 2 | FPS | Both | | <1 |
| 73 | 2004 | SKY Proleague 2004 | ProLeague | South Korea | - | 19 | 1 | Korea | | $244,750 | StarCraft: BW | 1 | RTS | SP | | 10 |
| 74 | 2004 | World Cyber Games 2004 | WCG | San Fransisco, California, USA | - | 36 | 11 | Global | | $350,000 | Counter-Strike: Condition Zero, FIFA 04, Halo: Combat Evolved, Need for speed: Underground, Project Gotham Racing 2, StarCraft: BW, Unreal Tournament 2004 and WarCraft III | 8 | FPS, Football, Racing and RTS | Both | | <1 |
| 75 | 2004 | QuakeCon 2004 | QuakeCon | Grapevine, Texas, USA | 3,000 | 19 | 5 | EU and NA | 5,000 | $150,000 | Doom 3 and Quake III Arena | 2 | FPS | Both | | <1 |
| 76 | 2004 | CPL Winter 2004 | CPL | Dallas, Texas, USA | - | 106 | 14 | Global | | $150,000 | CS, Painkiller, Day of Deafeat, Doom III | 4 | | Both | | <1 |

| | | | | | | | | | | | | | | | | | |
|---|---|---|---|---|---|---|---|---|---|---|---|---|---|---|---|---|---|
| 77 | 2004 | CPL Summer 2004: Extreme World Championship | CPL | Dallas, Texas, USA | - | 101 | 13 | Global | 52,000 | $275,000 | Call Of Duty, Counter-Strike,Halo: Combat Envolved, PainkillerUnreal Tournament | 5 | FPS | Both | | <1 | |
| 78 | 2004 | DreamHack Summer 2004 | DreamHack | Elmia, Jönköping, Sweden | 5,272 | 5 | 1 | EU | | $1,308 | Counter-Strike | 1 | FPS | TB | | <1 | |
| 79 | 2004 | DreamHack Winter 2004 | DreamHack | Elmia, Jönköping, Sweden | 5,852 | 5 | 1 | EU | | $3,000 | Counter-Strike | 1 | FPS | TB | | <1 | "WGR World largest Lan" |
| 80 | 2004 | Electronic Sports World Cup 2004 | ESCW | Futuroscope, Poitiers, France | 400 | 81 | 41 | Global | | $208,000 | Counter-Strike, Painkiller, Pro Evolution Soccer 3, Quake III Arena, Unreal Tournament 2004 and WarCraft III | 6 | FPS, Football and RTS | Both | BenQ, plantronics gaming, Nvidia and ubisoft | <1 | |
| 81 | 2004 | MLG 2004 Atlanta | MLG | Hyatt Regency, Atlanta, Georgia, USA | 128 | 16 | 1 | NA (USA) | | $10,000 | Halo: Combat Evolved | 1 | FPS | Both | | <1 | |
| 82 | 2004 | MLG 2004 Boston | MLG | Bayside Conference Center in Boston, Massachusetts | - | - | 1 | NA (USA) | | - | Halo: Combat Evolved, Soul Calibur II, Super Smash Bros. Melee and Gran Turismo III | 4 | FPS, Fighting and Racing | Both | | <1 | |
| 83 | 2004 | MLG 2004 Chicago | MLG | Chicago, Illinois, USA | - | - | 1 | NA (USA) | | - | Halo: Combat Evolved, Soul Calibur II, Super Smash Bros. Melee and Gran Turismo III | 4 | FPS, Fighting and Racing | Both | | <1 | |
| 84 | 2004 | MLG 2004 Dallas | MLG | Dallas, Texas, USA | - | 12 | 1 | NA (USA) | | $15,000 | Halo: Combat Evolved | 1 | FPS | Both | | <1 | |
| 85 | 2004 | MLG 2004 Los Angeles | MLG | Los Angeles, California, USA | - | - | 1 | NA (USA) | | - | Halo: Combat Evolved, Soul Calibur II, Super Smash Bros. Melee and Gran Turismo III | 4 | FPS, Fighting and Racing | Both | | <1 | |
| 86 | 2004 | MLG 2004 Philadelphia | MLG | Philadelphia, Pennsylvania, USA | - | 12 | 1 | NA (USA) | | $17,000 | Halo: Combat Evolved | 1 | FPS | Both | | <1 | |
| 87 | 2004 | MLG 2004 San Francisco | MLG | San Francisco, California, USA | - | - | 1 | NA (USA) | | - | Halo: Combat Evolved, Soul Calibur II, Super Smash Bros. Melee and Gran Turismo III | 4 | FPS, Fighting and Racing | Both | | <1 | |
| 88 | 2004 | MLG 2004 New York City | MLG | New York City, New York, USA | - | 12 | 1 | NA (USA) | | $5,800 | Halo: Combat Evolved | 1 | FPS | Both | | <1 | |

| | | | | | | | | | | | | | | | | | | |
|---|---|---|---|---|---|---|---|---|---|---|---|---|---|---|---|---|---|---|
| 89 | 2004 | MLG 2004 New York Championships | MLG | Horvath & Associates Studios in New York City, New York, USA | - | 32 | 1 | NA (USA) | | | $50,000 | Halo: Combat Evolved, Soul Calibur II, Super Smash Bros. Melee, Nokia Ngage - Tony Hawk and Gran Turismo III | 5 | FPS, Fighting, Racing and Skating (Sport simulation) | Both | | <1 | |
| 90 | 2005 | Electronic Sports World Cup 2005 | ESCW | The Carrousel du Lovre, Paris, France | - | 68 | 17 | Global | | | $236,000 | Counter-Strike, Gran Turismo 4, Pro Evolution Soccer 4, Quake III Arena, Unreal Torunument 2004 and WarCraft III | 6 | FPS, Racing, Football and RTS | Both | BenQ, plantronics gaming, Nvidia and ubisoft | <1 | |
| 91 | 2005 | SKY Proleague 2005 | ProLeague | Busan South Korea | - | 26 | 1 | Korea | 120,000 | | $70,189 | StarCraft: BW | 1 | RTS | SP | | 10 | Outdoor event at a Stadium in Busan, South Korea |
| 92 | 2005 | World e-sport Games (WEG): I | WEG | Various | - | 36 | 12 | Global | | Yes (Final at MTV Overdrive) | $132,000 | Counter-Strike and Warcraft 3 | 2 | FPS and RTS | Both | | <1 | |
| 93 | 2005 | World e-sport Games (WEG): II | WEG | Seoul, South Korea | - | 36 | 9 | Global | | Yes (Final at MTV Overdrive) | $137,000 | Counter-Strike and Warcraft 3 | 2 | FPS and RTS | Both | | <1 | |
| 94 | 2005 | World e-sport Games (WEG): III | WEG | Seoul, South Korea | - | 52 | 10 | Global | | Yes (Final at MTV Overdrive) | $143,000 | Counter-Strike and Warcraft 3 | 2 | FPS and RTS | Both | | <1 | |
| 95 | 2005 | MLG 2005 New York City | MLG | Secaucus, New Jersey, USA | - | 44 | 2 | NA | | | $88,000 | Halo 2 and Super Smash Bros. Melee | 2 | FPS and Fighting | Both | | <1 | |
| 96 | 2005 | MLG 2005 Chicago | MLG | Chicago, Illinois, USA | - | 32 | 2 | NA | | | $21,000 | Halo 2 and Super Smash Bros. Melee | 2 | FPS and Fighting | Both | | <1 | |
| 97 | 2005 | MLG 2005 Atlanta | MLG | Atlanta, Georgia, USA | - | 38 | 2 | NA | | | $27,000 | Halo 2 and Super Smash Bros. Melee | 2 | FPS and Fighting | Both | | <1 | |
| 98 | 2005 | MLG 2005 Los Angeles | MLG | Los Angeles, California, USA | - | 32 | 2 | NA | | | $21,000 | Halo 2 and Super Smash Bros. Melee | 2 | FPS and Fighting | Both | | <1 | |
| 99 | 2005 | MLG 2005 Seattle | MLG | Seattle, Washington, USA | - | 12 | 1 | USA (NA) | | | $5,000 | Halo 2, Tekken 5 and Super Smash Bros. Melee | 3 | FPS and Fighting | Both | | <1 | |
| 100 | 2005 | MLG 2005 Nashville | MLG | Nashville, Tennessee, USA | - | 12 | 2 | NA | | | $5,000 | Halo 2, Tekken 5 and Super Smash Bros. Melee | 3 | FPS and Fighting | Both | | <1 | |
| 101 | 2005 | MLG 2005 Las Vegas | MLG | Las Vegas, Nevada, USA | - | 12 | 1 | USA (NA) | | | $5,000 | Halo 2 and Super Smash Bros. Melee | 2 | FPS and Fighting | Both | | <1 | |
| 102 | 2005 | MLG 2005 Philadelphia | MLG | Philadelphia, Pennsylvania, USA | - | 22 | 1 | USA (NA) | | | $8,925 | Halo 2, Halo: Combat Evolved, Super Smash Bros. Melee and Tekken 5 | 4 | FPS and Fighting | Both | | <1 | |

| | | | | | | | | | | | | | | | |
|---|---|---|---|---|---|---|---|---|---|---|---|---|---|---|---|
| 103 | 2005 | MLG 2005 St. Louis | MLG | St. Louis, Missouri, USA | - | 18 | 1 | USA (NA) | | $6,250 | Halo 2, Tekken 5 and Super Smash Bros. Melee | 3 | FPS and Fighting | Both | <1 |
| 104 | 2005 | MLG 2005 Orlando | MLG | Orlando, Florida, USA | - | 13 | 1 | USA (NA) | | $5,500 | Halo 2, Tekken 5 and Super Smash Bros. Melee | 3 | FPS and Fighting | Both | <1 |
| 105 | 2005 | MLG 2005 Houston | MLG | Houston, Texas, USA | - | 18 | 3 | NA and EU | | $6,250 | Halo 2, Tekken 5 and Super Smash Bros. Melee | 3 | FPS and Fighting | Both | <1 |
| 106 | 2005 | MLG 2005 San Francisco | MLG | San Francisco, California, USA | - | 17 | 1 | USA (NA) | | $6,250 | Halo 2 and Super Smash Bros. Melee | 2 | FPS and Fighting | Both | <1 |
| 107 | 2005 | MLG 2005 Washington D.C. | MLG | Washington D.C., USA | - | 13 | 1 | USA (NA) | | $6,250 | Halo 2 and Super Smash Bros. Melee | 2 | FPS and Fighting | Both | <1 |
| 108 | 2005 | CPL Winter 2005 | CPL | Dallas, Texas, USA | - | 114 | 16 | Global | | $100,000 | CS, Quake IV, F.E.A.R | 3 | | Both | <1 |
| 109 | 2005 | CPL Summer 2005 | CPL | Dallas, Texas, USA | - | 80 | 13 | Global | | $200,000 | CS, Painkiller, CS:Source Warcraft III, Halo II, Day Of Defeat | 6 | | Both | <1 |
| 110 | 2005 | CPL Turkey 2005 | CPL | Istanbul, Turkey | - | 36 | 11 | Global | | $64,000 | Counter-Strike, Warcraft III and Painkiller | 3 | FPS and RTS | Both | <1 |
| 111 | 2005 | CPL Spain 2005 | CPL | Barcelona, Spain | - | 36 | 12 | Global | | $70,000 | Counter-Strike and Painkiller | 2 | FPS | Both | <1 |
| 112 | 2005 | CPL Brazil 2005 | CPL | Rio de Janeiro, Brazil | - | 36 | 10 | Global | | $75,000 | Counter-Strike and Painkiller | 2 | FPS | Both | <1 |
| 113 | 2005 | CPl Sweden 2005 | CPL | Jönköping, Sweden | - | 16 | 9 | Global | | $50,000 | Painkiller | 1 | FPS | SP | <1 |
| 114 | 2005 | CPL UK 2005 | CPL | Sheffield, UK | - | 31 | 9 | Global | | $60,000 | Counter-Strike and Painkiller | 2 | FPS | Both | <1 |
| 115 | 2005 | CPL Singapore 2005 | CPL | Singapore, Malaysia | - | 21 | 11 | Global | | $70,000 | Counter-Strike and Painkiller | 2 | FPS | Both | <1 |
| 116 | 2005 | CPL Italy 2005 | CPL | Milan, Italy | - | 16 | 10 | Global | | $50,000 | Painkiller | 1 | FPS | Both | <1 |
| 117 | 2005 | CPL Chile 2005 | CPL | Santiago, Chile | - | 6 | 11 | Global | | $60,000 | Counter-Strike and Painkiller | 2 | FPS | Both | <1 |
| 118 | 2005 | CPL World Tour Finals | CPL | New York City, New York, USA | - | 32 | 11 | Global | | $1,000,000 | Painkiller, Warcraft III, CS, Quake III | 4 | FPS and RTS | Both | <1 |
| 119 | 2005 | QuakeCon 2005 | QuakeCon | Mesquite Convention Center, Mesquite, Texas, USA | 3,200 | 9 | 5 | Global | 6,000 | $100,000 | Doom 3, Quake III Arena and Quake II | 3 | FPS | SP | <1 |
| 120 | 2005 | The World Cyber Games 2005 | WCG | Suntec City, Singapore | 800 | 39 | 14 | Global | | $325,000 | Counter-Strike Source, Dead or Alive Ultimate, FIFA 05, Halo 2, Need for speed: Ungerground 2, StarCraft: | 8 | FPS, Football, Racing and RTS | Both | <1 |

| | | | | | | | | | | | | | | | | | |
|---|---|---|---|---|---|---|---|---|---|---|---|---|---|---|---|---|---|
| | | | | | | | | | | | | BW, WarCraft III and Warhammmer 40K | | | | | |
| 121 | 2005 | Snickers All-Star League | - | Seoul, South Korea | - | 3 | 1 | Asia (Korea) | | | $40,000 | StarCraft: BW | 1 | RTS | SP | | <1 |
| 122 | 2005 | World Cyber Games 2005: Samsung Euro Championship | WCG | Hanover, Germany | - | 36 | 11 | EU | | | $79,475 | Counter-Strike, FIFA 05, Halo 2, Need for Speed: Underground 2, Project Gotham Racing 2, StarCraft: BW, Unreal Tournument 2004 and WarCraft III | 8 | FPS, Fotball, Racing and RTS | Both | Microsoft | <1 |
| 123 | 2005 | BlizzCon 2005 | BlizzCon | Anaheim, California, USA | - | 4 | 2 | Global | | | $26,000 | Warcraft III, Warcraft III: Defense of the Ancients Starcraft: BW | 3 | RTS and MOBA | Both | NVIDIA, Intel | <1 |
| 124 | 2005 | DreamHack Summer 2005 | DreamHack | Elmia, Jönköping, Sweden | 5,000 | 6 | 3 | EU | | | $5,263 | StarCraft: BW and WarCraft III | 2 | RTS | SP | | <1 |
| 125 | 2005 | DreamHack Winter 2005 | DreamHack | Elmia, Jönköping, Sweden | 7,538 | 7 | 2 | EU | | | $4,000 | Counter-Strike and WarCraft III | 2 | FPS and RTS | Both | | <1 |
| 126 | 2006 | Championship Gaming Invitational 2006 | CGI | San Fransisco, California, USA | - | 27 | 3 | NA and EU | | | $116,000 | Battlefield 2, Counter-Strike, DOA:4, Forza motorsport 2 and Halo 2 | 5 | FPS, Fighting and racing | Both | | <1 |
| 127 | 2006 | Electronic Sports World Cup 2006 | ESCW | Palais Omnisports de Paris-Bercy | - | 208 | 20 | Global | "Several Thousand" | 63,000 | $400,000 | Counter-Strike, Gran Turismo 4, Pro Evolution Soccer 5, Quake 4, TrackMania Nations and WarCraft III | 6 | FPS, Racing, Football and RTS | Both | BenQ, plantronics gaming, Nvidia and ubisoft | <1 |
| 128 | 2006 | DreamHack Summer 2006: WSCG 2006 | DramHack and WSVG | Elmia, Jönköping, Sweden | 5,000 | 21 | 4 | Global | | | $43,000 | Counter-Strike, Quake 4 and WarCraft III | 3 | FPS and RTS | Both | | <1 |
| 129 | 2006 | DreamHack Winter 2006 | DreamHack | Elmia, Jönköping, Sweden | 10,638 | 26 | 3 | EU | | | $12,633 | Warcraft III: DOTA, Starcraft II: BW, Warcraft III and Counter-Strike | 4 | MOBA, RTS and FPS | Both | | <1 |
| 130 | 2006 | QuakeCon 2006 | QuakeCon | Hilton Anatole Hotel, Dallas, Texas, USA. | 1,800 | 9 | 7 | Global | 4,000 | | $60,000 | Enemy Teritory: Quake Wars | 1 | FPS | Both | ID software, NVIDIA | <1 |

| | | | | | | | | | | | | | | | |
|---|---|---|---|---|---|---|---|---|---|---|---|---|---|---|---|
| 131 | 2006 | International E-Sport Festival 2006 | IEF | Shanghai, China | - | 8 | 2 | Asia | $55,000 | Counter-Strike, StarCraft: BW and WarCraft III | 3 | FPS and RTS | Both | | <1 |
| 132 | 2006 | MLG 2006 Las Vegas | MLG | Las Vegas, Nevada, USA | - | 21 | 2 | NA | $234,000 | Halo 2 and Super Smash Bros. Melee | 2 | FPS and Fighting | Both | | <1 |
| 133 | 2006 | MLG 2006 Orlando | MLG | Orlando, Flordia, USA | - | 39 | 1 | NA (USA) | $48,750 | Halo 2 and Super Smash Bros. Melee | 2 | FPS and Fighting | Both | | <1 |
| 134 | 2006 | MLG 2006 Chicago | MLG | Chicago, Illinois, USA | - | 43 | 2 | NA | $49,525 | Halo 2 and Super Smash Bros. Melee | 2 | FPS and Fighting | Both | | <1 |
| 135 | 2006 | MLG 2006 Anaheim | MLG | Anaheim, California, USA | - | 46 | 2 | NA | $49,525 | Halo 2 and Super Smash Bros. Melee | 2 | FPS and Fighting | Both | | <1 |
| 136 | 2006 | MLG 2006 Dallas | MLG | Dallas, Texas, USA | - | 34 | 2 | NA | $49,525 | Halo 2 and Super Smash Bros. Melee | 2 | FPS and Fighting | Both | | <1 |
| 137 | 2006 | MLG 2006 New York | MLG | Secaucus, New Jearsey, USA | - | 39 | 2 | NA | $48,750 | Halo 2 and Super Smash Bros. Melee | 2 | FPS and Fighting | Both | | <1 |
| 138 | 2006 | MLG 2006 New York Playoffs | MLG | New York City, New York, USA | - | 49 | 2 | NA | $123,775 | Halo 2 and Super Smash Bros. Melee | 2 | FPS and Fighting | Both | | <1 |
| 139 | 2006 | World Series of Video Games (WSVG): Intel Summer Championships | WSVG | Grapevine, Texas, USA | - | 96 | 13 | Global | $210,500 | Counter-Strike and Quake III Arena | 2 | FPS | Both | Intel | <1 |
| 140 | 2006 | World Series of Video Games (WSVG): London 2006 | WSVG | London, UK | - | 20 | 7 | NA and EU | $41,500 | Counter-Strike, Quake 4 and WarCraft III | 3 | FPS and RTS | Both | | <1 |
| 141 | 2006 | World Series of Video Games (WSVG): China 2006 | WSVG | Chengdu, China | - | 18 | 2 | Asia | $25,000 | Counter-Strike and WarCraft III | 2 | FPS and RTS | Both | | <1 |
| 142 | 2006 | World Series of Video Games (WSVG): Lanwar 2006 | WSVG | Louisville, Kentucky, USA | - | 22 | 3 | USA and EU | $43,000 | Counter-Strike, Quake 4 and WarCraft III | 3 | FPS and RTS | Both | | <1 |
| 143 | 2006 | CPL Winter 2006 | CPL | Dallas, Texas, USA | - | 112 | 15 | Global | $150,000 | Counter-Strike and Quake III Arena | 2 | FPS | Both | | <1 |
| 144 | 2006 | CPL Singapore 2006 | CPL | Singapore, Malaysia | - | 24 | 5 | Global | $31,000 | Counter-Strike and Quake III Arena | 2 | FPS | Both | | <1 |
| 145 | 2006 | CPL Brazil 2006 | CPL | São Paul, Brazil | - | 40 | 2 | Global | $21,000 | Counter-Strike | 1 | FPS | TB | | <1 |
| 146 | 2006 | CPL Italy 2006 | CPL | Verona, Italy | - | 34 | 7 | Global | $43,600 | Counter-Strike and Quake III Arena | 2 | FPS | Both | | <1 |

| | | | | | | | | | | | | | | | | | |
|---|---|---|---|---|---|---|---|---|---|---|---|---|---|---|---|---|---|
| 147 | 2006 | World Cyber Games 2006 | WCG | Monza, Italy | - | 39 | 16 | Global | | | $350,500 | Counter-Strike, Dead or Alive 4, FIFA 06, Need for speed: Most Wanted, Project Gotham Racing, Quake 4, StarCraft: BW, WarCraft III and WarHammer 40K | 9 | FPS, Fighting, Football, Racing, RTS | Both | | <1 |
| 148 | 2007 | International E-Sport Festival 2007 | IEF | Seoul, South Korea | - | 28 | 2 | Asia | | | $60,000 | Counter-Strike, StarCraft: BW and WarCraft III | 3 | FPS and RTS | Both | | <1 |
| 149 | 2007 | Sky Proleague Korea 2007 | ProLeague | South Korea | - | 27 | 1 | Asia (South Korea) | | | $184,338 | Starcraft: BW | 1 | RTS | SP | Shinhan Bank | 10 |
| 150 | 2007 | Electronic Sports World Cup 2007 | ESCW | Paris Expo Porte de Versailes, Paris, France | 750 | 27 | 51 | Global | | | $152,000 | Counter-Strike, Pro Evolution Soccer 6, Quake 4, Trackmania Nations and WarCraft III | 5 | FPS, Football, Racing and RTS | Both | BenQ, plantronics gaming, Nvidia and ubisoft | <1 |
| 151 | 2007 | World Cyber Games 2007 | WCG | Seattle, Washington, USA | - | 42 | 17 | Global | | | $300,000 | Age of Empires III, Carom 3D, Command & Conquer 3, Counter-Strike, FIFA 07, Gears of War, Need for speed: ProStreet, Project Gotham Racing 3, StarCraft: BW and WarCraft III | 11 | RTS, Football and pool (Sport simulation), FPS, Fighting and Raciing | Both | Microsoft | <1 |
| 152 | 2007 | World Cyber Games 2007 Samsung Euro Championship | WCG | Hanover, Germany | - | 30 | 11 | EU | | | $75,912 | Counter-Strike, Dead or Alive 4, FIFA 07, Need for speed Carbon, Project Gotham Racing and WarCraft III | 6 | FPS, Fighting, Football, Racing and RTS | Both | | <1 |
| 153 | 2007 | Championship Gaming Series 2007 by DirectTV | **CGS** | Culver City, California, USA | - | 117 | 16 | Global | 500-1500 | DirectTV | $1,025,000 | Counter-Strike: Source, Dead or Alive 4, FIFA 07, Project Gotham Racing and World of Warcraft | 5 | FPS, Fighting, Sports, MMORPG and Racing | Both | | <1 |
| 154 | 2007 | QuakeCon 2007 | QuakeCon | Hilton Anatole | 2,700 | 8 | 3 | NA and EU | 7,000 | | $50,000 | Quake 4 | 1 | FPS | SP | | <1 |

| | | | | | | | | | | | | | | | | |
|---|---|---|---|---|---|---|---|---|---|---|---|---|---|---|---|---|
| | | | | Hotel, Dallas, Texas, USA. | | | | | | | | | | | | |
| 155 | 2007 | CPL Italy 2007 | CPL | Verona, Italy | - | 22 | 7 | Global | $39,500 | F.E.A.R and World In Conflict | 2 | FPS and RTS | Both | | <1 | |
| 156 | 2007 | CPL USA 2007 | CPL | Dallas, Texas, USA | - | 8 | 3 | Global | $40,000 | F.E.A.R and World In Conflict | 2 | FPS and RTS | Both | | <1 | |
| 157 | 2007 | CPL UK 2007 | CPL | London, UK | - | 8 | 3 | Global | $38,000 | F.E.A.R and World In Conflict | 2 | FPS and RTS | Both | | <1 | |
| 158 | 2007 | CPL World Tour Finals 2007 | CPL | London, UK | - | 28 | 6 | Global | $300,000 | F.E.A.R and World In Conflict | 2 | FPS and RTS | Both | | <1 | |
| 159 | 2007 | CPL Extreme Winter Championships 2007 | CPL | Dallas, Texas, USA | - | 31 | 1 | NA (USA) | $101,250 | Counter-Strike, Counter-Strike:GØ and Halo 3 | 3 | FPS | Both | | <1 | |
| 160 | 2007 | MLG 2007 Las Vegas | MLG | Las Vegas. Nevada, USA | - | 32 | 1 | NA (USA) | $484,000 | Gears of War, Halo 2, RBS:V and Shadowrun | 4 | FPS and RPG | Both | | <1 | |
| 161 | 2007 | MLG 2007 Orlando | MLG | Orlando, Flordia, USA | - | 28 | 1 | NA (USA) | $96,300 | Gears of War, Halo 2, RBS:V and Shadowrun | 4 | FPS and RPG | Both | | <1 | |
| 162 | 2007 | MLG 2007 Chicago | MLG | Chicago, Illinois, USA | - | 27 | 1 | NA (USA) | $96,300 | Gears of War, Halo 2, RBS:V and Shadowrun | 4 | FPS and RPG | Both | | <1 | |
| 163 | 2007 | MLG 2007 Dallas | MLG | Dallas, Texas, USA | - | 27 | 1 | NA (USA) | $92,300 | Gears of War, Halo 2 and RBS:V | 3 | FPS | Both | | <1 | |
| 164 | 2007 | MLG 2007 Meadowlands | MLG | Meadowlands, New Jersey, USA | - | 30 | 1 | NA (USA) | $92,300 | Gears of War, Halo 2 and RBS:V | 3 | FPS | Both | | <1 | |
| 165 | 2007 | MLG 2007 Charlotte | MLG | Charlotte, North Carolina, USA | - | 28 | 1 | NA (USA) | $78,300 | Halo 2 and RBS:V | 2 | FPS | Both | | <1 | |
| 166 | 2007 | BlizzCon 2007 | BlizzCon | Anaheim, California, USA | - | 6 | 3 | Global | $208,000 | Warcraft III, Starcraft: BW, World Of Warcraft | 3 | RTS and MMORPG | Both | | <1 | |
| 167 | 2007 | Intel Challenge Cup 2007 | ICC | Moscow, Russia | - | 40 | 5 | EU | $50,000 | Counter-Strike | 1 | FPS | | Intel | <1 | |
| 168 | 2007 | IEM I Finals (Iintel Friday Night Games) | IEM (ESL) | Deutche Messe AG, Hanover, Germany | - | - | 24 | EU | $205,000 | Counter-Strike, WarCraft III | 2 | FPS and RTS | Both | Intel, Cemos, Logitech, Sennheiser, nVidia, Shuttle, Mindfactory, Snogard, Brunnen IT, Pandero and Upper Deck | <1 | Intel Racing Tour, Musiv event: Jan Hegenberg |
| 169 | 2007 | World Series of Video Games (WSVG): Toronto 2007 | WSVG | Toronto, Ontario, Canada | - | 8 | 6 | EU and NA | $38,100 | Counter-Strike and World of Warcraft | 2 | FPS and MMORPG Arena | TB | | <1 | |
| 170 | 2007 | World Series of Video Games | WSVG | Dallas, Texas, USA | - | 8 | 4 | EU and NA | $15,000 | Quake 4 | 1 | FPS | SP | | <1 | |

| | | | | | | | | | | | | | | | | | |
|---|---|---|---|---|---|---|---|---|---|---|---|---|---|---|---|---|---|
| | | (WSVG): Dallas 2007 | | | | | | | | | | | | | | | |
| 171 | 2007 | World Series of Video Games (WSVG): Lanwar 2007 | WSVG | Louisville, Kentucky, USA | - | 28 | 9 | EU and NA | | $75,000 | Counter-Strike, Quake 4 and World of WarCraft | 3 | FPS and MMORPG Arena | Both | | <1 | |
| 172 | 2007 | World Series of Video Games (WSVG): China 2006 | WSVG | Wuhan, China | - | 30 | 6 | Global | | $43,000 | Counter-Strike, WarCraft III, World of Warcraft | 3 | FPS, RTS and MMORPG Arena | Both | | <1 | |
| 173 | 2007 | PGL Season 1 | PGL | Bejing, Chinga | - | 6 | 3 | Global | | $23,000 | WarCraft III and StarCraft: BW | 2 | FPS and RTS | SP | | 2 | |
| 174 | 2007 | PGL Season 2 | PGL | Bejing, Chinga | - | 5 | 3 | Global | | $30,000 | WarCraft III and StarCraft: BW | 2 | FPS and RTS | SP | | 4 | |
| 175 | 2007 | DreamHack Summer 2007 | DreamHack | Elmia, Jönköping, Sweden | 6,000 | 29 | 7 | Global | | $22,596 | Counter-Strike, DOTA and WarCraft III | 3 | FPS, MOBA and RTS | Both | | <1 | |
| 176 | 2007 | DreamHack Winter 2007 | DreamHack and CPL Sweden | Elmia, Jönköping, Sweden | 13,000 | 37 | 7 | Global | | $44,603 | Counter-Strike, DOTA and WarCraft III | 3 | FPS, MOBA and RTS | Both | | <1 | |
| 177 | 2008 | DreamHack Summer 2008 | DreamHack | Elmia, Jönköping, Sweden | 7,500 | 57 | 8 | Global | | $38,223 | Counter-Strike, Counter-Strike Source, DOTA, StarCraft: BW and WarCraft III | 5 | FPS, MOBA and RTS | Both | | <1 | |
| 178 | 2008 | DreamHack Winter 2008 | DreamHack | Elmia, Jönköping, Sweden | 11,000 | 63 | 7 | Global | | $46,802 | Counter-Strike, Counter-Strike Source, DOTA, Quake III Arena, Quake World, Trackmania Nation Forever and WarCraft III | 8 | FPS, MOBA, Racing and RTS | Both | | <1 | |
| 179 | 2008 | IEM II Finals (Intel Friday Night Games) | IEM (ESL) | Deutche Messe AG, Hanover, Germany | - | - | - | Global | ESL TV and CeBIT | $285,000 | Counter-Strike, WarCraft III | 2 | FPS and RTS | Both | Intel, BenQ and Raidcall | <1 | |
| 180 | 2008 | Championship Gaming Series 2008 | CGS | Santa Monica, California, USA | - | 80 | 14 | Global | | $960,000 | Counter-Strike: Source, Dead or Alive 4, FIFA 08 and ForzaMotorsport 2 | 4 | FPS, Fighting, Sports and Racing | Both | | <1 | Kim Kardashian attended the event |
| 181 | 2008 | Electronic Sports World Cup 2008 | ESWC | Sane Jose, USA | - | 49 | 13 | Global | | $192,000 | Counter-Strike, WarCraft 3, DOTA Quake 3 and Trackmania ESCW | 5 | FPS, RTS, Racing and MOBA | TB | BenQ, plantronics gaming, Nvidia and ubisoft | <1 | |

| | | | | | | | | | | | | | | | | |
|---|---|---|---|---|---|---|---|---|---|---|---|---|---|---|---|---|
| 182 | 2008 | International e-Sport Festival 2008 | IEF | Wuhan, China | - | 28 | 5 | Global | | $77,000 | Counter-Strike, StarCraft: BW and WarCraft III | 3 | FPS and RTS | Both | | <1 |
| 183 | 2008 | World e-Sport Masters 2008 | WEG | Hangzhou, China | - | 48 | 10 | Global | | $100,000 | Counter-Strike and WarCraft III | 2 | FPS and RTS | Both | | <1 |
| 184 | 2008 | QuakeCon 2008 | QuakeCon | Hilton Anatole Hotel, Dallas, Texas, USA. | - | 19 | 4 | NA and EU | 6,000 | $25,000 | Quake Live | 1 | FPS | Both | Intel | <1 |
| 185 | 2008 | BlizzCon 2008 | BlizzCon | Anaheim, California, USA | - | 15 | 7 | Global | | $208 000 | Warcraft III, Starcraft: BW, World Of Warcraft | 3 | RTS and MMORPG Arena | Both | | <1 |
| 186 | 2008 | PGL Season 3 | PGL | Chengdu, China | - | 16 | 4 | Global | | $55,000 | WarCraft III and StarCraft: BW | 2 | RTS | SP | | <1 |
| 187 | 2008 | MLG 2008 Las Vegas | MLG | Las Vegas, Nevada, USA | - | 36 | 1 | NA (USA) | | $363,000 | Call of Duty 4: Modern Warfare, Gears of War, Halo 3 and RB6:V2 | 4 | FOS | TB | | <1 |
| 188 | 2008 | MLG 2008 Dallas | MLG | Dallas, Texas, Usa | - | 38 | 1 | NA (USA) | | $93,500 | Gears of War, Halo 3, RB6:V2, WoW | 3 | FPS and MMORPG Arena | TB | | <1 |
| 189 | 2008 | MLG 2008 Toronto | MLG | Toronto, Ontario, Canada | - | 32 | 1 | NA (USA) | | $71,000 | Halo 3, Gears of War and RB6:V2 | 3 | FPS | TB | | <1 |
| 190 | 2008 | MLG 2008 Orlando | MLG | Orlando, Flordia, USA | - | 35 | 1 | NA (USA) | | $93,500 | Gears of War, Halo 3, RB6:V2, WoW | 4 | FPS | TB | | <1 |
| 191 | 2008 | MLG 2008 San Diego | MLG | San Diego, California, USA | - | 38 | 3 | NA and EU | | $86,000 | Halo 3, RB6:V2 and WOW | 3 | FPS and MMORPG Arena | TB | | <1 |
| 192 | 2008 | MLG 2008 Meadowlands | MLG | Meadowlands, New Jersey, USA | - | 32 | 2 | NA | | $63,500 | Gears of War and Halo 3 | 2 | FPS | TB | | <1 |
| 193 | 2008 | The World Cyber Games (WCG 2008) | WCG | Cologne, Germany | 850 | 60 | 78 | Global | | $408 000 | Age of Empires III, Carom 3D, Command & Conquer 3, Counter-Strike, FIFA 08, Halo 3, Need for speed: ProStreet, Project Gotham Racing 4, Red Stone, StarCraft: BW, Virtua Fighter 5 and WarCraft III | 12 | | RTS, Pool and Football (Sport Simulation), FPS, Racing, MMORPG and Fighting | Microsoft | <1 |
| 194 | 2009 | Electronic Sports World Cup 2009 | ESWC | Cheonan, Korea | - | 21 | 9 | Global | | $54,000 | Counter-Strike, StarCraft: BW and WarCraft III | 3 | FPS and RTS | TB | | <1 |
| 195 | 2009 | PGL Season 4 | PGL | Chengdu, China | - | 16 | 4 | Global | | $55,000 | WarCraft III and | 2 | RTS | SP | | <1 |

| | | | | | | | | | | | | | | | | | |
|---|---|---|---|---|---|---|---|---|---|---|---|---|---|---|---|---|---|
| | | | | | | | | | | | | StarCraft: BW | | | | | |
| 196 | 2009 | QuakeCon 2009 | QuakeCon | Gaylord Texan Resort Hotel and Convention Center | - | 15 | 8 | NA and EU | 7,000 | | $47,500 | Quake Live | 1 | FPS | Both | | <1 |
| 197 | 2009 | World e-sport masters 2009 | WEG | Hangzhou, China | - | 48 | 10 | Global | | | $72,000 | Counter-Strike and WarCraft III | 2 | FPS and RTS | Both | | <1 |
| 198 | 2009 | International e-Sports festival 2009 | IEF | Suwon, South Korea | - | 21 | 4 | Global | | | $51,000 | Counter-Strike, StarCraft: BW and WarCraft III | 3 | FPS and RTS | Both | | <1 |
| 199 | 2009 | MLG 2009 Orlando | MLG | Orlando, Flordia, USA | - | 32 | 2 | NA | | | $370,000 | Halo 3, Gears of War 2 and World of Warcraft | 3 | FPS and MMORPG Arena | TB | | <1 |
| 200 | 2009 | MLG 2009 Anaheim | MLG | Anaheim, California, USA | - | 57 | 7 | Global | | | $116,750 | Call of Duty 4: Modern Warfare, Gear of War 2, Halo 2, Rainbow Six: Vegas 2, World of Warcraft | 5 | FPS and MMORPG Arena | TB | | <1 |
| 201 | 2009 | MLG 2009 Dallas | MLG | Dallas, Texas, Usa | - | 32 | 1 | NA | | | $86,750 | Gear of War 2, Halo 3 and World of WarCraft | 3 | FPS and MMORPG Arena | TB | | <1 |
| 202 | 2009 | MLG 2009 Columbus | MLG | Columbus, Ohio, USA | - | 41 | 4 | Global | | | $86,750 | Gear of War 2, Halo 3 and World of WarCraft | 3 | FPS and MMORPG Arena | TB | | <1 |
| 203 | 2009 | MLG 2009 Meadowlands | MLG | Meadowlands, New Jersey, USA | - | 32 | 2 | NA | | | $71,000 | Gears of War 2 and Halo 3 | 2 | FPS | TB | | <1 |
| 204 | 2009 | World Cyber Games 2009 | WCG | Chengdu, Sichuan, China | - | 45 | 10 | Global | | | $177,000 | Carom 3D, Counter-Strike, FIFA 09, Red Stone, StarCraft: BW, Tracmania Nations Forever, Virtua Fighter 5 and WarCraft III | 8 | Football and pool (Sport simulation), MMORPG, RTS, Fighting and Racing | Both | Microsoft | <1 |
| 205 | 2009 | BlizzCon 2009 | BlizzCon | Anaheim, California, USA | - | 15 | 4 | Global | | | $266 000 | Warcraft III, Starcraft: BW, World Of Warcraft | 3 | RTS and MMORPG Arena | | | <1 |
| 206 | 2009 | IEM III - World Championship Finals | IEM (ESL) | Deutche Messe AG, Hanover, Germany | - | - | - | - | | ESL TV and CeBIT | $200,000 | Counter-Strike and Warcraft III | 2 | FPS and RTS | Both | | <1 |
| 207 | 2009 | IEM III - Asian Championship Finals | IEM (ESL) | Heroes Digital Entertainment Arena, Chengdu, China | - | - | - | - | | ESL TV and CeBIT | $60,000 | Counter-Strike and World of Warcraft | 2 | FPS and MMORPG Arena | TB | | <1 |

| | | | | | | | | | | | | | | | |
|---|---|---|---|---|---|---|---|---|---|---|---|---|---|---|---|
| 208 | 2009 | IEM III - European Championship Finals | IEM (ESL) | Deutche Messe AG, Hanover, Germany | - | - | - | - | ESL TV and CeBIT | $80,000 | Counter-Strike and Warcraft III | 2 | FPS and RTS | Both | <1 |
| 209 | 2009 | IEM IV - American Championship Finals | IEM (ESL) | West Edmonton Mall, Edmonton, Canada | - | 48 | 4 | NA | ESL TV and CeBIT | $52,650 | Counter-Strike, Quake Live and World of Warcraft | 3 | FPS and MMORPG Arena | Both | <1 |
| 210 | 2009 | IEM IV - Global Challenge Dubai | IEM (ESL) | Gitex Shopper & Consumer Electronics Airport Expo, Dubai, United Arab Emirate | - | 68 | 18 | Global | ESL TV and CeBIT | $31,500 | Counter-Strike and Quake Live | 2 | FPS | Both | <1 |
| 211 | 2009 | IEM IV - Global Challenge Chengdu | IEM (ESL) | Heroes Digital Entertainment Arena, Chengdu, China | - | 84 | 12 | Global | ESL TV and CeBIT | $40,000 | Counter-Strike, Warcraft III: DOTA and Warcraft III | 3 | FPS, MOBA and RTS | Both | <1 |
| 212 | 2009 | IEM IV - Global Challenge Gamescom | IEM (ESL) | Koehlnmessen, Cologne, Germany | - | 21 | 5 | NA and EU | ESL TV and CeBIT | $170,000 | Warcraft III, Starcraft: BW, World Of Warcraft | 3 | RTS and MMORPG Arena | Both | <1 |
| 213 | 2009 | DreamHack Summer 2009 | DreamHack | Elmia, Jönköping, Sweden | 9,000 | 41 | 7 | EU | | $34,230 | Counter-Strike, DOTA, FIFA 09, StarCraft: BW, Street Fighter IV, TrackMania Nations Forever, Warcraft III and World of Warcraft | 8 | FPS, RTS, Sports, Fighting, Racing and MMORPG Arena | Both | <1 |
| 214 | 2009 | DreamHack Winter 2009 | DreamHack | Elmia, Jönköping, Sweden | 11,500 | 46 | 10 | EU and NA | | $42,031 | Counter-Strike, Call of Duty: Modern Warfare 2, Counter-Strike: Source, Quake Live and TrackMania Nations Forever | 5 | FPS and Racing | Both | <1 |
| 215 | 2010 | World E-sport Masters 2010 | WEG | Hangzhou, China | - | 48 | 8 | Global | | $100,000 | WarCraft III and Counter-Strike | 2 | FPS and RTS | Both | <1 |
| 216 | 2010 | International e-Sport Festival 2010 | IEF | Wuhan, China | | 21 | 2 | Global | | $58,000 | Counter-Strike, StarCraft: BW and WarCraft III | 3 | FPS and RTS | Both | <1 |

| | | | | | | | | | | | | | | | | |
|---|---|---|---|---|---|---|---|---|---|---|---|---|---|---|---|---|
| 217 | 2010 | World Cyber Games 2010 | WCG | Los Angeles, California, USA | - | 51 | 18 | Global | | $127,000 | Carom 3D, Counter-Strike, FIFA 10, Forza Motorsport 3, League of Legends, StarCraft: BW, Tekken 6, Trackmania Nations Forever and WarCraft III | 9 | Pool, Football (Sport Simulation), Driving, MOBA, RTS, Fighting | Both | Microsoft | <1 |
| 218 | 2010 | MLG 2010 Dallas | MLG | Dallas, Texas, Usa | - | 72 | 7 | Global | 1,800,000 | $404,500 | Call of Duty: Modern Warfare 2, Halo 3, Halo: Reach, StarCraft II, Super Smash Bros Brawl and Tekken 6 | 6 | FPS, RTS and Fighting | Both | | <1 |
| 219 | 2010 | MLG 2010 D.C. | MLG | Washington, D.C., USA | 128 | 56 | 3 | NA and Asia | 1,800,000 | $92,750 | Halo 3, StarCraft II, Super Smash Bros Brawl, Tekken 6 and World of Warcraft | 5 | FPS, RTS, Fighting and MMORPG Arena | Both | | <1 |
| 220 | 2010 | MLG 2010 Raleigh | MLG | Raleigh, North Carolina, USA | - | 56 | 5 | Global | 1,800,000 | $92,750 | Halo 3, StarCraft II, Super Smash Bros Brawl, Tekken 6 and World of Warcraft | 5 | FPS, RTS, Fighting and MMORPG Arena | Both | | <1 |
| 221 | 2010 | MLG 2010 Columbus | MLG | Columbus, Ohio, USA | - | 57 | 4 | NA and Asia | 1,800,000 | $85,750 | Halo 3, Super Smash Bros Brawl, Tekken 6 and World of Warcraft | 4 | FPS, Fighting and MMORPG Arena | Both | | <1 |
| 222 | 2010 | MLG 2010 Orlando | MLG | Orlando, Flordia, USA | - | 48 | 3 | NA and Asia | 1,800,000 | $70,000 | Halo 3, Super Smash Bros Brawl and Tekken 6 | 3 | FPS and Fighting | Both | | <1 |
| 223 | 2010 | Electronic Sports World Cup 2010 | ESWC | Disneyland, Paris, France | - | 73 | 15 | Global | | $203,000 | Counter-Strike, DOTA, FIFA 10, Need for Speed: Shift, Quake Live, Super Street Figher IV, Trackmania Nations Forever, WarCraft III | 8 | FPS, MOBA, Racing, Fighting and RTS | Both | BenQ, plantronics gaming, Nvidia and ubisoft | <1 |
| 224 | 2010 | Global Starcraft II League Open Season 1 | GSL | Seoul, South Korea | - | 64 | 3 | Global | | $172,784 | Starcraft II | 1 | RTS | SP | | 1.5 |
| 225 | 2010 | Global Starcraft II League Open Season 2 | GSL | Seoul, South Korea | - | 64 | 4 | Global | | $176,942 | Starcraft II | 1 | RTS | SP | | 2 |

| | | | | | | | | | | | | | | | | | | |
|---|---|---|---|---|---|---|---|---|---|---|---|---|---|---|---|---|---|---|
| 226 | 2010 | Global Starcraft II League Open Season 3 | GSL | Seoul, South Korea | - | 64 | 5 | Global | | | $176,621 | Starcraft II | 1 | RTS | SP | | 1 | |
| 227 | 2010 | QuakeCon 2010 | QuakeCon | Hilton Anatole Hotel, Dallas, Texas, USA. | - | 11 | 9 | Global | | | $50,500 | Quake Live | 1 | FPS | SP | | <1 | |
| 228 | 2010 | BlizzCon 2010 | BlizzCon | Anaheim, California, USA | - | 16 | 10 | Global | 8,000 | | $243,000 | Warcraft III, Starcraft II and World of Warcraft | 3 | RTS and MOBA Arena | Both | | <1 | |
| 229 | 2010 | IEM V - American Championship | IEM (ESL) | New York Comic Con, New York, USA | - | 56 | 6 | NA | | ESL TV and CeBIT | $44,000 | Counter-Strike, Quake Live and StarCraft II | 3 | FPS and RTS | Both | | <1 | |
| 230 | 2010 | IEM V - Cologne | IEM (ESL) | Koehlnmessen, Cologne, Germany | - | 30 | 13 | Global | | ESL TV and CeBIT | $25,000 | Quake Live and StarCraft II | 2 | FPS and RTS | Both | | <1 | |
| 231 | 2010 | IEM V - Shangai | IEM (ESL) | Shangai, China | - | 68 | 6 | Global | | ESL TV and CeBIT | $57,000 | Counter-Strike, WarCraft III and DOTA | 3 | FPS, MOBA and RTS | Both | | <1 | |
| 232 | 2010 | IEM IV - World Championship | IEM (ESL) | Deutche Messe AG, Hanover, Germany | - | 72 | 16 | Global | | ESL TV and CeBIT | $170,000 | Counter-Strike, Warcraft 3: DOTA and Warcraft III | 3 | FPS, MOBA and RTS | Both | | <1 | |
| 233 | 2010 | IEM IV - Asian Championship | IEM (ESL) | Tapei Game Show 2010, Tapei, Taiwan | - | 59 | 11 | Asia | | ESL TV and CeBIT | $49,000 | Counter-Strike, Quake Live and World of Warcraft | 3 | FPS and MMORPG Arena | Both | | <1 | |
| 234 | 2010 | IEM IV - European Championship | IEM (ESL) | Password Culture, Cologne, Germany | - | 85 | 16 | EU | | ESL TV and CeBIT | $74,500 | Counter-Strike, Quake Live and World of Warcraft | 3 | FPS and MMORPG Arena | Both | | <1 | |
| 235 | 2010 | DreamHack Summer 2010 | DreamHack | Elmia, Jönköping, Sweden | 8,000 | 33 | 8 | Global | | | $38,940 | Counter-Strike, Heroes of Newearth, Quake Live and Street Fighter IV | 4 | FPS, MOBA and Fighting | Both | | <1 | WR participants |
| 236 | 2010 | DreamHack Winter 2010 | DreamHack | Elmia, Jönköping, Sweden | 13,608 | 31 | 11 | Global | | | $67,171 | Counter-Strike, Heroes of Newearth, Quake Live and StarCraft II | 4 | FPS, MOBA and RTS | Both | | <1 | |
| 237 | 2011 | Electronic Sports World Cup 2011 | ESWC | Hall 3 Porte de Versailles, Paris, France | - | 109 | 26 | Global | | | $200,000 | Counter-Strike, Counterstrike: Source, DOTA 2, FIFA 11, StarCraft II and Trackmania Nations Forever | 6 | FPS, MOBA, Football, RTS and Racing | Both | BenQ, plantronics gaming, Nvidia and ubisoft | <1 | |

| | | | | | | | | | | | | | | | | | | | |
|---|---|---|---|---|---|---|---|---|---|---|---|---|---|---|---|---|---|---|---|
| 238 | 201 1 | DreamHack Summer: League of Legends event 2011 | League of Legends World Championship Season 1 | Elmia, Jönköping, Sweden | 16,000 | 98 | 21 | Global | | (900000 | 160000 0 (180000 ) | | $185,775 | Bloodline Champions, Counter-Strike, Heroes of The Newerth, Quake Live, StarCraft II, Super Street Fighter IV Arcade Edition, League of Legends and StarCraft II | 8 | MOBA, FPS, RTS and Fighting | Both | | <1 |
| 239 | 201 1 | DreamHack Winter 2011 | DreamHack | Elmia, Jönköping, Sweden | 20,984 | 94 | 26 | Global | | 6,755,72 8 | 1,673,27 0 | 94,932 | $176,223 | Bloodline Champions, Counter-Strike, DOTA 2, Heroes of Newerth, Quake Live, Super Street Fighter IV Arcade Edition and StarCraft II | 7 | MOBA, FPS, Fighting and RTS | Both | | <1 |
| 240 | **201 1** | **The Internation al 2011** | **The International** | Gamescom, Cologne, Germany | 80 | 42 | 13 | Global | | | 1,500,00 0 | | $1,600,00 0 | DOTA 2 | 1 | MMORP G | TB | | <1 |
| 241 | 201 1 | Internation al e-Sport Festival 2011 | IEF | Yongin City, South Korea | - | 5 | 2 | Global | | | | | $32,000 | StarCraft:BW and WarCraft III | 2 | RTS | SP | | <1 |
| 242 | 201 1 | World Cyber Games 2011 | WCG | Busan, South Korea | - | 66 | 12 | Global | | | | | $195,000 | Counter-Strike, CrossFire, FIFA 11, League of Legends, StarCraft II, Tekken 6, WarCraft III and World of WarCraft | 8 | FPS, Football, MOBA, Fighting, RTS and MMORP GA | Both | Microsoft | <1 |
| 243 | 201 1 | MLG 2011 Providence | MLG | Providence, Rhode Island, USA | - | 88 | 4 | Global | | | | | $577,000 | Call of Duty: Black Ops, Halo:Reach, League of Legends and StarCraft II | 4 | FPS, RTS and MOBA | Both | | <1 |
| 244 | 201 1 | MLG Global Invitention al | MLG | Online | - | 13 | 6 | Global | | | | | $15,250 | StarCraft II | 1 | RTS | SP | | 2 |
| 245 | 201 1 | MLG 2011 Orlando | MLG | Orlando, Flordia, USA | - | 64 | 3 | Global | | 3,500,00 0 | 181,000 | 241,000 | $98,000 | Call of Duty: Black Ops, Halo:Reach and StarCraft II | 3 | FPS and RTS | Both | | <1 |
| 246 | 201 1 | MLG 2011 Raleigh | MLG | Raleigh, North Carolina, USA | - | 78 | 4 | Global | 15,000 | 3,200,00 0 | 138,000 | | $118,000 | Call of Duty: Black Ops, League of Legends, Halo:Reach and StarCraft II | 4 | FPS, RTS and MOBA | Both | | <1 |
| 247 | 201 1 | MLG 2011 Anaheim | MLG | Anaheim, California, USA | - | 72 | 4 | Global | | 3,500,00 0 | | 241,000 | $97,600 | Call of Duty: Black Ops, Halo:Reach and StarCraft II | 3 | FPS and RTS | Both | | <1 |

| | | | | | | | | | | | | | | | | | | |
|---|---|---|---|---|---|---|---|---|---|---|---|---|---|---|---|---|---|---|
| 248 | 2011 | MLG 2011 Columbus | MLG | Columbus, Ohio, USA | - | 68 | 5 | Global | 3,500,000 | 241,000 | $98,000 | Call of Duty: Black Ops, Halo:Reach and StarCraft II | 3 | FPS and RTS | Both | | <1 | |
| 249 | 2011 | MLG 2011 Dallas | MLG | Dallas, Texas, Usa | - | 72 | 5 | Global | 3,500,000 | 241,000 | $98,000 | Call of Duty: Black Ops, Halo:Reach and StarCraft II | 3 | FPS and RTS | Both | | <1 | |
| 250 | 2011 | Global Starcraft II League 2011 | GSL | Various | - | 64 | 7 | Global | | | $940,677 | StarCraft II | 1 | RTS | SP | | 12 | |
| 251 | 2011 | GSL World Championship Seoul 2011 | GSL | Seoul, South Korea | - | 16 | 6 | Global | | | $97,048 | StarCraft II | 1 | RTS | SP | | <1 | |
| 252 | 2011 | GSL Blizzard Cup 2011 | GSL | South Korea | - | 10 | 3 | Global | | | $58,696 | StarCraft II | 1 | RTS | SP | | <1 | |
| 253 | 2011 | GSL Super Tournament | GSL | South Korea | - | 32 | 2 | Global | | | $185,832 | StarCraft II | 1 | RTS | SP | | 1 | |
| 254 | 2011 | BlizzCon 2011 | BlizzCon | Anaheim, California, USA | - | 12 | 7 | Global | | | $205,000 | Starcraft II, World Of Warcraft | 2 | RTS and MOBA Arena | Both | | <1 | |
| 255 | 2011 | QuakeCon 2011 | QuakeCon | Hilton Anatole Hotel, Dallas, Texas, USA. | - | 13 | 10 | Global | | | $42,000 | Quake Live | 1 | FPS | SP | | <1 | |
| 256 | 2011 | CPL Invitational 2011 | CPL | Shenyang City, China | - | 21 | 5 | Global | | | $26 056 | Warcraft III, Starcraft II, Warcraft III: DOTA | 3 | MOBA and RTS | Both | | <1 | |
| 257 | 2011 | Call of Duty: Experience 2011 | COD WC | Los Angeles, California, USA | - | 28 | 3 | Global | | | $1,000,000 | Call of Duty: Modern Warfare 3 | 1 | FPS | TB | Activision | <\| | Live Music: Dropkick murphys and Kanye West |
| 258 | 2011 | IEM V - World Championships | IEM (ESL) | Deutche Messe AG, Hanover, Germany | - | 104 | 18 | Global | | | $143,500 | Counter-Strike, Quake Live, Starcraft II and League of Legends | 4 | FPS, MOBA and RTS | Both | | <1 | |
| 259 | 2011 | IEM V - European Championshps | IEM (ESL) | CyberSport Arena, Kiev, Ukraine | - | 84 | 14 | Global | | | $79,000 | Counter-Strike, Quake Live and StarCraft II | 3 | FPS and RTS | Both | | <1 | |
| 260 | 2011 | IEM VI - New York | IEM (ESL) | New York Comic Con, New York, USA | - | 96 | 14 | Global | | | $93,000 | Counter-Strike, League of Legends and StarCraft II | 3 | FPS, MOBA and RTS | Both | | <1 | |
| 261 | 2011 | IEM VI - Guangzhou | IEM (ESL) | Jinhan Exhibition Center, Guangzhou, China | - | 96 | 18 | Global | | | $93,000 | Counter-Strike, League of Legends and StarCraft II | 3 | FPS, MOBA and RTS | Both | | <1 | |
| 262 | 2011 | IEM VI - Cologne | IEM (ESL) | Koehlnmessen, Cologne, Germany | - | 56 | 15 | Global | | | $53,000 | League of Legends and Starcraft II | 2 | MOBA and RTS | Both | | <1 | |
| 263 | 2012 | IEM VI - World Championship | IEM (ESL) | Deutche Messe AG, Hanover, Germany | - | 144 | 20 | Global | | | $283,000 | Counter-Strike, League of Legends | 3 | FPS, MOBA and RTS | Both | | <1 | |

| | | | | | | | | | | | | | | | | | | | |
|---|---|---|---|---|---|---|---|---|---|---|---|---|---|---|---|---|---|---|---|
| | | | | | | | | | | | | | | and StarCraft II | | | | | |
| 264 | 201 2 | IEM VI - Sao Paulo | IEM (ESL) | Anhembi Parque, Sao Paulo, Santana | - | 16 | 11 | Global | | | | | $21,000 | StarCraft II | 1 | RTS | SP | | <1 |
| 265 | 201 2 | IEM VI - Kiev | IEM (ESL) | CyberSport Arena, Kiev, Ukraine | - | 91 | 19 | Global | | | | | $93,000 | Counter-Strike, League of Legends and StarCraft II | 3 | FPS, MOBA and RTS | Both | | <1 |
| 266 | 201 2 | IEM VII - Cologne | IEM (ESL) | ESL Arena, Cologne, Germany | - | 40 | 12 | Global | | | | | $50,000 | League of Legends | 1 | MOBA | TB | | <1 |
| 267 | 201 2 | IEM VII - Singapore | IEM (ESL) | SITEX, Singapore | - | 39 | 12 | Global | | | | | $82,000 | League of Legends and Starcraft II | 2 | MOBA and RTS | Both | | <1 |
| 268 | 201 2 | IEM VII - Gamescom | IEM (ESL) | Cologne, Germany | - | 24 | 14 | Global | | | | | $182,000 | StarCraft II and League Of Legends | 2 | RTS and MOBA | Both | | <1 |
| 269 | 201 2 | Electronic Sports World Cup 2012 | ESWC | Hall 1 Porte de Versailles, Paris, France | - | 74 | 23 | Global | | | | | $155,000 | Counter-Strike: GØ, DOTA 2, Tekken Tag Tournement 2, FIFA 13, Shootmania Storm, StarCraft II and Trackmania Nations Forever | 7 | FPS, MOBA, Fighting, Football, RTS and Racing | Both | BenQ, plantronics gaming, Nvidia and ubisoft | <1 |
| 270 | 201 2 | 2012 MLG Winter Champions hip Columbus | MLG | Columbus, Ohio, USA | - | 72 | 9 | Global | 11,700,0 00 | 3,500,00 0 | | MLG.TV | $198,800 | Halo: Reach, King of Fighters XIII, Mortal Kombat 9, Soul Calibur V and StarCraft II | 5 | FPS, Fighting and RTS | | | |
| 271 | 201 2 | 2012 MLG Spring Champions hip Anaheim | MLG | Anaheim, California, USA | - | 56 | 8 | Global | 11,700,0 00 | 4,700,00 0 | 437,000 | MLG.TV | $164,400 | League of Legends, King of Fighters XIII, Mortal Kombat 9, Soul Calibur V, StarCraft II | 5 | MOBA, Fighting and RTS | Both | | <1 |
| 272 | 201 2 | 2012 MLG Summer Champions hip Raleigh | MLG | Raleigh, NC, USA | - | 62 | 8 | Global | 11,700,0 00 | 3,500,00 0 | | MLG.TV | $149,400 | League of Legends, Mortal Combat 9, Soul Calibur 4, StarCraft II | 4 | MOBA, RTS and Fighting | Both | | <1 |
| 273 | 201 2 | 2012 MLG Fall Champions hip Dallas | MLG | Dallas, Texas, Usa | - | 68 | 7 | Global | 11,700,0 00 | 3,500,00 0 | | MLG.TV | $177,700 | Halo 4, League of Legends, StarCraft II, Mortal Kombat 9, Tekken Tag 2 | 5 | FPS, MOBA, RTS and Fighting | Both | | <1 |
| 274 | 201 2 | MLG Prizefights | MLG | Online | - | 34 | 9 | Global | | | | | $14,000 | StarCraft II and League Of Legends | 2 | FPS and MOBA | Both | | 8 |
| 275 | 201 2 | CPL Champions hip 2012 | CPL | Shenyang City, China | - | 18 | 2 | Asia | | | | | $11,750 | League of Legends, Starcraft II | 2 | MOBA and RTS | Both | | <1 |

| | | | | | | | | | | | | | | | | | |
|---|---|---|---|---|---|---|---|---|---|---|---|---|---|---|---|---|---|
| 276 | 2012 | World E-Sport Masters 2010 | WEG | Hangzhou, China | - | 38 | 7 | Global | | | $100,000 | League of Legends, Starcraft II | 2 | MOBA and RTS | Both | | <1 |
| 277 | 2012 | International e-Cultural Festival 2012 | IEF | Wuhan, China | - | 21 | 3 | Global | | | $58,000 | League of Legends, StarCraft II and WarCraft III | 3 | MOBA and RTS | Both | | <1 |
| 278 | 2012 | DreamHack Summer 2012 | DreamHack | Elmia, Jönköping, Sweden | 15,531 | 87 | 24 | Global | | 12,522,319 | $180,839 | Battlefield 3, Bloodline Champions, Counter-Strike, DOTA 2, Heroes of Newerth, League of legends, Quake Live, Street Fighter X Tekken, Super Street Fighter IV Arcade Edition and StarCraft II | 10 | MOBA, FPS, RTS and Fighting | Both | Eizo, AMD, Sapphire, Logitech, Vengance, Corsair, Alienware, MSI, Beat it , Steelseries, Intel, Platonics, Twitch, Rocket Blast, Madcatz, Capcom, Energizer, Gamecom, Paradox, Funcom, Stunlcok. Telia and Kingston Hyper X | <1 |
| 279 | 2012 | DreamHack Winter 2012 (Including LCS Season 2) | DreamHack and LCS | Elmia, Jönköping, Sweden | 20,113 | 106 | 20 | Global | | | $286,334 | DOTA 2, Battlefield 3, Counter-Strike: GØ, Heroes of Newerth, Super Street Fighter IV Arcade Edition, StarCraft II, League of Legends, Quake Live | 8 | MOBA, FPS, RTS and Fighting | Both | Eizo, AMD, Sapphire, Roccat, Logitech, Hyper X, Thor, XMG, Steelseries, Energizer, Twitch, Capcom, Madcatz, Vengence, Corsair, Alienware, Rocket Jump AB, Twitch, Gamescom, Telia | <1 |
| 280 | 2012 | DreamHack Bucharest 2012 | DreamHack | Bucharest, Romania | - | 24 | 6 | Global | | | $41,000 | Counter-Strike, League of Legends and StarCraft II | 3 | FPS, MOBA and RTS | Both | | <1 |
| 281 | 2012 | DreamHack Valencia 2012 | DreamHack | Valencia, Spain | - | 65 | 11 | Global | | | $47,526 | Call of duty: Modern Warfare 3, Counter-Strike: GØ, DOTA 2, League of Legends and StarCraft II | 5 | FPS and MOBA | Both | | <1 |
| 282 | 2012 | QuakeCon 2012 | QuakeCon | Hilton Anatole Hotel, Dallas, Texas, USA. | 2,488 | 17 | 4 | Global | 7,652 | | $29,000 | Quake Live | 1 | FPS | SP | | <1 |
| 283 | 2012 | 2012 Battle.Net World | WCS | Shangai, China | - | 32 | 18 | Global | | | $250,000 | StarCraft II | 1 | RTS | SP | | <1 |

| | | | | | | | | | | | | | | | |
|---|---|---|---|---|---|---|---|---|---|---|---|---|---|---|---|
| | | Championship | | | | | | | | | | | | | |
| 284 | 2012 | WCS 2012 Asia | WCS | Shangai, China | - | 32 | 6 | Asia | $60,000 | StarCraft II | 1 | RTS | SP | | <1 |
| 285 | 2012 | WCS 2012 North America | WCS | Raleigh, NC | - | 32 | 3 | NA | $60,000 | StarCraft II | 1 | RTS | SP | | <1 |
| 286 | 2012 | WCS 2012 Europe | WCS | Stockholm, Sweden | - | 32 | 16 | EU | $60,000 | StarCraft II | 1 | RTS | SP | | <1 |
| 287 | 2012 | WCS 2012 Global Tour | WCS | Various | - | 375 | - | Global | $311,185 | StarCraft II | 1 | RTS | SP | | 4 |
| 288 | 2012 | 2012-2013 SK Telecom Proleague | ProLeague | Seoul, South Korea | - | 93 | 6 | Korea | $266,785 | StarCraft II | 1 | RTS | SP | SK Telecom | 10 |
| 289 | 2012 | Global StarCraft League Season 1 2012 | GSL | Seoul, South Korea | - | 72 | 6 | Global | $154,956 | StarCraft II | 1 | RTS | SP | | 3 |
| 290 | 2012 | Global StarCraft League Season 2 2012 | GSL | Seoul, South Korea | - | 72 | 3 | Global | $147,780 | StarCraft II | 1 | RTS | SP | | 2 |
| 291 | 2012 | Global StarCraft League Season 3 2012 | GSL | Seoul, South Korea | - | 72 | 2 | Global | $151,777 | StarCraft II | 1 | RTS | SP | | 2 |
| 292 | 2012 | Global StarCraft League Season 4 2012 | GSL | Seoul, South Korea | - | 72 | 3 | Global | $156,316 | StarCraft II | 1 | RTS | SP | | 2 |
| 293 | 2012 | Global StarCraft League Season 5 2012 | GSL | Seoul, South Korea | - | 73 | 1 | Global | $160,727 | StarCraft II | 1 | RTS | SP | | 2 |
| 294 | 2012 | Global StarCraft League Blizzard Cup 2012 | GSL | Seoul, South Korea | - | 10 | 1 | Global | $63,285 | StarCraft II | 1 | RTS | SP | | <1 |
| 295 | 2012 | Garena Premier League Opening Event | GPL | Tapei World Center II, Taiwan | - | 15 | 3 | Asia | $6,000 | League of legends | 1 | MOBA | TB | | <1 |
| 296 | 2012 | Garena Premier League Season 1 | GPL | Singapore, Malaysia | - | 53 | 7 | Asia | $40,000 | League of legends | 1 | MOBA | TB | | 6 |
| 297 | 2012 | World Cyber Games 2012 | WCG | Kunshan, China | - | 75 | 13 | Global | $210,000 | CrossFire, DOTA, FIFA 12, StarCraft II, WarCraft III and World of Tanks | 6 | FPS, Football, RTS and MMORPGA | Both | Microsoft | <1 |
| 298 | 2012 | Star Ladder Season I | Star Ladder | Online | - | 78 | 11 | Global | $54,850 | Bloodline Champions, DOTA 2, Point Blank and World of Tanks | 4 | MOBA FPS and MMOA | Both | | 2 |
| 299 | 2012 | Star Ladder Season II | Star Ladder | Online | - | 70 | 11 | Global | $45,000 | Dota 2, Poink Blank and World of Tanks | 3 | MOBA, FPS and MMOA | Both | | 2 |
| 300 | 2012 | Star Ladder Season III | Star Ladder | Online | - | 67 | 12 | Global | $45,000 | Dota 2 and Point Blank | 2 | MOBA and FPS | TB | | 1.5 |

| | | | | | | | | | | | | | | | | | | | |
|---|---|---|---|---|---|---|---|---|---|---|---|---|---|---|---|---|---|---|---|
| 301 | 2012 | Star Ladder Season IV | Star Ladder | Online | - | 60 | 12 | Global | | | | | $43,500 | CS:OG, Dota 2 and Point Blank | 3 | FPS and MOBA | TB | | 1.5 |
| 302 | 2012 | League of Legends World Championship Season 2 | League of Legends World Championship Season 2 | USC's Galen Center, Los Angeles, USA | - | 60 | 15 | Global | 8,000 | 8,200,000 | 1,100,000 | | $2,000,000 | League of Legends | 1 | MOBA | TB | | <1 |
| 303 | 2012 | The International 2012 | The International | Benaroya Hall, Seattle, Washington, USA | - | 40 | 8 | Global | | | | | $1,600,000 | DOTA 2 | 1 | MOBA | TB | | <1 |
| 304 | 2012 | OGN The Champions Spring 2012 | OGN | Seoul, South Korea | - | 80 | 7 | Global | | | | | $170,539 | League of Legends | 1 | MOBA | TB | | 3 |
| 305 | 2012 | OGN The Champions Summer 2012 | OGN | Seoul, South Korea | - | 80 | 10 | Global | | | | | $132,684 | League of Legends | 1 | MOBA | TB | | 3 |
| 306 | 2012 | OGN The Champions Winter 2012 | OGN | Seoul, South Korea | - | 55 | 1 | Asia (Korea) | | | | | $233,442 | League of Legends | 1 | MOBA | TB | | 5 |
| 307 | 2013 | OGN The Champions Spring 2013 | OGN | Seoul, South Korea | - | 60 | 1 | Asia (Korea) | | | | | $239,425 | League of Legends | 1 | MOBA | TB | | 2,5 |
| 308 | 2013 | OGN The Champions Summer 2013 | OGN | Seoul, South Korea | - | 80 | 1 | Asia (Korea) | | | | | $243,265 | League of Legends | 1 | MOBA | TB | | 2 |
| 309 | 2013 | Electronic Sports World Cup 2013 | ESWC | Hall 7 Porte de versailes, Paris, France | - | 62 | 15 | Global | | | | | $111,000 | Call of Duty: Black Ops 2, Counter Strike: GØ, DOTA 2, FIFA 14, TrackMania 2: Stadium, ShootMania Storm | 6 | FPS, MOBA, Football and Racing | Both | Plantronics, BenQ Numéricable, LDLC Nvidia, Roccat, Dailymotion, Paris Games Week, Jeuxvideo.com, DX Racer and Ubisoft | <1 |
| 310 | 2013 | MLG 2013 Winter Championship Dallas | MLG | Dallas, Texas, Usa | - | 78 | 9 | Global | | 54,000,000 | 6,000,000 | Various | $170,000 | Call of Dyty: Black Ops 2, League of Legends and StarCraft II | 3 | FPS, RTS and MOBA | Both | | <1 |
| 311 | 2013 | MLG 2013 Spring Championship Anaheim | MLG | Anaheim, California, USA | - | 24 | 4 | Global | | 54,000,000 | 6,000,000 | Various | $68,000 | Call of Dyty: Black Ops 2, League of Legends and StarCraft II | 3 | FPS, RTS and MOBA | Both | | <1 |
| 312 | 2013 | MLG 2013 Fall Championship Columbus | MLG | Columbus, Ohio, USA | - | 72 | 14 | Global | | 54,000,000 | 6,000,000 | Various | $184,025 | Call of Duty: Ghosts and Dota 2 | 2 | FPS and MOBA | TB | | <1 |
| 313 | 2013 | MLG Rising stars invitational | MLG | Online | - | 33 | 3 | NA | | | | | $6,000 | League of legends | 1 | MOBA | TB | | 3 |
| 314 | 2013 | Global StarCraft League 2013: Season 1 | GSL | Seoul, South Korea | - | 72 | 3 | Global | | | | | $158,823 | StarCraft II | 1 | RTS | SP | | 2 |
| 315 | 2013 | Global StarCraft League | GSL | Seoul, South Korea | - | 8 | 1 | Global | | | | | $41,623 | StarCraft II | 1 | RTS | SP | | <1 |

| | | | | | | | | | | | | | | | | |
|---|---|---|---|---|---|---|---|---|---|---|---|---|---|---|---|---|
| | | Hot6ix 2013 | | | | | | | | | | | | | | |
| 316 | 2013 | International E-Cultural Festival 2013 | IEF | Seongnam City, South Korea | - | 18 | 6 | Global | $42,000 | League of Legends and Starcraft II | 2 | MOBA and RTS | Both | | <1 | |
| 317 | 2013 | 2012-2013 SK Planet ProLeague | ProLeague | South Korea | - | 93 | 6 | Global | $266,785 | StarCraft II | 1 | RTS | TB | | 8 | |
| 318 | 2013 | DreamHack Summer 2013 | DreamHack | Elmia, Jönköping, Sweden | 18,361 | 147 | 24 | Global | $187,643 | Counter-Strike, GØ, DOTA 2, Heroes of Newerth, LOL, StarCraft II and Super Street Fighter IV Arcade Edition | 6 | FPS, RTS, MOBA and Fighting | Both | Eizo, Logitech, HyperX, ASUS, Steelseries, Intel and BenQ | <1 | |
| 319 | 2013 | DreamHack Winter 2013 | DreamHack | Elmia, Jönköping, Sweden | 22,810 | 176 | 29 | Global | $481,757 | Battlefield 4, Counter-Strike: GØ, DOTA 2, Heroes of Newerth, League of Legends, Quake Live, Super Street Fighter IV: Arcade Edition and StarCraft II | 8 | FPS, RTS, MOBA and Fighting | Both | | <1 | |
| 320 | 2013 | DreamHack Bucharest 2013 | DreamHack | Bucharest, Romania | - | 43 | 11 | Global | $51,858 | Counter-Strike: GØ, League of Legends and StarCraft II | 3 | FPS, MOBA and RTS | TB | | <1 | |
| 321 | 2013 | Dreamhack Valencia 2013 | DreamHack | Valencia, Spain | - | 45 | 11 | Global | $46,818 | Call of Duty: Black Ops 2, Counter Strike: GØ, League of Legends, Super Street Fighter IV Arcade Edition and StarCraft II | 5 | FPS, MOBA and RTS | Both | | <1 | |
| 322 | 2013 | Halo 4 Global Championship | - | PAX Prime, Banaroya Hall, Seattle | - | 8 | 1 | NA | $300,000 | Halo 4 | 1 | FPS | SP | | 1 | |
| 323 | 2013 | Crossfire Stars Season 1 | - | Shangai, China | - | 80 | 13 | Global | $180,000 | CrossFire | 1 | FPS | TB | | <1 | |
| 324 | 2013 | Turbo Racing League $1,000,000 Shell-Out | - | Los Angeles, CA, USA | - | - | - | - | $1,000,000 | Turbo Racing League | 1 | Racing | SP | Dreamworks Animation | 2 | Promotional contest for the movie Turbo |
| 325 | 2013 | BlizzCon 2013: World Of Warcraft Finals | World of Warcraft Arena World Championship | Anaheim, California, USA | - | 12 | 5 | Global | $189,000 | World Of Warcraft Arena | 1 | MMORPG | TB | | <1 | Music event: Blink-182 |
| 326 | 2013 | BlizzCon 2013: StarCraft II | WCS | Anaheim, California, USA | - | 16 | | | $250,000 | StarCraft II | 1 | RTS | SP | | <1 | Music event: Blink-182 |

| | | | | | | | | | | | | | | | |
|---|---|---|---|---|---|---|---|---|---|---|---|---|---|---|---|
| | | WCS Global Finals | | | | | | | | | | | | | |
| 327 | 2013 | WCS 2013 Season 1 Korea | WCS | Online | - | 73 | 1 | Asia (Korea) | $136,053 | StarCraft II | 1 | RTS | SP | | 3 |
| 328 | 2013 | WCS 2013 Season 1 America | WCS | Online | - | 43 | 12 | NA | $104,800 | StarCraft II | 1 | RTS | SP | | 3 |
| 329 | 2013 | WCS 2013 Season 1 Europe | WCS | Online | - | 41 | 14 | EU | $104,800 | StarCraft II | 1 | RTS | SP | | 3 |
| 330 | 2013 | WCS 2013 Season 1 Finals | WCS | Online | - | 16 | 4 | Global | $15,000 | StarCraft II | 1 | RTS | SP | | 3 |
| 331 | 2013 | WCS 2013 Season 2 Korea | WCS | Online | - | 72 | 1 | Asia (Korea) | $136,702 | StarCraft II | 1 | RTS | SP | | 3 |
| 332 | 2013 | WCS 2013 Season 2 America | WCS | Online | - | 39 | 9 | NA | $104,800 | StarCraft II | 1 | RTS | SP | | 3 |
| 333 | 2013 | WCS 2013 Season 2 Europe | WCS | Online | - | 44 | 15 | EU | $104,800 | StarCraft II | 1 | RTS | SP | | 3 |
| 334 | 2013 | WCS 2013 Season 2 Finals | WCS | Online | - | 16 | 5 | Global | $150,000 | StarCraft II | 1 | RTS | SP | | <1 |
| 335 | 2013 | WCS 2013 Season 3 Korea | WCS | Online | - | 73 | 1 | Asia (Korea) | $138,460 | StarCraft II | 1 | RTS | SP | | 3 |
| 336 | 2013 | WCS 2013 Season 3 America | WCS | Online | - | 36 | 8 | NA | $104,800 | StarCraft II | 1 | RTS | SP | | 3 |
| 337 | 2013 | WCS 2013 Season 3 Europe | WCS | Online | - | 42 | 16 | EU | $104,800 | StarCraft II | 1 | RTS | SP | | 3 |
| 338 | 2013 | WCS 2013 Season 3 Finals | WCS | Online | - | 16 | 2 | Global | $150,000 | StarCraft II | 1 | RTS | SP | | <1 |
| 339 | 2013 | WPC 2013 | WPC ACE | Shanghai, China | - | 15 | 3 | Asia | $231,328 | DOTA 2 | 1 | MOBA | TB | | 3 |
| 340 | 2013 | Garena Premier League Spring 2013 | GPL | Singapore, Malaysia | - | 59 | 7 | Asia | $100,000 | League of Legends | 1 | MOBA | TB | | 4.5 |
| 341 | 2013 | Garena Premier League Summer 2013 | GPL | Singapore, Malaysia | - | 52 | 7 | Asia | $100,000 | League of Legends | 1 | MOBA | TB | | 3 |
| 342 | 2013 | 2013 Call Of Duty Championships | COD WC | Los Angeles, CA, USa | - | 32 | 5 | Global | $1,000,000 | Call of Duty: Black Ops 2 | 1 | FPS | TB | | <1 |
| 343 | 2013 | IEM VII World Championship | IEM (ESL) | Deutche Messe AG, Hanover, Germany | - | 79 | 16 | Global | $250,000 | League of Legends and Starcraft II | 2 | MOBA and RTS | Both | | <1 |
| 344 | 2013 | IEM VII Sao Paulo (Brazil) | IEM (ESL) | Campus Party, Sao Paulo, Brazil | - | 20 | 6 | Global | $50,000 | League of Legends | 1 | MOBA | TB | | <1 |
| 345 | 2013 | IEM VII Katowice | IEM (ESL) | Spodek, Katowice, Poland | - | 79 | 16 | Global | $80,000 | League of Legends and Starcraft II | 2 | MOBA and RTS | Both | Hyper+HD, Dziennik Zachodni, Naszemiasto, metro and Gryonline.pl | <1 |

| | | | | | | | | | | | | | | | | | |
|---|---|---|---|---|---|---|---|---|---|---|---|---|---|---|---|---|---|
| 346 | 2013 | IEM VIII Singapore | IEM (ESL) | SITEX, Singapore | - | 47 | 8 | Global | | | $95,000 | League of Legends and Starcraft II | 2 | MOBA and RTS | Both | Intel, BenQ, HyperX, Gigabyte and Roccat | <1 |
| 347 | 2013 | IEM VIII Cologne (1) | IEM (ESL) | ESL Arena, Cologne, Germany | - | 50 | 17 | Global | | 280,000 | $80,000 | League of Legends | 1 | MOBA | TB | Intel, BenQ, HyperX, Gigabyte and Roccat | <1 |
| 348 | 2013 | IEM VIII New York | IEM (ESL) | New York Comic Con, New York, USA | - | 16 | 5 | Global | | | $25,000 | StarCraft II | 1 | RTS | SP | Intel, BenQ, HyperX, Gigabyte and Roccat | <1 |
| 349 | 2013 | IEM VIII Shangai | IEM (ESL) | Shanghai New International Expo Center, Shangai, China | - | 36 | 6 | Global | | | $55,000 | League of Legends and Starcraft II | 2 | MOBA and RTS | Both | Intel, BenQ, HyperX, Gigabyte and Roccat | <1 |
| 350 | 2013 | The International 2013 | The International | Benaroya Hall, Seattle, Washington, USA | - | 40 | 11 | Global | | 1,012,600 | $2,874,407 | DOTA 2 | 1 | MOBA | TB | | <1 |
| 351 | 2013 | QuakeCon 2013 | QuakeCon | Hilton Anatole Hotel, Dallas, Texas, USA. | 2,800 | 12 | 4 | Global | 10,000 | | $25,000 | Doom II and Quake Live | 1 | FPS | Both | | <1 |
| 352 | 2013 | CPL Championship 2013 | CPL | Shenyang City, China | - | 3 | 3 | Global | | | $6,500 | Starcraft II | 1 | RTS | SP | | <1 |
| 353 | 2013 | Tencent LPL Summer 2013 | LPL | Taicang, China | - | 40 | 2 | Asia (China and Hong Kong) | | | $237,974 | League of Legends | 1 | MOBA | TB | | 3 |
| 354 | 2013 | Tencent LPL Spring 2013 | LPL | Taicang, China | - | 46 | 3 | Asia (China, Hong Kong and Taiwan) | | | $236,341 | League of Legends | 1 | MOBA | TB | | 3 |
| 355 | 2013 | World Cyber Games 2013 | WCG | Kunshan, China | - | 63 | 10 | Global | | | $283,500 | CrossFire, FIFA 14, League of Legends, StarCraft II, Super Street Fighter IV, WarCraft III and World of Tanks | 7 | FPS, Football, MOBA, RTS, MMORPGA and Fighting | Both | Microsoft | <1 |
| 356 | 2013 | Star Ladder Season V | Star Ladder | Online | - | 60 | 10 | Global | | | $43,500 | CS:Global Offensive, DOTA 2, Point Blank | 3 | FPS and MOBA | TB | | 3 |
| 357 | 2013 | Star Ladder Season VI | Star Ladder | Online | - | 60 | 8 | Global | | | $51,000 | CS:Global Offensive, DOTA 2, Point Blank | 3 | FPS and MOBA | TB | | 2 |
| 358 | 2013 | Star Ladder Season VII | Star Ladder | Online | - | 60 | 9 | Global | | | $55,000 | CS:Global Offensive, DOTA 2, Point Blank | 3 | FPS and MOBA | TB | | 2 |

| | | | | | | | | | | | | | | | | | | |
|---|---|---|---|---|---|---|---|---|---|---|---|---|---|---|---|---|---|---|
| 359 | 201 3 | Star Ladder Season VIII | Star Ladder | Online | - | 60 | 16 | Global | | | | $155,858 | CS:Global Offensive , DOTA 2, Point Blank | 3 | FPS and MOBA | TB | | 2 |
| 360 | 201 3 | Perfect World's Dota 2 Super league | - | Shangai, China | - | 51 | 8 | Global | | | | $171,207 | DOTA 2 | 1 | MOBA | TB | | 2 |
| 361 | 201 3 | League of Legends Champions hip Series 2014: Spring Season 3 | LCS | EU: Berlin NA: Los Angeles | 40 | 40 | 13 | NA and EU | | | | $200,000 | League of Legends | 1 | MOBA | TB | | 4 |
| 362 | 201 3 | League of Legends Champions hip Series 2014: Summer Season 3 | LCS | EU: Cologne, Germany NA: Los Angeles, California, USA | 40 | 40 | 15 | NA and EU | | | | $20,000 | League of Legends | 1 | MOBA | TB | | 4 |
| 363 | 201 3 | League of Legends World Champions hip S3 | League of Legends World Championship Season 3 | Staple Center, Los Angelses, USA | - | 70 | 18 | Global | 13,000 | 32,000,0 00 | 8,500,0 00 | $2,050,00 0 | League of Legends | 1 | MOBA | TB | | <1 |
| 364 | 201 4 | DreamHack 2014 Summer | DreamHack | Elmia, Jönköping, Sweden | 21642 | 100 | 26 | Global | | | | $116,985 | Counter Strike GO, Hearthstone, League of Legends, StarCraft II and Ultra Street Fighter IV | 5 | FPS, OCC, MOBA, RTS and Fighting | Both | Eizo, Logitech, HyperX, Asus, Roccat, Steelseries, Platonics AMD, XMG, Astro and Saphire | <1 |
| 365 | 201 4 | DreamHack 2014 Winter | DreamHack | Elmia, Jönköping, Sweden | 26672 | 145 | 20 | Global | | | | $491,272 | Counter-Strike GO, Hearthstone, League of Legends, StarCraft II and DOTA 2 | 5 | FPS, OCC, MOBA and RTS | Both | HyperX, Roccat, Razer, Monster, Asus Nvidia Gforce, Alienware, MSI Beat it, Corsair, Plantronics, Zotac, Energizer and Sapphire AMD | <1 |
| 366 | 201 4 | DreamHack 2014 Bucharest | DreamHack | Bucharest, Romania | - | 36 | 15 | Global | | | | $67,790 | DOTA 2, Heartstone, League of Legends and StarCraft II | 4 | MOBA, OCC and RTS | Both | | <1 |
| 367 | 201 4 | DreamHack 2014 Valencia | DreamHack | Valencia, Span | - | 60 | 12 | Global | | | | $70,116 | Counter-Strike: GO, Hearthstone, StarCraft II, Ultra Street Fighter IV, League of Legends | 5 | MOBA, RTS, OCC and Fighting | Both | | <1 |
| 368 | 201 4 | DreamHack Invitational I | DreamHack | Stockholm, Sweden | - | 40 | 8 | Global | | | | $10,000 | CS:GO | 1 | FPS | TB | | 3 |
| 369 | 201 4 | DreamHack Invitational II | DreamHack | Stockholm, Sweden | - | 40 | 8 | Global | | | | $30,000 | CS:GO | 1 | FPS | TB | HyperX | <1 |

| | | | | | | | | | | | | | | | | | | | | | |
|---|---|---|---|---|---|---|---|---|---|---|---|---|---|---|---|---|---|---|---|---|---|
| 370 | 2014 | Electronic Sports World Cup 2014 | ESCW | Porte de Versailes, Paris, France | - | 65 | 13 | Global | | | | | | $116,000 | Call of Duty: Ghosts, Counter Strike: GØ, FIFA 15, Just Dance 2014, ShootMania Storm and TrackMania 2: Stadium | 6 | FPS, Football, Dancing and Racing | Both | Plantronics, BenQ Numéricable, LDLC Nvidia, Roccat, Dailymotion Paris Games Week Jeuxvideo.com, DX Racer and Ubisoft | <1 | |
| 371 | 2014 | CrossFire Stars Season 2 | CFS | Chengdu, China | - | 80 | 13 | Global | | | | | | $180,000 | Crossfire | 1 | FPS | TB | | <1 | |
| 372 | 2014 | CrossFire Stars 2014 Grand Final | CFS | Seoul, South Korea | - | 20 | 5 | Global | | | | | | $105,000 | Crossfire | 1 | FPS | TB | | <1 | |
| 373 | 2014 | 2014 SK Telecom ProLeague | ProLeague | Online | - | 41 | 1 | Korea | | | | | | $154,559 | Starcraft II | 1 | RTS | SP | | 10 | |
| 374 | 2014 | Garena Premier League Winter 2014 | GPL | Singapore, Malaysia | - | 60 | 7 | Asia | | | | | Yes | $200,000 | League of Legends | 1 | MOBA | TB | | 3 | |
| 375 | 2014 | Garena Premier League Spring 2014 | GPL | Singapore, Malaysia | - | 60 | 6 | ASia | | | | | Yes | $200,000 | League of Legends | 1 | MOBA | TB | | 2 | |
| 376 | 2014 | Garena Premier League Summer 2014 | GPL | Singapore, Malaysia | - | 60 | 6 | Asia | | | | | Yes | $1,920,000 | League of Legends | 1 | MOBA | TB | | 2 | |
| 377 | 2014 | Global StarCraft II League Championship 2014 | GSL | Seoul, South Korea | - | 8 | 1 | Asia (Korea) | | | | | | $18,900 | Starcraft II | 1 | RTS | SP | | <1 | |
| 378 | 2014 | GSL Hot6ix Cup 2014 | GSL | Seoul, South Korea | - | 16 | 1 | Asia (Korea) | | | | | | $38,786 | Starcraft II | 1 | RTS | SP | | 1 | |
| 379 | 2014 | OGN The Champions Spring 2014 | OGN | Yongsan E-sports Stadium, Seoul, South Korea | - | 80 | 1 | Asia (Korea) | 12,000 | 27,000,000 | | 1,300,000 | | $249,930 | League of Legends | 1 | MOBA | TB | | 2 | |
| 380 | 2014 | OGN The Champions Summer 2014 | OGN | Yongsan E-sports Stadium, Seoul, South Korea | - | 80 | 1 | Asia (Korea) | 7,000 | | 2,000,000 | 600,000 | | $263,879 | League of Legends | 1 | MOBA | TB | | 2 | |
| 381 | 2014 | Smite Launch Events | - | Center Stage Theater Atlanta, Georgia, USA | - | 40 | 9 | Global | 1,050 | | | | | $217,909 | Smite | 1 | MOBA | TB | | <1 | Meet the proes event |
| 382 | 2014 | 2014 MLG Championship Anaheim | MLG | Anaheim, California, USA | - | 56 | 6 | Global | | | | | | **$131,000** | **Call of Duty: Ghost, Injustice: God Among us, Killer Instinct (2013), StarCraft II and Super** | **5** | **FPS, Fighting and RTS** | **Both** | | **<1** | |

| | | | | | | | | | | | | | | | |
|---|---|---|---|---|---|---|---|---|---|---|---|---|---|---|---|
| | | | | | | | | | | **Smash Bros. Melee** | | | | | |
| 383 | 2014 | World Cyber Arena 2014 | WCA | Yinchuan, China | - | 64 | 14 | Global | $1,052,300 | CrossFire, Dota 2, Hearthstone, WarCraft III and World of Tanks | 5 | | | <1 | |
| 384 | 2014 | 2014 Call Of Duty Championships | COD WC and MLG | Los Angeles, California, USA | - | 30 | 3 | Global | **$1,000,000** | **Call of Duty: Ghosts** | **1** | **FPS** | **TB** | **<1** | |
| 385 | 2014 | Wargaming World of Tanks Grand Finals | - | Warsaw, Poland | - | 21 | 3 | Global | $300,000 | World Of Tanks | 1 | MMOA | TB | <1 | |
| 386 | 2014 | BlizzCon 2014: Finals (WoW AWC) (WCS) (HSC) | WCS, WOW AWC and HSC | Anaheim, California, USA | - | 56 | 16 | Global | $750,000 | Hearthstone, StarCraft II and World Of Warcraft Arena | 3 | OCCG, RTS and MMORPG | Both | <1 | Music event: Metallica |
| 387 | 2014 | WCS 2014 Season 1 Korea | WCS | Online | - | 55 | 1 | Asia (Korea) | $166,759 | Starcraft II | 1 | RTS | SP | 3 | |
| 388 | 2014 | WCS 2014 Season 1 Noth America | WCS | Online | - | 56 | 9 | NA | $145,400 | Starcraft II | 1 | RTS | SP | 3 | |
| 389 | 2014 | WCS 2014 Season 1 Europe | WCS | Online | - | 56 | 18 | EU | $145,400 | Starcraft II | 1 | RTS | SP | 3 | |
| 390 | 2014 | WCS 2014 Season 2 Korea | WCS | Online | - | 56 | 1 | Asia (Korea) | $174,194 | Starcraft II | 1 | RTS | SP | 3 | |
| 391 | 2014 | WCS 2014 Season 2 Noth America | WCS | Online | - | 48 | 7 | NA | $140,600 | Starcraft II | 1 | RTS | SP | 3 | |
| 392 | 2014 | WCS 2014 Season 2 Europe | WCS | Online | - | 48 | 16 | EU | $140,600 | Starcraft II | 1 | RTS | SP | 3 | |
| 393 | 2014 | WCS 2014 Season 3 Korea | WCS | Online | - | 56 | 1 | Asia (Korea) | $170,982 | Starcraft II | 1 | RTS | SP | 3 | |
| 394 | 2014 | WCS 2014 Season 3 Noth America | WCS | Online | - | 48 | 7 | NA | $140,600 | Starcraft II | 1 | RTS | SP | 3 | |
| 395 | 2014 | WCS 2014 Season 3 Europe | WCS | Online | - | 48 | 14 | EU | $140,600 | Starcraft II | 1 | RTS | SP | 3 | |
| 396 | 2014 | WPC 2014 | WPC ACE | Shanghai, China | - | 42 | 7 | Global | $245,549 | DOTA 2 | 1 | MOBA | TB | 3 | |
| 397 | 2014 | World E-sport Championship | WEC | Hangzhou, China | - | 64 | 13 | Global | $317,563 | DOTA 2, Hearthstone, StarCraft II and WarCraft III | 4 | MOBA, OCC and RTS | Both | <1 | |
| 398 | 2014 | Starladder Season IX | Starseries | Various | - | 110 | 18 | Global | $269,512 | CS:Global Offensive, League of Legends, Point Blank and Dota 2 | 4 | FPS and MOBA | TB | 2.5 | |

| | | | | | | | | | | | | | | | | | | | | |
|---|---|---|---|---|---|---|---|---|---|---|---|---|---|---|---|---|---|---|---|---|
| 399 | 2014 | Starladder Season X | Starseries | Various | - | 115 | 16 | Global | | | | | $320,604 | CS:Global Offensive, League of Legends, Point Blank and Dota 2 | 4 | FPS and MOBA | TB | | 5 | |
| 400 | 2014 | Starladder Season XI | Starseries | Various | - | 100 | 13 | Global | | | | | $171,227 | CS:Global Offensive, League of Legends, Point Blank and Dota 2 | 4 | FPS and MOBA | TB | | 5 | |
| 401 | 2014 | The International 2014 | The International | Key Arena, Seattle, Washington, USA | - | 70 | 16 | Global | 17,000 | 20,000,000 | 2,000,000 | ESPN 3 | $10,931,100 | DOTA 2 | 1 | MOBA | TB | | <1 | |
| 402 | 2014 | ESL One Frankfurt 2014 | IEM (ESL) | Frankfurt, Germany | - | 40 | 13 | Global | 12,500 | | 500,000 | | $210,900 | DOTA 2 | 1 | MOBA | TB | | < 1 | First ever esport event held in a world cup stadium |
| 403 | 2014 | ESL One New York 2014 | IEM (ESL) | Madison Square Garden Theater, New York, USA | - | 40 | 13 | Global | | | | | $113,388 | DOTA 2 | 1 | MOBA | TB | Madcatz, iBuypower, T Mobile, Pizza Hut, Twitch, reedpop and joinDOTA | < 1 | 60 Different events |
| 404 | 2014 | ESL One Cologne | IEM (ESL) | Cologne, Germany | - | 80 | 15 | Global | | 409,368 | 348,638 | | $250,000 | CS:GO | 1 | FPS | TB | | < 1 | |
| 405 | 2014 | XMG Captain Draft Invitentional | XMG CD | Online | - | 20 | 8 | Global | | | | | $40,170 | DOTA 2 | 1 | MOBA | TB | | < 1 | |
| 406 | 2014 | XMG Captain Draft 2.0 | XMG CD | Online | - | 39 | 12 | Global | | | | | $275,998 | DOTA 2 | 1 | MOBA | TB | | < 1 | |
| 407 | 2014 | HoN Tour Season 2 World Finals | - | BITEC Convention Center, Bangkok, Thailand | - | 40 | 4 | Global | | | | | $181,888 | Heroes of Newerth | 1 | MOBA | TB | | <1 | |
| 408 | 2014 | ASUS ROG DreamLeague Season 1 | DreamLeague | Online and jönköping, Sweeden | - | 59 | 13 | Global | | | | | $259,996 | DOTA 2 | 1 | MOBA | TB | ASUS, ROG | < 1 | |
| 409 | 2014 | QuakeCon 2014 | QuakeCon | Hilton Anatole Hotel, Dallas, Texas, USA. | 2,200 | 18 | 8 | Global | 9,000 | | | | $42,000 | Quake Live | 1 | FPS | SP | | < 1 | |
| 410 | 2014 | i-league 2014 Season 1 | ImbaTV (i-league) | Shangai, China | - | 40 | 9 | Global | | | | | $308,200 | Dota 2 | 1 | MOBA | TB | | < 1 | |
| 411 | 2014 | i-league 2014 Season 2 | ImbaTV (i-league) | Shangai, China | - | 40 | 6 | Global | | | | | $129,269 | Dota 2 | 1 | MOBA | TB | | 3 | |
| 412 | 2014 | i-league 2014 Season 3 | ImbaTV (i-league) | Shangai, China | - | 48 | 5 | Global | | | | | $448,969 | Dota 2 and Hearthstone | 2 | MOBA and OCCG | Both | | 2 | |
| 413 | 2014 | Tencent LPL Spring 2014 | LPL | China | - | 40 | 2 | Asia (China) | | | | | $231,839 | League of Legends | 1 | MOBA | TB | | 3 | |
| 414 | 2014 | Tencent LPL Summer 2014 | LPL | China | - | 40 | 2 | Asia (China) | | | | | $235,868 | League of Legends | 1 | MOBA | TB | | 3 | |
| 415 | 2014 | IEM VIII - World Championship | IEM (ESL) | Spodek, Katowice, Poland | - | 45 | 12 | Global | 73,000 | 23,164,454 | | 643,362 | $508,000 | League of Legends, StarCraft II, Hearthstone, Counter-Strike: GO and FIFA 14 | 3 | MOBA, RTS, OCCG, FPS and Football | TB | Intel, BenQ, HyperX, Gigabyte and Roccat | <1 | |
| 416 | 2014 | IEM VIII - Cologne (2) | IEM (ESL) | ESL Arena, Cologne, Germany | - | 16 | 4 | Global | | | | | $25,000 | StarCraft II | 1 | RTS | SP | Intel, BenQ, HyperX, Gigabyte and Roccat | <1 | |
| 417 | 2014 | IEM VIII - Sao Paulo | IEM (ESL) | Campus Party, Sao Paulo, Brazil | - | 56 | 12 | Global | | | | | $75,000 | StarCraft II and League Of Legends | 2 | RTS and MOBA | Both | Intel, BenQ, HyperX, Gigabyte and Roccat | <1 | |
| 418 | 2014 | IEM IX - San Jose | IEM (ESL) | San Jose SAP Center, California, USA | - | 51 | 13 | Global | | | 4,000,000 | | $75,000 | League of Legends, StarCraft II | 2 | RTS and MOBA | Both | Intel, BenQ, HyperX, Gigabyte and Roccat | <1 | ARAM Showmatch |
| 419 | 2014 | IEM IX - Toronto | IEM (ESL) | FAN EXPO Canada, Toronto, Canada | - | 16 | 4 | Global | | | | | $25,000 | StarCraft II | 1 | RTS | SP | Intel, BenQ, HyperX, Gigabyte and Roccat | <1 | |
| 420 | 2014 | IEM IX - Shenzen | IEM (ESL) | Shenzen Cartoon and Animation Festival, Shenzen, China | - | 34 | 7 | Global | | | | | $40,000 | Leage of Legends, Starcraft II and Hearthstone | 3 | MOBA, RTS and OCCG | Both | Intel, BenQ, HyperX and Gigabyte | <1 | |
| 421 | 2014 | League of Legends Championship Series 2014: Spring Season 4 | LCS | EU: Berlin NA: Los Angeles | 40 | 10 | 12 | NA and EU | | | | | $200,000 | League of Legends | 1 | MOBA | TB | | 4 | |
| 422 | 2014 | League of Legends Championship Series 2014: Summer Season 4 | LCS | EU: Cologne, Germany NA: Los Angeles, California, USA | 40 | 10 | 14 | NA and EU | | | | | $200,000 | League of Legends | 1 | MOBA | TB | | 4 | |
| 423 | 2014 | League of Legends World Championship S4 | League of Legends World Championship Season Four | Sangam Stadium, Seoul, South Korea | 82 | 10 | 16 | Global | 65,000 | 27,000,000 | 11,200,000 | | $2,130,000 | League of Legends | 1 | MOBA | TB | | <1 | Music event: Imagine Dragons |
| 424 | 2014 | League of Legends All-Star 2014 | - | Le Zenith Theater, Paris, France | 35 | 10 | - | Global | | | 18,000,000 | | $50,000 | League of Legends | 1 | MOBA | TB | American Express Serve and Coke Zero | | All Star Event Showmatches |
| 425 | 2014 | The Summit 1 | The | Los angeles, CA, USA | - | 30 | 12 | Global | | | | | $102,358 | Dota 2 | 1 | MOBA | TB | G2A | <1 | |
| 426 | 2014 | The Summit 2 | The Sumit | Los angeles, CA, USA | - | 35 | 12 | Global | | | | | $320,589 | Dota 2 | 1 | MOBA | TB | G2A | <1 | |
| 427 | 2015 | The Summit 3 | The Sumit | Los angeles, CA, USA | - | 40 | 14 | Global | | | | | $271,685 | Dota 2 | 1 | MOBA | TB | G2A | <1 | |
| 428 | 2015 | League of Legends Championship Series 2015: Spring Season 5 | LCS | EU: Berlin, Germany NA: Santa Monica, CA, | 50 | 50 | 17 | NA and EU | | | | 290,000 | $200,000 | League of Legends | 1 | MOBA | TB | | 4 | |
| 429 | 2015 | Garena Premier League Spring 2015 | GPL | | - | 80 | 6 | Asia | | | | | $100,000 | League of legends | 1 | MOBA | TB | | 5 | |

| | | | | | | | | | | | | | | | | | | | | | | |
|---|---|---|---|---|---|---|---|---|---|---|---|---|---|---|---|---|---|---|---|---|---|---|
| 430 | 2015 | Lol Masters Series Taiwan Sping 2015 | LMS | | - | 49 | 5 | Taiwan | | | | | | | $187,637 | League of legends | 1 | MOBA | TB | | 3 | |
| 431 | 2015 | Starladder Season XII | Starseries | Various | - | 112 | 21 | Global | | | | | | | $233,040 | CS:GO, League of Legends, Point Blank and Dota 2 | 4 | MOBA and FPS | TB | | 5 | |
| 432 | 2015 | Global Starcraft League 2015 Season 1 | GSL | | - | 32 | 1 | Asia (Korea) | | | | | | | $84,041 | StarCraft II | 1 | RTS | SP | | 3 | |
| 433 | 2015 | Global Starcraft League 2015 Season 2 | GSL | | - | 24 | 1 | Asia (Korea) | | | | | | | $8,774 | StarCraft II | 1 | RTS | SP | | <1 | |
| 434 | 2015 | IEM IX - Taipei | IEM (ESL) | Tapei Game Show 2015, Tapei, Taiwan | - | 46 | 9 | Global | | | | | | | $75,000 | Leauge of Legends, StarCraft II | 2 | RTS and MOBA | Both | Intel, BenQ, HyperX, G1 Gaming and Roccat | <1 | |
| 435 | 2015 | IEM IX - Cologne | IEM (ESL) | ESL Studios, Cologne, Germany | - | 30 | 13 | Global | | | | | | | $30,000 | League of Legends | 1 | MOBA | TB | Intel, BenQ, HyperX, Gigabyte and Roccat | <1 | |
| 436 | 2015 | OGN Champion Spring 2015 | OGN | Yongsan E-sports Stadium, Seoul, South Korea | - | 52 | 1 | Asia | | | | | | | $250,152 | League of Legends | 1 | MOBA | TB | SBENU, Sonokong IB, Corsair gaming, Azubu, Maximum Gear | 5 | |
| 437 | 2015 | Wargaming World of Tanks Grand Finals 2015 | | Warsaw, Poland | - | 60 | 9 | Global | | | | | | | $300,000 | World Of Tanks | 1 | MMOA | SP | | <1 | |
| 438 | 2015 | 2015 Call Of Duty Championship | COD WC | Los angeles, CA, USA | - | 32 | 3 | Global | | | | | | | $1,000,000 | Call of Duty: Advanced Warfare | 1 | FPS | TB | | <1 | |
| 439 | 2015 | ESL One Katowice 2015: IEM IX - World Championship | IEM (ESL) | Spodek, Katowice, Poland | - | 138 | 24 | Global | 100,000 | 75,000,000 | | 8,780,000 | 1,012,742 | | $550,000 | Starcraft II, League of Legends and CS:GO | 3 | RTS, MOBA and FPS | Both | Intel, BenQ, HyperX, Gigabyte and Roccat | <1 | |
| 440 | 2015 | Heroes of the Dorm | HotD | 665 West Jefferson Boulevard, Los Angeles, CA 90007 | 3500 | 20 | 1 | NA | | | 96,000 | | | ESPN 2 | $450,000 | Heroes of the Storm | 1 | MOBA | TB | HyperX, Rosewill, Gigabyte, Steelseries, Intel, Cyberpower | 1 | Live on ESPN 2 he first time an esport has ever been broadcast on a major American television network |
| 441 | 2015 | Starcraft II World Championship Season 1: Premier League | | Cobb Energy Centre Ballroom, Atlanta, GA | - | 32 | 17 | Global | | | | | | | $217,000 | StarCraft II | 1 | RTS | SP | | 3 | |
| 442 | 2015 | Smite World | SWC | Atlanta, GA, USA | - | 40 | 12 | Global | | | | 2,926,660 | | | $2,612,971 | Smite | 1 | MOBA | TB | | <1 | |

| | | | | | | | | | | | | | | | |
|---|---|---|---|---|---|---|---|---|---|---|---|---|---|---|---|
| | | Champions hip | | | | | | | | | | | | | |
| 443 | 201 5 | Tencent LPL Spring 2015 | LPL | China | - | 82 | 82 | Asia | | $378,782 | League of Legends | 1 | MOBA | TB | 3 |
| 444 | 201 5 | Dota 2 Asian Champions hip 2015 | DAC | Shanghai, China | - | 100 | 18 | Global | 465,000 | $3,057,00 0 | DOTA 2 | 1 | MOBA | TB | 1 |

## Appendix B Thematic Analysis: References

*As explained in previous points, a separate reference list were created for the thematic analysis, as it included over 1,000 references.* This is done separately from the main reference list to make a better and an easier overview for the read and legibility. As stated in the analysis, some events were not used due of lack of data or outliners, these events are marked with (*) and greyed out.

**Event 1 - First National Space Invaders Competition**

**Event 2 - National Video Game Championship (Atari Coin-Op $50 000 World Championship)**

**Event 3 - North America Video Game Olympics 1982 - That's Incredible Ms. Pac-Man Tournament**

**Event 4 - "North America Video Game Olympics 1983 That's Incredible!"**

**Event 77 - CPL Summer 2004: Extreme World Championship**

**Event 78 - DreamHack Summer 2004**

**Event 79 - DreamHack Winter 2004**

**Event 80 - Electronic Sports World Cup 2004**

**Event 81 - MLG 2004 Atlanta**

**Event 82 - MLG 2004 Boston**

**Event 83 - MLG 2004 Chicago**

**Event 84 - MLG 2004 Dallas**

**Event 100 - MLG 2005 Nashville**

**Event 101 - MLG 2005 Las Vegas**

**Event 102 - MLG 2005 Philadelphia**

**Event 103 - MLG 2005 St. Louis**

**Event 104 - MLG 2005 Orlando**

**Event 105 - MLG 2005 Houston**

**Event 106 - MLG 2005 San Francisco**

**Event 127 - Electronic Sports World Cup 2006**

**Event 128 - DreamHack Summer 2006: WSCG 2006**

**Event 129 - DreamHack Winter 2006**

**Event 130 - QuakeCon 2006**

**Event 131 - International E-Sport Festival 2006**

**Event 132 - MLG 2006 Las Vegas**

**Event 133 - MLG 2006 Orlando**

**Event 264 - IEM VI - Sao Paulo**

**Event 265 - IEM VI – Kiev**

**Event 266 - IEM VII – Cologne**

**Event 267 - IEM VII – Singapore**

**Event 268 - IEM VII – Gamescom**

**Event 403 - ESL One New York 2014**

**Event 404 - ESL One Cologne**

**Event 405 - XMG Captain Draft Invitational**

**Event 406 - XMG Captain Draft 2.0**

**Event 407 - HoN Tour Season 2 World Finals**

**Event 408 - ASUS ROG DreamLeague Season 1**

**Event 409 - QuakeCon 2014**

**Event 410 - i-league 2014 Season 1**

## Appendix C Thematic analysis and Business analysis: Various findings not presented

All models attached in Appendix C is based on data collected from Appendix B. The subsequent models is various figures and tables not used in the thesis, but has relevance to the content.

### Appendix C I: Single Player or Team Based Numbers

**SPTB Single player (SP) or team based (TB)?**

| | | Frequency | Percent | Valid Percent | Cumulative Percent |
|---|---|---|---|---|---|
| Valid | SP | 114 | 25.6 | 26.5 | 26.5 |
| | TB | 110 | 24.7 | 25.6 | 52.1 |
| | Both | 206 | 46.3 | 47.9 | 100.0 |
| | Total | 430 | 96.6 | 100.0 | |
| Missing | System | 15 | 3.4 | | |
| Total | | 445 | 100.0 | | |

### Appendix C II: Event Location Percentage

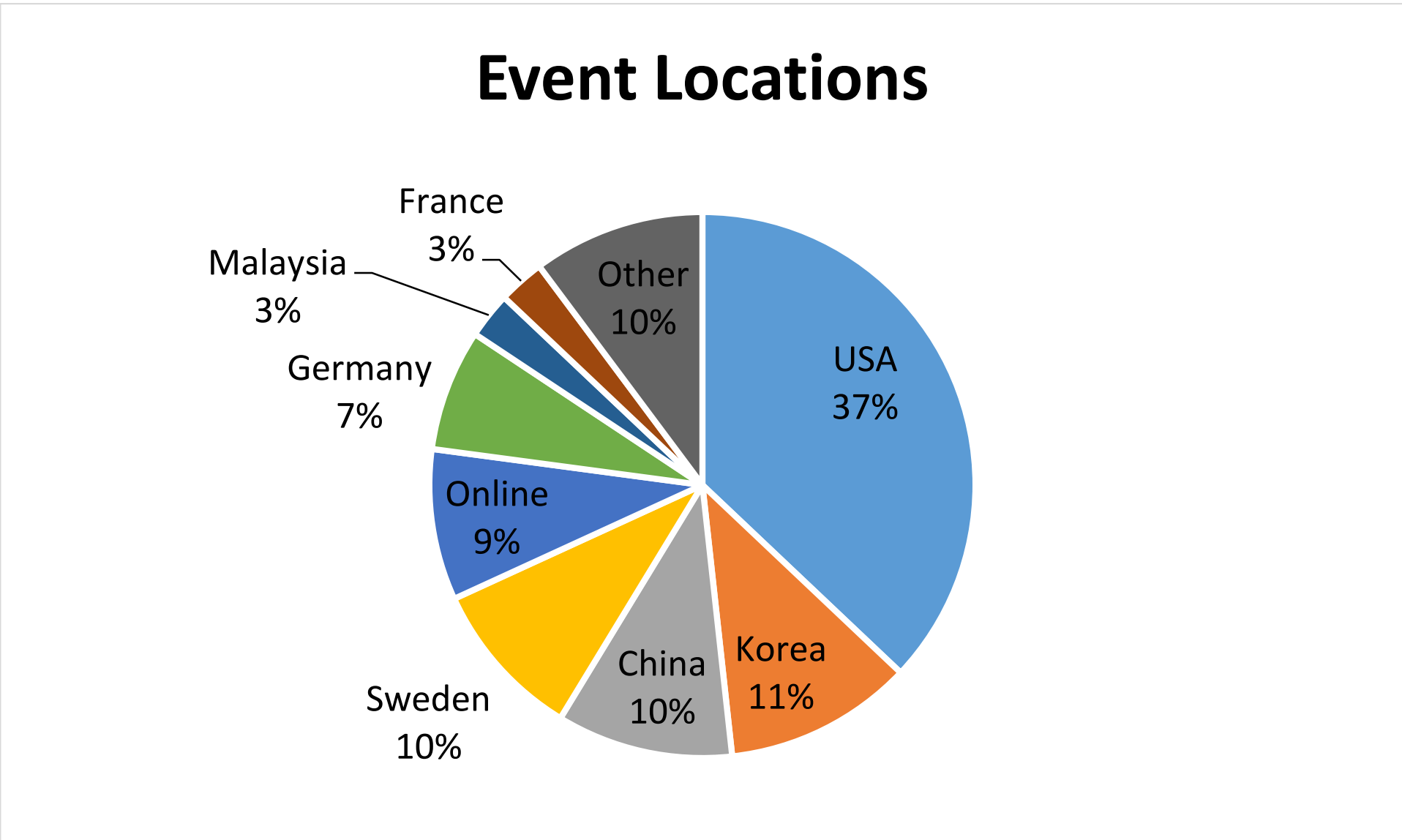

## Appendix C III: Games Percentage

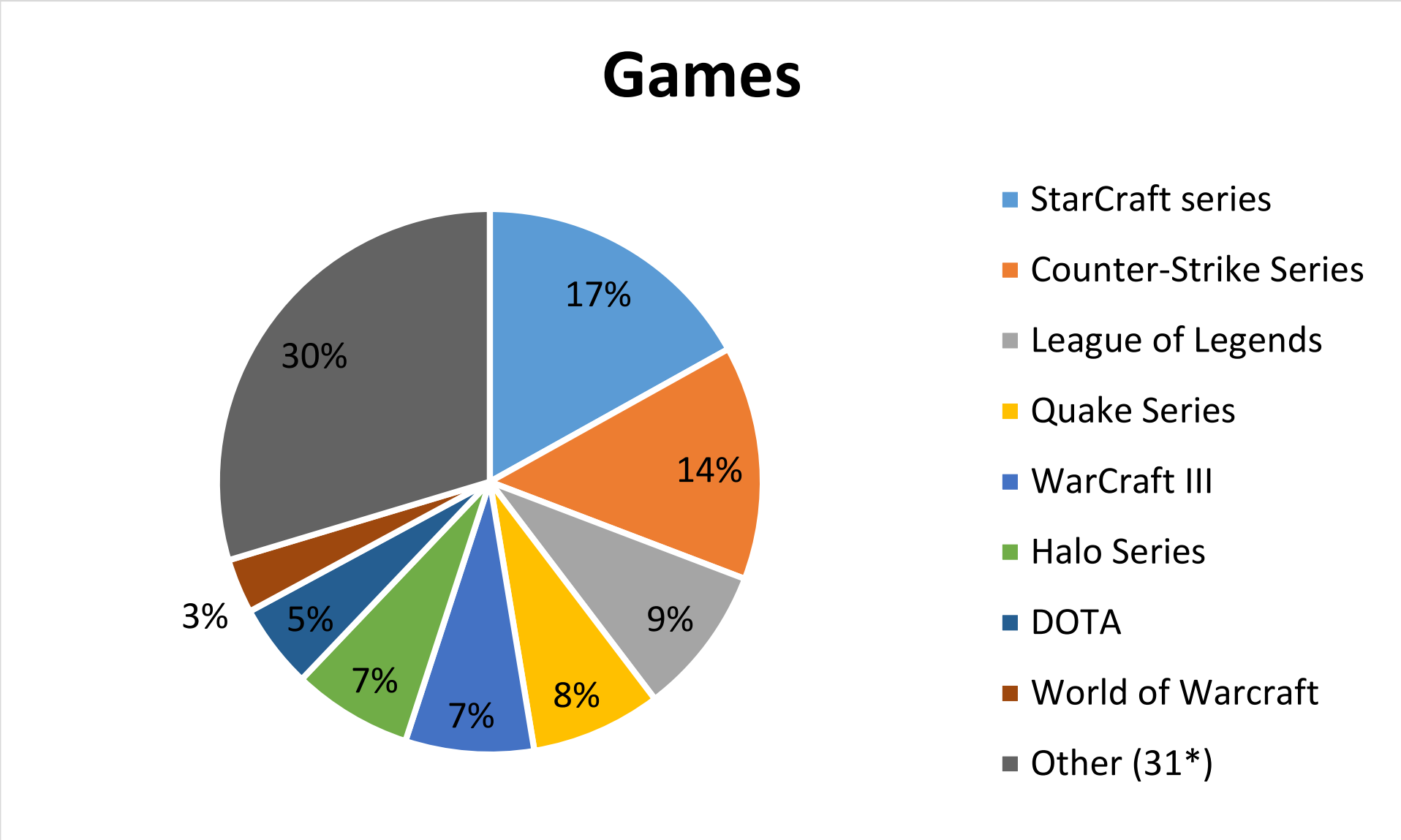

## Appendix C IV: Game Genre Percentage

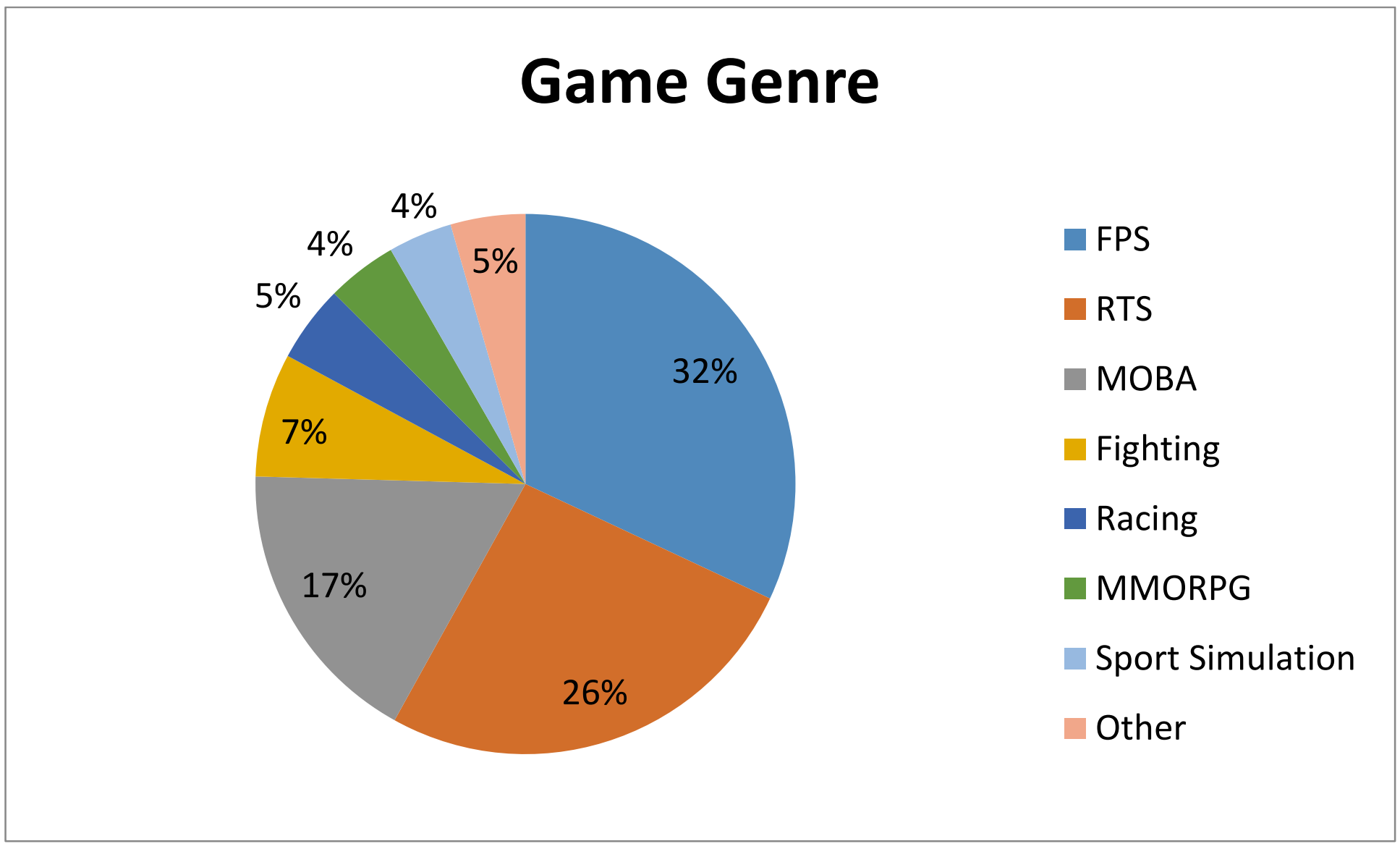

## Appendix C V: Plot analysis [X] Game Genre [Y] Year

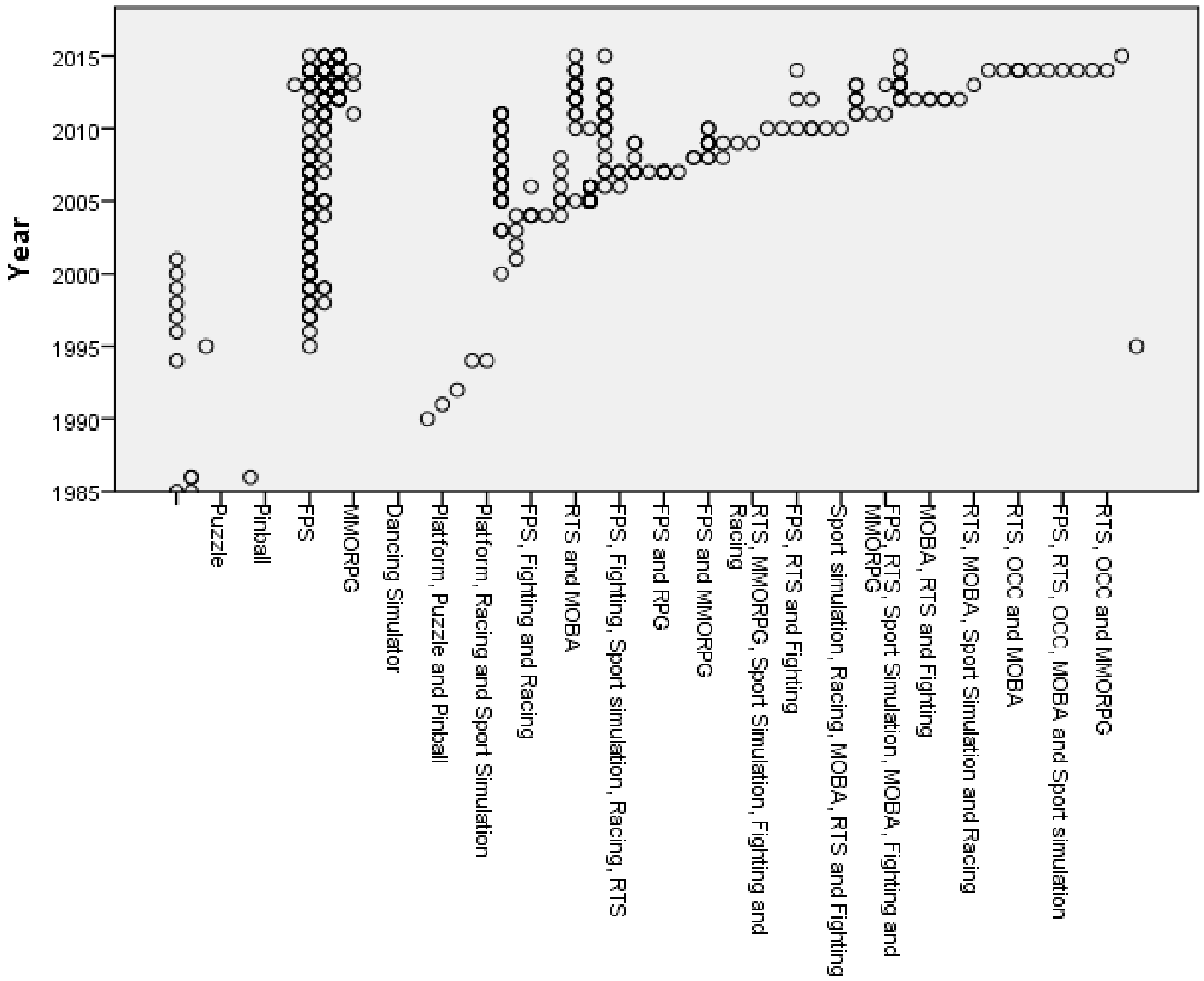

## Appendix C VI: Prize pool Numerical numbers

| Years | Total | Avarage | Events |
|---|---|---|---|
| 1980 | | | |
| 1981 | 50000 | | |
| 1982 | 500 | | |
| 1983 | | | |
| 1984 | | | |
| 1985 | | | |
| 1986 | | | |
| 1987 | | | |
| 1988 | | | |
| 1989 | | | |
| 1990 | 10000 | | 1 |
| 1991 | | | |
| 1992 | 15000 | | 1 |
| 1993 | | | |
| 1994 | | | |
| 1995 | | | |
| 1996 | | | |
| 1997 | 7000 | 3500 | 2 |
| 1998 | 46200 | 15400 | 4 |
| 1999 | 173815 | 34783 | 5 |
| 2000 | 265000 | 132500 | 6 |
| 2001 | 605370 | 100895 | 6 |
| 2002 | 646202 | 107700 | 6 |
| 2003 | 707318 | 78591 | 9 |
| 2004 | 1329858 | 110822 | 13 |
| 2005 | 1982352 | 61949 | 35 |
| 2006 | 1656583 | 82829 | 22 |
| 2007 | 3851799 | 137564 | 29 |
| 2008 | 2549525 | 169968 | 17 |
| 2009 | 1898161 | 94908 | 21 |
| 2010 | 2579208 | 117237 | 22 |
| 2011 | 5481601 | 228400 | 26 |
| 2012 | 8445352 | 201800 | 44 |
| 2013 | 11268837 | 220958 | 57 |
| 2014 | 14507406 | 268656 | 63 |
| 2015 | 8055615 | 575401 | 18 |

## Appendix C VII: Duration Numerical Data

**Duration**

| | | Frequency | Percent | Valid Percent | Cumulative Percent |
|---|---|---|---|---|---|
| Valid | | 25 | 5.6 | 5.6 | 5.6 |
| | <1 | 322 | 72.4 | 72.4 | 78.0 |
| | 1 | 6 | 1.3 | 1.3 | 79.3 |
| | 1.5 | 3 | .7 | .7 | 80.0 |
| | 10 | 5 | 1.1 | 1.1 | 81.1 |
| | 12 | 1 | .2 | .2 | 81.3 |
| | 2 | 21 | 4.7 | 4.7 | 86.1 |
| | 2,5 | 1 | .2 | .2 | 86.3 |
| | 2.5 | 1 | .2 | .2 | 86.5 |
| | 3 | 42 | 9.4 | 9.4 | 96.0 |
| | 4 | 8 | 1.8 | 1.8 | 97.8 |
| | 4.5 | 1 | .2 | .2 | 98.0 |
| | 5 | 6 | 1.3 | 1.3 | 99.3 |
| | 6 | 1 | .2 | .2 | 99.6 |
| | 8 | 2 | .4 | .4 | 100.0 |
| | Total | 445 | 100.0 | 100.0 | |

## Appendix C VIII: Live Audience Numerical Data

| AVERAGE VALUES | | | | |
|---|---|---|---|---|
| | Live Audience | Streaming Total views | Unique veiwers | Viewing Peak |
| 2015 | | | 4,507,000 | 651,371 |
| 2014 | 24,549 | 26,721,000 | 9,268,000 | 898,667 |
| 2013 | 11,500 | 48,500,000 | 6,000,000 | 3,264,200 |
| 2012 | 7,826 | 11,000,000 | 4,803,720 | |
| 2011 | | 3,550,811 | 1,018,454 | 211,786 |
| 2010 | | 1,800,000 | | |
| 2009 | | | | |
| 2008 | | | | |

| | |
|---|---|
| 2007 | 4,000 |
| 2006 | |
| 2005 | |
| 2004 | 63,000 |
| 2003 | |
| 2002 | |
| 2001 | |
| 2000 | 1567 |

| TOTAL VALUES | | | | |
|---|---|---|---|---|
| | Live Audience | Streaming Total views | Unique viewers | Viewing Peak |
| 2015 | 100,000 | 96,000 | 12,171,660 | 1,302,742 |
| 2014 | 196,550 | 77,164,454 | 55,609,368 | 5,392,000 |
| 2013 | 23,000 | 194,000,000 | 18,000,000 | 9,792,600 |
| 2012 | 15,652 | 55,000,000 | 28,882,319 | |
| 2011 | 15,000 | 24,885,728 | 5,092,270 | 1,058,932 |
| 2010 | 8,000 | 9,000,000 | | |
| 2009 | 7,000 | | | |
| 2008 | 6,000 | | | |
| 2007 | 8,000 | | | |
| 2006 | 4,000 | | | |
| 2005 | 126,000 | | | |
| 2004 | 5,000 | | | |
| 2003 | 4,000 | | | |
| 2002 | 3,250 | | | |

| | |
|---|---|
| 200 | |
| 1 | 3,000 |
| 200 | |
| 0 | 4,700 |
| 199 | |
| 9 | 1,100 |
| 199 | |
| 8 | |
| 199 | |
| 7 | 650 |

## Appendix C IX: Prize Pool Matrix Plot [X] Prize Pool [Y] Year

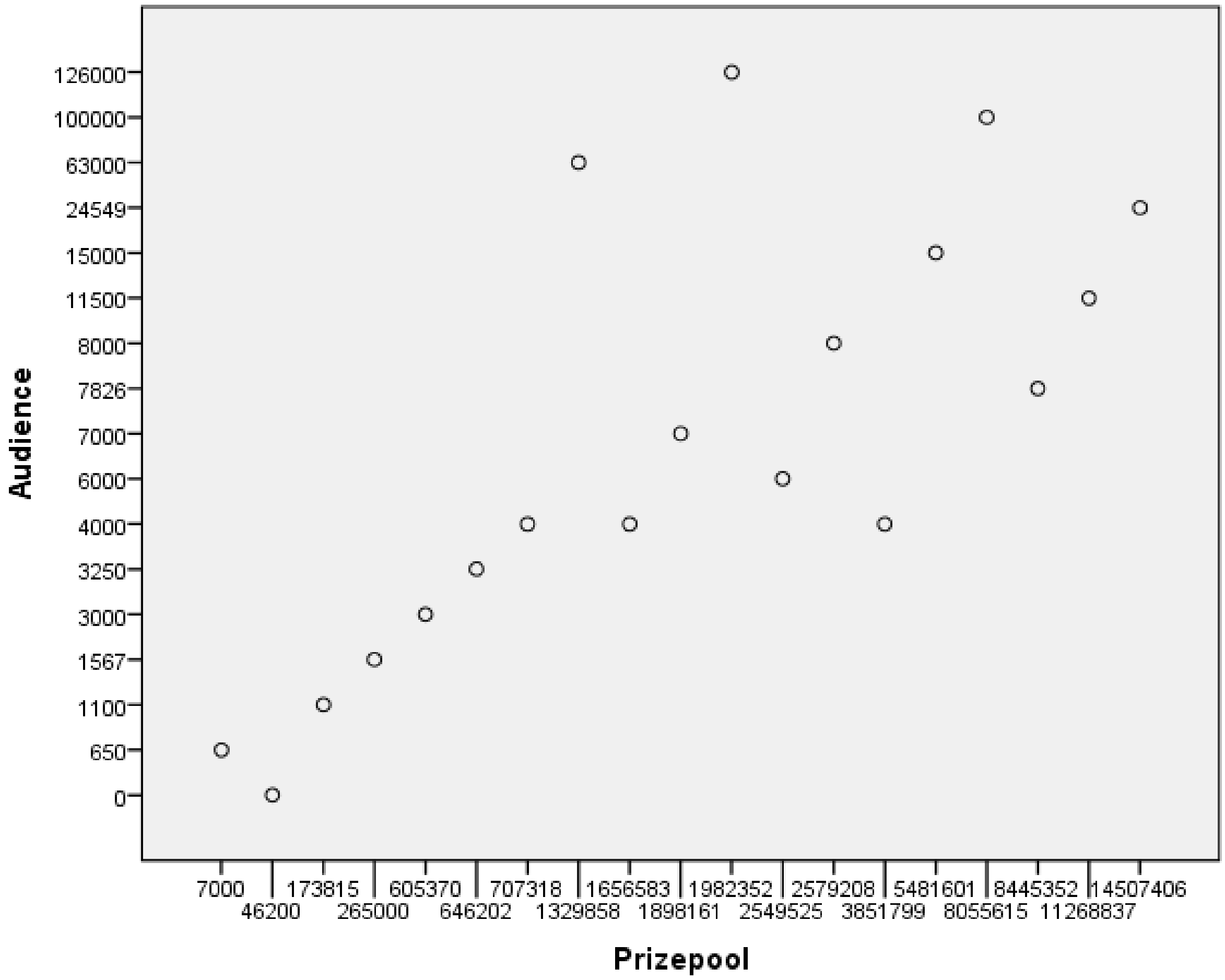

## Appendix D Business analytics: Trend analysis

All models attached in Appendix D is based on data collected from Appendix B. The subsequent models is trend analyses used and referred to in Chapter 5.

### Appendix D I: Unique views

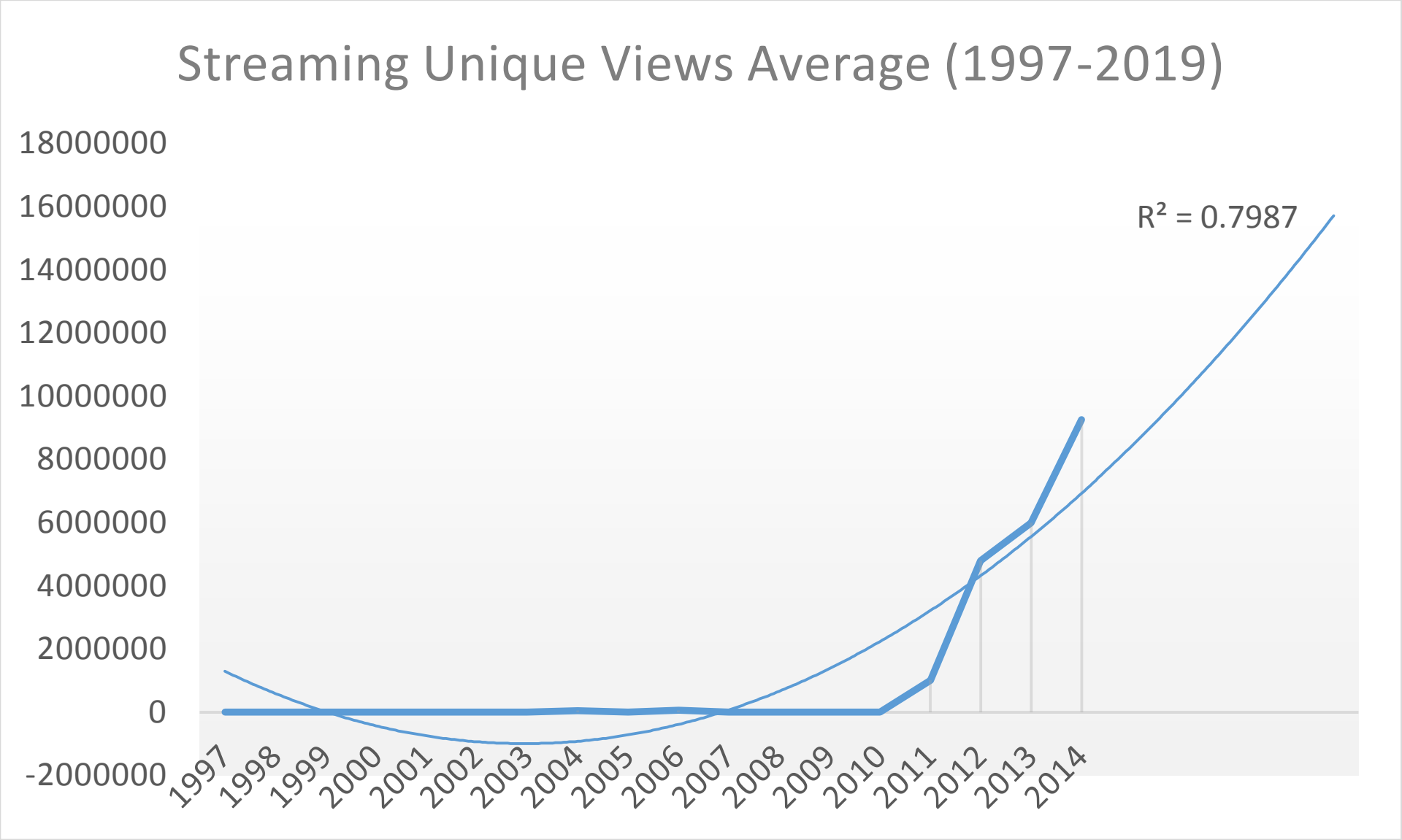

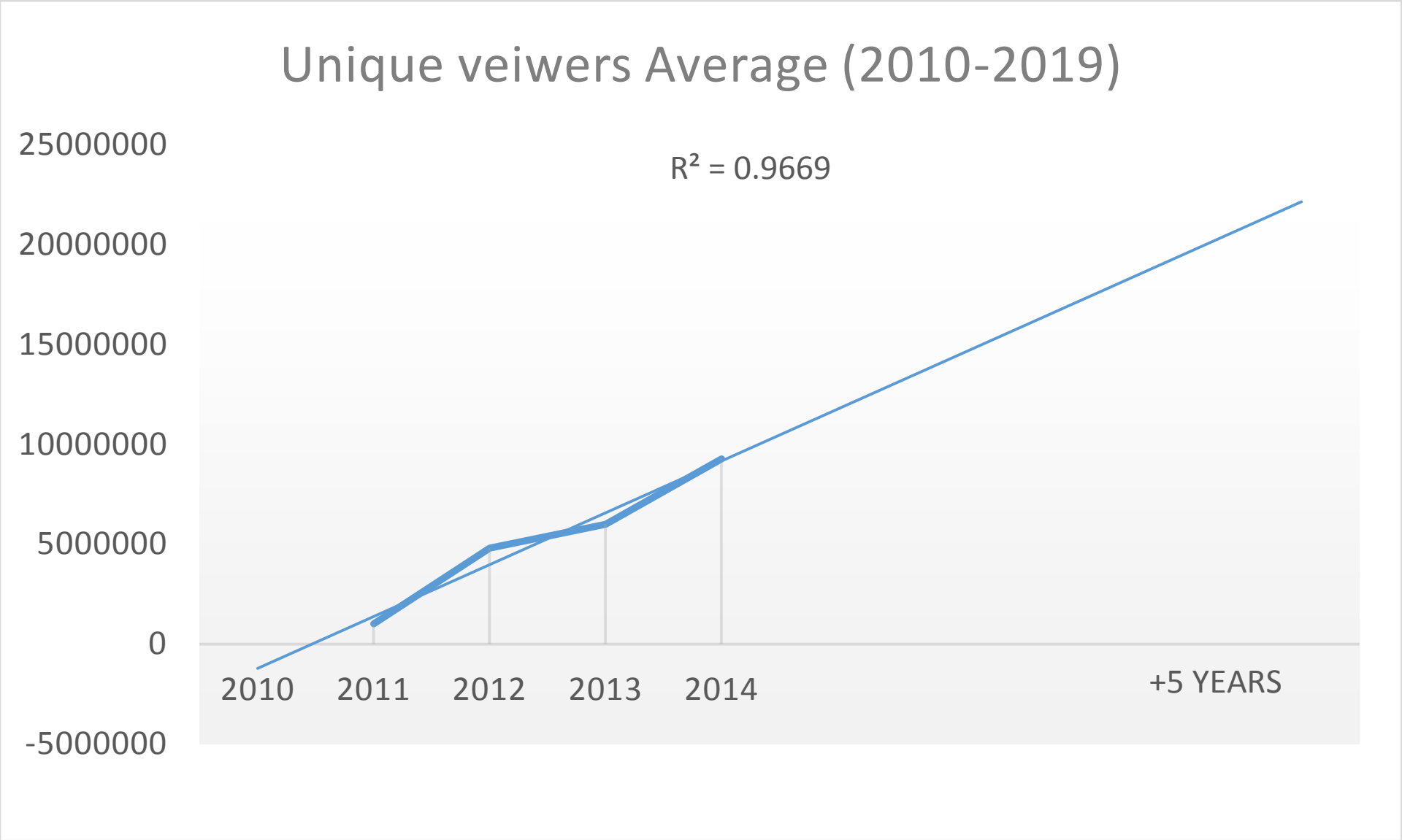

## Appendix D II: Audience

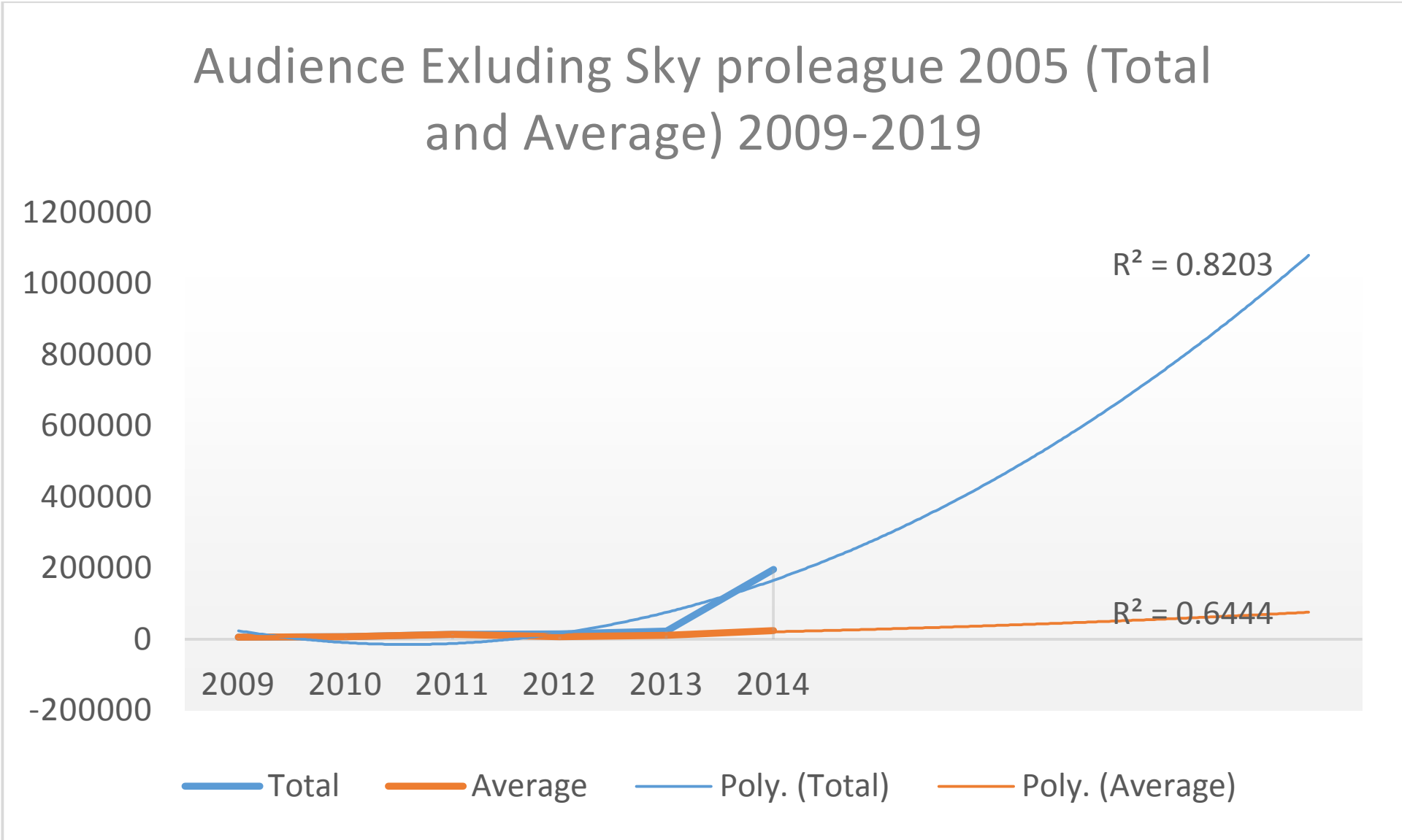

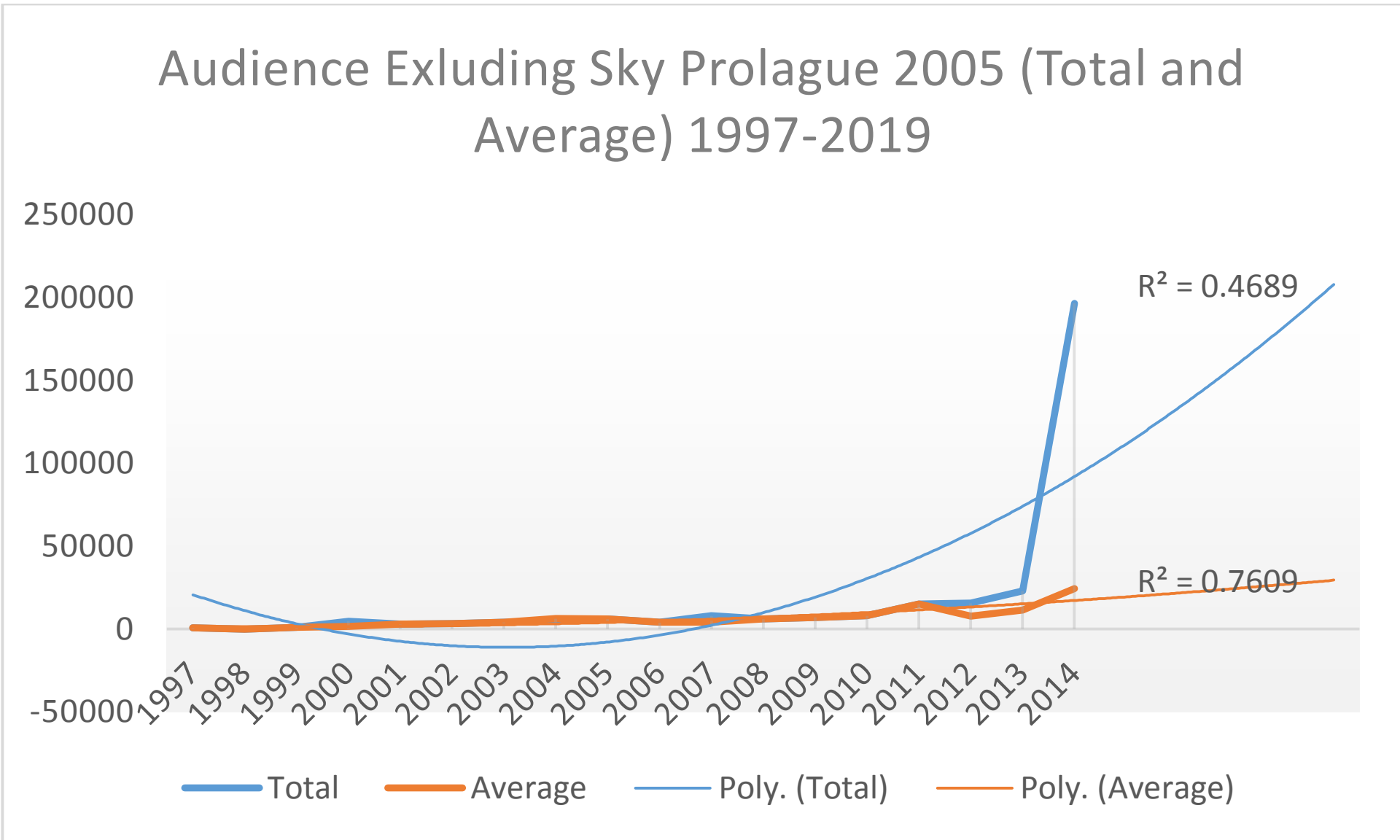

## Appendix D III: SP TB

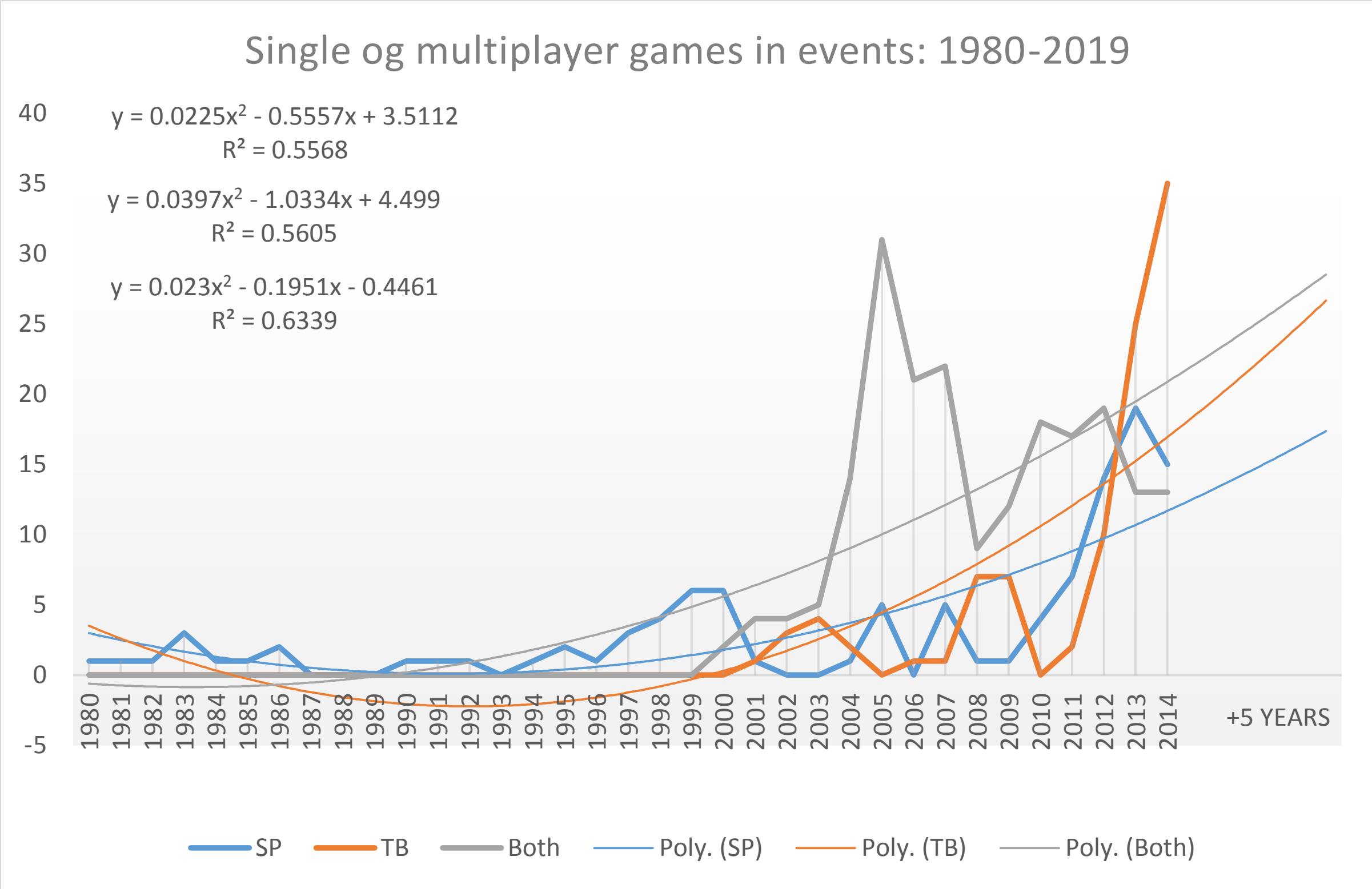

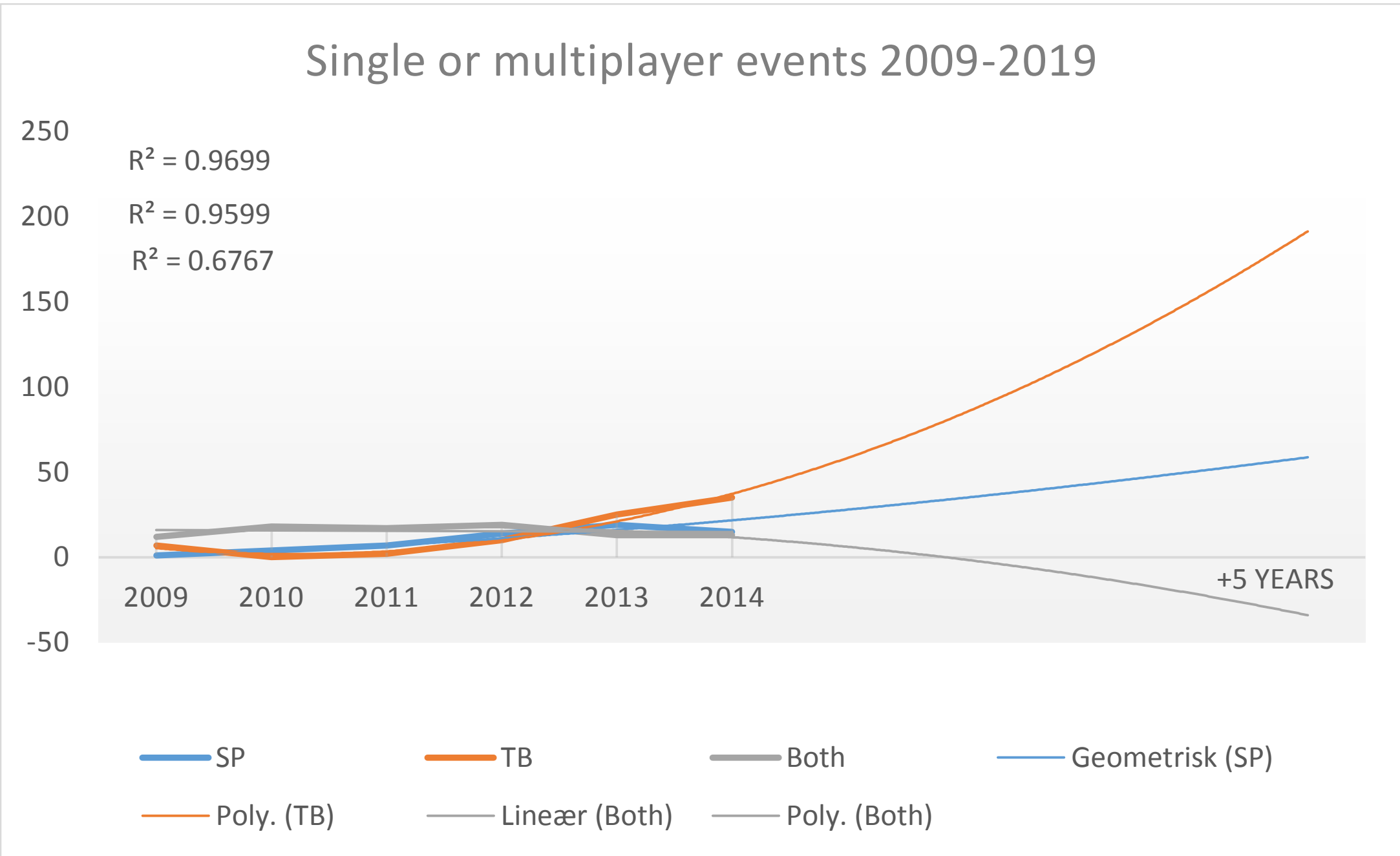

## Appendix D IV: Participants total and average

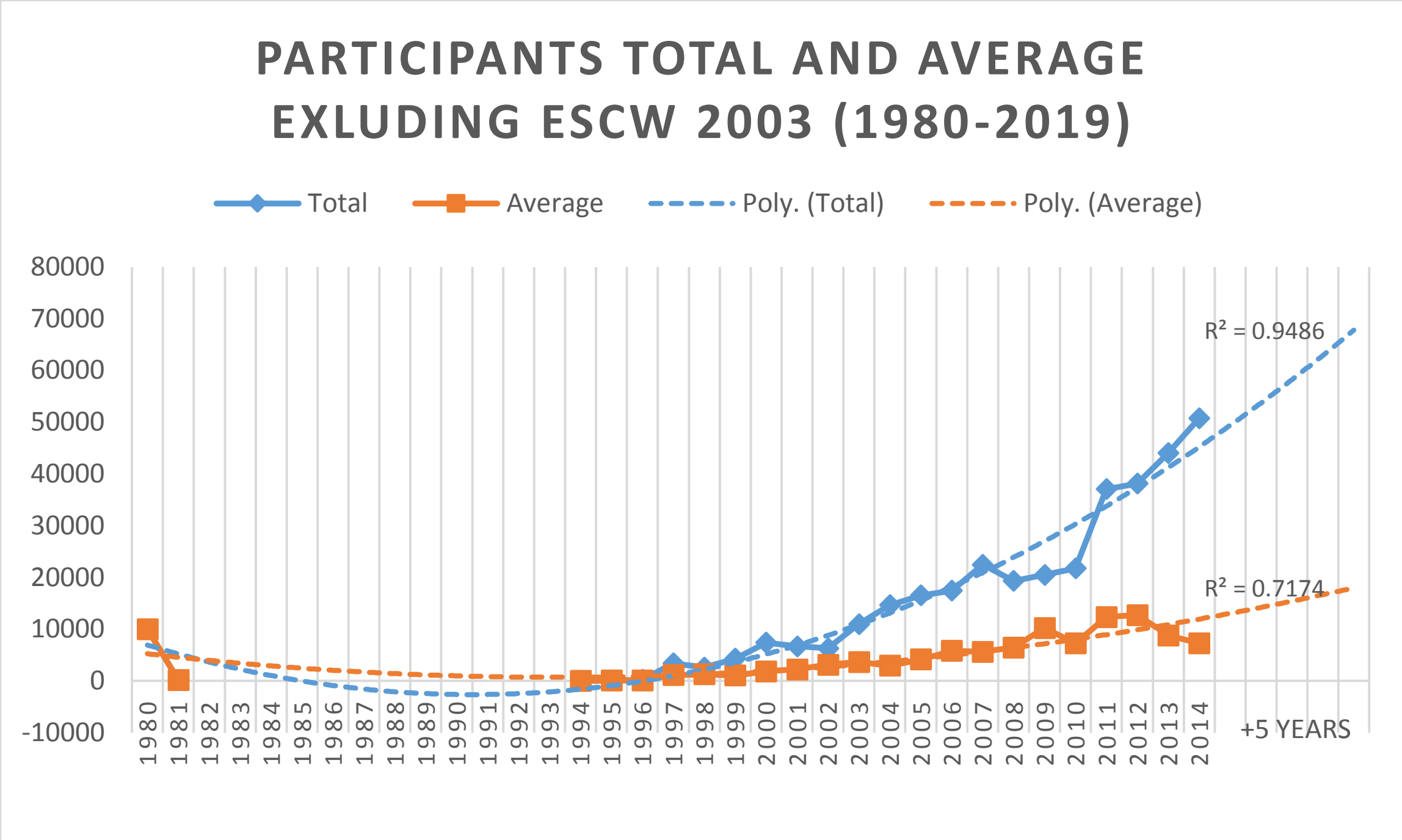

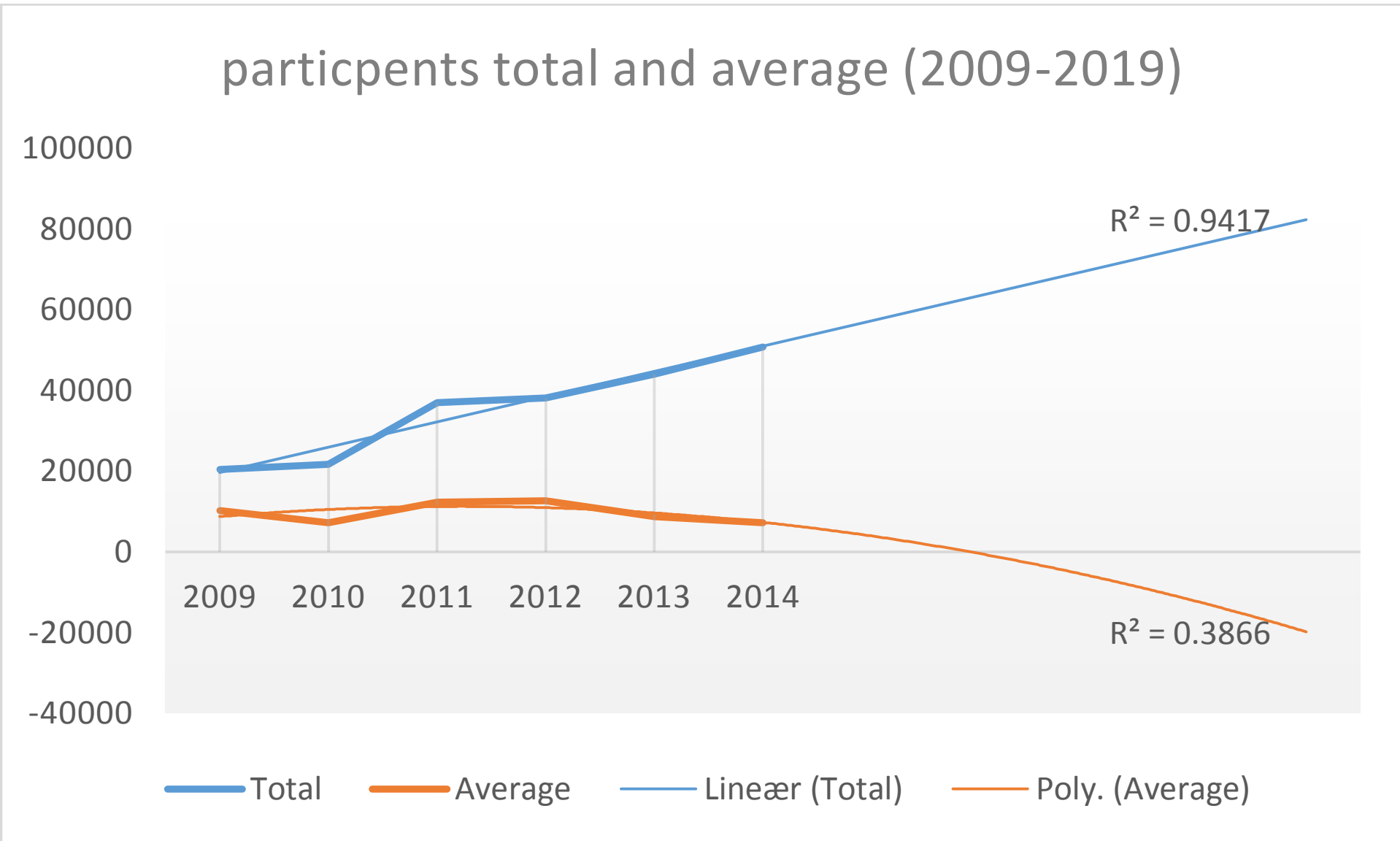

## Appendix D V: Participant countries

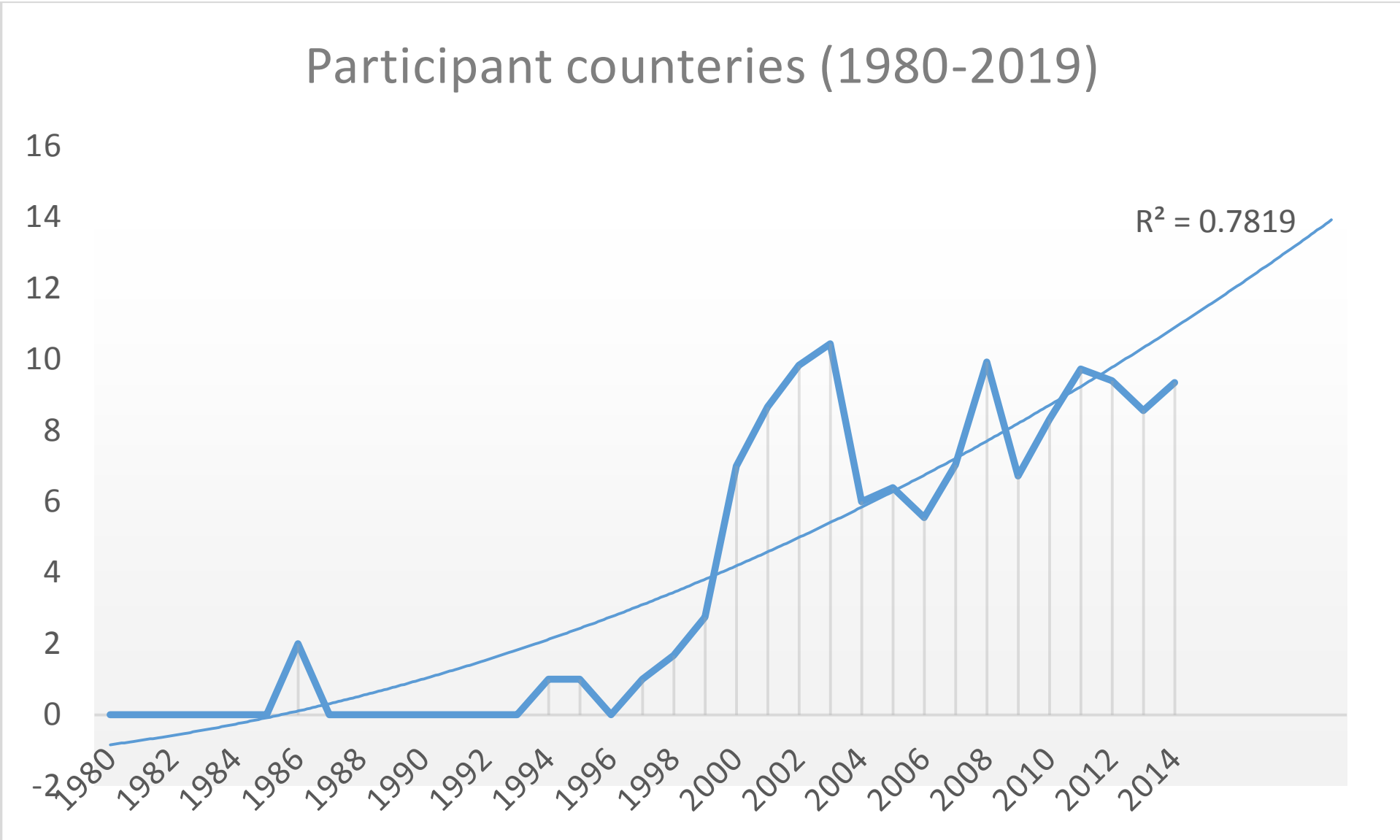

## Appendix D VI: Participants total

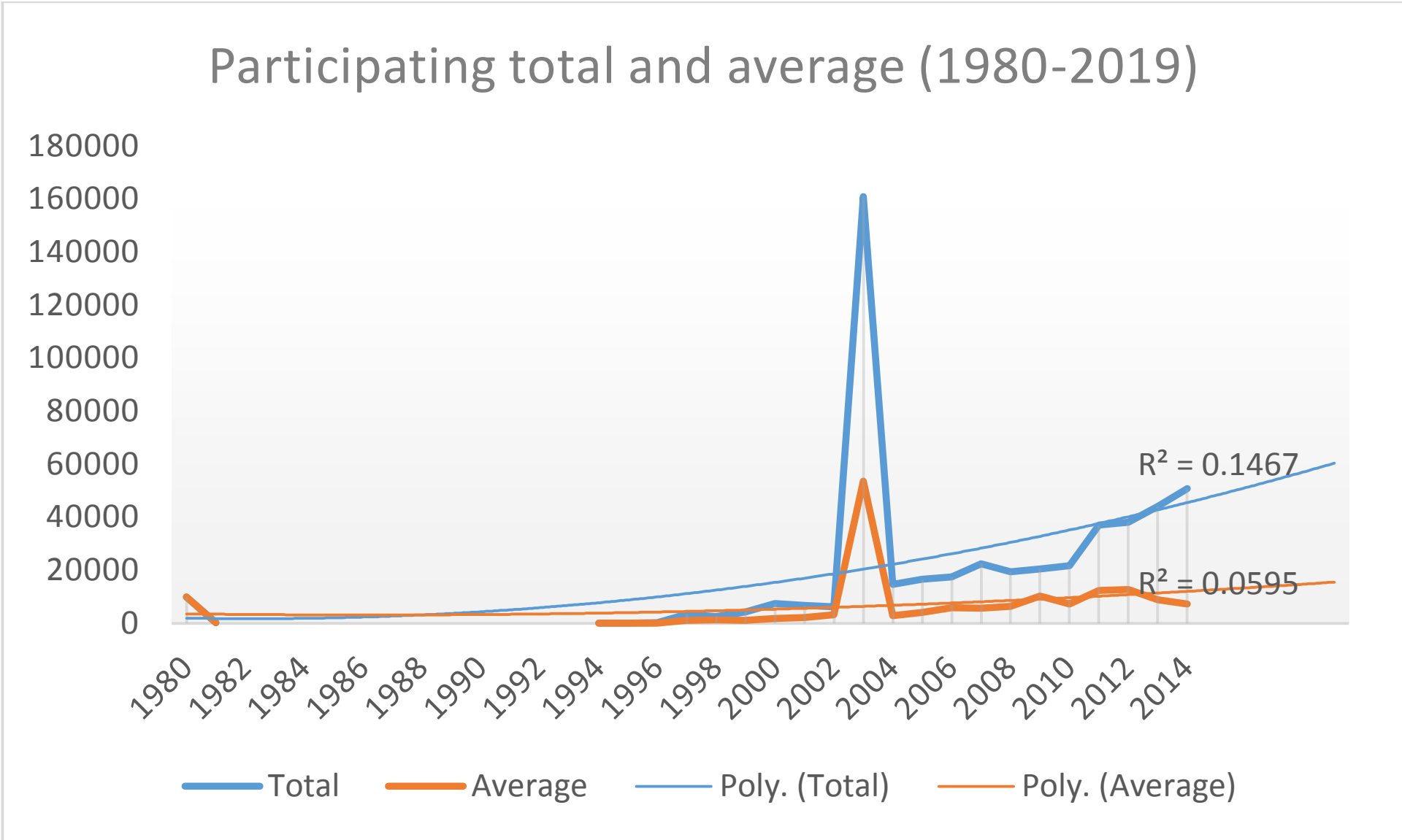

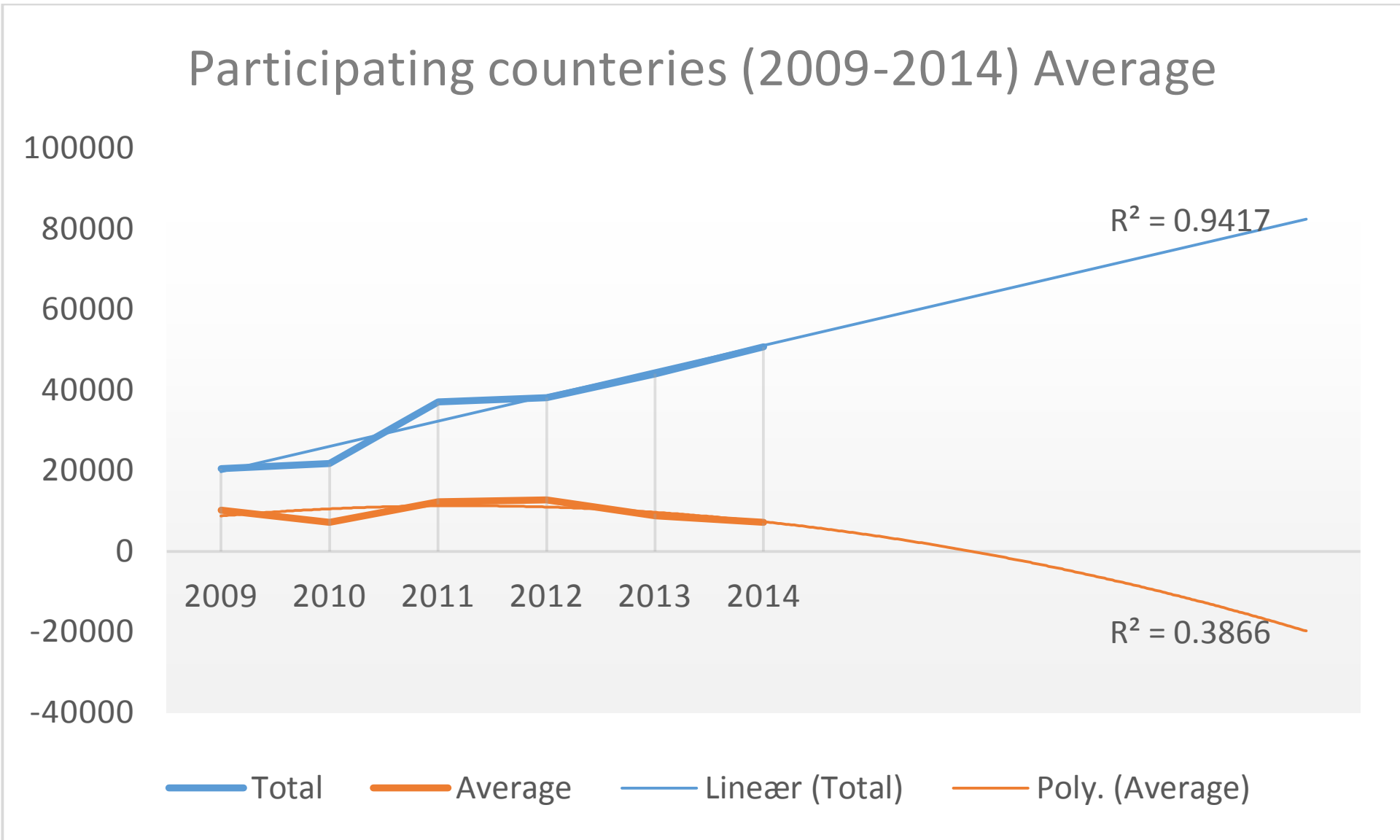

## Appendix D VII: Participants finalists

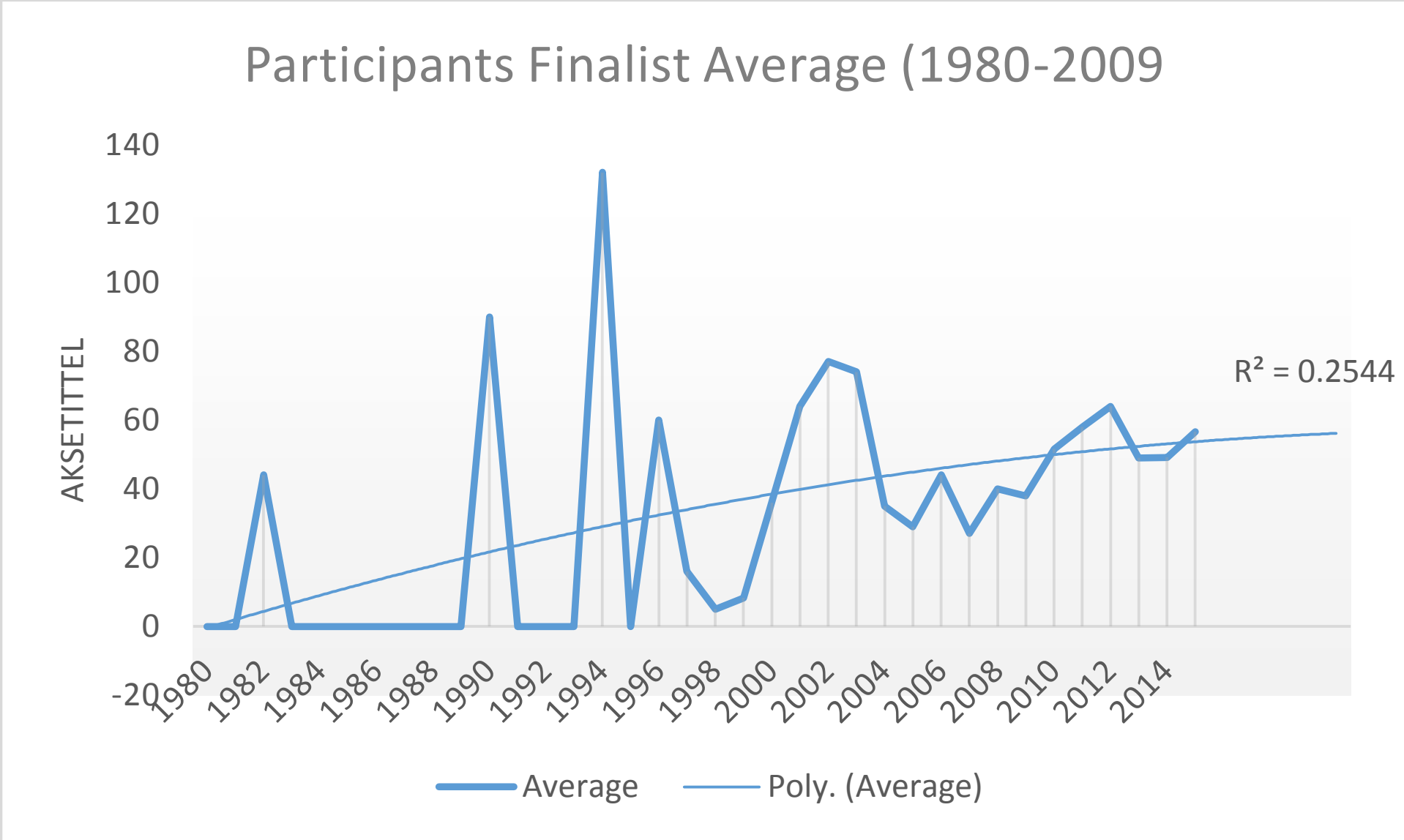

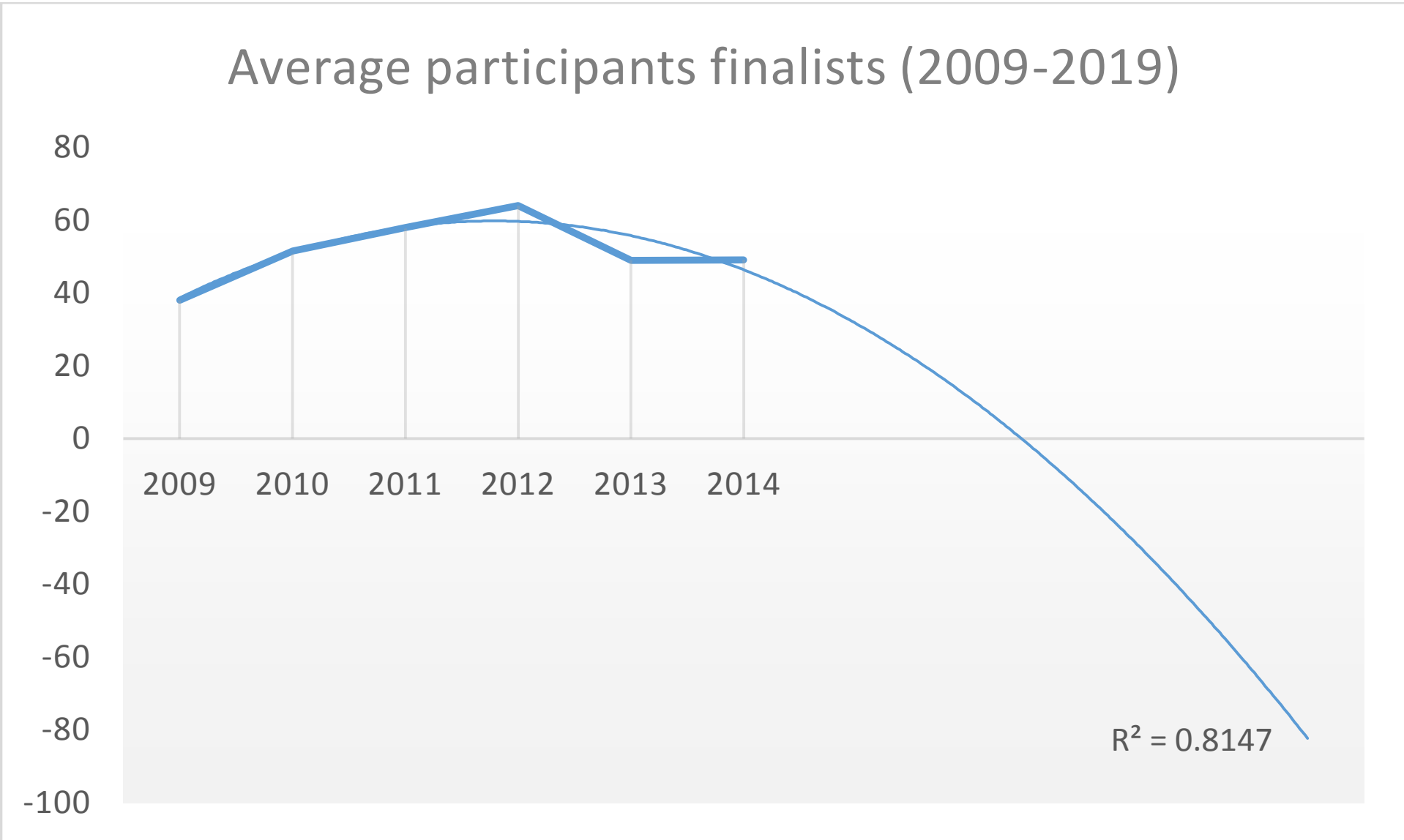

## Appendix D VIII: Duration

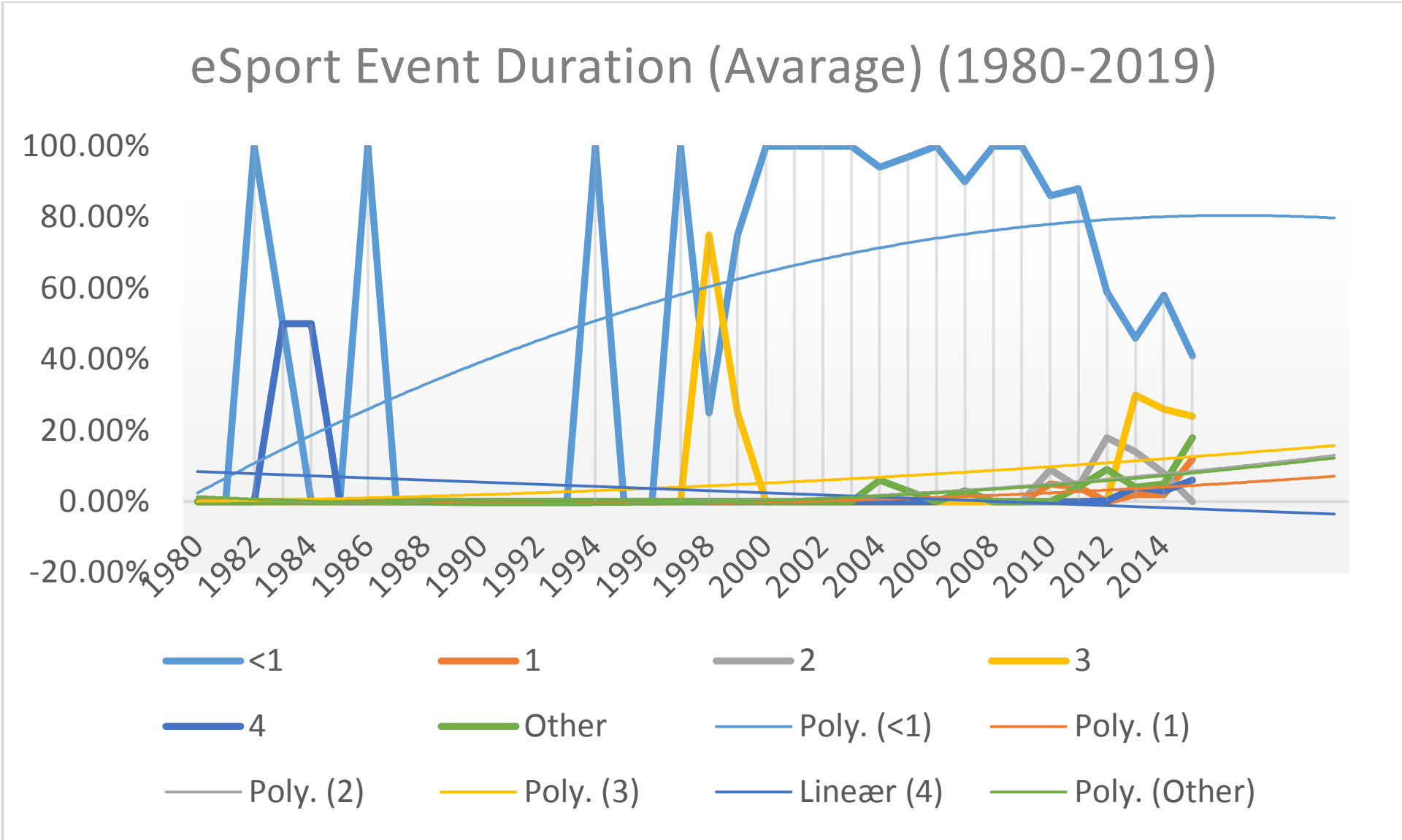

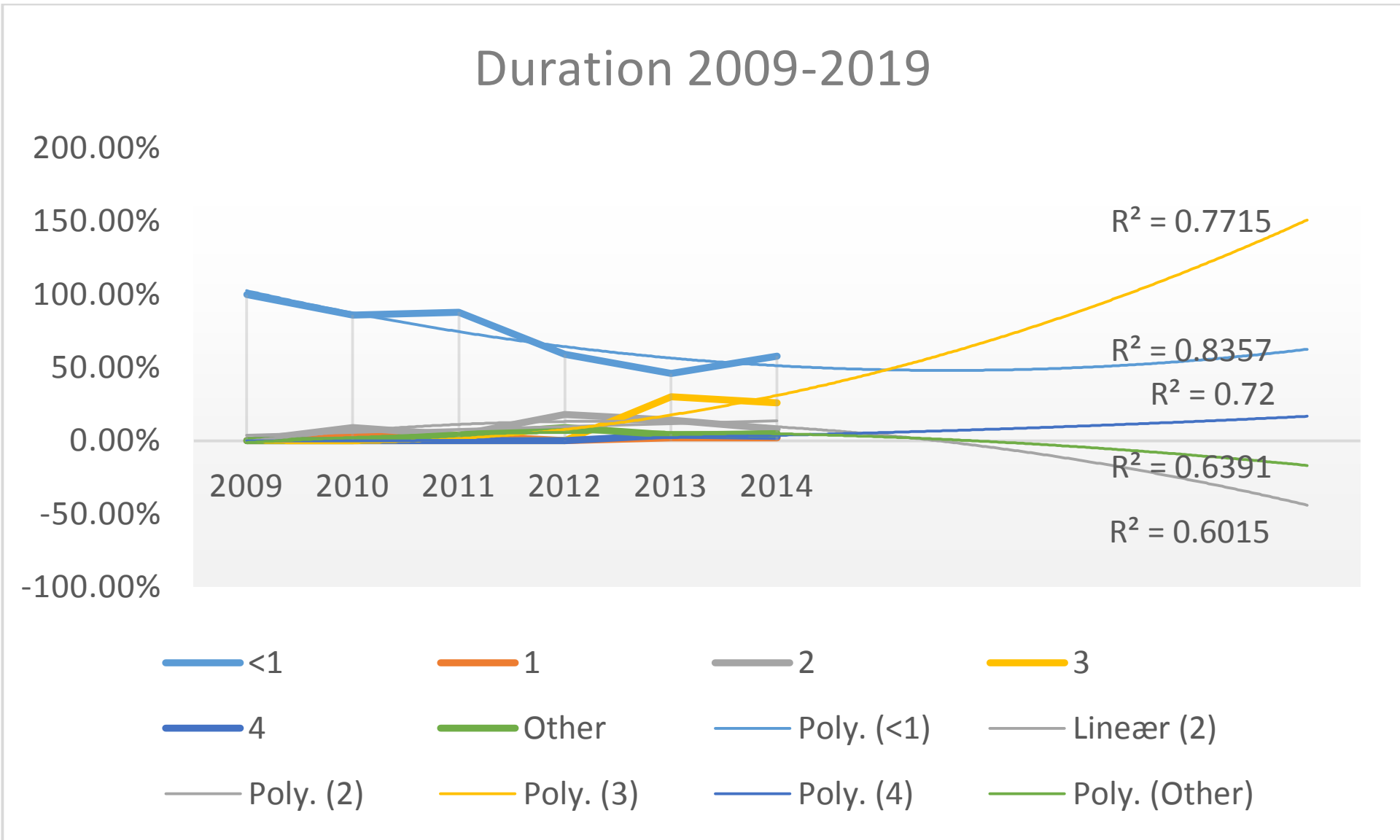

## Appendix D IX: Prize pool

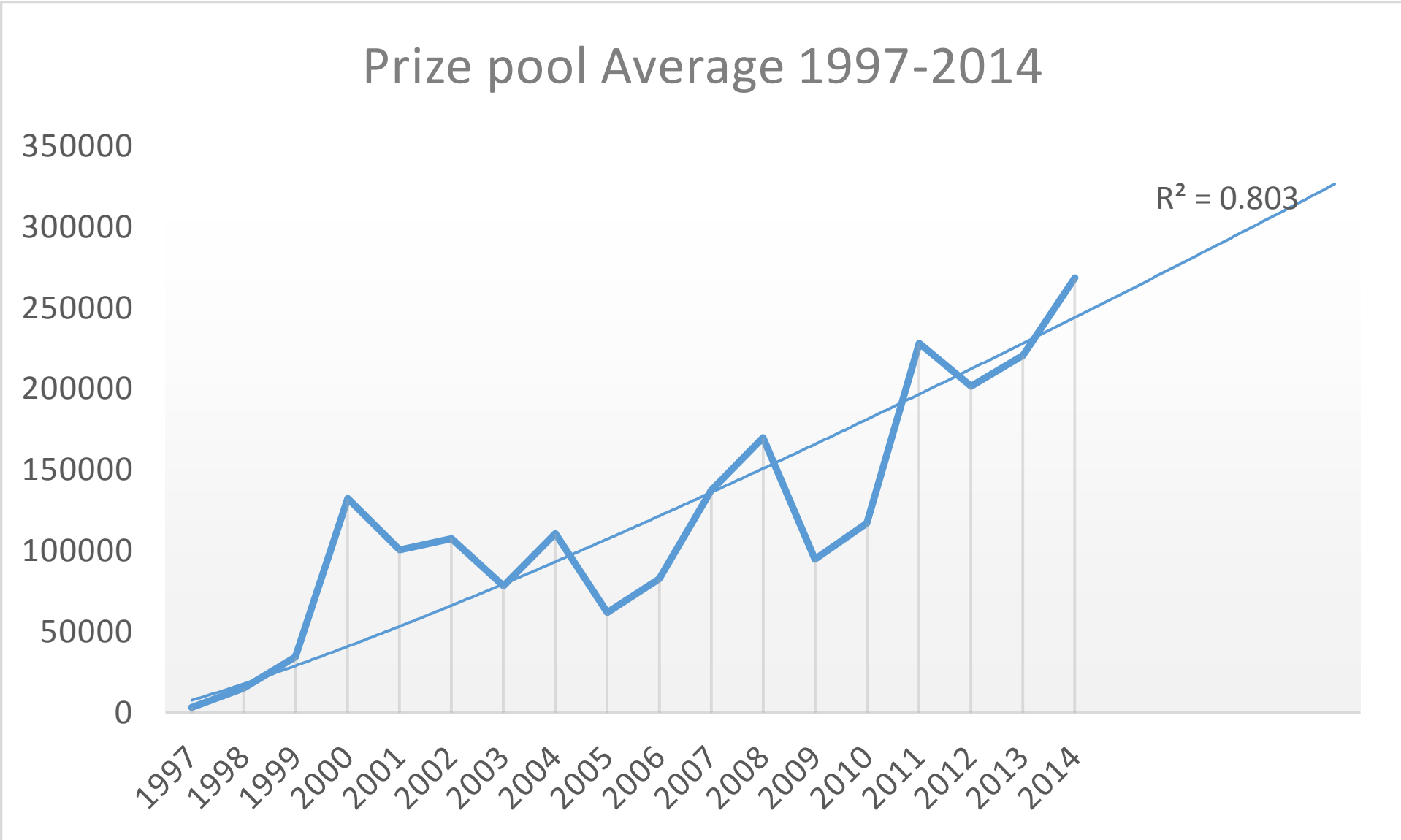

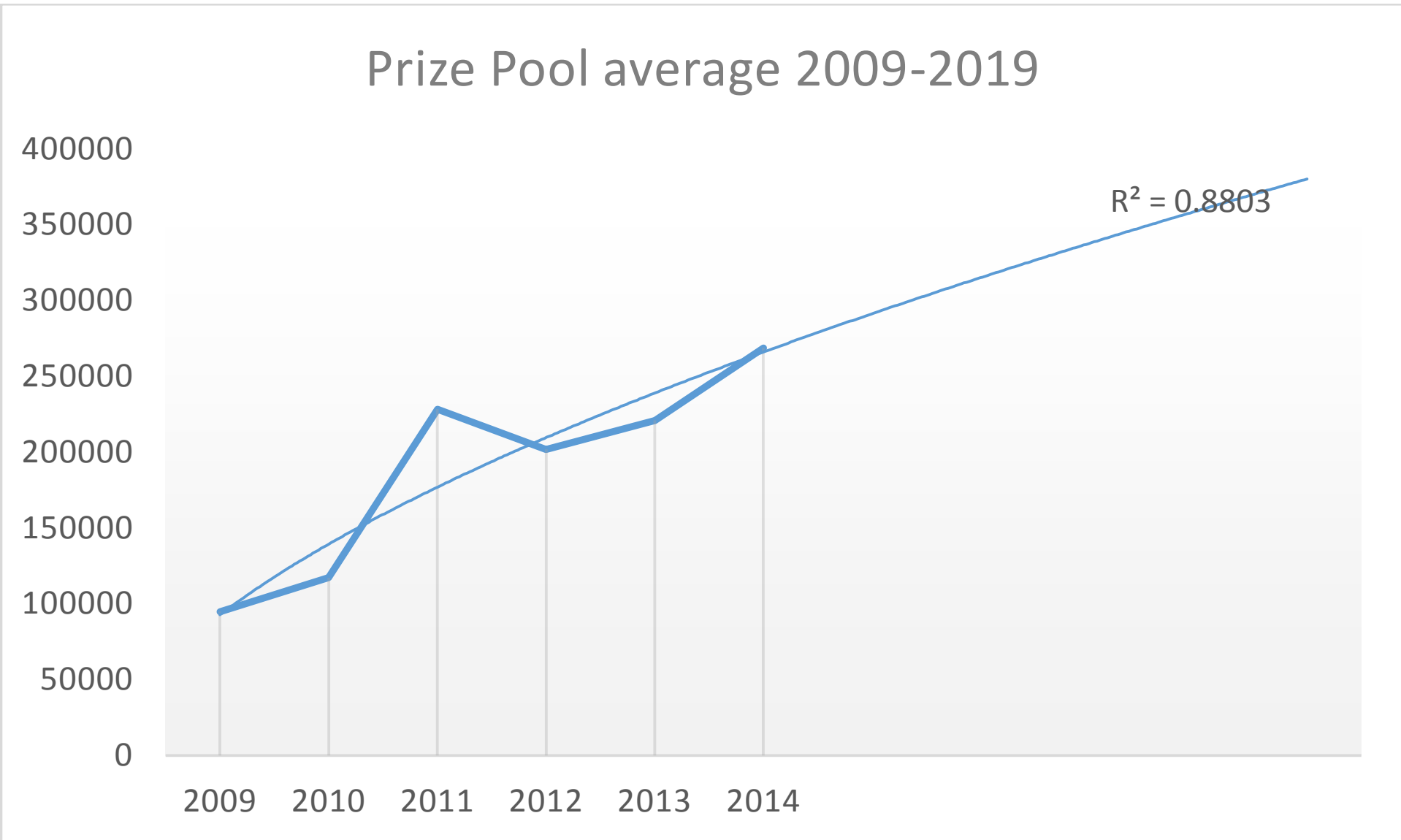

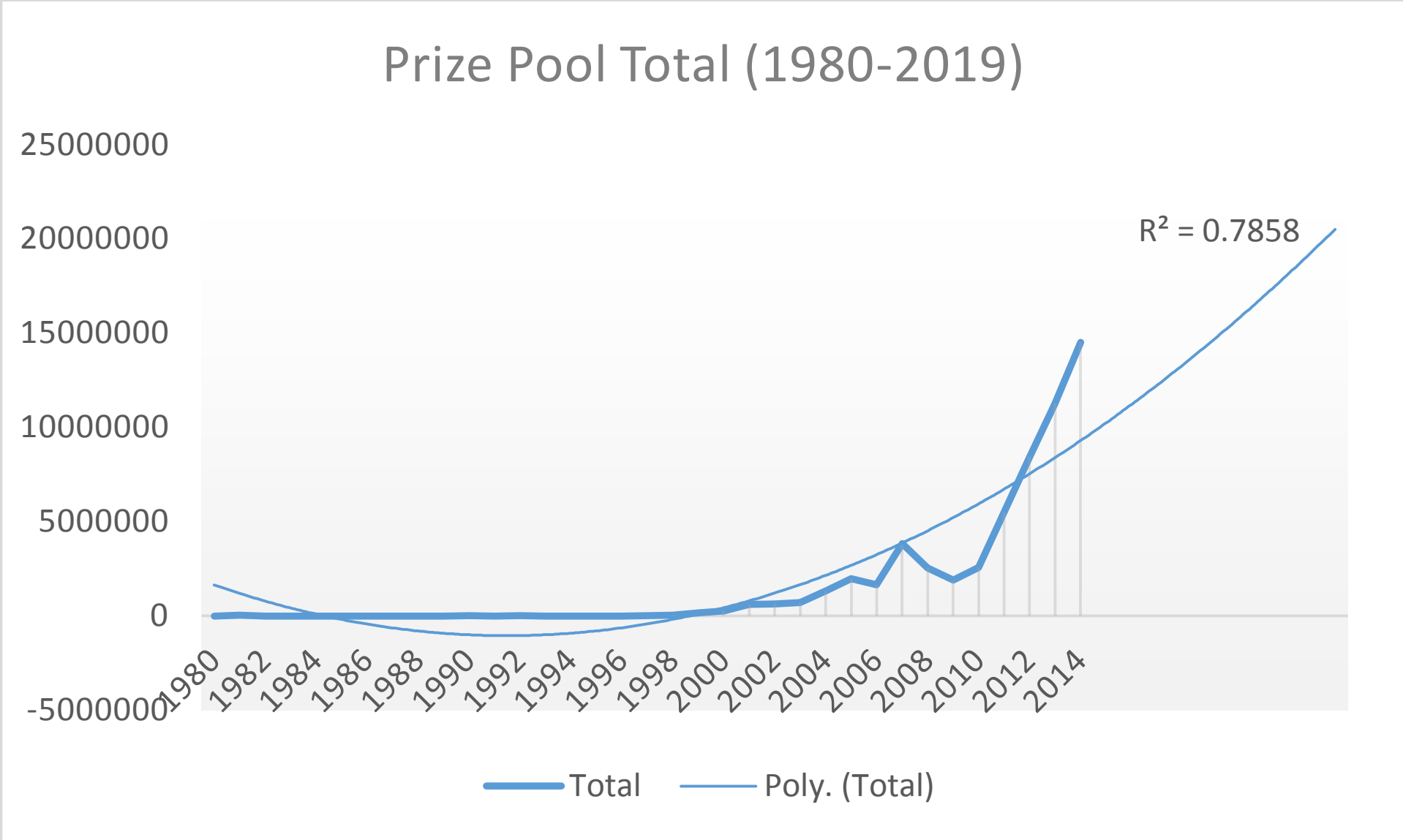

Prize Pool Total (1980-2019)
25000000
20000000
15000000
10000000
5000000
0
-5000000
1980
1982
1984
1986
1988
1990
1992
1994
1996
1998
2000
2002
2004
2006
2008
2010
2012
2014
$R^2 = 0.7858$
Total
Poly. (Total)

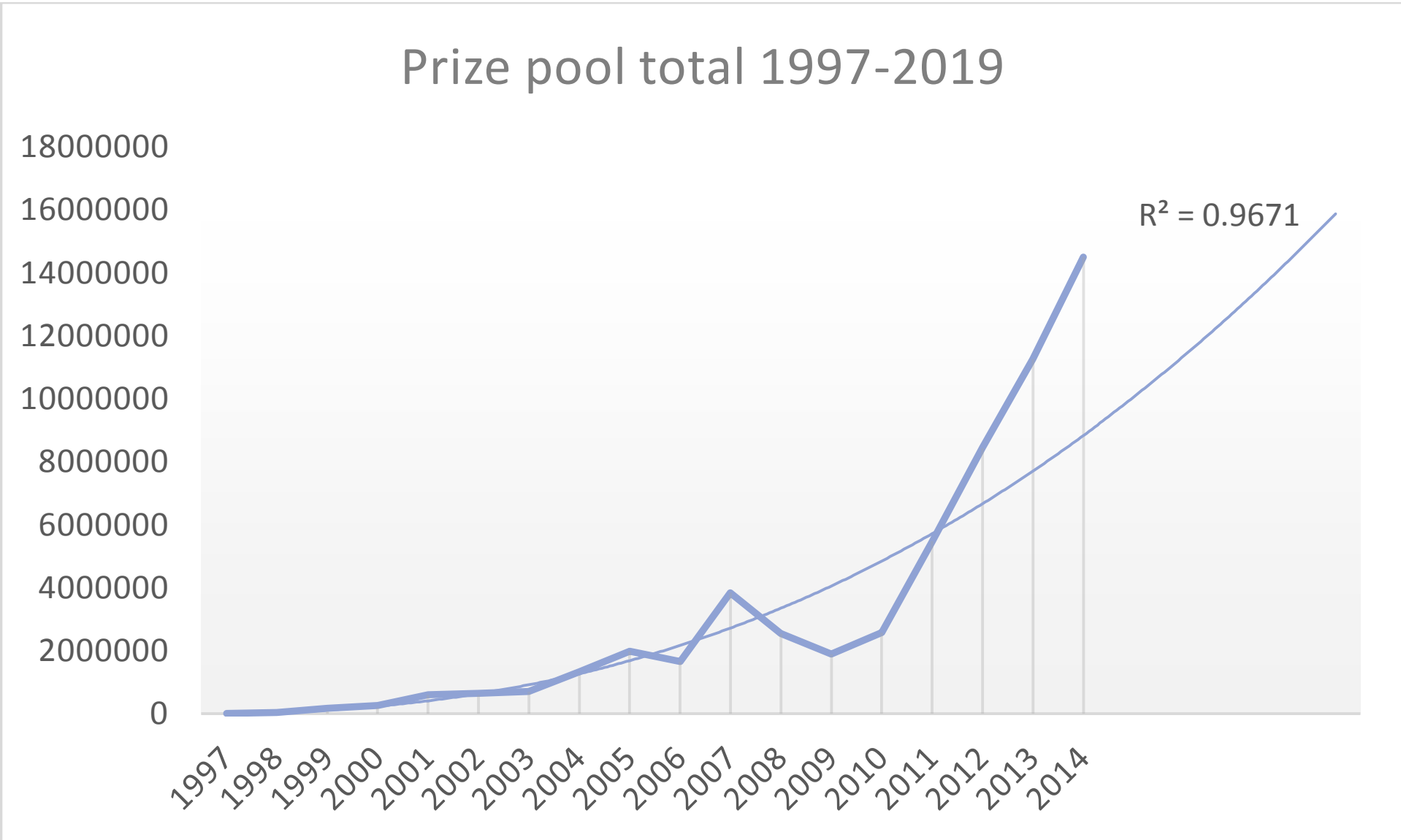

Prize pool total 1997-2019
18000000
16000000
14000000
12000000
10000000
8000000
6000000
4000000
2000000
0
1997
1998
1999
2000
2001
2002
2003
2004
2005
2006
2007
2008
2009
2010
2011
2012
2013
2014
$R^2 = 0.9671$

## Appendix E Piloting of the Data

### Appendix E I: Piloting of the thematic analysis

| Games Played | 1980-1989 | 1990-1999 | 2000-2009 | 2010-2015 |
|---|---|---|---|---|
| Space Invanders | 1 | | | |
| Centipede | 1 | | | |
| Joust | 1 | | | |
| Defender | 1 | | | |
| Congo Bongo | 1 | | | |
| Centipede | 1 | | | |
| Q-Bert | 1 | | | |
| Ms Pac-man | 1 | | | |
| Galage | 1 | | | |
| Overall score in 60 different Arcade games | 1 | | | |
| Overall score in 97 different Arcade games | 1 | | | |
| Overall score in 129 different Arcade games | 1 | | | |
| Super Mario Bros | | 1 | | |
| Rad Racers | | 1 | | |
| Tetris | | 1 | | |
| Quake | | 1 | | |
| Quake II | | 1 | | |
| Quake World | | 1 | 1 | |
| StarCraft: BW | | 3 | 12 | |
| Quake III Arena | | 2 | 21 | |
| Quake IV | | | 3 | |
| Quake Live | | | 3 | |
| Enemy Teritory: Quake Wars | | | 1 | |
| Age of Empires | | | 1 | |
| Age of Empires II | | | 1 | |
| FIFA 2000 | | | 1 | |
| FIFA 2001 | | | 1 | |
| FIFA 2007 | | | 2 | |
| FIFA 2008 | | | 1 | |
| FIFA 2009 | | | 1 | |
| Counter-Strike 1.6 | | | 63 | |
| Counter-Strike: Source | | | 4 | |
| Counter-Strike: Global Offensive | | | 1 | |
| Doom II | | | 1 | |
| Doom III | | | 1 | |
| Unreal Tournament | | | 1 | |
| Age of Empires II | | | 1 | |
| Alien Versus Predator 2 | | | 1 | |
| Unreal Tournument 2003 | | | 2 | |
| Unreal Tournument 2004 | | | 2 | |
| WarCraft III | | | 28 | |
| Painkiller | | | 14 | |
| Call of Duty | | | 2 | |
| Halo | | | 2 | |
| Halo 2 | | | 2 | |
| Halo 3 | | | 3 | |
| Day of Defeat | | | 2 | |
| Pro Evolution Soccer 3 | | | 1 | |
| Pro Evolution Soccer 4 | | | 1 | |
| Pro Evolution Soccer 5 | | | 1 | |
| Pro Evolution Soccer 6 | | | 1 | |
| Gran Turismo 4 | | | 2 | |
| F.E.A.R | | | 6 | |
| Warcraft III: Defense of the Ancients (DOTA) | | | 8 | |
| TrackMania Nations Series | | | 6 | |
| Dead or Alive 4 | | | 3 | |
| Project Gotham Racing 3 | | | 2 | |
| World in Conflict | | | 5 | |
| World of Warcraft | | | 8 | |
| Street Fighter IV | | | 1 | |

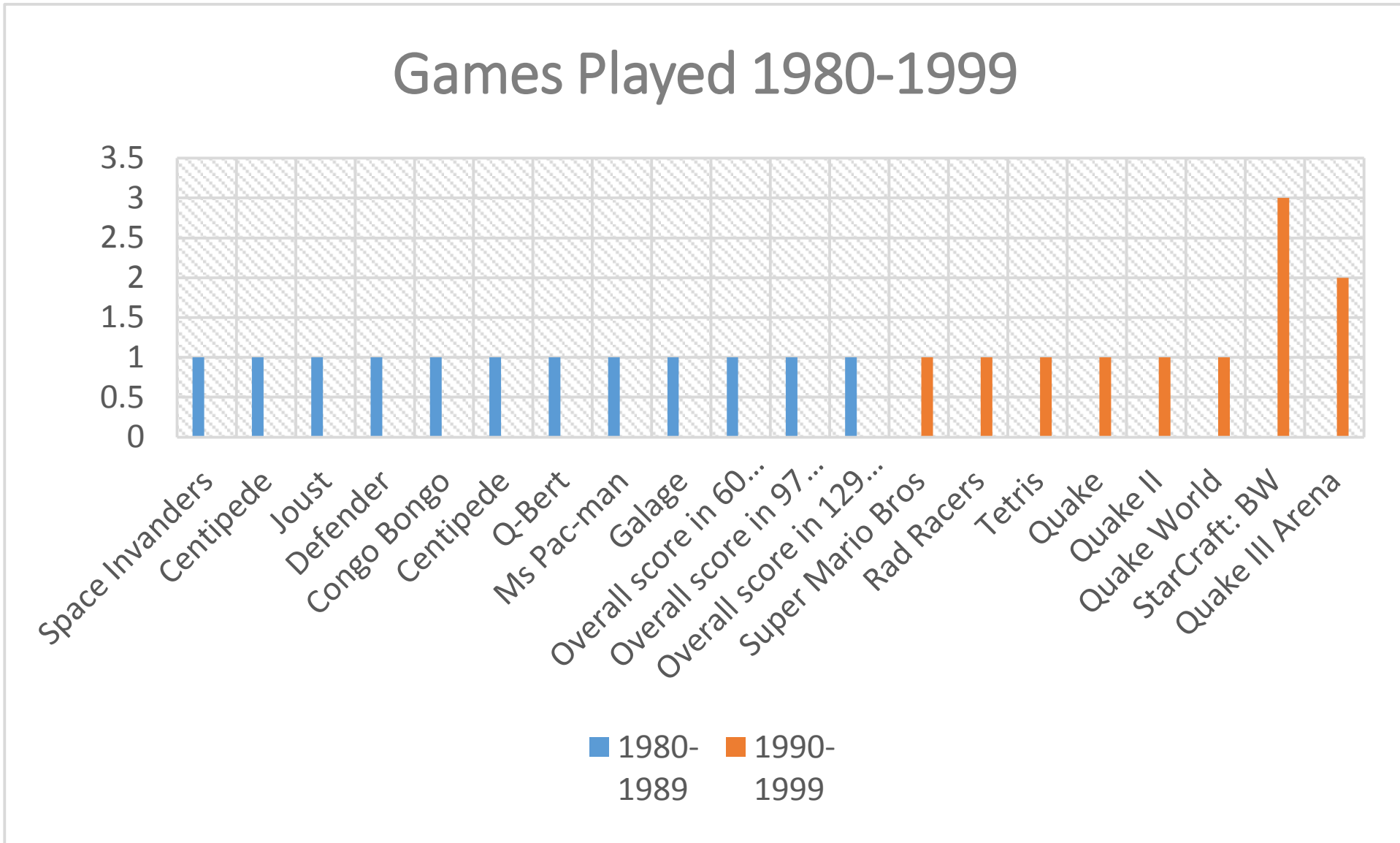

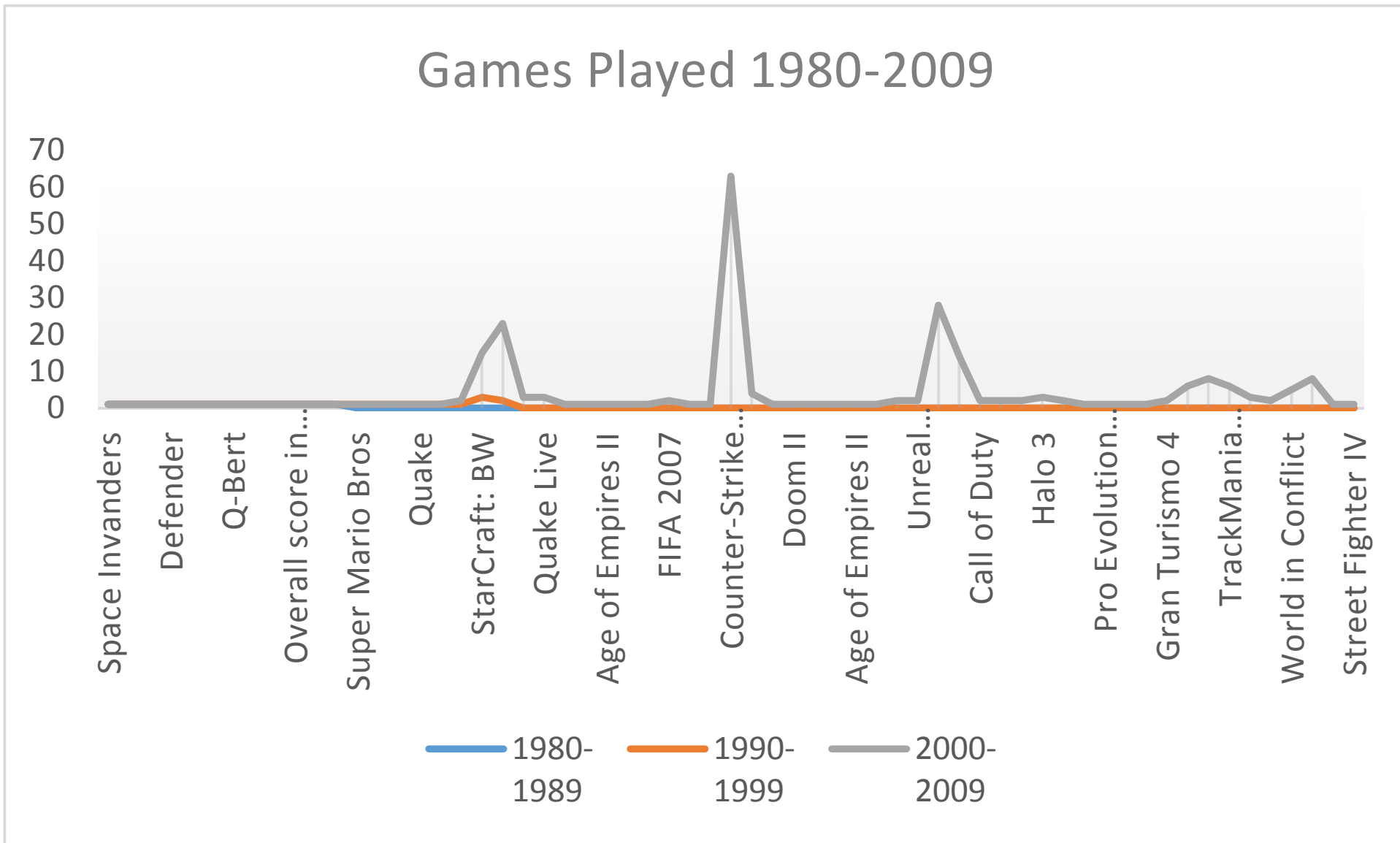

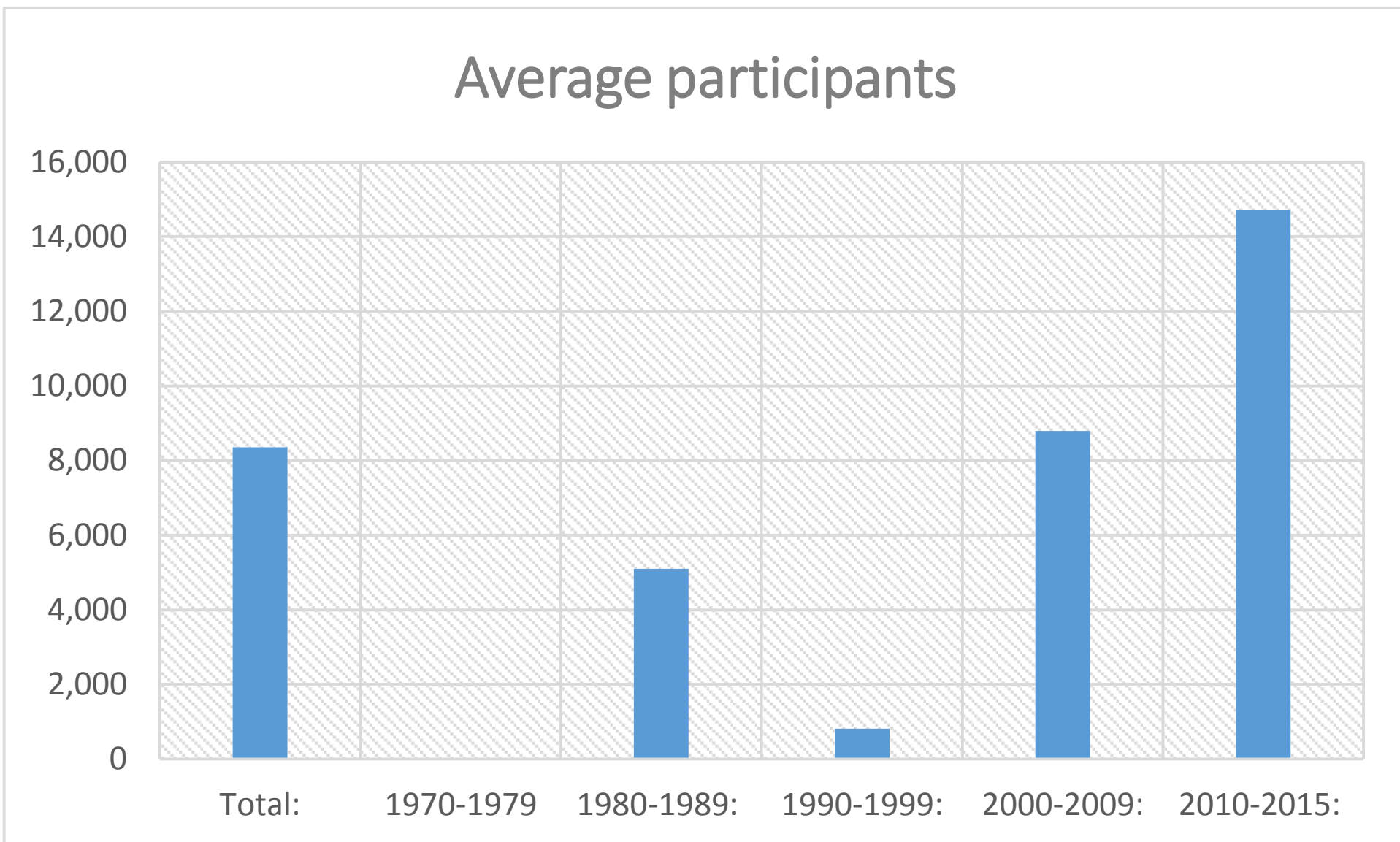

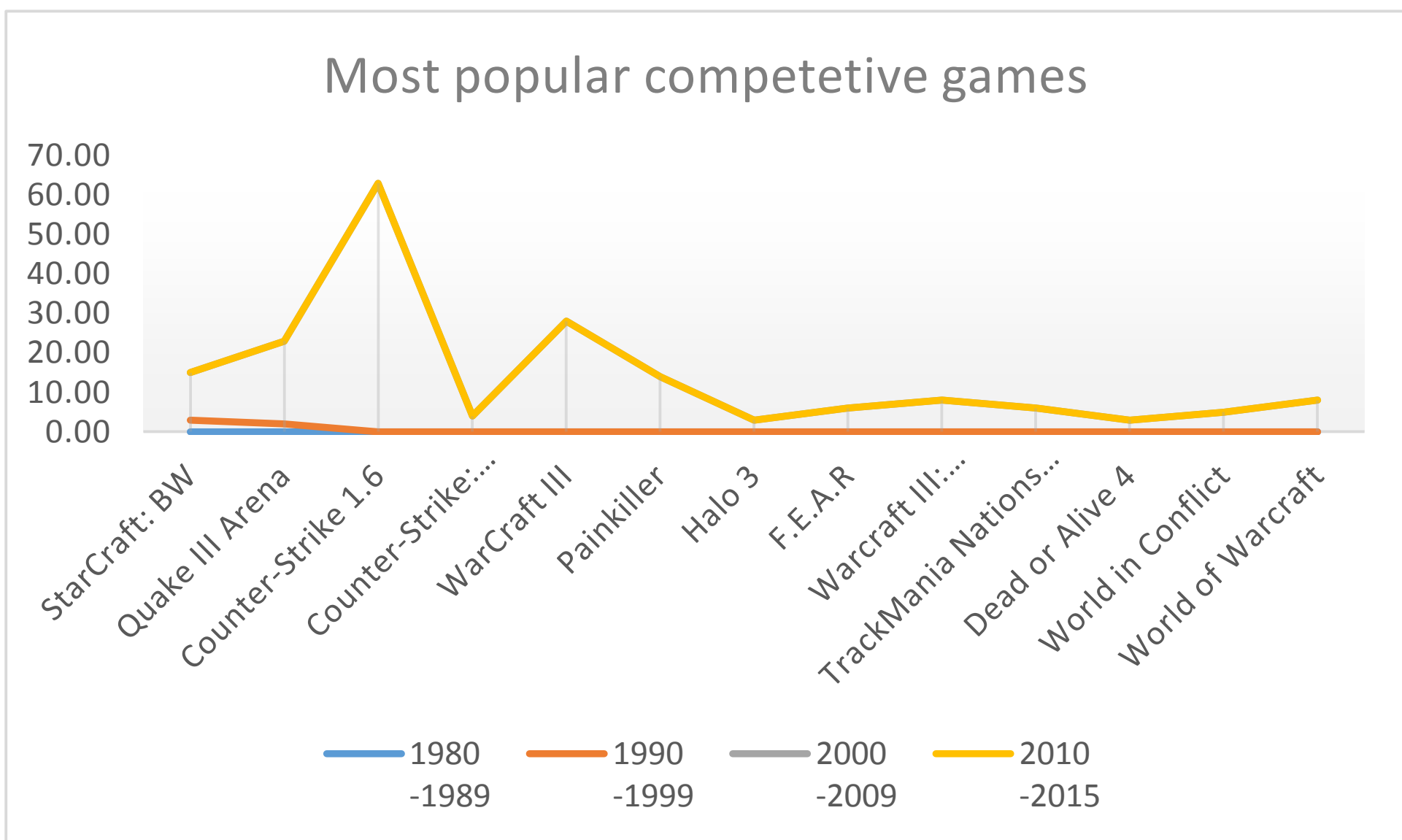

| **Average Participants** | |
|---|---|
| Total: | 8,355 |
| 1970-1979 | 0 |
| 1980-1989: | 5,100 |
| 1990-1999: | 817 |
| 2000-2009: | 8,790 |
| 2010-2015: | 14,708 |

| **Games Played (Top)** | **1980 -1989** | **1990 -1999** | **2000 -2009** | **2010 -2015** |
|---|---|---|---|---|
| StarCraft: BW | 0.00 | 3.00 | 12.00 | |
| Quake III Arena | 0.00 | 2.00 | 21.00 | |
| Counter-Strike 1.6 | 0.00 | 0.00 | 63.00 | |
| Counter-Strike: Source | 0.00 | 0.00 | 4.00 | |
| WarCraft III | 0.00 | 0.00 | 28.00 | |
| Painkiller | 0.00 | 0.00 | 14.00 | |
| Halo 3 | 0.00 | 0.00 | 3.00 | |
| F.E.A.R | 0.00 | 0.00 | 6.00 | |
| Warcraft III: Defense of the Ancients (DOTA) | 0.00 | 0.00 | 8.00 | |
| TrackMania Nations Series | 0.00 | 0.00 | 6.00 | |
| Dead or Alive 4 | 0.00 | 0.00 | 3.00 | |
| World in Conflict | 0.00 | 0.00 | 5.00 | |
| World of Warcraft | 0.00 | 0.00 | 8.00 | |

## Appendix E II: Piloting of the interview Results

Introduction (Question 1: Yourself)

- Could you start by telling a bit about yourself, and what experience you have and your responsibilities within the company?
  - More specific. "Never make a combination of two questions"
- How would you, personally, describe eSports?
  - OK
- Do you watch eSports yourself?
  - OK

Pre-topic (Question 2: Around eSports)

- What event would you say defined eSports, as we know it today?
  - ESL, IEM, LCS, LCK, etc.
  - OK
- Who would you say are the major communities and teams within eSports?
  - OK
- Who watches eSport?
  - OK

Topic (Question 3: eSport today)

- What would you say is the biggest difference between eSports today, since it began in the 1980s?
  - OK
  - Answer: International
  - Growth of Internet
- Why is eSport seeing an increase in viewership now, and not earlier?
  - OK
  - Internet
  - Because of the interactions?
- Of the following factors, which would you rank the highest and the lowest in the essence of an eSport event, and why?
  - Participants, Live audience, Viral Audience, Television coverage, Prize pool, Genre of games, Specific games, Number of games or Extraordinary events such as live music
    - ASK him to prepare him for these questions, as its way to long.
    - Do you think high price money effect the viewership?
- How important is social media for the growth of eSports?
  - Twitter, Facebook and Reddit mainly.
  - OK
  - Be more accurate on what you ask

Company specific (Question 4: Your company)

- Who is your target audience?
  - OK

- Why do you sponsor eSports?
    - OK
- What company specific goals do you set that influence the field of eSports?
    - OK

Off/end-topic (Question 5: Future of eSports)

- Would you say there is a lot of difference between eSport events and sport events?
    - If yes, why?
    - If no, why?
        - OK
- How would you say the future for eSport look like economically?
    - OK
- What is essential for the future for eSports?
    - If eSports were to improve, what would it improve? [Company spefics]
        - OK
        - Infrastructure and marketing
- How does eSport difference itself in NA, EU and Asia?
    - OK
- Would you give any credit to a specific game or company regarding the growth of eSports?
    - ESPN, ESL
    - DOTA, League of Legends, Starcraft, LOLoo

## Appendix F Gant Chart

### Appendix F I: GANTT Summarised

| ID | Task Mode | Task Name | Duration | Start | Finish | Predecessors |
|---|---|---|---|---|---|---|
| 1 | | **Task: Choosing the topic** | **24 days** | **Mon 01/12/14** | **Thu 01/01/15** | |
| 2 | | Brainstorming | 5 days | Mon 01/12/14 | Fri 05/12/14 | |
| 3 | | Concept Mapping | 5 days | Mon 08/12/14 | Fri 12/12/14 | 2 |
| 4 | | Choose Topic | 5 days | Mon 15/12/14 | Fri 19/12/14 | 3 |
| 5 | | Gather Information About Topic | 5 days | Mon 15/12/14 | Fri 19/12/14 | 3 |
| 6 | | Define draft Aim, Objectives and deliverables | 9 days | Mon 22/12/14 | Thu 01/01/15 | 5 |
| 7 | | **Chapter 1-2: First Draft (40%)** | **28 days** | **Wed 01/04/15** | **Fri 08/05/15** | |
| 8 | | Final Aim, Objectives ar | 18 days | Wed 01/04/15 | Fri 24/04/15 | |
| 9 | | Litterature Review | 5 days | Mon 27/04/15 | Fri 01/05/15 | 8 |
| 10 | | Write backround and m | 6 days | Mon 27/04/15 | Mon 04/05/15 | 8 |
| 11 | | Make headers and structure | 4 days | Tue 05/05/15 | Fri 08/05/15 | 9,10 |
| 12 | | **Chapter 3-5: Second Draft (70%)** | **50 days** | **Mon 11/05/15** | **Fri 17/07/15** | |
| 13 | | Book and do interveiws | 5 days | Tue 16/06/15 | Mon 22/06/15 | 11 |
| 14 | | Define main KPI's | 15 days | Mon 11/05/15 | Fri 29/05/15 | 11 |
| 15 | | Research methodology | 15 days | Mon 11/05/15 | Fri 29/05/15 | 11 |
| 16 | | Write Interpretations | 20 days | Mon 01/06/15 | Fri 26/06/15 | 14,15 |
| 17 | | Collect data (Primary) | 9 days | Tue 23/06/15 | Fri 03/07/15 | 13 |
| 18 | | Collect data (Secondary | 9 days | Tue 23/06/15 | Fri 03/07/15 | 9 |
| 19 | | Moddeling and tests | 5 days | Mon 06/07/15 | Fri 10/07/15 | 17,18 |
| 20 | | Write Analysis | 5 days | Mon 13/07/15 | Fri 17/07/15 | 19 |
| 21 | | **Chapter 5-7: Final Draft (100%)** | **23 days** | **Mon 13/07/15** | **Wed 12/08/15** | |
| 22 | | Compare findings | 5 days | Mon 13/07/15 | Fri 17/07/15 | 19 |
| 23 | | Vizualise data | 5 days | Mon 20/07/15 | Fri 24/07/15 | 22 |
| 24 | | Write Recommedations | 8 days | Mon 27/07/15 | Wed 05/08/15 | 23 |
| 25 | | Write Conclusiion | 5 days | Thu 06/08/15 | Wed 12/08/15 | 24 |
| 26 | | Do references | 5 days | Thu 06/08/15 | Wed 12/08/15 | 24 |
| 27 | | **Task: Ethics** | **32 days** | **Tue 30/06/15** | **Wed 12/08/1!** | |
| 28 | | Do ethic form | 1 day | Tue 30/06/15 | Tue 30/06/15 | |
| 29 | | Continously do ethic checks | 31 days | Wed 01/07/15 | Wed 12/08/15 | 28 |
| 30 | | Remove ethic keyword: | 31 days | Wed 01/07/15 | Wed 12/08/15 | 28 |

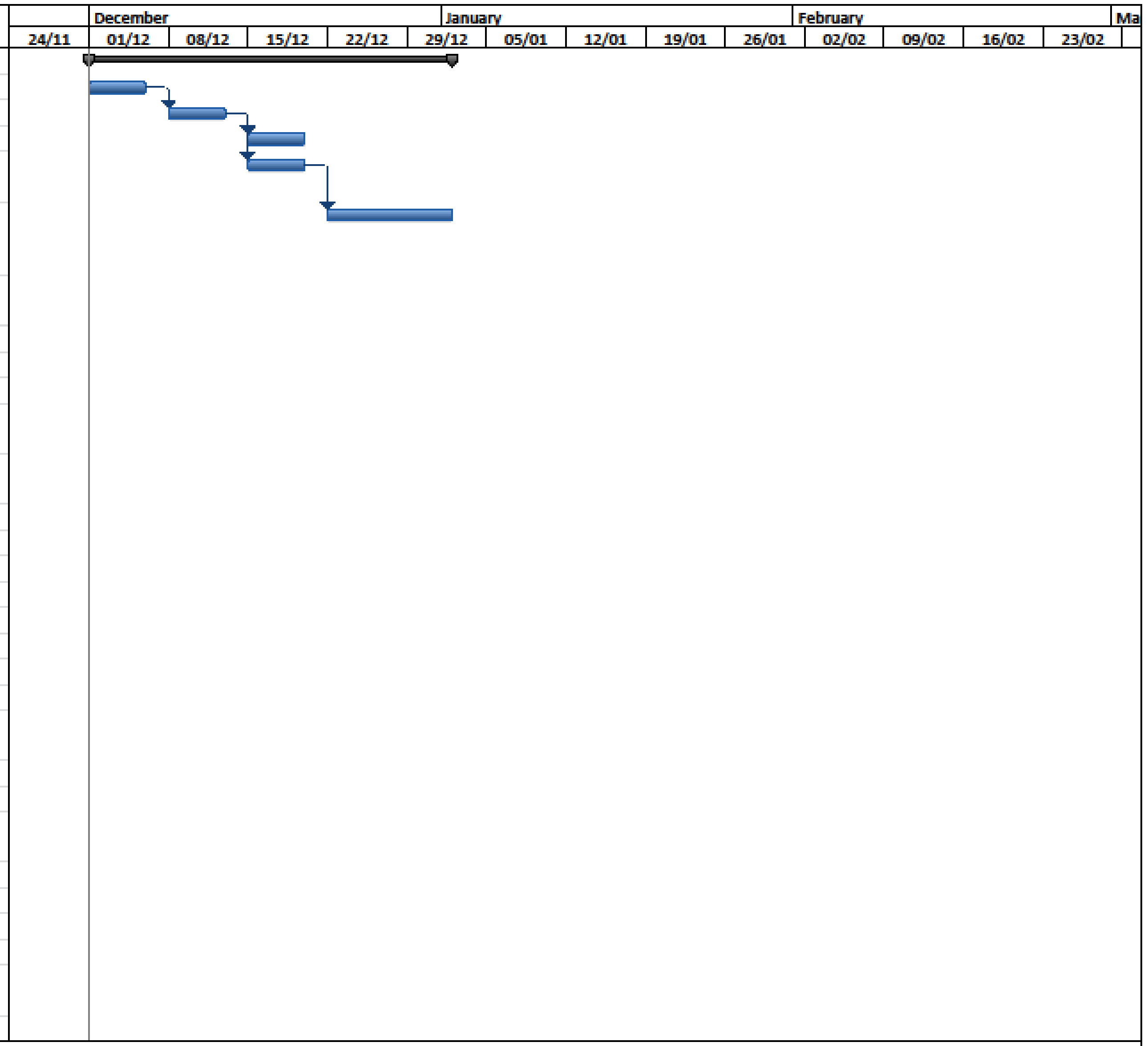

Project: Master summerized
Date: Wed 06/05/15

| | | | | | | | |
|---|---|---|---|---|---|---|---|
| Task | Project Summary | Inactive Milestone | Manual Summary Rollup | Deadline | | | |
| Split | External Tasks | Inactive Summary | Manual Summary | Progress | | | |
| Milestone | External Milestone | Manual Task | Start-only | | | | |
| Summary | Inactive Task | Duration-only | Finish-only | | | | |

## Appendix F II: Gantt based upon objectives and deliverables

| ID | Task Mode | Task Name | Duration | Start | Finish | Predecessors |
|---|---|---|---|---|---|---|
| 1 | | **Objective 1: Discover previous research within the field, and summarize their importance for this study** | **10 days** | **Mon 20/04/15** | **Fri 01/05/15** | |
| 2 | | A complete literature review by week 19 | 5 days | Mon 20/04/15 | Fri 24/04/15 | |
| 3 | | A summary of the major findings within the e-Sport community by week 20 | 5 days | Mon 27/04/15 | Fri 01/05/15 | 2 |
| 4 | | **Objective 2: Define what e-sports is, and outline its major historic milestones** | **30 days** | **Mon 06/04/15** | **Fri 15/05/15** | |
| 5 | | A Complete and synthesised definition of e-sports by week 16. | 5 days | Mon 06/04/15 | Fri 10/04/15 | |
| 6 | | A summary with the milestones of e-sport's history by week 19. | 19 days | Tue 14/04/15 | Fri 08/05/15 | 5 |
| 7 | | A visual timeline of e-sports' historic data by week 20. | 5 days | Mon 11/05/15 | Fri 15/05/15 | 6 |
| 8 | | A visual timeline of e-sports' historic data by week 20. | 5 days | Mon 11/05/15 | Fri 15/05/15 | 6 |
| 9 | | A report on the major e-sport events by week 20 | 5 days | Mon 11/05/15 | Fri 15/05/15 | 6 |
| 10 | | **Objective 3: Arrange and compare the main financial factors within e-sport** | **10 days** | **Mon 18/05/15** | **Fri 29/05/15** | |
| 11 | | A list of the main financial factors within e-sports by week 21 | 5 days | Mon 18/05/15 | Fri 22/05/15 | 5,7,8,9 |
| 12 | | A visual comparison of financial factors to differentiate their impact by week 22 | 5 days | Mon 25/05/15 | Fri 29/05/15 | 11 |
| 13 | | A brief explanation of the financial factors by week 22 | 5 days | Mon 25/05/15 | Fri 29/05/15 | 11 |
| 14 | | **Objective 4: Review previously successful e-sport events, and use business analytics tools to visualize the common patterns** | **25 days** | **Mon 01/06/15** | **Fri 03/07/15** | |

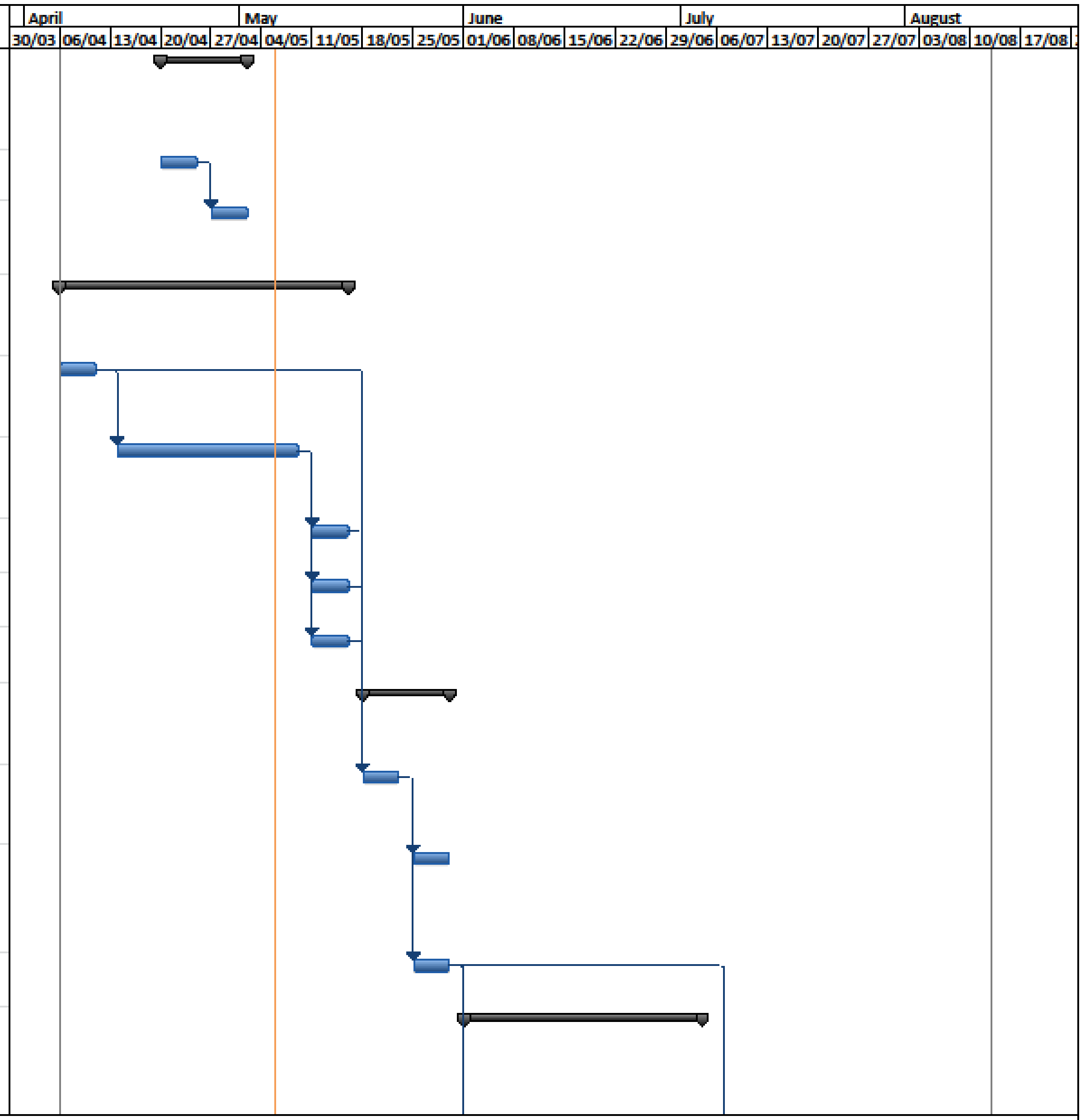

Project: Objectivess GANTT
Date: Wed 06/05/15

Task | Split | Milestone | Summary | Project Summary | External Tasks | External Milestone | Inactive Task | Inactive Milestone | Inactive Summary | Manual Task | Duration-only | Manual Summary Rollup | Manual Summary | Start-only | Finish-only | Deadline | Progress

| ID | | Task Mode | Task Name | Duration | Start | Finish | Predecessors |
|---|---|---|---|---|---|---|---|
| 15 | | | A critical analysis of the most important success factors in previous e-sport events by week 25. | 11 days | Mon 01/06/15 | Mon 15/06/15 | 13 |
| 16 | | | A set of visualised data to compare e-sport events and other sport events by week | 14 days | Tue 16/06/15 | Fri 03/07/15 | 15 |
| 17 | | | A statistic correlation test to see if there is any common patterns between the teams by week 25. | 2 days | Tue 16/06/15 | Wed 17/06/15 | 15 |
| 18 | | | **Objective 5: Discover the features of e-sport that defines the essence of the e-sport business** | **26 days** | **Mon 01/06/15** | **Mon 06/07/15** | |
| 19 | | | A interview with some of the important actors within e-sports by week 25 | 2 days | Tue 16/06/15 | Wed 17/06/15 | 15 |
| 20 | | | A synthesised report of the comparison of the financial factors by week 26 | 20 days | Mon 01/06/15 | Fri 26/06/15 | 13 |
| 21 | | | A juxtaposed summary of the most important features within e-sport by week 26 | 1 day | Mon 29/06/15 | Mon 29/06/15 | 19,20 |
| 22 | | | A juxtaposed summary of the causing factors of the recent growth within e-Sport by week 27 | 5 days | Tue 30/06/15 | Mon 06/07/15 | 21 |
| 23 | | | **Objective 6: Examine the overall awareness of e-sports** | **15 days** | **Mon 15/06/15** | **Fri 03/07/15** | |
| 24 | | | An online survey of the overall awareness of e-sports by week 26 | 10 days | Mon 15/06/15 | Fri 26/06/15 | |
| 25 | | | A report based upon a quantitative analysis of the quantitative data by week 27 | 3 days | Mon 29/06/15 | Wed 01/07/15 | 24 |
| 26 | | | A summary of the e-sport awareness, differenced by age segments by week 27 | 2 days | Thu 02/07/15 | Fri 03/07/15 | 25 |

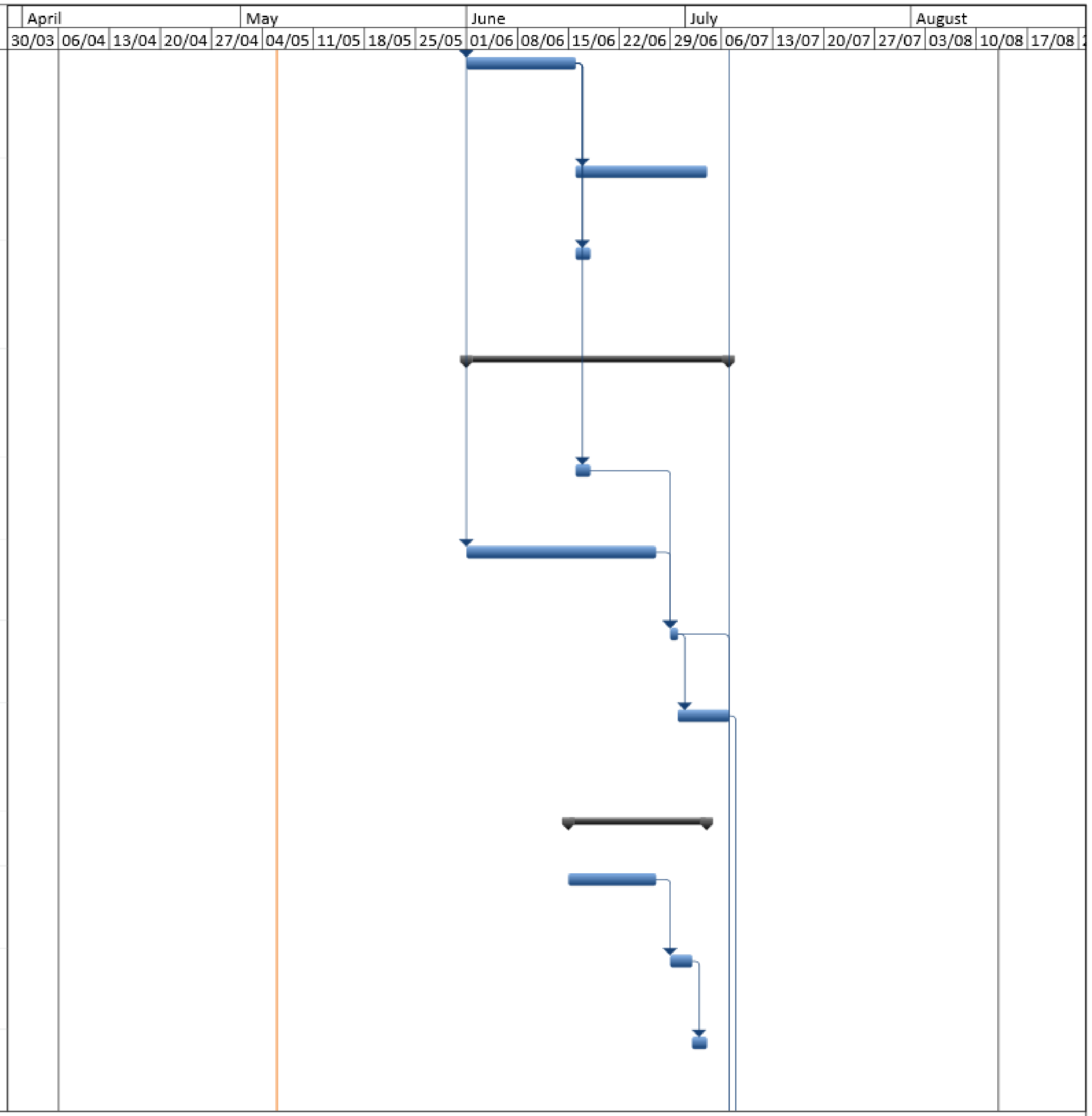

Project: Objectivess GANTT
Date: Wed 06/05/15

| | | | | | |
|---|---|---|---|---|---|
| Task | Project Summary | Inactive Milestone | Manual Summary Rollup | Deadline | |
| Split | External Tasks | Inactive Summary | Manual Summary | Progress | |
| Milestone | External Milestone | Manual Task | Start-only | | |
| Summary | Inactive Task | Duration-only | Finish-only | | |

| ID | | Task Mode | Task Name | Duration | Start | Finish | Predecessors |
|---|---|---|---|---|---|---|---|
| 27 | | | **Objective 7: Evaluate the benefits and disbenefits of implementing ITM tools in e-Sport and synthesise which ITM tools would be most effective** | **36 days** | **Wed 06/05/15** | **Wed 24/06/15** | |
| 28 | | | A complete definition of Information technology, and its tools, by week 21 | 13 days | Wed 06/05/15 | Fri 22/05/15 | |
| 29 | | | A critical evaluation of the difference between Information Technology tools, business analytic tools and organizational tools by week 22 | 5 days | Mon 25/05/15 | Fri 29/05/15 | 28 |
| 30 | | | A critical analysis of the most common usage of Information Technology Management tools within businesses and other sports by week 23 | 10 days | Mon 25/05/15 | Fri 05/06/15 | 28 |
| 31 | | | An evaluation of which ITM-tools implementations that can be done within e-sport by week 26 | 13 days | Mon 08/06/15 | Wed 24/06/15 | 30 |
| 32 | | | **Objective 8: Evaluate the effect of implementing well-justified ITM tools to increase the growth of e-sport** | **7 days** | **Thu 25/06/15** | **Fri 03/07/15** | |
| 33 | | | A list of benefits and disbenefits of the implementation of ITM-tools in e-sports by week 26 | 2 days | Thu 25/06/15 | Fri 26/06/15 | 31 |

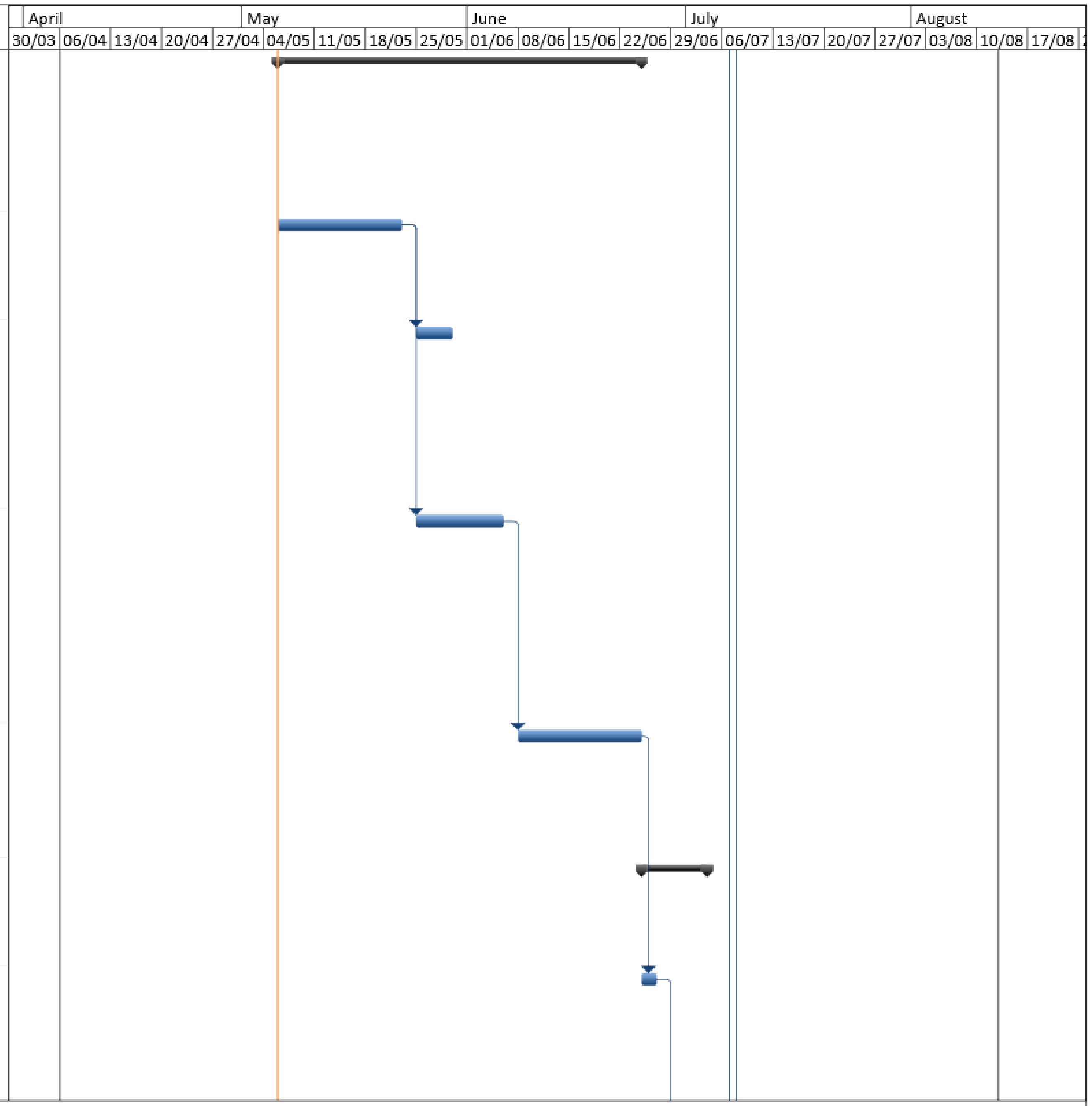

Project: Objectivess GANTT
Date: Wed 06/05/15

| | | | | |
|---|---|---|---|---|
| Task | Project Summary | Inactive Milestone | Manual Summary Rollup | Deadline |
| Split | External Tasks | Inactive Summary | Manual Summary | Progress |
| Milestone | External Milestone | Manual Task | Start-only | |
| Summary | Inactive Task | Duration-only | Finish-only | |

| ID | | Task Mode | Task Name | Duration | Start | Finish | Predecessors |
|---|---|---|---|---|---|---|---|
| 34 | | | A critical evaluation of how various ITM tools might improve the growth of e-Sport, and its effects by week 27 | 3 days | Mon 29/06/15 | Wed 01/07/15 | 33 |
| 35 | | | A set of graphs visualising the effect ITM-tools might bring by week 27 | 2 days | Thu 02/07/15 | Fri 03/07/15 | 34 |
| 36 | | | **Objective 9: Predict the financial future of E-sports** | **14 days** | **Tue 07/07/15** | **Fri 24/07/15** | |
| 37 | | | An in-depth analysis of the economic future of e-sports based upon both statistical data made from business analytics and the literature review by week 28 | 3 days | Tue 07/07/15 | Thu 09/07/15 | 13,22,21 |
| 38 | | | A set of milestones of the predication of e-sports by week 28 | 1 day | Fri 10/07/15 | Fri 10/07/15 | 37 |
| 39 | | | A critical evaluation of the financial effect of the implementation of Information Technology Management tools by week 29 | 6 days | Fri 10/07/15 | Fri 17/07/15 | 37 |
| 40 | | | A brief recommendation on how E-sport can improve its financial value by week 30 | 5 days | Mon 20/07/15 | Fri 24/07/15 | 38,39 |
| 41 | | | **Objective 10: Recommend alterations the e-sport business might commence to uphold and increase its growth.** | **8 days** | **Mon 27/07/15** | **Wed 05/08/15** | |
| 42 | | | A summary of the data gathered in the dissertation by week 31 | 3 days | Mon 27/07/15 | Wed 29/07/15 | 40 |

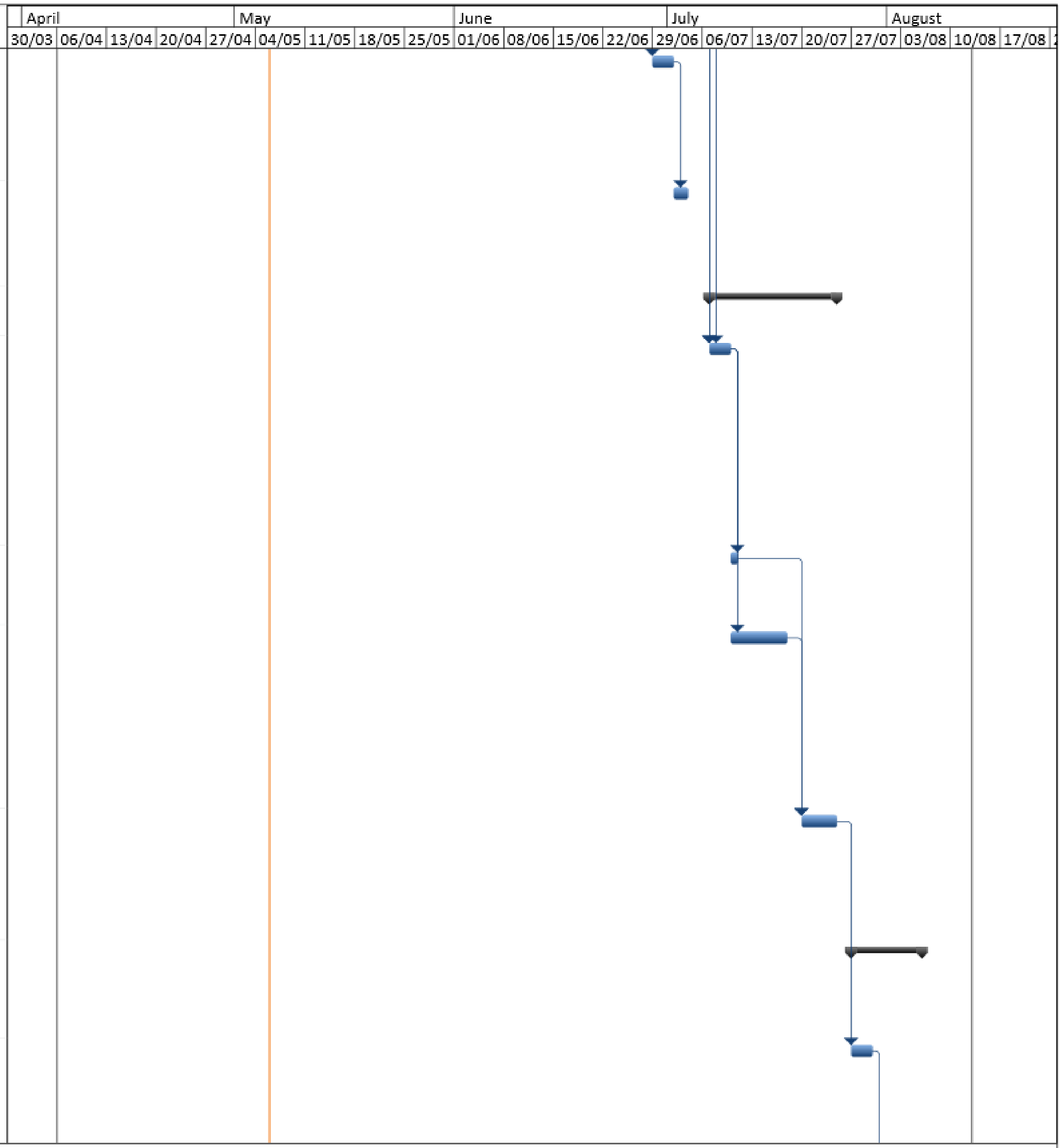

Project: Objectivess GANTT
Date: Wed 06/05/15

Task; Split; Milestone; Summary; Project Summary; External Tasks; External Milestone; Inactive Task; Inactive Milestone; Inactive Summary; Manual Task; Duration-only; Manual Summary Rollup; Manual Summary; Start-only; Finish-only; Deadline; Progress

| ID | | Task Mode | Task Name | Duration | Start | Finish | Predecessors |
|---|---|---|---|---|---|---|---|
| 34 | | | A critical evaluation of how various ITM tools might improve the growth of e-Sport, and its effects by week 27 | 3 days | Mon 29/06/15 | Wed 01/07/15 | 33 |
| 35 | | | A set of graphs visualising the effect ITM-tools might bring by week 27 | 2 days | Thu 02/07/15 | Fri 03/07/15 | 34 |
| 36 | | | **Objective 9: Predict the financial future of E-sports** | **14 days** | **Tue 07/07/15** | **Fri 24/07/15** | |
| 37 | | | An in-depth analysis of the economic future of e-sports based upon both statistical data made from business analytics and the literature review by week 28 | 3 days | Tue 07/07/15 | Thu 09/07/15 | 13,22,21 |
| 38 | | | A set of milestones of the predication of e-sports by week 28 | 1 day | Fri 10/07/15 | Fri 10/07/15 | 37 |
| 39 | | | A critical evaluation of the financial effect of the implementation of Information Technology Management tools by week 29 | 6 days | Fri 10/07/15 | Fri 17/07/15 | 37 |
| 40 | | | A brief recommendation on how E-sport can improve its financial value by week 30 | 5 days | Mon 20/07/15 | Fri 24/07/15 | 38,39 |
| 41 | | | **Objective 10: Recommend alterations the e-sport business might commence to uphold and increase its growth.** | **8 days** | **Mon 27/07/15** | **Wed 05/08/15** | |
| 42 | | | A summary of the data gathered in the dissertation by week 31 | 3 days | Mon 27/07/15 | Wed 29/07/15 | 40 |

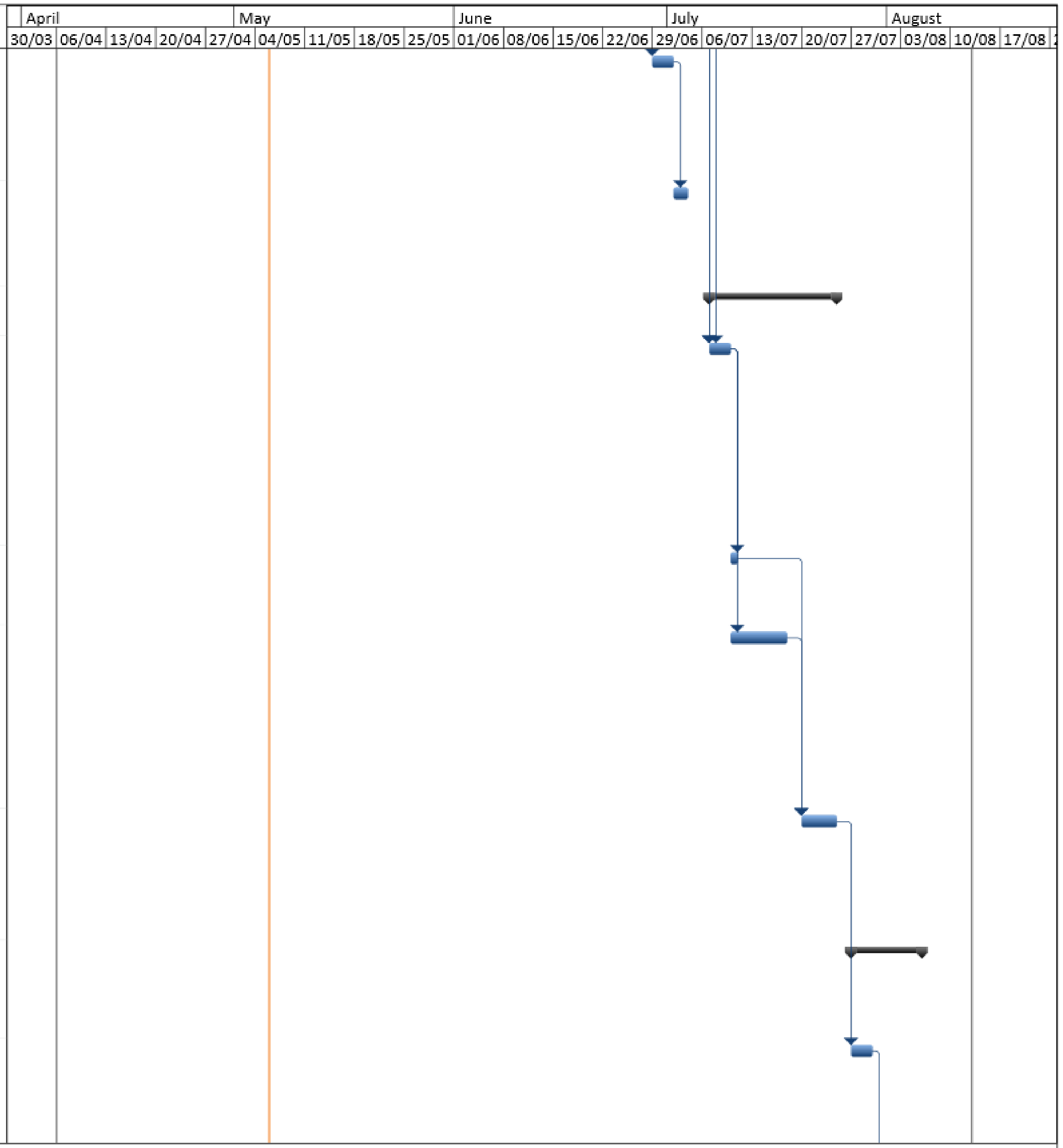

Project: Objectivess GANTT
Date: Wed 06/05/15

| | | | | | | | | | |
|---|---|---|---|---|---|---|---|---|---|
| Task | | Project Summary | | Inactive Milestone | | Manual Summary Rollup | | Deadline | |
| Split | | External Tasks | | Inactive Summary | | Manual Summary | | Progress | |
| Milestone | | External Milestone | | Manual Task | | Start-only | | | |
| Summary | | Inactive Task | | Duration-only | | Finish-only | | | |

## Appendix G Ethical Documents

### Appendix G I: Participant Information Sheet

# Participant Information Sheet

**Purpose of project**

This project aims to get an understanding of the past of eSports, and its future. This is done by looking at the main eSports events through time, and getting an understanding of what made them successful nor a failure. This information, together with other collected data, will be used to get an understanding of the eSports future. This research will be divided in two different fields; firstly this research will use business analytics tools to critical analyse what momentums and factors that has led to the increase of e-sport consumption; furthermore this research will analyse the effect om implementing Information Technology Management theories to uphold the growth of eSports. This information will be measured up with qualitative data mainly consisting of interviews to get both a better understanding, and a higher validity on the data.

**Why you have been chosen**

To gain a deeper understanding of what eSport is, and the future within it, various companies, communities and names within the field of eSport has been chosen. It is mainly the experience and the knowledge of the chosen names that is interesting of the research paper. You have specifically been chosen to give your view and insight of the past, and the future of eSports. Whatever you say will be analysed with the data already collected, and further play a big part in the research itself.

**Do you have to take part?**

You are not forced to participate, but your help would greatly be appreciated.

**What you have to do**

You have to attend to an interview – either personally or over phone/skype. During the interview, I will ask several open questions concerning the research topic (these questions will be sent to you before the interview, to give you an insight of what we will talk about and give you some time to prepare). All I ask for is for you to answer these questions, in addition to follow-up questions, as honestly you can. The interview itself will not take longer than 30 minutes.

**Risk associated with the project**

There is very low risk associated with this project. All information you provide will be used to get an understanding of general trends – not isolate your personal views, which ultimately is the goal for this research. This is further done correctly in the data protection laws, meaning everything you say will be written anonymously. However, you have the ultimate voice in the interview, so if any question feels uncomfortable or risky, you do not have to answer it.

**Your benefits for taking apart**

This master dissertation involves a field that is yet not heavily researched, and I believe the result of this research will be interesting for you as a person or a company – as it will be interesting for the eSport community. Including you with the research will ensure the projects validity, which will be appreciated within the text. Additionally, if you want me to specifically include a view or research something within the research field, I would be more than happy to do it.

**Withdrawal options**

If you at any point feel uncomfortable doing this, please let me know, as you are allowed to withdraw at any time.

**Data protection & confidentiality**

All data collected will be stored within the Coventry Universities data protection guidelines, which is based upon the Data Protection Act 1998. Secure measurements will be done to ensure that the data will not be lost or damaged. When the research is finished, all collected data will be deleted.

**Contact persons**

If something were to go wrong, you can firstly talk to me, the researcher, and we can try to solve it out. If you want to take this to someone above me, you can contact my supervisor named Stella-Maris Orim.

**After the study**

The results of the study will be used to get an understanding of the future of eSports. These results will further be implemented with Information Management Technology Tools to get an understanding of what the results might bring the future of eSports. If it were to be published, or used in any other way that was indicated, you would be contacted. A copy will also be sent to you personally, if this is preferable.

**Who has reviewed this study?**

My supervisor, Stella-Maris Orim, and my second marker, Owen Richards.

**Key contact details of researcher and supervisor**

Researcher:

| | |
|---|---|
| Name: | Anders Hval Olsen |
| Email: | PURPOSELY LEFT BLANK FOR ETHICAL REASONS |
| Telephone UK: | PURPOSELY LEFT BLANK FOR ETHICAL REASONS |
| Telephone Norway: | PURPOSELY LEFT BLANK FOR ETHICAL REASONS |
| Address: | PURPOSELY LEFT BLANK FOR ETHICAL REASONS |

Supervisor:

| | |
|---|---|
| Name: | Stella-Maris Orim |
| Email: | PURPOSELY LEFT BLANK FOR ETHICAL REASONS |
| Address: | PURPOSELY LEFT BLANK FOR ETHICAL REASONS |

## Appendix G II: Informed Consent Form

# Informed Consent Form

**Summary of the research**

This project aims to get an understanding of the past of eSports, and its future. This is done by looking at the main eSports events through time, and getting an understanding of what made them successful nor a failure. This information, together with other collected data, will be used to get an understanding of the eSports future. This research will be divided in two different fields; firstly this research will use business analytics tools to critical analyse what momentums and factors that has led to the increase of e-sport consumption; furthermore this research will analyse the effect om implementing Information Technology Management theories to uphold the growth of e-sports. This information will be measured up with qualitative data mainly consisting of semi-structured interviews to get both a better understanding, and a higher validity on the data.

**Consent Form**

Please tick each box, if you agree, and further state your name, signature an date.

| | **Please tick** |
|---|---|
| 1. I confirm that I have read and understood the participant information sheet for the above study and have had the opportunity to ask questions. | ☐ |
| 2. I understand that my participation is voluntary and that I am free to withdraw at anytime without giving a reason. | ☐ |
| 3. I understand that all the information I provide will be treated in confidence | ☐ |
| 4. I understand that I also have the right to change my mind about participating in the study for a short period after the study has concluded (17.08.2015). | ☐ |
| 5. I agree to be recorded as part of the research project | ☐ |
| 6. I agree to take part in the research project | ☐ |

**Signature**

Name of participant: ...........................................................................................

Signature of participant: ...................................................................................

Date: ...........................................................................................................

Witnessed by (if appropriate): ............................................................................

Name of witness: ................................................................................................

Signature of witness:..........................................................................................

Name of Researcher: ...........................................................................................

Signature of researcher: .....................................................................................

Date:..............................................................................................................

Appendix G III Certificate of Ethical Approval

# Certificate of Ethical Approval

Applicant:

Anders Hval Olsen

Project Title:

The past and the future of eSports: An analysis of its evolution, and the implementation of information technology management tools to gain further growth.

This is to certify that the above named applicant has completed the Coventry University Ethical Approval process and their project has been confirmed and approved as Medium Risk

Date of approval:

04 June 2015

Project Reference Number:

P34072

## Appendix H Time line of the 23 most important eSport events

The 22 most important eSport events (Appendix A; Appendix B) [Explanation found in 6.2.2]

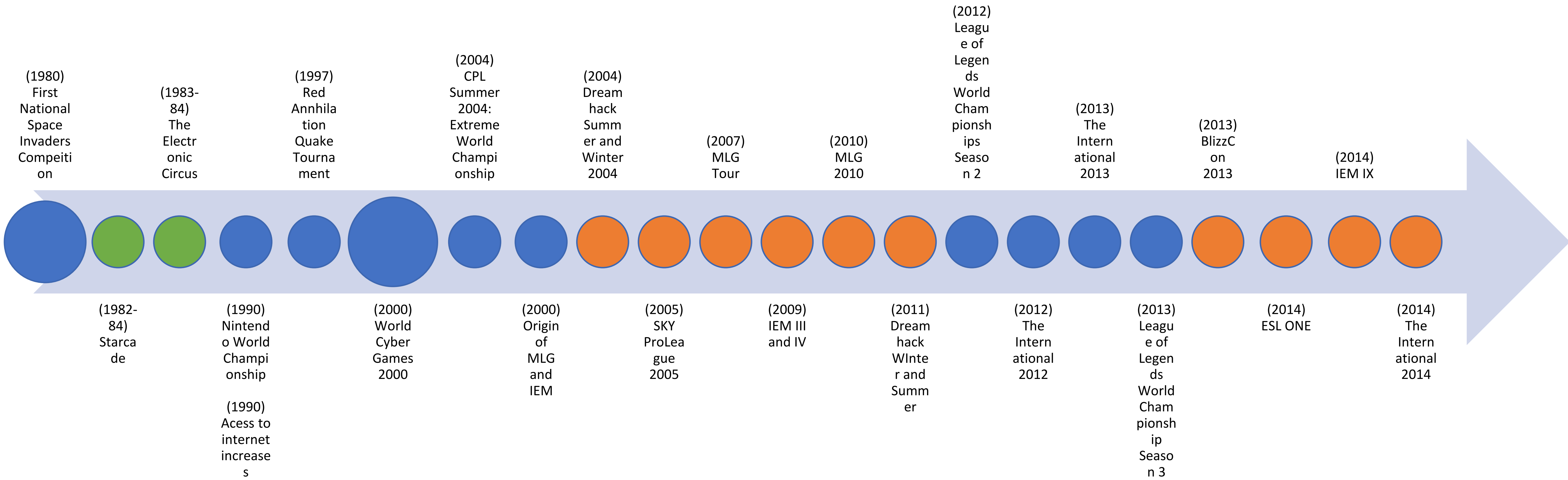

## Appendix I Interview guide (After pilot)

**Introduction**

The purpose of this interview is to understand the eSports position today, what defines it, and what its future looks like. All information you provide will be used to get an understanding of general trends – not isolate your personal views, which ultimately is the goal for this research. I will record the interviews we do; the data will be stored in accordance with the guidelines set by Coventry University and the Data Protection Act 1998, to prevent incorrect citations and more. If you are not comfortable with this and want to pull out, please let me know now. The entire interview will take approximate 30 minutes, depending on how detailed you choose to respond.

Do you have any questions before we start?

**Interview questions**

Introduction

- Could you start by telling a bit about yourself, what working experience you have and your responsibilities within the company/community?
- Do you watch eSports yourself?
- How would you describe eSports?

eSports: Past (History)

- What are the most important events that defined eSports as we know it today? And which of these would you rank as the most important milestone?
- Who are the major communities and teams within the history of eSports?
  - Is there a specific game or company that has had a major importance in the growth of eSports?
- Who (which segments) watches eSports?
- Is eSports different in the NA, EU and Asia?

eSports: Today (Present)

- What is the biggest difference between eSports today, since it really began in the 1980s?
- Why is eSport seeing an increase in viewership now, and not earlier?

- Of the following factors, which would you rank the highest and the lowest in the importance of an eSport event?
  - Total Number of participants
  - Number of live audience
  - Number of Viral Audience (Streaming)
  - Television Coverage
  - Total Prize pool
  - Genre of games played at the event

- Specific games (such as Counter-strike of League of legends) played at a event
  - Number of games played at the event
  - Having extraordinary events, such as live music

    - Is there any other factors that highly affect the success of a eSport event?

eSports: Organisations and communities

- How important are organisations and communities for eSports?
- Does the increase of organisations and communities affect the growth of eSports?
  - If yes, why and how?
  - If no, why?
- Do these organisations and communities give more structure and a more business approach, making it more realistic to grow?
  - If yes, why and how?
  - If no, why?

eSport: Future

- How does the future for eSport look like – both economically and in size?
- What is essential for the future for eSports?
  - If eSports were to improve, what would it improve?
  - How important are the structure – in the sense of management and strategy – for the future of eSports?
- Would you say there is a lot of difference between eSport events and sport events?
  - If yes, why?
  - If no, why?
- Would you recommend – for the future of eSports – to increase the usage of business aspects and its management tools to enhance the growth of eSports seen today?

**Ending**

That was all the questions I had for this time. Do you have something you wish to ask or add, after hearing my questions?